\documentclass[aps,prd,superscriptaddress,floatfix,nofootinbib,notitlepage,twocolumn]{revtex4-1}

\usepackage{graphicx}
\usepackage{amsmath}
\usepackage{longtable}
\usepackage{xcolor}
\usepackage{color}
\usepackage{colortbl}
\definecolor{mygray}{gray}{0.8}

\usepackage[sort&compress]{natbib}
\begin{document}

\title{The angular power spectrum of the diffuse gamma-ray emission as measured by the \emph{Fermi} Large Area Telescope and constraints on its Dark Matter interpretation}

\date{\today}

\author{Mattia Fornasa}
\email{fornasam@gmail.com}
\affiliation{GRAPPA, University of Amsterdam, Science Park, 1098 XH Amsterdam, Netherlands}

\author{Alessandro Cuoco}
\email{cuoco@physik.rwth-aachen.de}
\affiliation{Institute for Theoretical Particle Physics and Cosmology (TTK), RWTH Aachen University, D-52056 Aachen, Germany}

\author{Jes\'{u}s Zavala}
\email{jzavala@dark-cosmology.dk}
\thanks{Marie Curie Fellow}
\affiliation{Dark Cosmology Centre, Niels Bohr Institute, University of Copenhagen, Juliane Maries Vej 30, 2100 Copenhagen, Denmark}
\affiliation{Center for Astrophysics and Cosmology, Science Institute, University of Iceland, Dunhagi 5, 107 Reykjavik, Iceland}

\author{Jennifer M.~Gaskins}
\affiliation{GRAPPA, University of Amsterdam, Science Park, 1098 XH Amsterdam, Netherlands}
\affiliation{California Institute of Technology, Pasadena, California, 91125, United States of America}

\author{Miguel A. S\'{a}nchez-Conde}
\affiliation{The Oskar Klein Centre for Cosmoparticle Physics, AlbaNova, SE-106 91 Stockholm, Sweden}
\affiliation{Department of Physics, Stockholm University, AlbaNova, SE-106 91 Stockholm, Sweden}

\author{German Gomez-Vargas}
\affiliation{Instituto de Astrofis\'{i}ca, Pontificia Universidad Catolica de Chile, Avenida Vicuna Mackenna 4860, Santiago, Chile}

\author{Eiichiro Komatsu}
\affiliation{Max-Planck-Institut f\"ur Astrophysik, 85740 Garching bei M\"unchen, Germany}
\affiliation{Kavli Institute for the Physics and Mathematics of the Universe (Kavli IPMU, WPI), Todai Institutes for Advanced Study, The University of Tokyo, Kashiwa 277-8583, Japan}

\author{Tim Linden}
\affiliation{University of Chicago, Kavli Institute for Cosmological Physics, Chicago, Illinois, 60637, United States of America}
\affiliation{Ohio State University, Center for Cosmology and AstroParticle Physcis (CCAPP), Columbus, Ohio, 43210, United States of America}

\author{Francisco Prada}
\affiliation{Instituto de F\'{i}sica Te\'{o}rica, (UAM/CSIC), Universidad Aut\'{o}noma de Madrid, Cantoblanco, E-28049 Madrid, Spain}
\affiliation{Campus of International Excellence UAM+CSIC, Cantoblanco, E-28049 Madrid, Spain}
\affiliation{Instituto de Astrof\'{i}sica de Andaluc\'{i}a (IAA-CSIC), Glorieta de la Astronom\'{i}a, E-18008, Granada, Spain}

\author{Fabio Zandanel}
\affiliation{GRAPPA, University of Amsterdam, Science Park, 1098 XH Amsterdam, Netherlands}

\author{Aldo Morselli}
\affiliation{Istituto Nazionale di Fisica Nucleare, Sezione di Roma ''Tor Vergata'', I-00133 Roma, Italy}

\begin{abstract}
The isotropic gamma-ray background arises from the contribution of unresolved 
sources, including members of confirmed source classes and proposed gamma-ray 
emitters such as the radiation induced by dark matter annihilation and decay. 
Clues about the properties of the contributing sources are imprinted in the 
anisotropy characteristics of the gamma-ray background. We use 81 months of 
Pass 7 Reprocessed data from the {\it Fermi} Large Area Telescope to perform a 
measurement of the anisotropy angular power spectrum of the gamma-ray 
background. We analyze energies between 0.5 and 500 GeV, extending the range 
considered in the previous measurement based on 22 months of data. We also 
compute, for the first time, the cross-correlation angular power spectrum 
between different energy bins. We find that the derived angular spectra are 
compatible with being Poissonian, i.e. constant in multipole. Moreover, the 
energy dependence of the anisotropy suggests that the signal is due to two 
populations of sources, contributing, respectively, below and above $\sim$2 
GeV. Finally, using data from state-of-the-art numerical simulations to model 
the dark matter distribution, we constrain the contribution from dark matter 
annihilation and decay in Galactic and extragalactic structures to the 
measured anisotropy. These constraints are competitive with those that can be 
derived from the average intensity of the isotropic gamma-ray background.
Data are available at https://www-glast.stanford.edu/pub\_data/552.
\end{abstract}

\maketitle

\pretolerance=10000

\section{Introduction}
\label{sec:intro}

In 2012, the \emph{Fermi} Large Area Telescope (LAT) Collaboration measured
for the first time the auto-correlation angular power spectrum (auto-APS)
of the diffuse gamma-ray emission detected far from the Galactic plane 
\cite{Ackermann:2012uf}. In that analysis, point sources in the first 
\emph{Fermi} LAT source catalog (1FGL) \cite{Abdo:2010ru} and a band along the
Galactic plane with Galactic latitude $|b|<30^\circ$ were masked in order to 
isolate the contribution to the auto-APS from the so-called Isotropic 
Gamma-Ray Background (IGRB). 

The IGRB is what remains of the gamma-ray sky after the subtraction of the
emission from resolved sources and from the Galactic diffuse foreground 
induced by cosmic rays \cite{Abdo:2010nz,Ackermann:2014usa}. It dominates the
gamma-ray sky at large Galactic latitudes and its intensity energy spectrum is 
found to be compatible with a power-law with a slope of $2.32 \pm 0.02$ 
between 100 MeV and $\sim$300 GeV, and with an exponential cut-off at higher 
energies \cite{Ackermann:2014usa}. These values for the spectral slope and for 
the energy cut-off are those found when ``model A'' from 
Ref.~\cite{Ackermann:2014usa} is used to describe the Galactic diffuse 
foreground emission. A different foreground model for the Galaxy would lead
to a slightly different energy spectrum for the IGRB. Deviations can be as 
large as 20-30\% depending on energy.

The IGRB is interpreted as the cumulative emission of sources (e.g., blazars,
star-forming and radio galaxies) that are too faint to be detected 
individually (see Ref.~\cite{Fornasa:2015qua} for a recent review and the 
references therein)\footnote{Instead, the sum of the emission from the 
resolved and unresolved sources is generally referred to as the Extragalactic 
Gamma-ray Background.}. Yet, its exact composition remains unknown. It is 
expected to be isotropic on large angular scales but it can still contain 
anisotropies on small angular scales. Indeed, the contribution to the IGRB 
from unresolved sources imprints anisotropies in the diffuse emission which 
can be used to infer the properties of the contributing sources (see 
Refs.~\cite{Zhang:2004tj,Ando:2006mt,Ando:2009nk,SiegalGaskins:2010mp,
Cuoco:2012yf,Harding:2012gk,Ando:2013ff,DiMauro:2013xta,Zandanel:2014pva,
Hooper:2016gjy} among others). For example, the detection of a significant
angular power in Ref.~\cite{Ackermann:2012uf} determined an upper limit to the 
contribution of unresolved blazars \cite{Cuoco:2012yf,Harding:2012gk,
DiMauro:2014wha} to the IGRB. Additional tools to reconstruct the nature of 
the IGRB are the study of its cross-correlation with catalogs of resolved 
galaxies \cite{Xia:2015wka,Shirasaki:2015nqp,Cuoco:2015rfa,Regis:2015zka,
Ando:2013xwa,Ando:2014aoa}, with gravitational lensing cosmic shear 
\cite{Shirasaki:2014noa} and with lensing of the cosmic microwave background 
radiation \cite{Fornengo:2014cya}. Complementary information can also be 
inferred by modeling its 1-point photon count distribution 
\cite{Malyshev:2011zi,Zechlin:2015wdz,Zechlin:2016pme}.

The detection of the auto-APS presented in Ref.~\cite{Ackermann:2012uf} was 
based on $\sim$22 months of data. Since then, \emph{Fermi} LAT has increased 
its statistics by approximately a factor of 4. Therefore, we expect that an 
updated measurement of the auto-APS will significantly improve our 
understanding of the IGRB. In the first part of this work, we perform this 
measurement by analyzing 81 months of \emph{Fermi} LAT data from 0.5 to 500 
GeV, extending the 1-50 GeV energy range considered in 
Ref.~\cite{Ackermann:2012uf}. This enables a more precise characterization of 
the energy dependence of the auto-APS. Indeed, looking for features in the 
so-called ``anisotropy energy spectrum'' is a powerful way to single out 
different components of the IGRB \cite{SiegalGaskins:2009ux}. We also compute, 
for the first time, the cross-correlation angular power spectrum (cross-APS) 
of the diffuse gamma-ray emission between different energy bins. The cross-APS 
additionally enhances our ability to break down the IGRB into its different 
components since it provides information about the degree of correlation of 
the emission at different energies, which is stronger if the emission 
originates from one single source population (see, e.g., 
Ref.~\cite{Zavala:2009zr,Cuoco:2010jb}).

In the second part of this paper, we focus our analysis on one possible 
contributor to the IGRB, namely the emission induced by dark matter (DM). If 
DM is a weakly-interacting massive particle (WIMP), its annihilation or decay 
could generate gamma rays. The radiation produced in extragalactic and 
Galactic DM structures could contribute to the IGRB (see 
Ref.~\cite{Fornasa:2015qua} and references therein) and, therefore, the IGRB 
could be used to indirectly search for non-gravitational DM interactions. 
Indeed, both the measurement of the IGRB energy spectrum 
\cite{Ackermann:2014usa} and of its auto-APS \cite{Ackermann:2012uf} have been 
already used to set constraints on the possible DM-induced gamma-ray emission 
\cite{Ackermann:2015tah,Ando:2013ff,Gomez-Vargas:2014yla,Lange:2014ura}. 

In this work, we also update the predictions for the auto- and cross-APS 
expected from DM annihilation or decay with respect to 
Ref.~\cite{Fornasa:2012gu}. The distribution and properties of DM structures 
are modeled according to the results of state-of-the-art $N$-body cosmological 
simulations. We also employ well-motivated semi-analytical recipes to account 
for the emission of DM structures below the mass resolution of the simulations. 
The latter is a significant part of the expected signal, at least in the case 
of annihilating DM. We take special care to estimate the uncertainties 
introduced when modeling the clustering of DM, especially at the smallest 
scales. Our predicted DM signal is then compared to the updated {\it Fermi} 
LAT measurement of the auto- and cross-APS. In the most conservative scenario, 
this comparison provides an upper limit to the gamma-ray production rate by 
DM particles, i.e. an upper limit to its annihilation cross section or a lower 
limit to its decay lifetime, as a function of DM mass.

The paper is organized as follows: in Sec.~\ref{sec:data} we provide details
on the data set that will be used in Sec.~\ref{sec:analysis}, where we 
describe our data analysis pipeline. We validate the latter in 
Sec.~\ref{sec:clvalidation} on Monte Carlo (MC) simulations of the unresolved 
gamma-ray sky. In Sec.~\ref{sec:results_analysis}, we present our results for 
the auto- and cross-APS, and we describe the validation tests performed. 
Sec.~\ref{sec:interpretation} provides a phenomenological interpretation of 
our results in terms of one or multiple populations of gamma-ray sources. In 
Sec.~\ref{sec:simulations} we focus on DM-induced gamma-ray emission: we 
provide details on how this signal is simulated, distinguishing among different 
components and discussing the main uncertainties affecting its calculation. In 
Sec.~\ref{sec:limits} the auto- and cross-APS expected from DM are compared to 
the measurements and exclusion limits are derived. Finally, 
Sec.~\ref{sec:conclusion} summarizes our conclusions.

\section{Data selection and processing}
\label{sec:data}

The data analysis pipeline proceeds similarly to what is described in 
Ref.~\cite{Ackermann:2012uf}. We use Pass 7 Reprocessed \emph{Fermi} LAT data 
taken between August 4 2008 and May 25 2015 (MET Range: 239557417 -- 
454279160), and restrict ourselves to photons passing the ULTRACLEAN event 
selection. Thus, we use P7REP\_ULTRACLEAN\_V15 as the instrument response 
functions (IRFs). We place standard selection cuts on the \emph{Fermi} LAT 
data, removing events entering the detector with a zenith angle exceeding 
100$^\circ$, events recorded when the \emph{Fermi} LAT instrument was oriented 
at a rocking angle exceeding 52$^\circ$ and events recorded while the 
\emph{Fermi} LAT was passing through the South Atlantic anomaly, or when it 
was not in science survey mode. Since photons which pair-convert in the 
front of the \emph{Fermi} LAT detector have a better angular resolution,
we split our data set into front- and back-converting events, running each 
data set through the same data analysis pipeline. The front-converting events
will represent our default deta set, with the corresponding 
P7REP\_ULTRACLEAN\_V15 IRF. To produce flux maps we bin the resulting 
\emph{Fermi} LAT event counts and exposure maps into {\sc HEALPix}-format 
maps\footnote{http://healpix.jpl.nasa.gov} \cite{Gorski:2004by} with angular 
bins of size $\sim$0.06$^\circ$ ({\sc HEALPix} order 10, 
{\ttfamily Nside}=1024), as well as into 100 logarithmically-spaced energy 
bins spanning the energy range between 104.46 MeV and 1044.65 GeV. The 
conversion of the exposure maps into {\sc HEALPix}-format maps is performed 
with the GaRDiAn package \cite{Ackermann:2009zz}. Flux maps are, then, built 
by dividing the count map by the corresponding exposure map, in each energy 
bin. The flux maps obtained with the fine energy binning are later co-added 
into 13 larger bins spanning the energy range between 500 MeV and 500 GeV. 
This is done to ensure sufficient statistics within each energy bin. We use 
the smaller energy bins to calculate the beam window function and the photon 
noise within each larger energy bin, as described in Sec.~\ref{sec:analysis}.

\section{Anisotropy analysis}
\label{sec:analysis}

\subsection{Auto- and cross-correlation angular power spectra}
An intensity sky map can be decomposed into spherical harmonics as follows:
\begin{equation}
I(\psi)=\sum_{\ell m} a_{\ell,m} Y_{\ell,m}(\psi),
\end{equation}
where $I(\psi)$ is the intensity from the line-of-sight direction $\psi$ and 
$Y_{\ell,m}(\psi)$ are the spherical harmonic functions. The auto-APS $C_\ell$ 
of the intensity map is given by the $a_{\ell,m}$ coefficients as:
\begin{equation}
C_{\ell} = \frac{1}{2\ell + 1} \sum_{m=-\ell}^{\ell} |a_{\ell,m}|^{2}.
\label{eqn:auto_corr}
\end{equation}

Similarly, the cross-APS between two intensity maps $I_i$ and $I_j$ is
constructed from the individual $a_{\ell,m}^i$ and $a_{\ell,m}^j$ coefficients,
obtained from the decomposition in the two energy bins, independently:
\begin{equation}
C_{\ell}^{ij} = \frac{1}{2\ell + 1} \sum_{m=-\ell}^{\ell} a_{\ell,m}^{i} a_{\ell,m}^{j\star}.
\label{eqn:cross_corr}
\end{equation}

The auto- and cross-APS are computed with specific numerical tools as, e.g., 
{\sc HEALPix} and {\sc PolSpice} \cite{Challinor:2011}. However, before 
applying Eqs.~\ref{eqn:auto_corr} and \ref{eqn:cross_corr}, the data set must 
be prepared, accounting for possible masking, foreground subtraction and
pixelization. Additionally the calculations are complicated by the finite 
angular resolution of the instrument. In the following subsections, we 
summarize how these aspects are taken into consideration.

\subsection{Masking}
\label{sec:masks}
We apply a mask to the all-sky data to reduce contamination from Galactic 
diffuse foregrounds and from sources already detected in the third 
\emph{Fermi} LAT source catalog (3FGL) \cite{Acero:2015gva}. The mask applied 
in our default analysis excludes low Galactic latitudes ($|b|<30^\circ$). 
We also mask each point-like source in 3FGL with a disk whose radius depends
on the flux detected from the source between 0.1 and 100 GeV: for the 500 
brightest sources we consider a disk with a radius of 3.5$^\circ$, for the 
following 500 sources a disk with a radius of 2.5$^\circ$, a disk with a radius
of of 1.5$^\circ$ for the following 1000 sources, and, finally, a radius of 
1.0$^\circ$ for the remaining objects. Validation of the choice for the mask 
will be performed in Sec.~\ref{sec:validations}. The 3FGL catalog contains 
3 extended sources at moderate and high latitudes: Centaurus A and the Large 
and Small Magellanic Clouds. Centaurus A and the Large Magellanic Cloud are 
each masked excluding a 10$^\circ$-region from their center in the catalog. We 
employ a $5^\circ$-mask for the Small Magellanic Cloud. The fraction $f_{\rm sky}$ 
of the sky outside the mask is 0.275. 

We also consider an alternative mask that covers the same strip around the 
Galactic plane but only the sources in the second \emph{Fermi} LAT source 
catalog (2FGL) \cite{Fermi-LAT:2011yjw}. In this case, we mask all the sources 
with a 2$^\circ$-radius disk. The validation for this choice is performed in
Sec.~\ref{sec:validations} and, in this case, $f_{\rm sky}=0.309$.

\begin{figure*}
\includegraphics[width=0.38\textwidth,angle=90]{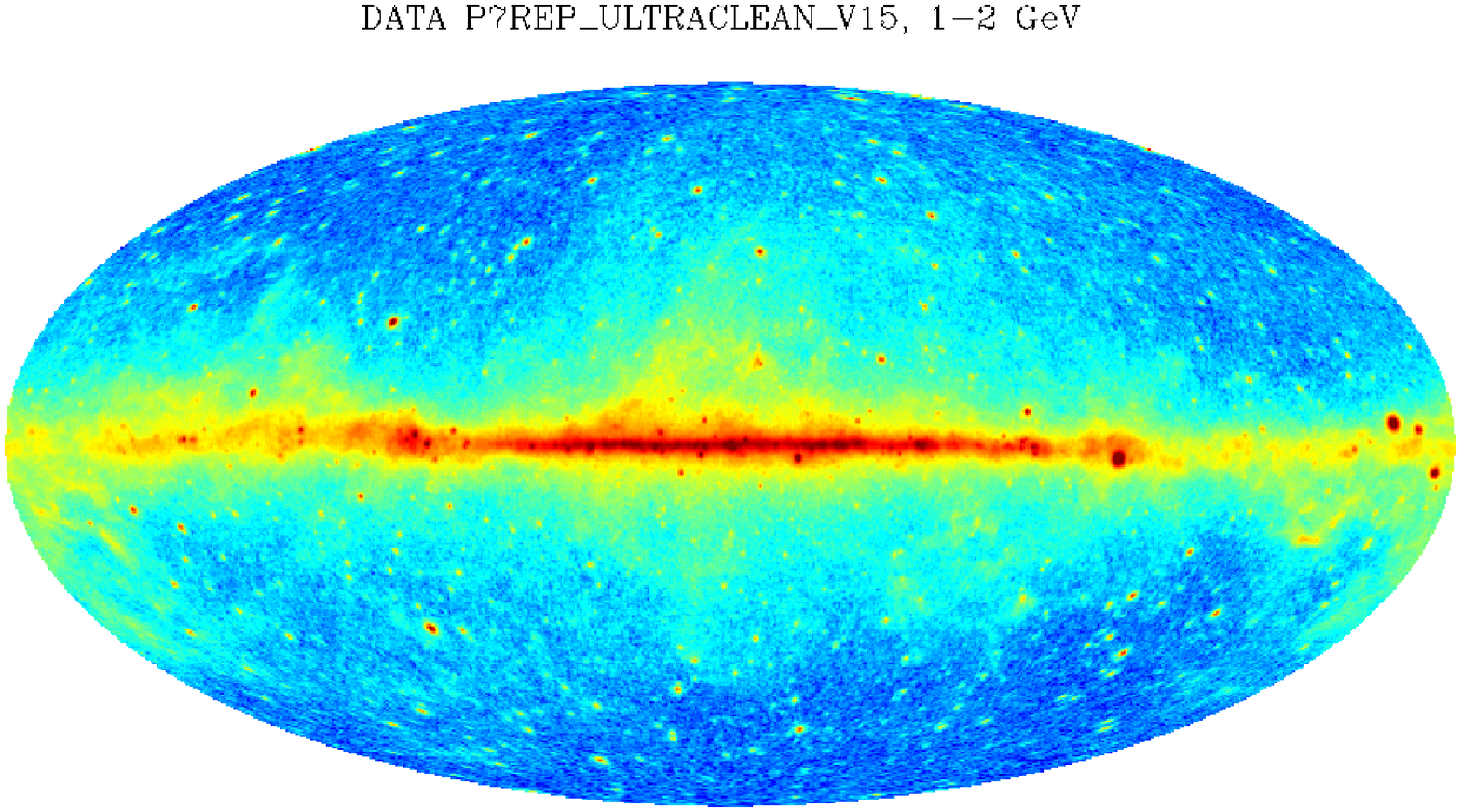}
\includegraphics[width=0.38\textwidth,angle=90]{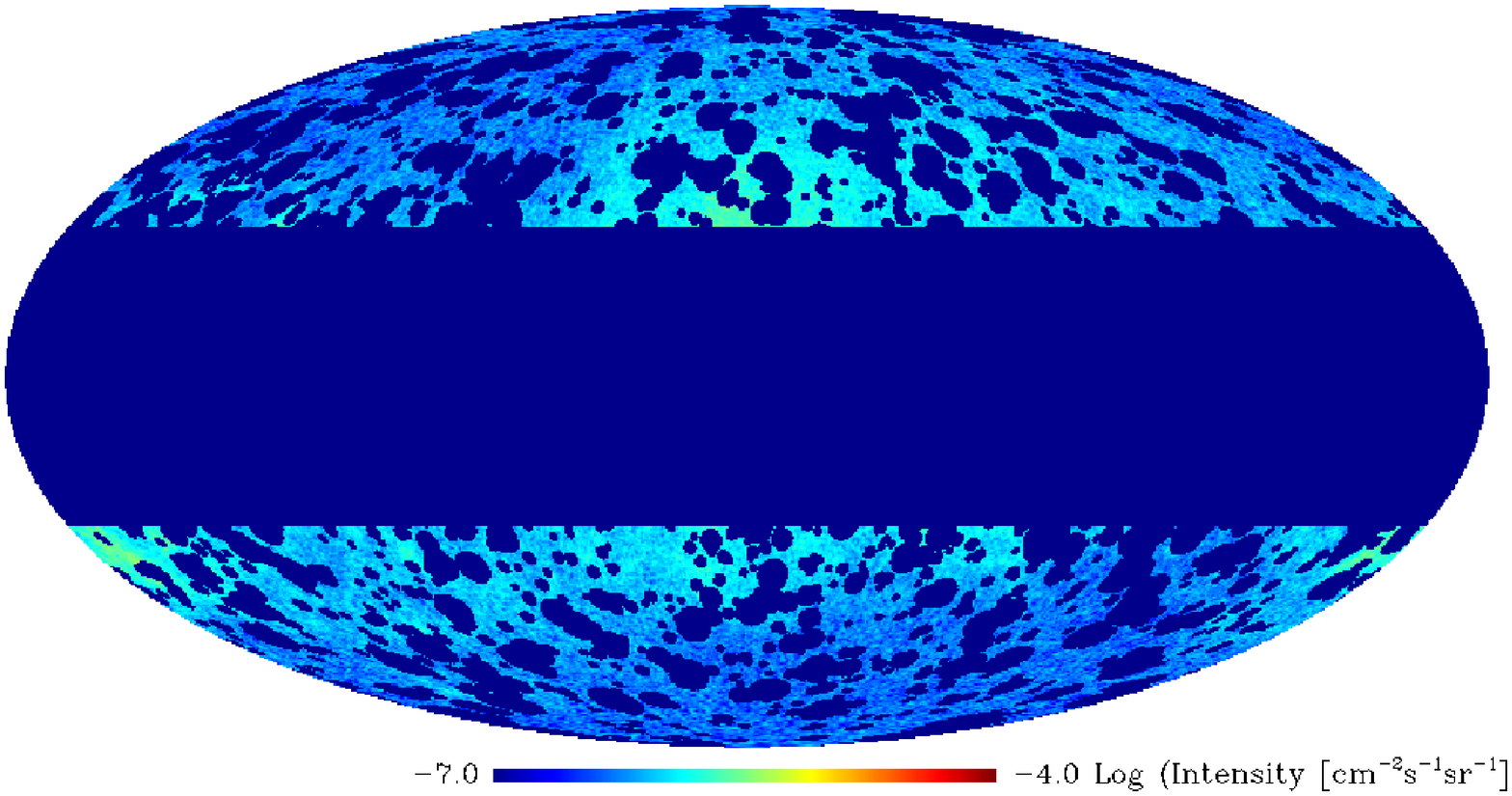}
\caption{\label{fig:maps} Intensity maps (in $\mbox{cm}^{-2}\mbox{s}^{-1}\mbox{sr}^{-1}$) in Galactic coordinates for energies between 1.0 and 2.0 GeV, shown unmasked ({\it top}) and after applying the default mask removing sources in 3FGL, as described in Sec.~\ref{sec:masks} ({\it bottom}). Data used here follow the default processing (see Sec.~\ref{sec:data}), but they include both front- and back-converting events. Both maps have been smoothed with a gaussian beam with $\sigma=0.5^\circ$ and their projection scheme is Mollweide.}
\vspace{-0.3cm}
\end{figure*}

As an illustrative example, the intensity sky maps of the data between 1.0 and 
2.0 GeV are shown in Fig.~\ref{fig:maps}, both unmasked (top panel) and with 
the default mask excluding sources in the 3FGL (bottom panel).

\begin{figure*}
\includegraphics[width=0.38\textwidth,angle=90]{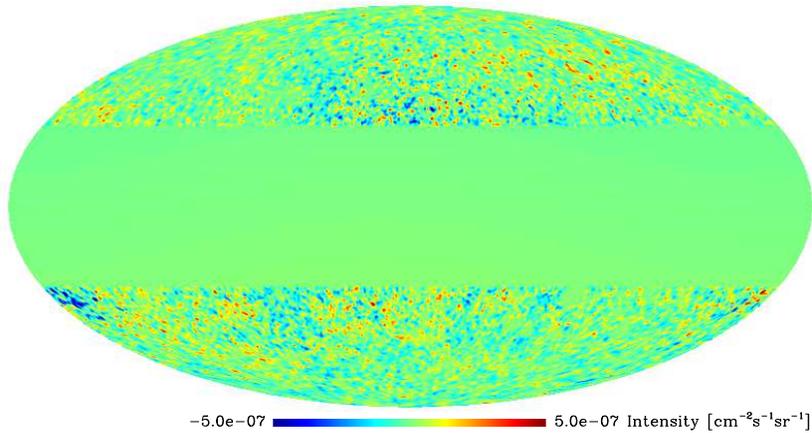}
\caption{\label{fig:mapresiduals} Same as the bottom panel of Fig.~\ref{fig:maps} but with our model for the Galactic foreground subtracted (see Sec.~\ref{sec:cleaning}). The residuals have been smoothed with a gaussian beam with $\sigma=1^\circ$. The projection scheme is Mollweide.}
\vspace{-0.3cm}
\end{figure*}

\subsection{Foreground cleaning}
\label{sec:cleaning}
Despite applying a generous cut in Galactic latitude, some Galactic diffuse 
emission remains visible in the unmasked area of the sky map, particularly at 
low energies (see Fig.~\ref{fig:maps}). To reduce this contamination further, 
we perform foreground cleaning by subtracting a model of the Galactic diffuse 
emission. We use the recommended model for Pass 7 Reprocessed data analysis, 
i.e. \texttt{gll\_iem\_v05\_rev1.fit}\footnote{http://fermi.gsfc.nasa.gov/ssc/data/access/lat/BackgroundMod\-els.html}. Details of the derivation of the 
model are described in Ref.~\cite{Acero:2016qlg}. This foreground model, 
together with an isotropic component, is fitted to the data in the unmasked 
region of the sky and in each one of the 13 coarser energy bins, using 
{\sc GaRDiAn}. The default mask is adopted when fitting the diffuse components.
The resulting best-fit model is then subtracted from the intensity maps in 
each energy bin to obtain residual intensity maps, on which the anisotropy 
measurements are performed. Fig.~\ref{fig:mapresiduals} shows an example of 
the residual intensity map for the data in the energy bin between 1 and 2 GeV.

We investigate the impact of foreground cleaning on the auto- and cross-APS
measurements in Sec.~\ref{sec:validations}.

\subsection{Noise and beam window functions}
\label{sec:noiseandbeam}
We calculate the auto- and cross-APS of the intensity maps using the 
{\sc PolSpice} package \cite{Challinor:2011} to deconvolve the effect of the 
mask on the spectra and to provide the covariance matrix for the estimated 
$C_{\ell}$. 

\begin{figure}
\includegraphics[width=0.49\textwidth]{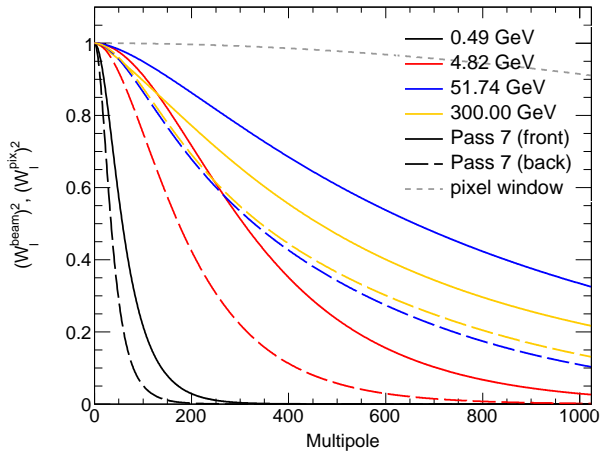}
\caption{\label{fig:beamwindow} Pixel and beam window functions as a function of energy and for different data selections. The gray short-dashed line shows the pixel window function of a {\sc HEALPix} map with \texttt{Nside}=1024. The pixel window function is independent of energy and IRFs. The solid and long-dashed lines show the beam window functions for P7REP\_ULTRACLEAN\_V15 IRFs. The solid lines are for front-converting events and the long-dashed ones for back-converting events. The different colors stand for 4 different representative energies.}
\end{figure} 

Both the finite angular resolution of the instrument (given by its 
point-spread function, PSF) and the finite angular resolution of the map 
(i.e., the pixelization scheme) suppress the measured auto- and cross-APS
at large multipoles (i.e. small angular scales). This effect is described 
using the beam window function $W_\ell^{\rm beam}$ and pixel window function 
$W_\ell^{\rm pix}$, respectively. We note that they affect the signal but not the
noise term $C_{\rm N}$ (see Ref.~\cite{Ackermann:2012uf}). We use the beam and 
pixel window functions to correct the suppression at large multipoles so that
our estimation for the auto- and cross-APS is as follows\footnote{In the 
remainder of the paper, we commonly refer to this estimator simply by $C_\ell$ 
instead of $C_\ell^{\rm signal}$.}:
\begin{equation}
C_\ell^{{\rm signal},ij} = 
\frac{C_{\ell}^{{\rm Pol},ij} - \delta^{ij} C_{\rm N}^{i}}
{(W_\ell^{{\rm beam},i} W_\ell^{{\rm beam},j}) (W_\ell^{\rm pix})^2},
\label{eq:clest}
\end{equation}
where the $i$ and $j$ indexes run from 1 to 13 and label emission in 
different energy bins. The case $i=j$ corresponds to the auto-APS and the 
one with $i \neq j$ to the cross-APS between energy bins. Also, 
$C_{\ell}^{{\rm Pol},ij}$ is the APS delivered by {\sc PolSpice}, which is already 
corrected for the effect of masking. The noise term $\delta^{ij} C_{\rm N}^i$ is 
equal to zero for the cross-APS since it is due to shot noise from the finite 
statistics of the gamma-ray events, which is uncorrelated between different 
energy bins. We compute $C_{\rm N}^i$ from the shot noise $C_{\rm N}^k$ of the 100 
\emph{finely-gridded} intensity maps, where
\begin{equation}
C_{\rm N}^k =  \frac{\langle n_{\gamma,\rm pix}^k / (A_{\rm pix}^k)^{2} \rangle}
{\Omega_{\rm pix}},
\label{eq:noise}
\end{equation} 
where $n_{\gamma,\rm pix}^k$ and $A_{\rm pix}^k$ are the number of observed events 
and the exposure, respectively, in each pixel and for the $k$-th 
finely-gridded energy bin. The averaging is done over the unmasked pixels. 
$\Omega_{\rm pix}$ is the pixel solid angle, which is the same for each pixel. 
See Appendix~\ref{sec:photon_noise} for a derivation of Eq.~\ref{eq:noise}.
The noise term $C_{\rm N}^{i}$ for the auto-APS in the $i$-th large energy bin 
is given by the sum of the noise terms in Eq.~\ref{eq:noise} of all the 
finely-gridded energy bins covered by the $i$-th bin. We note that 
Eq.~\ref{eq:noise} is more accurate than the shot noise used in 
Ref.~\cite{Ackermann:2012uf}, i.e. $C_{\rm N} = \langle n_{\gamma,\rm pix} \rangle / ( \Omega_{\rm pix} \langle A_{\rm pix}^{2} \rangle)$.

The beam window function is computed as follows:
\begin{equation}
W_{\ell}^{\rm beam}(E)=
2\pi \int_{-1}^1 d\cos\theta P_{\ell}(\cos(\theta)) \rm{PSF}(\theta; E),
\label{eq:beamwindow}
\end{equation}
where $P_{\ell}(\cos(\theta))$ are the Legendre polynomials and 
$\rm{PSF}(\theta; E)$ is the energy-dependent PSF for a given set of IRFs, 
with $\theta$ denoting the angular distance in the PSF. We use the 
\emph{gtpsf} tool in the {\sc Science Tools} package to calculate the 
effective PSF, as a function of energy, averaged over the actual pointing and 
live-time history of the LAT. The beam window functions are calculated 
separately for the P7REP\_ULTRACLEAN\_V15 front- and back-converting events.
Finally, the pixel window function $W_\ell^{\rm pix}$ is computed using the tools 
provided in the {\sc HEALPix} package for \texttt{Nside}=1024. Since we use 
the same map resolution for all maps, the pixel window function does not 
depend on the energy.

The pixel window function and the beam window function for front and back 
events are shown separately in Fig.~\ref{fig:beamwindow}, for the 
P7REP\_ULTRACLEAN\_V15 IRF at 4 representative energies. They are also 
available at https://www-glast.stanford.edu/pub\_data/552. Note that the pixel 
window function (short-dashed gray line) has a negligible effect up to 
multipoles of, at least, $\sim$500 and it is subdominant with respect to the 
beam window functions at all multipoles and energies. At energies below 
$\sim$0.5 GeV, the beam window function leads to a strong suppression of power 
for $\ell \gtrsim 100$, even with the front event selection.

Given that the PSF of the \emph{Fermi} LAT varies significantly over the 
energy range considered in this analysis, and in some cases within the 
individual energy bins used when computing the auto- and cross-APS, it is 
necessary to calculate an effective beam window function for each energy 
bin. Therefore, for the $i$-th energy bin, we define the average window 
function $\langle W_{\ell}^{{\rm beam},i} \rangle$ by weighting 
Eq.~\ref{eq:beamwindow} with the intensity spectrum of the events in that bin 
outside the mask:
\begin{equation}
\langle W_{\ell}^{{\rm beam},i} \rangle = 
\frac{1}{I_{\rm bin}} \int_{E_{{\rm min},i}}^{E_{{\rm max},i}}\! \! {\rm d}{E} \; 
W_{\ell}^{\rm beam} (E) \, \frac{{\rm d}I(E)}{{\rm d}E},
\end{equation}
where 
$I_{\rm bin} \equiv \int_{E_{{\rm min},i}}^{E_{{\rm max},i}} {\rm d}E \, ({\rm d}I/{\rm d}E)$ 
and $E_{{\rm min},i}$ and $E_{{\rm max},i}$ are the lower and upper bounds of the
$i$-th energy bin. We approximate the energy spectrum of the data by using the 
measured differential intensity ${\rm d}I/{\rm d}E$ outside the mask in each 
intensity map for the finely-binned energy bins.

\section{Monte Carlo validation of the binning of the APS and of its Poissonian fit}
\subsection{Auto-correlation angular power spectrum}
\label{sec:clvalidation}

In this section we describe in detail the procedure used to bin the auto-APS 
estimated in Eq.~\ref{eq:clest} into large multipole bins. Binning is required
in order to reduce the correlation among nearby $C_\ell$ due to the presence
of the mask.

In contrast with the analysis of Ref.~\cite{Ackermann:2012uf}, in the present
work the binned spectra $\overline{C_l}$ are taken to be the unweighted 
average of the individual $C_\ell$ in the bin. Also, the error 
$\overline{\sigma_\ell}$ on $\overline{C_l}$ is computed by averaging all the 
entries of the covariance matrix provided by {\sc PolSpice} in the block 
corresponding to the bin under consideration. A dedicated set of MC 
simulations of all-sky data are produced to validate these choices and to 
additionally test alternative binning schemes. The MC validation procedure is 
described below.

\subsubsection{Monte Carlo simulations}
The simulations are performed for a single energy bin from 1 to 10 GeV. We 
assume an underlying population of sources with a power-law source-count 
distribution, i.e., $dN/dS=A \ (S/S_0)^{-\alpha}$. The parameters $A$, $S_0$ and 
$\alpha$ are fixed to the values 
$3.8 \times 10^8 \mbox{ cm}^2 \, \mbox{s} \, \mbox{ sr}^{-1}$, 
$10^{-8} \mbox{ cm}^{-2} \mbox{s}^{-1}$ and 2.0, respectively, in agreement with 
the best-fit results of Ref.~\cite{Zechlin:2015wdz}. We consider sources 
with fluxes (in the energy range between 1 and 10 GeV) from 
$10^{-11} \, \mbox{cm}^{-2} \mbox{s}^{-1}$ to 
$10^{-10} \, \mbox{cm}^{-2} \mbox{s}^{-1}$. The upper value is roughly equal to 
the 3FGL sensitivity threshold. In this way, the level of anisotropy expected 
from these sources is roughly equal to that observed in the data when masking 
the 3FGL sources. The lower value is not crucial since the auto-APS is 
dominated by the sources just below the detection threshold. From the source 
count distribution $dN/dS$, we create a realization of the source population, 
producing about 40,000 objects and assigning them random positions on the sky. 
This creates a map with a Poissonian (i.e., constant in multipole) auto-APS, 
$C_{\rm P}$, whose value can be computed by summing together the squared flux, 
$\Phi_i^2$, of all the simulated sources divided by $4\pi$: 
$C_{\rm P}=\sum_i \Phi_i^2 / 4 \pi$. This is equivalent to the usual way of 
calculating $C_{\rm P}$ by integrating $S^2 dN/dS$ over the range in flux 
mentioned above. The resulting Poissonian auto-APS $C_{\rm P}$ is 
$3.42 \times 10^{-18} \, \mbox{cm}^{-4} \mbox{s}^{-2} \mbox{sr}^{-1}$. This is the 
nominal auto-APS that we want to recover by applying our analysis pipeline 
to the simulations.

We use the exposure (averaged in the energy range between 1 and 10 GeV) for
5 years of data-taking to convert the intensity map into a counts map. The map
is also convolved with the average PSF for the P7REP\_ULTRACLEAN\_V15 IRFs
for front-converting events (averaged in the 1-10 GeV range, assuming an 
energy spectrum $\propto E^{-2.3}$). The result is a {\sc HEALPix}-formatted 
map with resolution \texttt{Nside}=1024 containing the expected emission, in 
counts, from the simulated sources. Purely isotropic emission is also included 
by adding an isotropic template to the map, which was also convolved with 
the IRFs and normalized to give the number of counts expected from the 
IGRB measured in the 1-10 GeV energy range, including the contamination from 
residual cosmic rays. For simplicity we did not model the Galactic foregrounds. 
This final map is then Poisson-sampled pixel-by-pixel 200 times to yield 200 
different realizations of the expected counts. The auto-APS of each map is 
calculated with {\sc PolSpice}, after applying the default mask used in the 
analysis of the real data, i.e., excluding the region with $|b|<30^\circ$ and 
the sources in 3FGL, even though the simulation does not include those 
sources. Finally, noise subtraction and beam correction are also applied as 
described in Sec.~\ref{sec:noiseandbeam}.

\subsubsection{Binning validation}
We first validate our recipe to determine the binned auto-APS. In this case, 
the standard analytic error $\sigma_\ell$ on each $C_\ell$ (assuming that
$C_\ell$ follows a $\chi^2_{2\ell+1}$ distribution \cite{Knox:1995dq}) is:
\begin{equation}
\sigma_{\ell} = \sqrt{\frac{2}{(2\ell+1) \, f_{\rm sky}}}
\left( C_\ell + \frac{C_{\rm N}}{W_\ell^2} \right),
\label{eqn:clerr}
\end{equation}
with $W_\ell=W_\ell^{\rm beam}W_\ell^{\rm pix}$. We test three approaches to obtain
$\overline{C_{\ell}}$: $i)$ computing the weighted average of the $C_\ell$ in 
each multipole bin, using $w_\ell=\sigma_\ell^{-2}$ as weight, $ii)$ computing 
the weighted average of the $C_\ell$ in the bin, with a weight 
$w_\ell=\sigma_\ell^{-2}$, defining $\sigma_\ell$ as in Eq.~\ref{eqn:clerr} but 
{\it only with the noise term} $C_{\rm N}/W_\ell^2$ and $iii)$ computing the 
unweighted average of the $C_\ell$ in the bin. Note that in the first 
approach, the weight $w_\ell$ depends on the data via the $C_\ell$ term in 
Eq.~\ref{eqn:clerr}, while in the second and third methods there is no 
dependence on the estimated auto-APS. The first method is the one employed in 
Ref.~\cite{Ackermann:2012uf}.  

\begin{figure}
\includegraphics[width=0.49\textwidth]{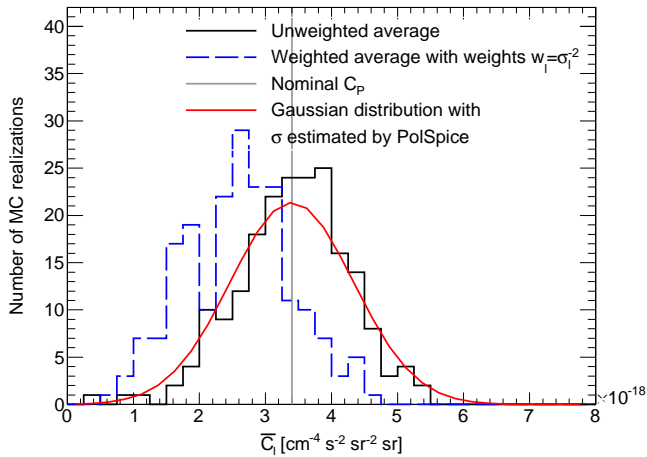}
\caption{\label{fig:clbintest} Comparison between different methods to bin the auto-APS measured in the bin between $\ell=243$ and 317 from the MC simulations described in the text. The nominal $C_{\rm P}$ is represented by the vertical grey line. The solid black histogram shows the distribution of the measured $\overline{C_\ell}$ for the 200 simulated realizations, where the binned auto-APS is computed by an unweighted average. The dashed blue histogram denotes the case of a weighted average with weights given in Eq.~\ref{eqn:clerr}. The solid red curves is a Gaussian distribution centered on the nominal $C_{\rm P}$ and with a standard deviation of $9.3 \times 10^{-19} \, \mbox{cm}^{-4} \mbox{s}^{-2} \mbox{sr}^{-1}$, as estimated with {\sc PolSpice}.}
\end{figure}

In Fig.~\ref{fig:clbintest}, we show a histogram of the binned 
$\overline{C_{\ell}}$ in the bin between $\ell=243$ and 317 for the 200 MC 
realizations. The nominal $C_{\rm P}$ is denoted by the grey vertical solid 
line. The solid black histogram refers to the case in which no weights are 
used (method $iii$), while the dashed blue histogram is for the weighted 
average with weights from Eq.~\ref{eqn:clerr} (method $i$). The results for 
method $ii$ (i.e., weighted average but with only the noise term in 
Eq.~\ref{eqn:clerr}) are not plotted but they are similar to the solid black 
histogram. It is clear that binning the data by means of a weighted average 
which includes the data $C_\ell$ itself gives a result which underestimates the 
nominal $C_{\rm P}$. On the other hand, using the unweighted average (as we do 
in the current analysis) or weighting using only the noise term gives results 
compatible with the input. The intuitive reason for this bias can be traced to 
the fact that method $i)$ uses the measured auto-APS in the estimation of the 
error: at each multipole the measured auto-APS fluctuates up and down 
significantly. If we use Eq.~\ref{eqn:clerr} with the measured $C_\ell$ to 
weight the data at each multipole, a downward fluctuation of $C_\ell$ is 
assigned a smaller error bar and, thus, a larger weight. This will lead to a 
downward biased $\overline{C_\ell}$. Finally, the histograms also show that the 
distribution of the $\overline{C_\ell}$ obtained from the MC realizations is, 
to a good approximation, Gaussian. Indeed, it agrees well with the solid red 
curve representing a Gaussian distribution centered on the nominal $C_{\rm P}$ 
and with a standard deviation of 
$9.3 \times 10^{-19} \mbox{cm}^{-4} \mbox{s}^{-2} \mbox{sr}^{-1}$ (see below).

To assign an error $\overline{\sigma_\ell}$ to the binned auto-APS 
$\overline{C_\ell}$ we also test three methods: $i)$ the unweighted average of 
$\sigma_\ell^2$ from Eq.~\ref{eqn:clerr} in the bin, $ii)$ the weighted average 
of $\sigma_\ell^2$ from Eq.~\ref{eqn:clerr} with weight $w_\ell=\sigma_\ell^{-2}$ 
and $iii)$ the average of the covariance matrix computed by {\sc PolSpice} in 
the bin\footnote{{\sc PolSpice} returns the covariance matrix of the 
beam-uncorrected $C_\ell$, denoted here by $V_{\ell\ell'}$. In method $iii)$ the
error $\overline{\sigma_\ell}^2$ is defined as 
$\sum_{\ell\ell'} V_{\ell\ell'} / (W_\ell^2 W_{\ell'}^2 \Delta\ell^2)$, where the sum 
runs over the $\ell,\ell'$ inside each multipole bin and $\Delta\ell$ is the 
width of the bin.}. Differently from the estimation of $\overline{C_\ell}$, the 
three methods for the estimation of $\overline{\sigma_\ell}$ produce similar 
results. Thus, we decide to choose method $iii)$ as our standard prescription. 
This has also the advantage that, by averaging different blocks of the 
covariance matrix provided by {\sc PolSpice}, one can build a covariance
matrix for the binned auto-APS. The average of $\overline{\sigma_\ell}$ from
method $iii)$ in the multipole bin between $\ell=243$ and 317 over the 200 MC 
realizations is $9.3 \times 10^{-19} \mbox{cm}^{-4} \mbox{s}^{-2} \mbox{sr}^{-1}$, 
i.e. the value considered in Fig.~\ref{fig:clbintest} for the standard 
deviation of the red curve. Our validation with MC simulations shows that our
estimate of the errors is realiable and that higher-order effects, e.g. those
related to the bispectrum and trispectrum discussed in 
Ref.~\cite{Campbell:2014mpa}, can be neglected. It remains interesting, 
nonetheless, to understand if a small bispectrum and trispectrum can be used to 
independently constrain the sources contributing to the IGRB.

\subsubsection{Poissonian fit validation}
Having validated the binning procedure for the measured auto-APS, we are now
interested in fitting the binned auto-APS $\overline{C_\ell}$ with a constant 
value. Indeed, a Poissonian APS $C_{\rm P}$ (i.e. an APS that is constant in
multipole) is a natural expectation for the anisotropies induced by 
unclustered unresolved point sources. One possibility is to infer $C_{\rm P}$ 
by minimizing the following $\chi^2$ function:
\begin{equation}
\chi^2(C_{\rm P}) = 
\sum_{\ell} \frac{(\overline{C_\ell}-C_{\rm P})^2}{\overline{\sigma_\ell}^2},
\label{eqn:chi2_CP}
\end{equation}
where $\overline{C_\ell}$ and $\overline{\sigma_\ell}$ are the the binned data 
and their errors, as described in the previous section.

A second possibility is to consider a likelihood function $\mathcal{L}$ that, 
up to a normalization constant, can be written as follows:
\begin{equation}
\log \mathcal{L}(C_{\rm P})  =
-\sum_{\ell}\log(\overline{\sigma_\ell}) - \frac{1}{2} 
\sum_{\ell} \frac{( \overline{C_\ell} - C_{\rm P})^2}{\overline{\sigma_\ell}^2}.
\label{eq:logLmethod}
\end{equation}
This expression for the likelihood takes into account the fact that 
$\overline{\sigma_\ell}$ also depends on $C_{\rm P}$, since 
$\overline{\sigma_\ell}^2$ in Eq.~\ref{eq:logLmethod} is defined as the 
average of 
\begin{equation}
\sigma_\ell^2= \frac{2}{(2\ell+1) f_{\rm sky}} \, 
\left( C_{\rm P} + \frac{C_{\rm N}}{W_\ell^2} \right)^2,
\label{eqn:error_logL}
\end{equation} 
over the specific multipole bin. In fact, for large multipoles, the expected  
$\chi^2_{2\ell+1}$ distribution of a given $C_\ell$ can be approximated by a 
Gaussian for which \emph{the mean and the standard deviation are not 
independent} but related as in Eq.~\ref{eqn:error_logL}. Thus, the main 
difference between the $\chi^2$ minimization (as in Eq.~\ref{eqn:chi2_CP}) and 
the likelihood method is that, in the latter, $\overline{\sigma_\ell}$ depends 
on $C_{\rm P}$. Ignoring such a dependence may bias the result of the fit.

\begin{figure}
\includegraphics[width=0.49\textwidth]{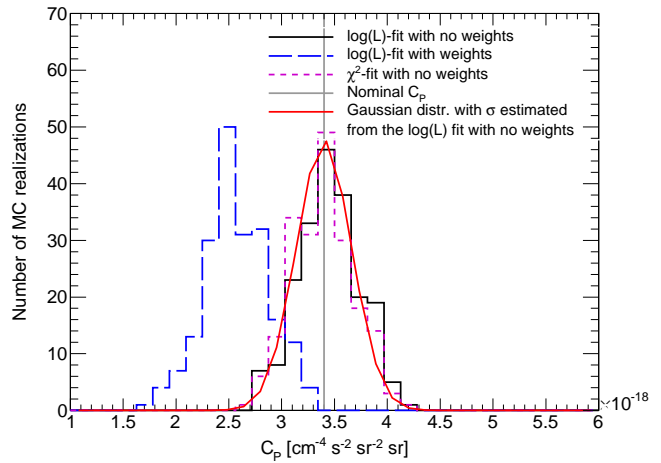}
\caption{\label{fig:cpbintest} Comparison between different methods to measure the Poissonian $C_{\rm P}$ in the MC simulations, given the binned $\overline{C_\ell}$. The nominal $C_{\rm P}$ is represented by the vertical grey line. The solid black histogram shows the distribution of the Poissonian $C_{\rm P}$ for the 200 simulated realizations obtained by maximazing the $\log\mathcal{L}$ in Eq.~\ref{eq:logLmethod} over the multipole range from 49 to 706. The binned $\overline{C_\ell}$ in Eq.~\ref{eq:logLmethod} are computed with no weights. If the weighted average is considered, the distribution of $C_{\rm P}$ is shown by the long-dashed blue histogram, which is clearly biased low. The short-dashed pink histogram shows the distribution of $C_{\rm P}$ computed by the minimization of the $\chi^2$ in Eq.~\ref{eqn:chi2_CP} from $\overline{C_\ell}$ binned with no weights. The solid red curve is a Gaussian distribution centered on the nominal $C_{\rm P}$ and with a standard deviation of $2.6 \times 10^{-19} \mbox{cm}^{-4} \mbox{s}^{-2} \mbox{sr}^{-1}$, as estimated from the $\log\mathcal{L}$ method.}
\end{figure}

The two methods described above are used to determine the best-fit $C_{\rm P}$ 
for the 200 MC realizations described above, by considering 10 
$\overline{C_\ell}$ in 10 bins in multipole uniformly spaced in $\log\ell$ 
between $\ell=49$ and 706. As we discuss in Sec.~\ref{sec:results_analysis}, 
this multipole range excludes the large angular scales where the reconstructed 
$C_{\ell}$ are most uncertain due to possible contamination of the Galactic 
foreground, and the high-multipole range where the effect of the window 
functions becomes too severe. The results are summarized in 
Fig.~\ref{fig:cpbintest}: the vertical grey line is the nominal $C_{\rm P}$, 
while the solid black histogram shows the distribution of the $C_{\rm P}$ 
determined by maximazing the $\log\mathcal{L}$ of Eq.~\ref{eq:logLmethod} if 
the binned $\overline{C_\ell}$ are computed with no weights. This approach 
produces a distribution that is approximately Gaussian and centered on the 
nominal $C_{\rm P}$. On the other hand, if the binned $\overline{C_\ell}$ are 
computed with the weights from Eq.~\ref{eqn:clerr}, then the maximization of 
$\log\mathcal{L}$ underestimates the Poissonian auto-APS (long-dashed blue 
histogram in Fig.~\ref{fig:cpbintest}). Making use of the $\chi^2$ function in 
Eq.~\ref{eqn:chi2_CP} instead of the $\log\mathcal{L}$ in 
Eq.~\ref{eq:logLmethod} gives similar results, i.e. an unbiased distribution 
for $C_{\rm P}$ if the binned $\overline{C_\ell}$ are computed without weights 
(short-dashed pink line) and an underestimation of the nominal $C_{\rm P}$ when 
weights are included (not shown in Fig.~\ref{fig:cpbintest})\footnote{Note 
that applying the $\log\mathcal{L}$ or the $\chi^2$ approach to the unbinned 
$C_\ell$ provided by {\sc PolSpice} also leads to an underestimation of 
$C_{\rm P}$.}. The error associated to the best-fit $C_{\rm P}$ corresponds to 
the 68\% confidence-level (CL) region. We note that the $\log\mathcal{L}$ 
approach yields slightly smaller errors and we decide to adopt this as our 
standard way to measure the Poissonian auto-APS in the following. The average 
of the error on the best-fit $C_{\rm P}$ over the 200 MC realizations is 
$2.6 \times 10^{-19} \, \mbox{cm}^{-4} \mbox{s}^{-2} \mbox{sr}^{-1}$, i.e. the 
value used as the standard deviation for the Gaussian function plotted as the 
solid red line in Fig.~\ref{fig:cpbintest}, which is centered on the nominal 
$C_{\rm P}$.

\subsection{Cross-correlation angular power spectrum}
Similar checks to what is described above for the auto-APS are performed for 
the cross-APS between two energy bins. In this case, the standard analytical
error is:
\begin{equation}
\sigma_{\ell}^2 = \frac{1}{(2\ell+1) \, f_{\rm sky}}
\left[ C_\ell^2 + \left( C_{1,\ell} + \frac{C_{1,{\rm N}}}{W_{1,\ell}^2} \right)
\left( C_{2,\ell} + \frac{C_{2,{\rm N}}}{W_{2,\ell}^2} \right) \right],
\label{eqn:clcrosserr}
\end{equation}
where $C_\ell$ is the cross-APS and $C_{1,\ell}$ and $C_{2,\ell}$ are the auto-APS 
for the two energy bins. Similarly, $W_{1,\ell}$ and $W_{2,\ell}$ are the window 
functions for the two energies considered and $C_{1,{\rm N}}$ and $C_{2,{\rm N}}$
are the two photon noises. After testing different averaging schemes, we 
decide to use the same method as for the auto-APS case, i.e. to bin the 
cross-APS with an unweighted average and to estimate $\overline{\sigma_\ell}$ 
by computing the block-average of the covariance matrix provided by 
{\sc PolSpice}.

\begin{figure*}
\includegraphics[width=0.49\textwidth]{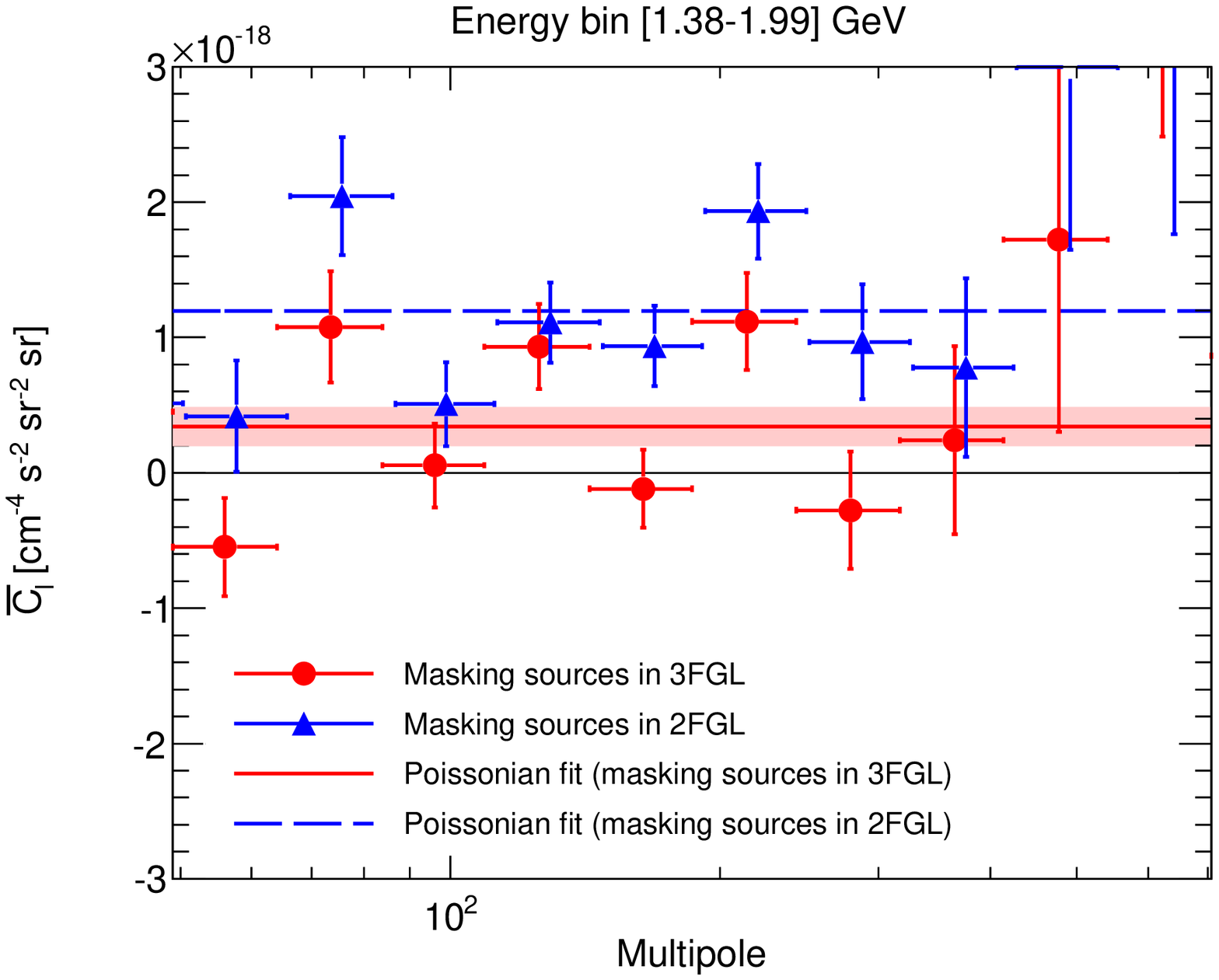}
\includegraphics[width=0.49\textwidth]{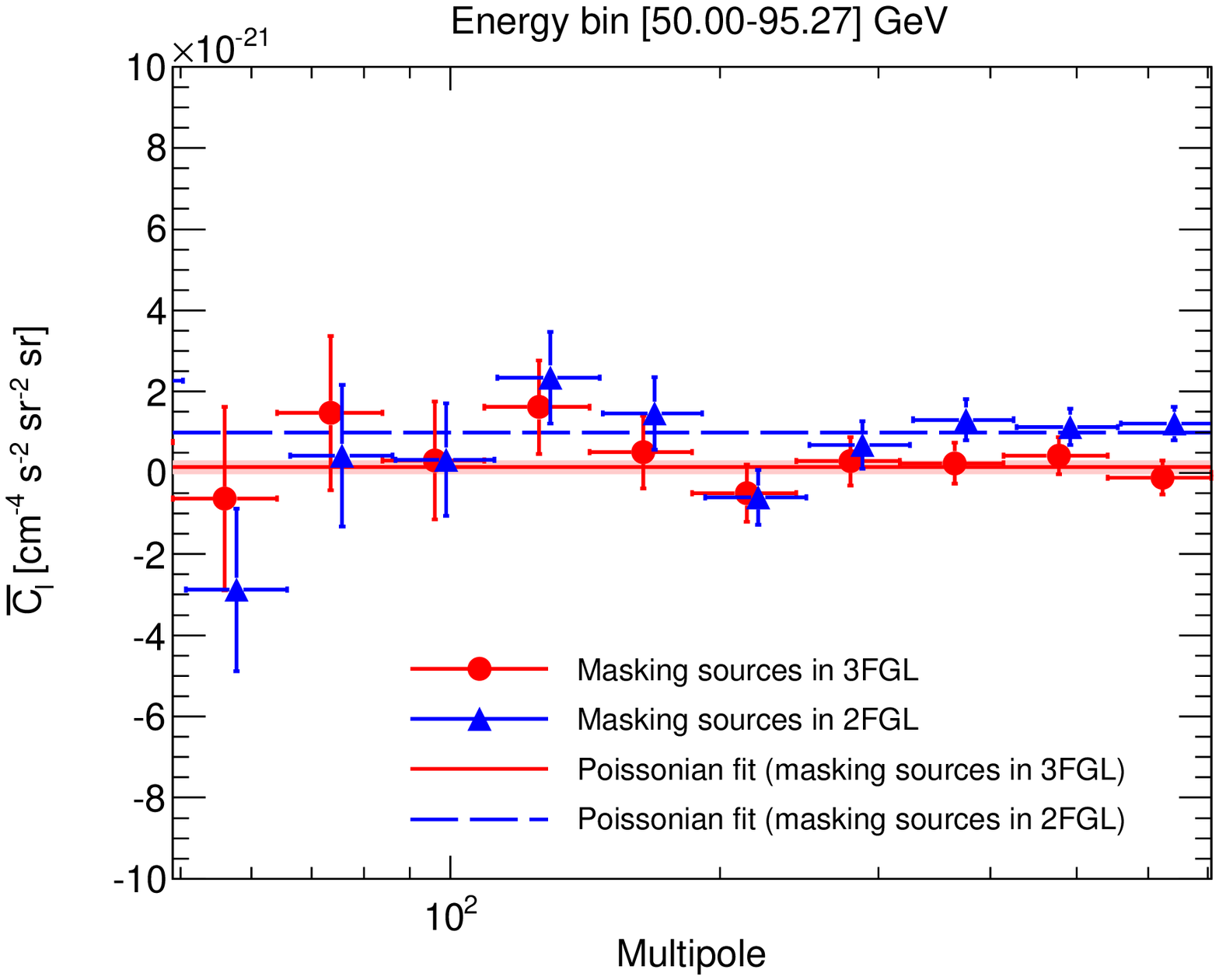}
\caption{\label{fig:clsauto} Auto-APS of the IGRB for 2 representative energy bins (between 1.38 and 1.99 GeV in the left panel and between 50.0 and 95.27 GeV in the right panel) and for the reference data set (P7REP\_ULTRACLEAN\_V15 front events) using the reference mask which excludes $|b|<30^\circ$ and 3FGL sources (red circles). The blue triangles show the same but masking the sources in 2FGL. Data have been binned as described in Sec.~\ref{sec:clvalidation}. The solid red line shows the best-fit $C_{\rm P}$ for the red data points, with the pink band indicating its 68\% CL error. The dashed blue line corresponds to the best-fit $C_{\rm P}$ for the blue data points. Note that only the results in our signal region (i.e. between $\ell=49$ and 706) are plotted and that the scale of the $y$-axis varies in the two panels. Also, the blue triangles have been slightly shifted horizontally with respect to the red circles to increase the readibility of the plots. This will happen also in many of the following plots.}
\end{figure*}

Similarly, we tested the likelihood and $\chi^2$ approach to derive the 
Poissonian best-fit $C_{\rm P}$ to the cross-APS data. For the likelihood 
approach, $\overline{\sigma_\ell}^2$ is now defined as the average of 
Eq.~\ref{eqn:clcrosserr} after having replaced $C_\ell$ by $C_{\rm P}$. As for 
the auto-APS, we find compatible results between the two methods, with the 
likelihood approach providing slightly smaller errors. Therefore, in the 
following, we will quote Poissonian cross-APS derived with this method. 

We end this section by noting that the proper way to estimate $C_{\rm P}$ for 
the cross-APS would be to use the likelihood method but replacing the
auto-APS $C_{1,\ell}$ and $C_{2,\ell}$ in Eq.~\ref{eqn:clcrosserr} with their
Poissonian estimates $C_{1,{\rm P}}$ and $C_{2,{\rm P}}$, and to perform a joint 
likelihood fit to all three quantities, i.e. $C_{\rm P}$, $C_{1,{\rm P}}$ and 
$C_{2,{\rm P}}$. However, this approach would not provide results that are
significantly different than the ones obtained as described above. In fact, 
at present, the noise terms in Eq.~\ref{eqn:clcrosserr} dominate over the 
signal terms, reducing the effect of covariance between energy 
bins\footnote{The noise terms in Eq.~\ref{eqn:clcrosserr} are a factor of 4-5
larger than $C_{\rm P}$, $C_{1,{\rm P}}$ and $C_{2,{\rm P}}$. Therefore, not 
performing the joint likelihood fit as described in the text generates an 
error of, at most, 10-20\% on $\sigma_{\ell}$. The effect on the estimated 
best-fit Poissonian auto- and cross-APS will be even smaller.}.

\section{Measured auto- and cross-correlation angular power spectra of the Isotropic Gamma-Ray Background}
\label{sec:results_analysis}
Following the analysis described in the previous section, we measure the 
auto- and cross-APS in 13 energy bins spanning the energy range between 500 
MeV and 500 GeV.

\subsection{Auto-correlation angular power spectra}
The auto-APS of the IGRB is shown in Fig.~\ref{fig:clsauto} for two 
representative energy bins. The auto-APS for all 13 energy bins 
considered is shown in Appendix~\ref{sec:appendixauto} and it is available at
https://www-glast.stanford.edu/pub\_data/552. The $y$-axis range of 
Fig.~\ref{fig:clsauto} and in Appendix~\ref{sec:appendixauto} has been 
chosen to better illustrate the auto-APS in the multipole range of interest, 
i.e. between $\ell=$49 and 706, divided into 10 bins equally spaced in 
$\log\ell$. The red circles indicate the auto-APS for our reference data set 
(i.e., P7REP\_ULTRACLEAN\_V15 front events) and the default mask covering the 
region with $|b|<30^\circ$ and 3FGL sources, as described in 
Sec.~\ref{sec:masks}. Instead, the blue triangles refer to the same data set 
but using the default mask covering only 2FGL sources (see 
Sec.~\ref{sec:masks}). Note that the blue triangles are systematically higher 
than the red circles, due to the anisotropy power associated with the sources 
that are present in 3FGL but still unresolved in 2FGL.

\begin{figure*}
\includegraphics[width=0.49\textwidth]{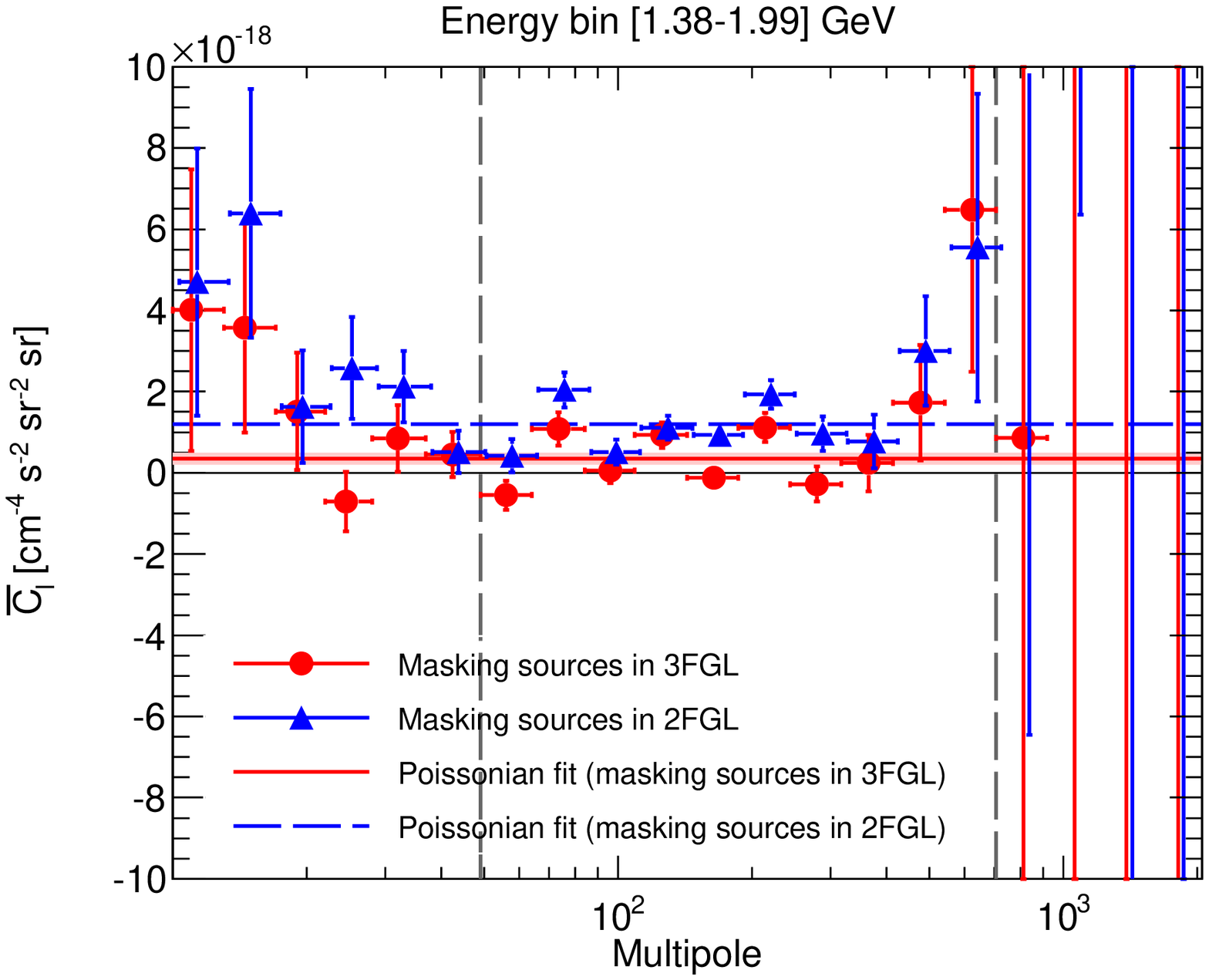}
\includegraphics[width=0.49\textwidth]{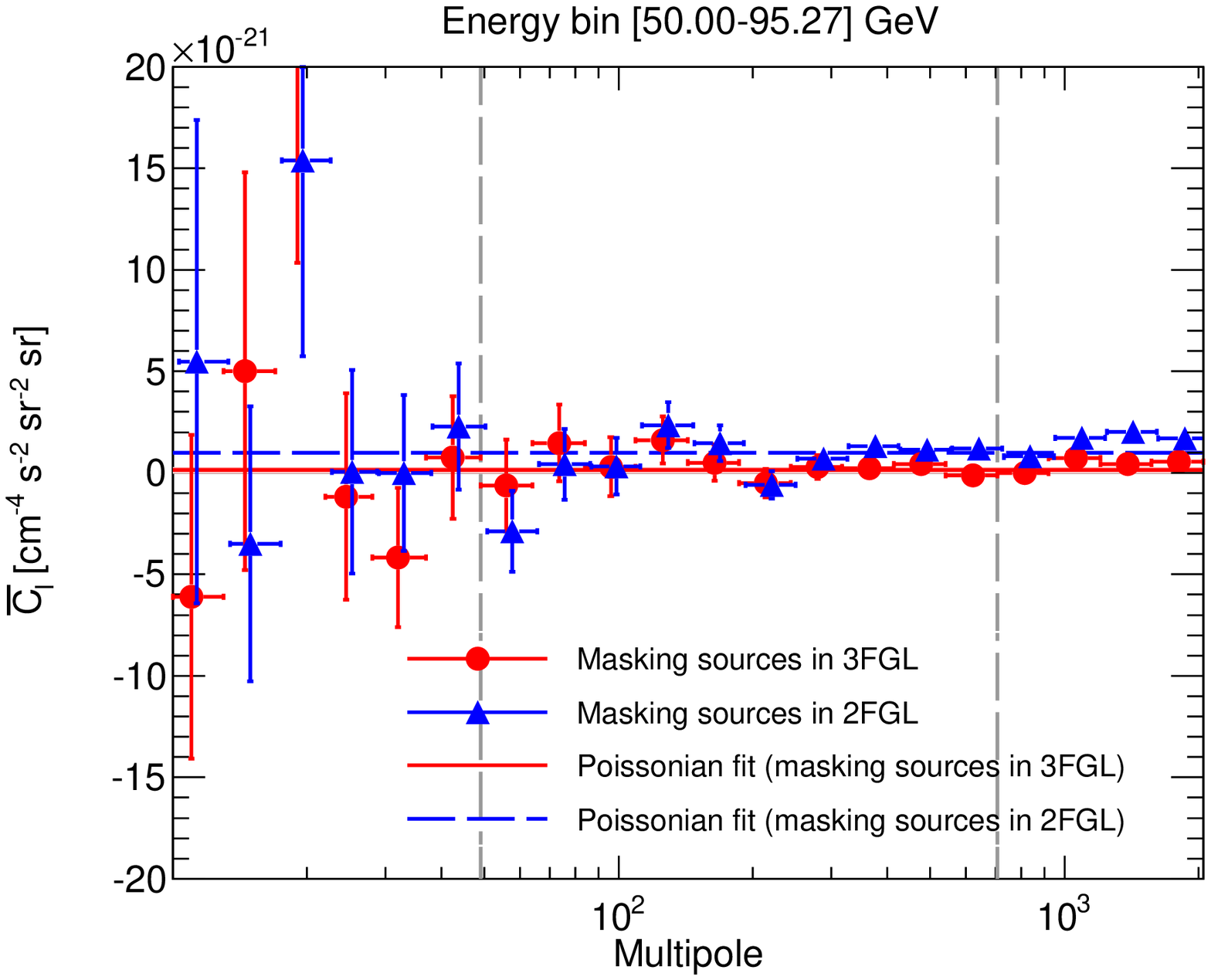}
\caption{\label{fig:clszoom} Same as Fig.~\ref{fig:clsauto}, but showing a wider range in multipole, going from $\ell=10$ to 2000. The two dashed grey vertical lines indicate the lower and upper bounds of the multipole range used for the present analysis. Note the different scale of the $y$-axis in each panel.}
\end{figure*}

Fig.~\ref{fig:clszoom} shows the auto-APS for the two same energy bins but
over a broader multipole range, i.e. from $\ell=10$ to 2000. This illustrates 
the behavior of the auto-APS above and below the signal region used in our 
analysis, i.e. between $\ell=49$ and 706. At large scales (i.e., low 
multipoles), there might be some residual contamination from the Galactic 
foregrounds. This motivates our choice of neglecting the APS below $\ell=49$. 
In Sec.~\ref{sec:validations} the effect of foreground contamination is 
discussed in more detail. On the other hand, at high multipoles and at low 
energies (left panel), the size of the error bars increases dramatically due 
to the strong signal suppression caused by the beam window functions. Our
signal region neglects any $\overline{C_\ell}$ above 706. At high energies 
(right panel), the effect of the beam window function is more modest, even up 
to $\ell=2000$ (see Fig.~\ref{fig:beamwindow}). In principle, for high 
energies, we could consider a signal region in multipole that extends to 
smaller scales. However, we prefer to work with a window in multipole that is 
independent of the energy bin and, therefore, we choose the value of 
$\ell=706$ as a reasonable compromise.

Note that each individual data point in Figs.~\ref{fig:clsauto} and 
\ref{fig:clszoom} can be negative, since our auto-APS estimator quantifies 
the excess of power with respect to the photon noise $C_{\rm N}$. We fit the 
auto-APS (between $\ell=49$ and 706) in each energy bin to a constant value, 
in order to determine the Poissonian $C_{\rm P}$ (the possibility of a 
non-constant $\overline{C_\ell}$ is considered later). The fit is performed as 
discussed in the previous section. The best-fit $C_{\rm P}$ are reported in 
Tabs.~\ref{tab:cp_3FGL} and \ref{tab:cp_2FGL} for the different energy bins 
and for the masks around 3FGL and 2FGL sources, respectively. They are also 
available at https://www-glast.stanford.edu/pub\_data/552 and they are reported 
as the solid red and dashed blue lines in Figs.~\ref{fig:clsauto}, 
\ref{fig:clszoom}, \ref{fig:cls1} and \ref{fig:cls2}, when masking sources 
in 3FGL and 2FGL, respectively. In the former case, we also show the 
estimated 68\% CL error on $C_{\rm P}$ as a pink band. The significance of the 
measured Poissonian auto-APS can be quantified by computing the Test 
Statistics (TS) of the best-fit $C_{\rm P}$, defined as the difference between 
the $-2\ln\mathcal{L}$ of the best fit and the $-2\ln\mathcal{L}$ of the null 
hypothesis. The latter is obtained from Eq.~\ref{eq:logLmethod} by setting 
$C_{\rm P}$ to zero. Assuming Wilks' theorem, TS is distributed as $\chi^2$ 
distribution with 1 degree of freedom and, thus, it can be used to estimate 
the significance associated to $C_{\rm P}$. For the default data set masking 
3FGL sources, the significance of the measured auto-APS $C_{\rm P}$ is larger 
than 3$\sigma$ for all energy bins up to 21.8 GeV, except between 5.00 and 
10.45 GeV. The significance of the detection is reported in italics in 
Tabs.~\ref{tab:cp_3FGL} and \ref{tab:cp_2FGL}. In the case of the mask around 
3FGL sources, the highest significance in the auto-APS is 6.3$\sigma$ and it 
is reached in the second energy bin, i.e. between 0.72 and 1.04 GeV.

The way the auto- and cross-APS depend on the energy (i.e. the so-called 
``anisotropy energy spectrum'') is an informative observable that can 
provide insight into the emission causing the anisotropic signal. In fact, 
in the case that the auto-APS is produced by a single population of sources, 
the anisotropy energy spectrum allows their energy spectrum to be reconstructed
\cite{SiegalGaskins:2009ux,Hensley:2009gh,Hensley:2012xj}\footnote{The 
anisotropy energy spectrum traces the intensity energy spectrum of the sources
responsible for the anisotropy signal only if the clustering of the source
population is independent of energy.}. If more than one class of objects are
responsible for the signal, then, by detecting features in the anisotropy 
energy spectrum, it may be possible to identify energy regimes where the 
different classes dominate the signal.

\begin{figure}
\includegraphics[width=0.49\textwidth]{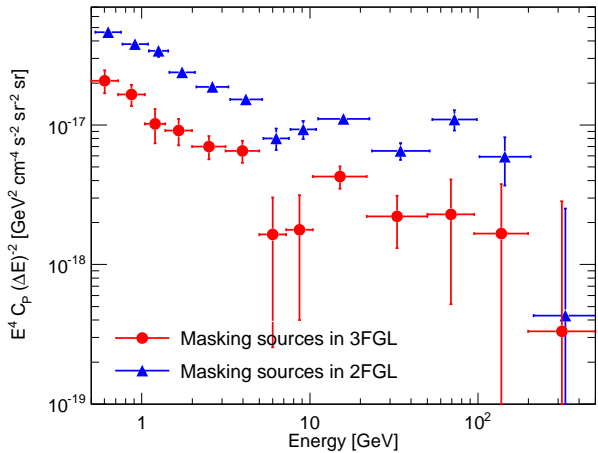}
\caption{\label{fig:autoCPvsE} Anisotropy energy spectra for the auto-APS using the reference data set with the default 3FGL mask (red circles) in comparison with the case in which we use the default mask around 2FGL sources (blue triangles).}
\end{figure}

\begin{figure*}
\includegraphics[width=0.49\textwidth]{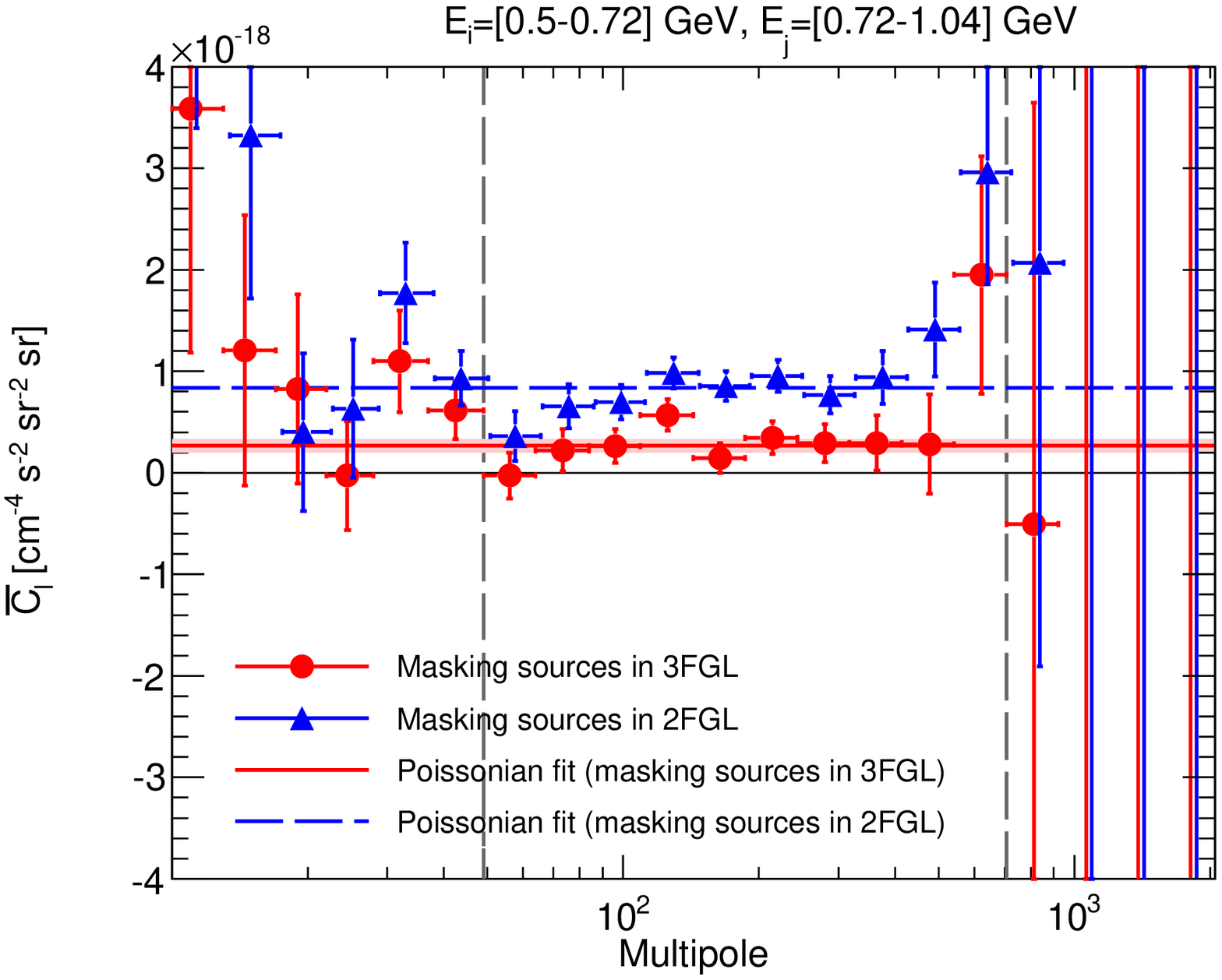}
\includegraphics[width=0.49\textwidth]{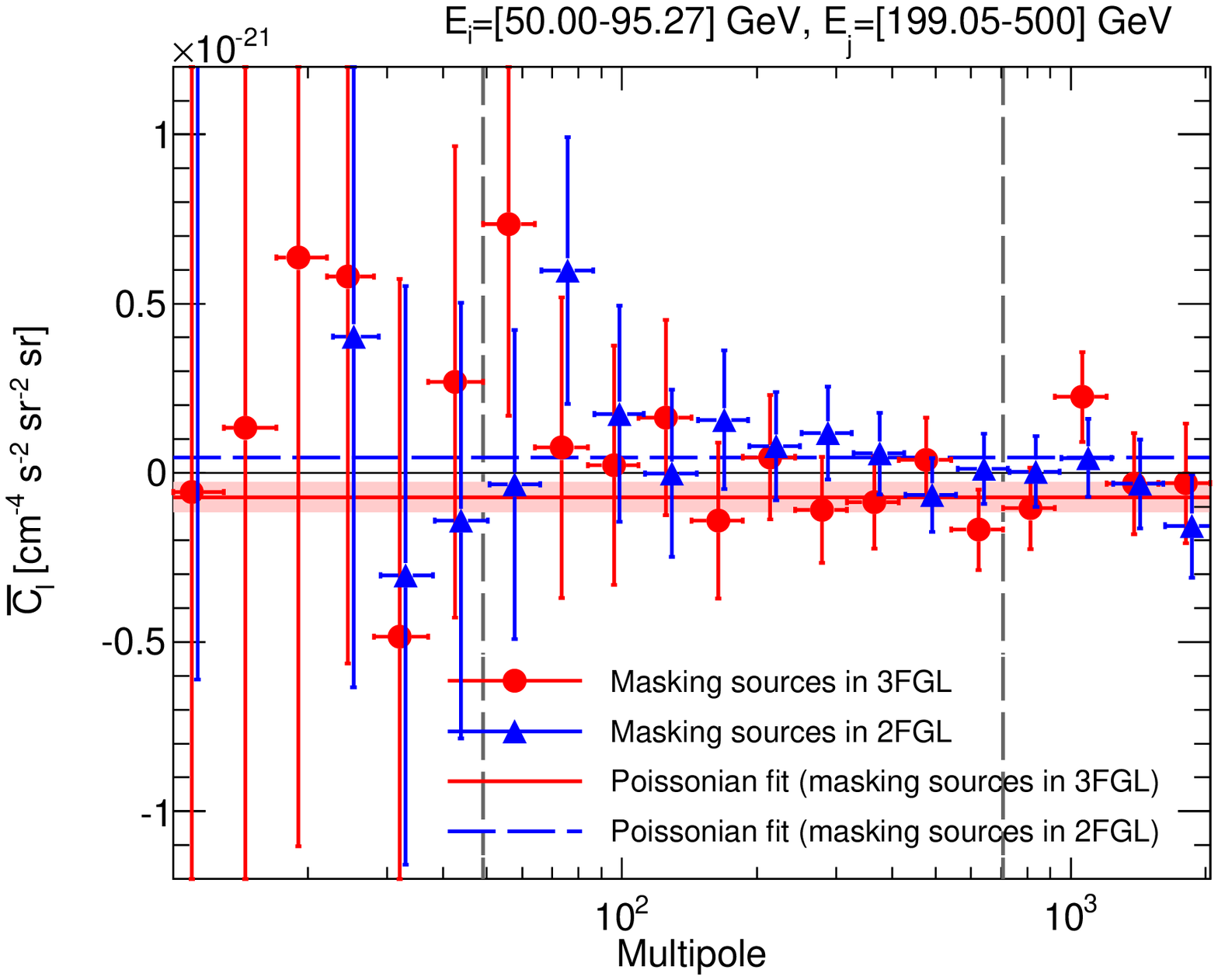}
\caption{\label{fig:crosscorrcl} Cross-APS for two representative combinations of energy bins as indicated above the two panels. The default data set and 3FGL mask are used to compute the red circles, while the blue triangles are for the mask excluding 2FGL sources. The vertical dashed grey lines denote the bounds of the multipole range used in the analysis. The best-fit $C_{\rm P}$ is shown as a solid red line in the case of the mask around 3FGL sources, with a pink band indicating its 68\% CL error. The dashed blue line corresponds to the best-fit $C_{\rm P}$ for the blue data points (i.e. masking 2FGL sources).}
\end{figure*}

The measured anisotropy energy spectrum for the auto-APS is shown in 
Fig.~\ref{fig:autoCPvsE}. In the figure, the data points are weighted by 
$E^4/\Delta E^2$ where $E$ is the log-center of the energy bin and $\Delta E$ 
is the width of the bin. This weighting is introduced in order to compare the 
anisotropy energy spectrum directly with the squared intensity energy spectrum 
of the sources responsible for the anisotropy signal. Fig.~\ref{fig:autoCPvsE} 
compares the auto-APS $C_{\rm P}$ for the case of the mask excluding 3FGL 
sources (red circles) to that of the mask excluding 2FGL sources (blue 
triangles). As already mentioned, the amplitude of the auto-APS is lower when 
we exclude the sources in 3FGL. In both data sets, the low-energy part of the 
spectrum appears generally consistent with a power law, while a feature is 
apparent around 7~GeV. We comment further on the structure of the anisotropy 
energy spectrum in Sec.~\ref{sec:interpretation}.

\begin{figure}
\includegraphics[width=0.49\textwidth]{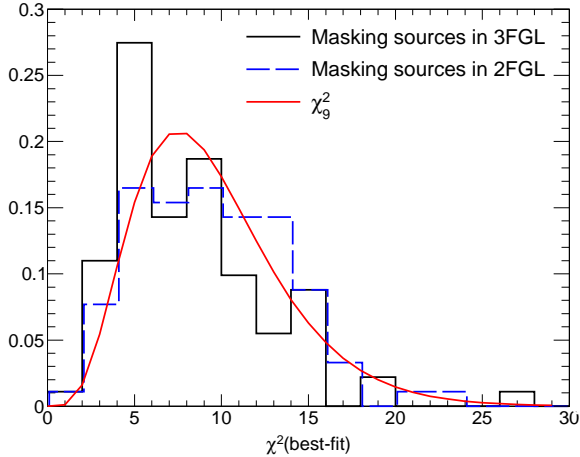}
\caption{\label{fig:chi2_distribution} Normalized distribution of the $\chi^2$ (defined as in Eq.~\ref{eqn:chi2_CP}) for the best-fit Poissonian $C_{\rm P}$ for all 91 independent combinations of energy bins. The solid black line is for the case when 3FGL sources are masked and the dashed blue line is for the mask covering 2FGL sources. The solid red curve is a $\chi^2$ distribution with 9 degrees of freedom.}
\end{figure}

\subsection{Cross-correlation angular power spectra}
\label{sec:cross_correlation}
Two examples of the cross-APS between energy bins are shown in  
Fig.~\ref{fig:crosscorrcl}. The left panel is for the cross-APS between bins
at low energies. A clear correlation is detected in the multipole range of 
interest (bounded by the vertical grey lines in the figure). Note the effect 
of the beam window function on the error bars at high multipoles, as in 
Fig.~\ref{fig:clszoom}. The right panel shows the cross-APS between two 
high-energy bins. This combination does not correspond to a significant 
detection, as the best-fit $C_{\rm P}$ is compatible with zero at a $2\sigma$
level.

The best-fit $C_{\rm P}$ for the cross-APS between the $i$-th and the $j$-th 
energy bins are shown in Appendix~\ref{sec:appendix}, multiplied by 
$E^2_i E^2_j / \Delta E_i \Delta E_j$ and for all the possible combinations of 
energy bins. Cross-APS $C_{\rm P}$ is detected in most combinations of energy 
bins, with the ones failing to yield a detection mainly involving the two 
highest energy bins. Tabs.~\ref{tab:cp_3FGL} and \ref{tab:cp_2FGL} report the 
detected cross-APS with their significance\footnote{Note that in some cases 
the best-fit $C_{\rm P}$ is negative. However, whenever that happens the 
estimated error is large and the measurement is compatible with zero.} The 
largest detection significance is 7.8$\sigma$ for the case of the cross-APS 
between the energy bin from 1.99 and 3.15 GeV and the energy bin between 3.15 
and 5.0 GeV.

The tables also report in bold the $\chi^2$ associated with the best-fit 
$C_{\rm P}$ according to the definition in Eq.~\ref{eqn:chi2_CP}. 
Fig.~\ref{fig:chi2_distribution} shows the distribution of the 91 $\chi^2$
of best-fit $C_{\rm P}$ in the 91 independent combinations of the 13 energy 
bins. The solid black line refers to the case when all sources in the 3FGL are 
masked and the dashed blue line when only sources in the 2FGL are masked. 
Both distributions are compatible with that of a $\chi^2$ distribution with 9 
degrees of freedom (i.e. the 10 data points inside the signal region in 
multipole minus 1 fitted parameter). The latter is represented by a solid red 
line in Fig.~\ref{fig:chi2_distribution}. Only 3 (4) combinations of energy 
bins have a $\chi^2$ larger than 16.9 (that would correspond to a $p$-value of 
0.05) when masking 3FGL (2FGL) sources.

\begin{figure*}
\includegraphics[width=0.49\textwidth]{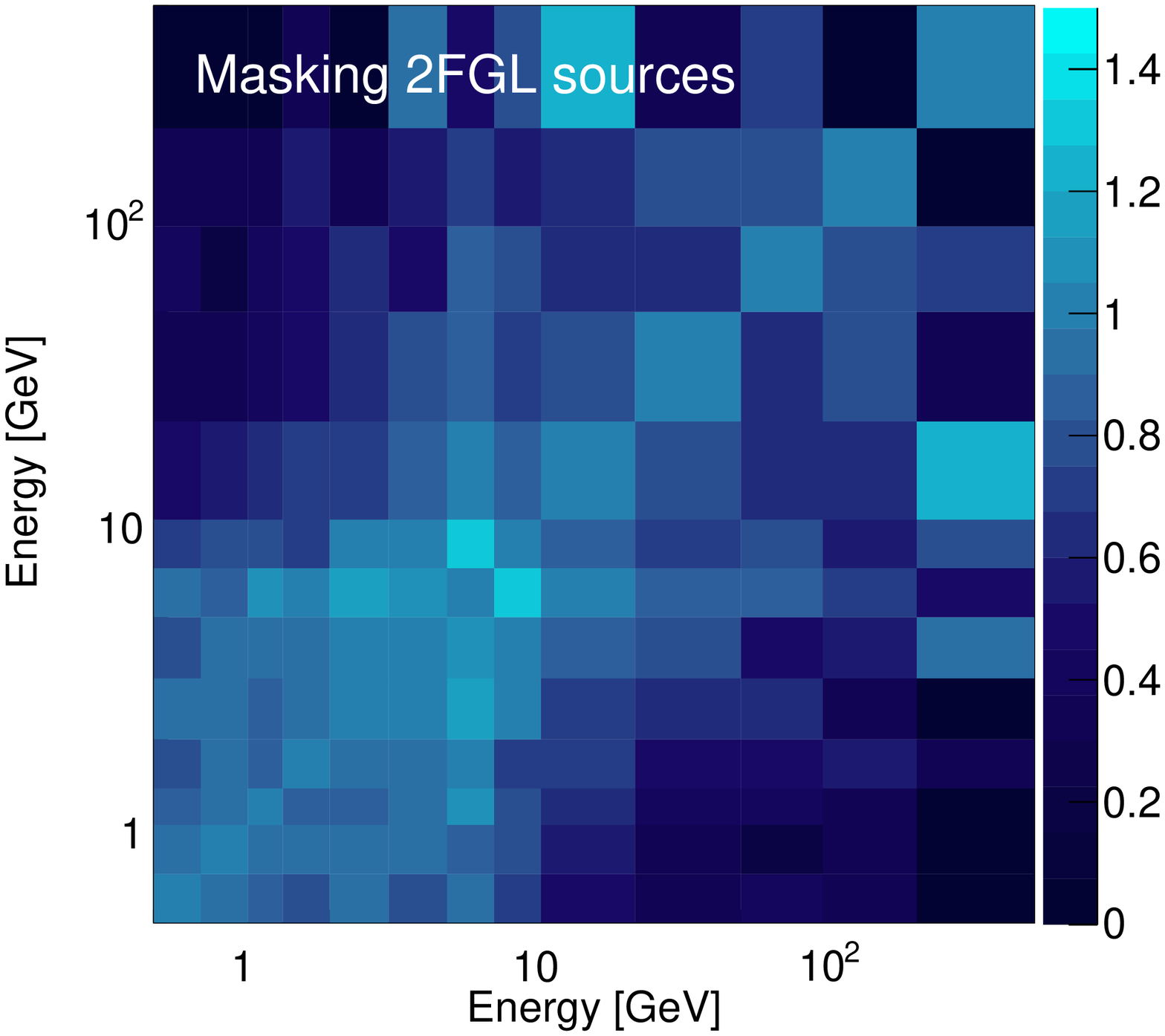}
\includegraphics[width=0.49\textwidth]{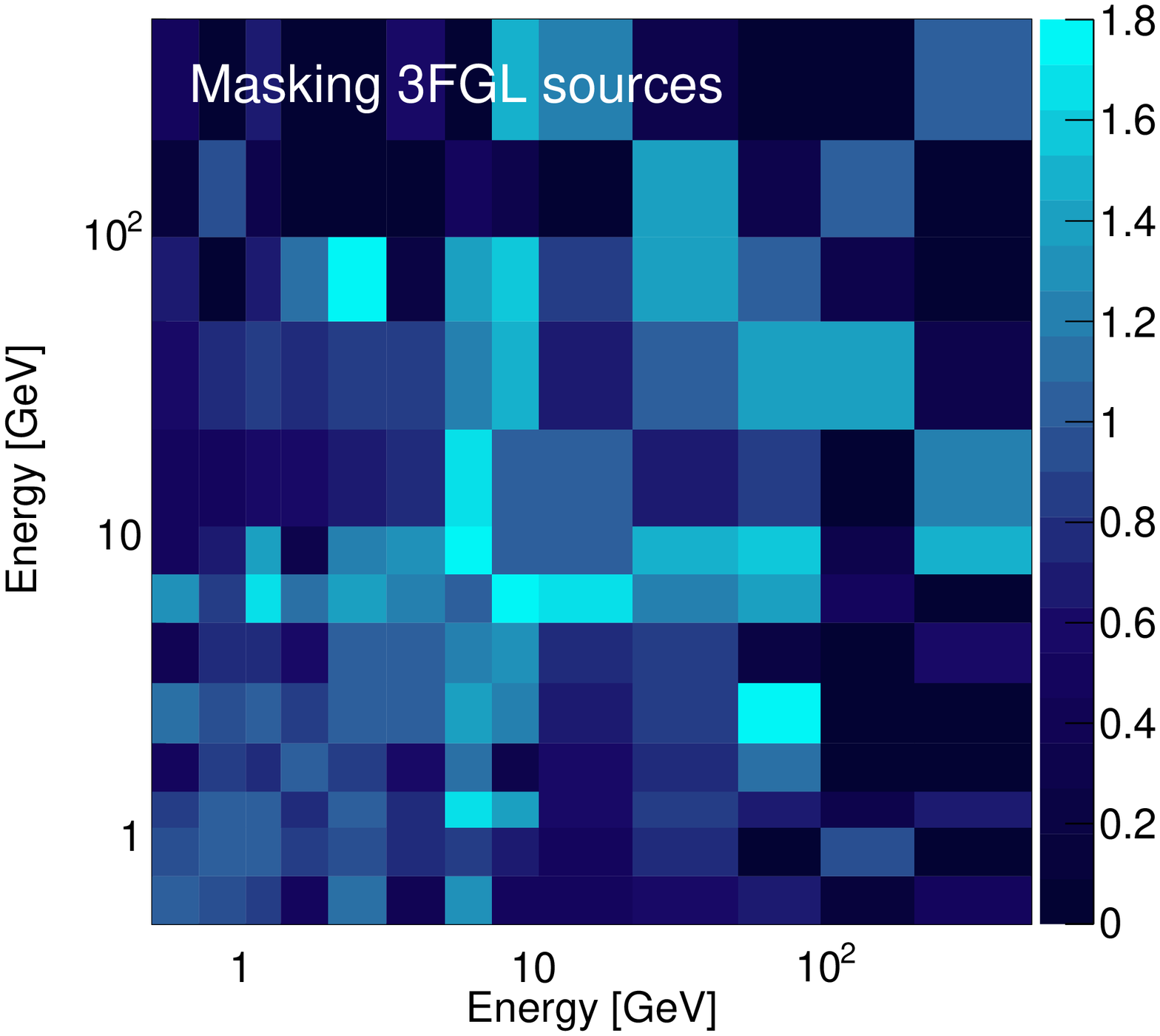}
\caption{\label{fig:corr_coefficents} Cross-correlation coefficients between energy bins. Each pixel in the panels corresponds to a pair $(i,j)$ of energy bins and it is colored according to the cross-correlation coefficent $r_{i,j}$. By construction the panels are symmetric with respect to the diagonal. The panel on the left refers to the default data set with a mask that covers the sources in 2FGL, while the one on the right is for the mask covering 3FGL sources. Cross-correlation coefficents below 1 indicate that the signal is due to multiple populations of sources.}
\end{figure*}

Together with the auto-APS in Fig.~\ref{fig:autoCPvsE}, the cross-APS provides 
an important handle to characterize the emission responsible for the 
anisotropy signal. In particular, if the latter is due to only one class of 
unresolved sources, the auto-APS $C_{\rm P}^{i,i}$ allows us to reconstruct their 
energy spectrum and the cross-APS can be {\it predicted} as 
$C_{\rm P}^{i,j} = \sqrt{C_{\rm P}^{i,i} C_{\rm P}^{j,j}}$. Alternatively, if we define 
the so-called cross-correlation coefficients $r_{i,j}$ as 
$C_{\rm P}^{i,j} / \sqrt{C_{\rm P}^{i,i} C_{\rm P}^{j,j}}$, any deviation from 1 when 
$i\neq j$ can be interpreted as an indication of multiple source classes 
contributing to the signal. In Fig.~\ref{fig:corr_coefficents}, we show the 
cross-correlation coefficients corresponding to the best-fit $C_{\rm P}^{i,j}$ 
for the data set obtained masking 2FGL sources (left panel) and masking 3FGL 
sources (right panel). In the former case, it is clear that the 
cross-correlation coefficients of low-energy bins are systematically smaller 
than 1, when correlated with high-energy bins. This is in qualitative
agreement with the findings of Ref.~\cite{DiMauro:2014wha}, in which the
auto-APS measured in Ref.~\cite{Ackermann:2012uf} was explained by the sum of
two different populations of unresolved blazars at low energies, while, above
$\sim$10 GeV, the signal was compatible with only one source class. 
Figs.~\ref{fig:cross_corr_coeff_1} and \ref{fig:cross_corr_coeff_2} in 
Appendix~\ref{sec:cross_corr_coefficents} show, for each energy bin $i$, how
the cross-correlation coefficents $r_{i,j}$ depend on energy $E_{j}$.

When 3FGL sources are masked (right panel) the situation is less clear as 
errors are larger (especially at high energies) and the estimated $C_{\rm P}$ 
more uncertain. We further discuss about the nature of our auto- and cross-APS 
in Sec.~\ref{sec:interpretation}. Note that in some cases the coefficients 
$r_{i,j}$ shown in Fig.~\ref{fig:corr_coefficents} are larger than 1, since only 
the best-fit values are plotted. They are, however, compatible with 1, within 
their uncertainty. Also, some coefficients are negative (and they are 
associated with a black pixel). Although within the error bars these negative 
$r_{i,j}$ are actually compatible with 0, we note that negative values are 
allowed, in the case of anti-correlations between two energy bins.

We finish this section by studying whether the binned auto- and cross-APS
$\overline{C_\ell}$ are better described by an APS that changes with multipole, 
instead of a constant value. We fit the binned $\overline{C_\ell}$ with a power 
law\footnote{The fit is performed with {\sc Minuit2} v5.34.14, 
http://lcgapp.cern.ch/project/cls/work-packages/mathlibs/minuit/index.html.}, 
i.e., $C_\ell=A (\ell/\ell_0)^{-\alpha}$, with $\ell_0=100$. We leave the 
normalization $A$ free to vary independently in all the 91 combinations of 
energy bins but we consider one common slope\footnote{If the auto- and 
cross-APS are interpreted as produced by a population of unresolved sources, 
they can be expressed in terms of the 3-dimensional power spectrum of the 
density field associated with the sources of the gamma-ray emission 
\cite{Ando:2005xg,Fornengo:2013rga,Fornasa:2015qua}. The latter determines the 
dependence on multipole, hence the shape of the auto- and cross-APS. Normally, 
the 3-dimensional power spectrum is only mildly dependent on the gamma-ray 
energy, which is encoded in the ``window function''. Therefore, the APS 
associated with different combinations of energy bins is expected to have 
approximately an energy-independent shape. It is therefore reasonable to 
assume a constant $\alpha$.}, i.e. $\alpha$. The best-fit value of $\alpha$ is 
$-0.06 \pm 0.08$ and it corresponds to a $\chi^2$ per degree of freedom of 
0.91. This should be compared to the global (i.e. for all the 91 combinations 
of energy bins) Poissonian fit, with also has a $\chi^2$ per degree of freedom 
of 0.91. Therefore, we cannot deduce any preference for the power-law scenario. 

\begin{turnpage}
\begin{table*}
\caption{\label{tab:cp_3FGL} Best-fit Poisson auto- and cross-APS $C_{\rm P}$ for the default data set and for the mask covering all sources in 3FGL, in units of $\mbox{cm}^{-4} \mbox{s}^{-2} \mbox{sr}^{-1}$. The numbers in italics indicate the significance of the detection in units of standard deviation (see text), while the numbers in bold give the $\chi^2$ associated with the corresponding $C_{\rm P}$. The entries marked in grey correspond to the auto-APS.}
\begin{ruledtabular}
\tiny
\begin{tabular}{c|ccccccccccccc}
Energy & 0.50- & 0.72- & 1.04- & 1.38- & 1.99- & 3.15- & 5.00- & 7.23- & 10.45- & 21.83- & 50.00- & 95.27- & 199.05- \\
bin [GeV] & 0.72 & 1.04 & 1.38 & 1.99 & 3.15 & 5.00 & 7.23 & 10.45 & 21.83 & 50.00 & 95.27 & 199.05 & 500.00 \\
\hline
0.50- & \cellcolor{mygray} $(7.88 \pm 1.47)$ & & & & & & & & & & & & \\

0.72 & \cellcolor{mygray} $\times 10^{-18}$ & & & & & & & & & & & & \\

     & \cellcolor{mygray} {\it 6.2}, {\bf 5.76} & & & & & & & & & & & & \\

\hline
0.72- & $(4.42 \pm 0.62)$ & \cellcolor{mygray} $(3.00 \pm 0.52)$ & & & & & & & & & & & \\

1.04 & $\times 10^{-18}$ & \cellcolor{mygray} $\times 10^{-18}$ & & & & & & & & & & & \\

     & {\it 7.1}, {\bf 4.12} & \cellcolor{mygray} {\it 6.3}, {\bf 8.52} & & & & & & & & & & & \\

\hline
1.04- & $(1.76 \pm 0.34)$ & $(1.28 \pm 0.21)$ & \cellcolor{mygray} $(5.45 \pm 1.51)$ & & & & & & & & & & \\

1.38 & $\times 10^{-18}$ & $\times 10^{-18}$ & \cellcolor{mygray} $\times 10^{-19}$ & & & & & & & & & & \\

     & {\it 5.1}, {\bf 14.43} & {\it 6.2}, {\bf 18.03} & \cellcolor{mygray} {\it 3.7}, {\bf 5.36} & & & & & & & & & & \\

\hline
1.38- & $(9.92 \pm 2.86)$ & $(1.02 \pm 0.16)$ & $(3.95 \pm 0.88)$ & \cellcolor{mygray} $(4.56 \pm 0.98)$ & & & & & & & & & \\

1.99 & $\times 10^{-19}$ & $\times 10^{-18}$ & $\times 10^{-19}$ & \cellcolor{mygray} $\times 10^{-19}$ & & & & & & & & & \\

     & {\it 3.5}, {\bf 15.34} & {\it 6.2}, {\bf 13.19} & {\it 4.5}, {\bf 2.58} & \cellcolor{mygray} {\it 4.9}, {\bf 26.21} & & & & & & & & & \\

\hline
1.99- & $(1.51 \pm 0.21)$ & $(8.12 \pm 1.18)$ & $(3.83 \pm 0.62)$ & $(2.96 \pm 0.48)$ & \cellcolor{mygray} $(2.41 \pm 0.46)$ & & & & & & & & \\

3.15 & $\times 10^{-18}$ & $\times 10^{-19}$ & $\times 10^{-19}$ & $\times 10^{-19}$ & \cellcolor{mygray} $\times 10^{-19}$ & & & & & & & & \\

     & {\it 7.4}, {\bf 3.91} & {\it 6.9}, {\bf 5.68} & {\it 6.2}, {\bf 8.85} & {\it 6.2}, {\bf 8.40} & \cellcolor{mygray} {\it 5.6}, {\bf 2.17} & & & & & & & & \\

\hline
3.15- & $(3.51 \pm 1.32)$ & $(3.89 \pm 0.75)$ & $(1.65 \pm 0.39)$ & $(1.25 \pm 0.29)$ & $(1.51 \pm 0.19)$ & \cellcolor{mygray} $(8.94 \pm 1.60)$ & & & & & & & \\

5.00 & $\times 10^{-19}$ & $\times 10^{-19}$ & $\times 10^{-19}$ & $\times 10^{-19}$ & $\times 10^{-19}$ & \cellcolor{mygray} $\times 10^{-20}$ & & & & & & & \\

     & {\it 2.7}, {\bf 8.39} & {\it 5.2}, {\bf 15.75} & {\it 4.3}, {\bf 8.83} & {\it 4.3}, {\bf 8.84} & {\it 7.8}, {\bf 7.62} & \cellcolor{mygray} {\it 5.8}, {\bf 9.21} & & & & & & & \\

\hline
5.00- & $(2.81 \pm 0.82)$ & $(1.22 \pm 0.46)$ & $(9.96 \pm 2.35)$ & $(6.02 \pm 1.73)$ & $(5.29 \pm 1.15)$ & $(2.88 \pm 0.64)$ & \cellcolor{mygray} $(6.23 \pm 5.26)$ & & & & & & \\

7.23 & $\times 10^{-19}$ & $\times 10^{-19}$ & $\times 10^{-20}$ & $\times 10^{-20}$ & $\times 10^{-20}$ & $\times 10^{-20}$ & \cellcolor{mygray} $\times 10^{-21}$ & & & & & & \\

     & {\it 3.4}, {\bf 8.25} & {\it 2.7}, {\bf 6.64} & {\it 4.3}, {\bf 1.79} & {\it 3.5}, {\bf 14.43} & {\it 4.7}, {\bf 5.55} & {\it 4.5}, {\bf 10.33} & \cellcolor{mygray} {\it 1.2}, {\bf 9.48} & & & & & & \\

\hline
7.23- & $(7.86 \pm 6.06)$ & $(6.79 \pm 3.36)$ & $(5.77 \pm 1.69)$ & $(1.09 \pm 1.26)$ & $(3.47 \pm 0.81)$ & $(2.21 \pm 0.46)$ & $(1.36 \pm 0.25)$ & \cellcolor{mygray} $(3.22 \pm 2.50)$ & & & & & \\

10.24 & $\times 10^{-20}$ & $\times 10^{-20}$ & $\times 10^{-20}$ & $\times 10^{-20}$ & $\times 10^{-20}$ & $\times 10^{-20}$ & $\times 10^{-20}$ & \cellcolor{mygray} $\times 10^{-21}$ & & & & & \\

      & {\it 1.3}, {\bf 15.77} & {\it 2.0}, {\bf 4.47} & {\it 3.4}, {\bf 7.64} & {\it 0.9}, {\bf 6.29} & {\it 4.3}, {\bf 2.67} & {\it 4.9}, {\bf 15.62} & {\it 5.3}, {\bf 8.11} & \cellcolor{mygray} {\it 1.3}, {\bf 4.61} & & & & & \\

\hline
10.24- & $(1.51 \pm 0.58)$ & $(8.63 \pm 3.18)$ & $(4.15 \pm 1.60)$ & $(4.37 \pm 1.16)$ & $(3.39 \pm 0.75)$ & $(2.43 \pm 0.41)$ & $(1.38 \pm 0.23)$ & $(5.85 \pm 1.56)$ & \cellcolor{mygray} $(1.06 \pm 0.19)$ & & & & \\

21.83 & $\times 10^{-19}$ & $\times 10^{-20}$ & $\times 10^{-20}$ & $\times 10^{-20}$ & $\times 10^{-20}$ & $\times 10^{-20}$ & $\times 10^{-20}$ & $\times 10^{-21}$ & \cellcolor{mygray} $\times 10^{-20}$ & & & & \\

21.83 & {\it 2.6}, {\bf 6.15} & {\it 2.7}, {\bf 4.63} & {\it 2.6}, {\bf 4.90} & {\it 3.8}, {\bf 6.66} & {\it 4.6}, {\bf 6.90} & {\it 5.9}, {\bf 12.80} & {\it 6.0}, {\bf 4.40} & {\it 3.7}, {\bf 11.35} & \cellcolor{mygray} {\it 5.6}, {\bf 13.87} & & & & \\

\hline
21.83- & $(6.13 \pm 3.35)$ & $(4.90 \pm 1.82)$ & $(2.47 \pm 0.91)$ & $(2.03 \pm 0.66)$ & $(1.53 \pm 0.42)$ & $(1.01 \pm 0.23)$ & $(3.61 \pm 1.31)$ & $(3.17 \pm 0.88)$ & $(2.56 \pm 0.79)$ & \cellcolor{mygray} $(1.47 \pm 0.60)$ & & & \\

50.00 & $\times 10^{-20}$ & $\times 10^{-20}$ & $\times 10^{-20}$ & $\times 10^{-20}$ & $\times 10^{-20}$ & $\times 10^{-20}$ & $\times 10^{-21}$ & $\times 10^{-21}$ & $\times 10^{-21}$ & \cellcolor{mygray} $\times 10^{-21}$ & & & \\

      &  {\it 1.8}, {\bf 14.23} & {\it 2.7}, {\bf 5.22} & {\it 2.7}, {\bf 11.01} & {\it 3.0}, {\bf 4.13} & {\it 3.6}, {\bf 8.44} & {\it 4.4}, {\bf 9.82} & {\it 2.8}, {\bf 6.50} & {\it 3.6}, {\bf 4.52} & {\it 3.3}, {\bf 18.78} & \cellcolor{mygray} {\it 2.5}, {\bf 3.08} & & & \\

\hline
50.00- & $(2.90 \pm 1.69)$ & $(-1.34 \pm 0.93)$ & $(7.50 \pm 4.63)$ & $(1.08 \pm 0.34)$ & $(1.23 \pm 0.22)$ & $(0.97 \pm 1.18)$ & $(1.59 \pm 0.66)$ & $(1.28 \pm 0.44)$ & $(1.21 \pm 0.39)$ & $(7.82 \pm 2.22)$ & \cellcolor{mygray} $(2.07 \pm 1.60)$ & & \\

95.27 & $\times 10^{-20}$ & $\times 10^{-20}$ & $\times 10^{-21}$ & $\times 10^{-20}$ & $\times 10^{-20}$ & $\times 10^{-21}$ & $\times 10^{-21}$ & $\times 10^{-21}$ & $\times 10^{-21}$ & $\times 10^{-22}$ & \cellcolor{mygray} $\times 10^{-22}$ & & \\

95.27 & {\it 1.7}, {\bf 10.92} & {\it 1.4}, {\bf 11.02} & {\it 1.6}, {\bf 10.31} & {\it 3.2}, {\bf 5.82} & {\it 5.7}, {\bf 2.86} & {\it 0.8}, {\bf 9.59} & {\it 2.4}, {\bf 9.41} & {\it 2.9}, {\bf 13.96} & {\it 3.0}, {\bf 5.61} & {\it 3.6}, {\bf 1.29} & \cellcolor{mygray} {\it 1.3}, {\bf 4.07} & & \\

\hline
95.27- & $(0.32 \pm 1.04)$ & $(1.16 \pm 0.58)$ & $(1.49 \pm 2.88)$ & $(0.17 \pm 2.10)$ & $(0.23 \pm 1.35)$ & $(1.09 \pm 7.48)$ & $(2.58 \pm 4.15)$ & $(1.18 \pm 2.84)$ & $(-1.17 \pm 2.52)$ & $(3.72 \pm 1.38)$ & $(3.63 \pm 7.09)$ & \cellcolor{mygray} $(4.98 \pm 6.30)$ & \\

199.05 & $\times 10^{-20}$ & $\times 10^{-20}$ & $\times 10^{-21}$ & $\times 10^{-21}$ & $\times 10^{-21}$ & $\times 10^{-22}$ & $\times 10^{-22}$ & $\times 10^{-22}$ & $\times 10^{-22}$ & $\times 10^{-22}$ & $\times 10^{-23}$ & \cellcolor{mygray} $\times 10^{-23}$ & \\

      & {\it 0.3}, {\bf 5.10} & {\it 2.0}, {\bf 8.35} & {\it 0.5}, {\bf 5.55} & {\it 0.1}, {\bf 1.16} & {\it 0.2}, {\bf 7.96} & {\it 0.2}, {\bf 6.23} & {\it 0.6}, {\bf 11.13} & {\it 0.4}, {\bf 7.34} & {\it 0.5}, {\bf 2.59} & {\it 2.7}, {\bf 4.39} & {\it 0.5}, {\bf 4.25} & \cellcolor{mygray} {\it 0.7}, {\bf 14.18} & \\

\hline
199.05 & $(2.54 \pm 6.16)$ & $(-4.13 \pm 3.43)$ & $(0.91 \pm 1.71)$ & $(-1.77 \pm 1.26)$ & $(-6.13 \pm 8.02)$ & $(2.87 \pm 4.44)$ & $(-0.97 \pm 2.51)$ & $(1.43 \pm 1.71)$ & $(2.11 \pm 1.52)$ & $(1.85 \pm 8.33)$ & $(-4.93 \pm 4.32)$ & $(-1.37 \pm 2.74)$ & \cellcolor{mygray} $(0.30 \pm 2.30) $ \\

500.0 & $\times 10^{-21}$ & $\times 10^{-21}$ & $\times 10^{-21}$ & $\times 10^{-21}$ & $\times 10^{-22}$ & $\times 10^{-22}$ & $\times 10^{-22}$ & $\times 10^{-22}$ & $\times 10^{-22}$ & $\times 10^{-23}$ & $\times 10^{-23}$ & $\times 10^{-23}$ & \cellcolor{mygray} $\times 10^{-23}$ \\

      & {\it 0.4}, {\bf 7.64} & {\it 1.2}, {\bf 5.81} & {\it 0.5}, {\bf 6.99} & {\it 1.4}, {\bf 9.91} & {\it 0.8}, {\bf 9.07} & {\it 0.6}, {\bf 11.64} & {\it 0.4}, {\bf 5.83} & {\it 0.8}, {\bf 5.30} & {\it 1.4}, {\bf 3.42} & {\it 0.2}, {\bf 3.61} & {\it 1.1}, {\bf 4.72} & {\it 0.5}, {\bf 4.82} & \cellcolor{mygray} {\it 0.1}, {\bf 3.64} \\
\end{tabular}
\end{ruledtabular}
\end{table*}

\begin{table*}
\caption{\label{tab:cp_2FGL} Same as Tab.~\ref{tab:cp_3FGL} but with the mask covering the sources in 2FGL. Data are available at https://www-glast.stanford.edu/pub\_data/552.}
\begin{ruledtabular}
\tiny
\begin{tabular}{c|ccccccccccccc}
Energy & 0.50- & 0.72- & 1.04- & 1.38- & 1.99- & 3.15- & 5.00- & 7.23- & 10.45- & 21.83- & 50.00- & 95.27- & 199.05- \\
bin [GeV] & 0.72 & 1.04 & 1.38 & 1.99 & 3.15 & 5.00 & 7.23 & 10.45 & 21.83 & 50.00 & 95.27 & 199.05 & 500.00 \\
\hline
0.50- & \cellcolor{mygray} $(1.76 \pm 0.15)$ & & & & & & & & & & & & \\

0.72 & \cellcolor{mygray} $\times 10^{-17}$ & & & & & & & & & & & & \\

     & \cellcolor{mygray} {\it 14.7}, {\bf 8.81} & & & & & & & & & & & & \\

\hline
0.72- & $(1.02 \pm 0.07)$ & \cellcolor{mygray} $(6.92 \pm 0.56)$ & & & & & & & & & & & \\

1.04 & $\times 10^{-17}$ & \cellcolor{mygray} $\times 10^{-18}$ & & & & & & & & & & & \\

     & {\it 15.2}, {\bf 7.52} & \cellcolor{mygray} {\it 15.5}, {\bf 14.61} & & & & & & & & & & & \\

\hline
1.04- & $(4.66 \pm 0.38)$ & $(3.27 \pm 0.22)$ & \cellcolor{mygray} $(1.81 \pm 0.17)$ & & & & & & & & & & \\

1.38 & $\times 10^{-18}$ & $\times 10^{-18}$ & \cellcolor{mygray} $\times 10^{-18}$ & & & & & & & & & & \\

     & {\it 12.6}, {\bf 5.82} & {\it 14.9}, {\bf 12.07} & \cellcolor{mygray} {\it 12.7}, {\bf 11.35} & & & & & & & & & & \\

\hline
1.38- & $(3.61 \pm 0.30)$ & $(2.78 \pm 0.19)$ & $(1.25 \pm 0.09)$ & \cellcolor{mygray} $(1.19 \pm 0.11)$ & & & & & & & & & \\

1.99 & $\times 10^{-18}$ & $\times 10^{-18}$ & $\times 10^{-18}$ & \cellcolor{mygray} $\times 10^{-18}$ & & & & & & & & & \\

     & {\it 12.0}, {\bf 14.57} & {\it 15.8}, {\bf 8.59} & {\it 13.4}, {\bf 4.76} & \cellcolor{mygray} {\it 13.2}, {\bf 20.84} & & & & & & & & & \\

\hline
1.99- & $(3.11 \pm 0.22)$ & $(2.04 \pm 0.13)$ & $(9.40 \pm 0.66)$ & $(8.39 \pm 0.49)$ & \cellcolor{mygray} $(6.44 \pm 0.48)$ & & & & & & & & \\

3.15 & $\times 10^{-18}$ & $\times 10^{-18}$ & $\times 10^{-19}$ & $\times 10^{-19}$ & \cellcolor{mygray} $\times 10^{-19}$ & & & & & & & & \\

     & {\it 14.3}, {\bf 3.85} & {\it 16.4}, {\bf 9.66} & {\it 14.5}, {\bf 8.88} & {\it 16.8}, {\bf 12.20} & \cellcolor{mygray} {\it 15.3}, {\bf 9.72} & & & & & & & & \\

\hline
3.15- & $(1.45 \pm 0.14)$ & $(1.12 \pm 0.08)$ & $(5.56 \pm 0.40)$ & $(4.70 \pm 0.30)$ & $(3.65 \pm 0.20)$ & \cellcolor{mygray} $(2.08 \pm 0.17)$ & & & & & & & \\

5.00 & $\times 10^{-18}$ & $\times 10^{-18}$ & $\times 10^{-19}$ & $\times 10^{-19}$ & $\times 10^{-19}$ & \cellcolor{mygray} $\times 10^{-19}$ & & & & & & & \\

     & {\it 10.5}, {\bf 7.55} & {\it 14.4}, {\bf 23.99} & {\it 13.9}, {\bf 14.04} & {\it 15.7}, {\bf 7.33} & {\it 18.5}, {\bf 4.52} & \cellcolor{mygray} {\it 14.0}, {\bf 14.77} & & & & & & & \\

\hline
5.00- & $(6.86 \pm 0.87)$ & $(3.98 \pm 0.47)$ & $(2.61 \pm 0.25)$ & $(1.91 \pm 0.18)$ & $(1.66 \pm 0.12)$ & $(8.63 \pm 0.68)$ & \cellcolor{mygray} $(3.05 \pm 0.53)$ & & & & & & \\

7.23 & $\times 10^{-19}$ & $\times 10^{-19}$ & $\times 10^{-19}$ & $\times 10^{-19}$ & $\times 10^{-19}$ & $\times 10^{-20}$ & \cellcolor{mygray} $\times 10^{-20}$ & & & & & & \\

     & {\it 8.1}, {\bf 13.94} & {\it 8.4}, {\bf 4.62} & {\it 10.9}, {\bf 4.61} & {\it 10.7}, {\bf 11.95} & {\it 14.4}, {\bf 5.02} & {\it 13.2}, {\bf 11.26} & \cellcolor{mygray} {\it 6.0}, {\bf 9.03} & & & & & & \\

\hline
7.23- & $(4.06 \pm 0.63)$ & $(2.59 \pm 0.35)$ & $(1.33 \pm 0.18)$ & $(1.01 \pm 0.13)$ & $(1.03 \pm 0.08)$ & $(6.10 \pm 0.45)$ & $(2.91 \pm 0.25)$ & \cellcolor{mygray} $(1.69 \pm 0.25)$ & & & & & \\

10.24 & $\times 10^{-19}$ & $\times 10^{-19}$ & $\times 10^{-19}$ & $\times 10^{-19}$ & $\times 10^{-19}$ & $\times 10^{-20}$ & $\times 10^{-20}$ & \cellcolor{mygray} $\times 10^{-20}$ & & & & & \\

      & {\it 6.5}, {\bf 14.30} & {\it 7.5}, {\bf 9.69} & {\it 7.6}, {\bf 12.06} & {\it 7.9}, {\bf 2.92} & {\it 12.6}, {\bf 9.20} & {\it 13.4}, {\bf 13.67} & {\it 11.4}, {\bf 7.19} & \cellcolor{mygray} {\it 7.0}, {\bf 5.45} & & & & & \\

\hline
10.24- & $(3.39 \pm 0.60)$ & $(2.30 \pm 0.33)$ & $(1.49 \pm 0.17)$ & $(1.30 \pm 0.12)$ & $(9.07 \pm 0.77)$ & $(6.57 \pm 0.42)$ & $(2.99 \pm 0.23)$ & $(1.80 \pm 0.16)$ & \cellcolor{mygray} $(2.76 \pm 0.19)$ & & & & \\

21.83 & $\times 10^{-19}$ & $\times 10^{-19}$ & $\times 10^{-19}$ & $\times 10^{-19}$ & $\times 10^{-20}$ & $\times 10^{-20}$ & $\times 10^{-20}$ & $\times 10^{-20}$ & \cellcolor{mygray} $\times 10^{-20}$ & & & & \\

      & {\it 5.6}, {\bf 13.66} & {\it 7.0}, {\bf 15.13} & {\it 9.0}, {\bf 1.53} & {\it 10.8}, {\bf 13.47} & {\it 11.9}, {\bf 2.82} & {\it 15.6}, {\bf 12.72} & {\it 12.8}, {\bf 13.79} & {\it 11.4}, {\bf 13.18} & \cellcolor{mygray} {\it 14.5}, {\bf 11.56} & & & & \\

\hline
21.83- & $(9.00 \pm 3.42)$ & $(6.08 \pm 1.89)$ & $(3.52 \pm 0.94)$ & $(3.58 \pm 0.69)$ & $(3.19 \pm 0.42)$ & $(2.36 \pm 0.23)$ & $(9.54 \pm 1.27)$ & $(6.07 \pm 0.85)$ & $(8.79 \pm 0.75)$ & \cellcolor{mygray} $(4.35 \pm 0.61)$ & & & \\

50.00 & $\times 10^{-20}$ & $\times 10^{-20}$ & $\times 10^{-20}$ & $\times 10^{-20}$ & $\times 10^{-20}$ & $\times 10^{-20}$ & $\times 10^{-21}$ & $\times 10^{-21}$ & $\times 10^{-21}$ & \cellcolor{mygray} $\times 10^{-21}$ & & & \\

      & {\it 2.6}, {\bf 14.20} & {\it 3.2}, {\bf 6.29} & {\it 3.8}, {\bf 8.80} & {\it 5.3}, {\bf 4.51} & {\it 7.5}, {\bf 6.94} & {\it 10.1}, {\bf 11.18} & {\it 7.4}, {\bf 12.72} & {\it 7.0}, {\bf 6.34} & {\it 11.2}, {\bf 17.72} & \cellcolor{mygray} {\it 7.4}, {\bf 9.31} & & & \\

\hline
50.00- & $(5.66 \pm 1.77)$ & $(1.46 \pm 0.97)$ & $(1.65 \pm 0.48)$ & $(1.65 \pm 0.35)$ & $(1.67 \pm 0.22)$ & $(6.58 \pm 1.21)$ & $(4.79 \pm 0.67)$ & $(3.33 \pm 0.45)$ & $(3.48 \pm 0.41)$ & $(1.29 \pm 0.23)$ & \cellcolor{mygray} $(9.88 \pm 1.66)$ & & \\

95.27 & $\times 10^{-20}$ & $\times 10^{-20}$ & $\times 10^{-20}$ & $\times 10^{-20}$ & $\times 10^{-20}$ & $\times 10^{-21}$ & $\times 10^{-21}$ & $\times 10^{-21}$ & $\times 10^{-21}$ & $\times 10^{-21}$ & \cellcolor{mygray} $\times 10^{-22}$ & & \\

      & {\it 3.2}, {\bf 10.32} & {\it 1.5}, {\bf 7.70} & {\it 3.4}, {\bf 11.91} & {\it 4.7}, {\bf 4.67} & {\it 7.6}, {\bf 5.76} & {\it 5.4}, {\bf 15.15} & {\it7.1}, {\bf 9.64} & {\it 7.3}, {\bf 10.32} & {\it 8.5}, {\bf 7.88} & {\it 5.8}, {\bf 12.85} & \cellcolor{mygray} {\it 6.1}, {\bf 12.39} & & \\

\hline
95.27- & $(2.07 \pm 1.10)$ & $(1.12 \pm 0.61)$ & $(6.20 \pm 3.02)$ & $(8.01 \pm 2.22)$ & $(3.99 \pm 1.42)$ & $(3.41 \pm 0.77)$ & $(1.67 \pm 0.43)$ & $(1.02 \pm 0.29)$ & $(1.46 \pm 0.26)$ & $(7.23 \pm 1.43)$ & $(3.31 \pm 0.74)$ & \cellcolor{mygray} $(1.77 \pm 0.67)$ & \\

199.05 & $\times 10^{-20}$ & $\times 10^{-20}$ & $\times 10^{-21}$ & $\times 10^{-21}$ & $\times 10^{-21}$ & $\times 10^{-21}$ & $\times 10^{-21}$ & $\times 10^{-21}$ & $\times 10^{-21}$ & $\times 10^{-22}$ & $\times 10^{-22}$ & \cellcolor{mygray} $\times 10^{-22}$ & \\

       & {\it 1.9}, {\bf 9.32} & {\it 1.8}, {\bf 7.93} & {\it 2.0}, {\bf 16.59} & {\it 3.6}, {\bf 4.21} & {\it 2.8}, {\bf 8.62} & {\it 4.4}, {\bf 5.85} & {\it 3.9}, {\bf 4.06} & {\it 3.5}, {\bf 11.83} & {\it 5.6}, {\bf 3.72} & {\it 5.0}, {\bf 7.81} & {\it 4.4}, {\bf 2.94} & \cellcolor{mygray} {\it 2.6}, {\bf 14.31} & \\

\hline
199.05 & $(-1.32 \pm 5.74)$ & $(-2.19 \pm 3.14)$ & $(0.07 \pm 1.57)$ & $(0.70 \pm 1.17)$ & $(-5.67 \pm 7.44)$ & $(8.39 \pm 4.08)$ & $(1.73 \pm 2.27)$ & $(2.05 \pm 1.54)$ & $(4.14 \pm 1.38)$ & $(4.59 \pm 7.58)$ & $(4.51 \pm 3.95)$ & $(-3.57 \pm 2.54)$ & \cellcolor{mygray} $(0.39 \pm 1.91)$ \\

500.0 & $\times 10^{-21}$ & $\times 10^{-21}$ & $\times 10^{-21}$ & $\times 10^{-21}$ & $\times 10^{-22}$ & $\times 10^{-22}$ & $\times 10^{-22}$ & $\times 10^{-22}$ & $\times 10^{-22}$ & $\times 10^{-23}$ & $\times 10^{-23}$ & $\times 10^{-23}$ & \cellcolor{mygray} $\times 10^{-23}$ \\

      & {\it 0.2}, {\bf 4.73} & {\it 0.7}, {\bf 6.78} & {\it $<$0.1}, {\bf 11.42} & {\it 0.6}, {\bf 14.07} & {\it 0.8}, {\bf 10.48} & {\it 2.1}, {\bf 17.24} & {\it 0.8}, {\bf 6.79} & {\it 1.3}, {\bf 6.92} & {\it 3.0}, {\bf 5.18} & {\it 0.6}, {\bf 9.32} & {\it 1.1}, {\bf 3.97} & {\it 1.4}, {\bf 9.04} & \cellcolor{mygray} {\it $<$0.1}, {\bf 4.17} \\
\end{tabular}
\end{ruledtabular}
\end{table*}
\end{turnpage}

\subsection{Validation studies}
\label{sec:validations}
We note that the uncertainties reported in the last section only include 
statistical errors. It is therefore important to estimate any systematic 
errors as, e.g., those related to the analysis (such as the foreground 
cleaning and the use of the mask) or to the characterization of the 
instrument, which may affect the effective area and beam window functions. We
discuss possible sources of systematic errors in the following sections.

\begin{figure*}
\includegraphics[width=0.49\textwidth]{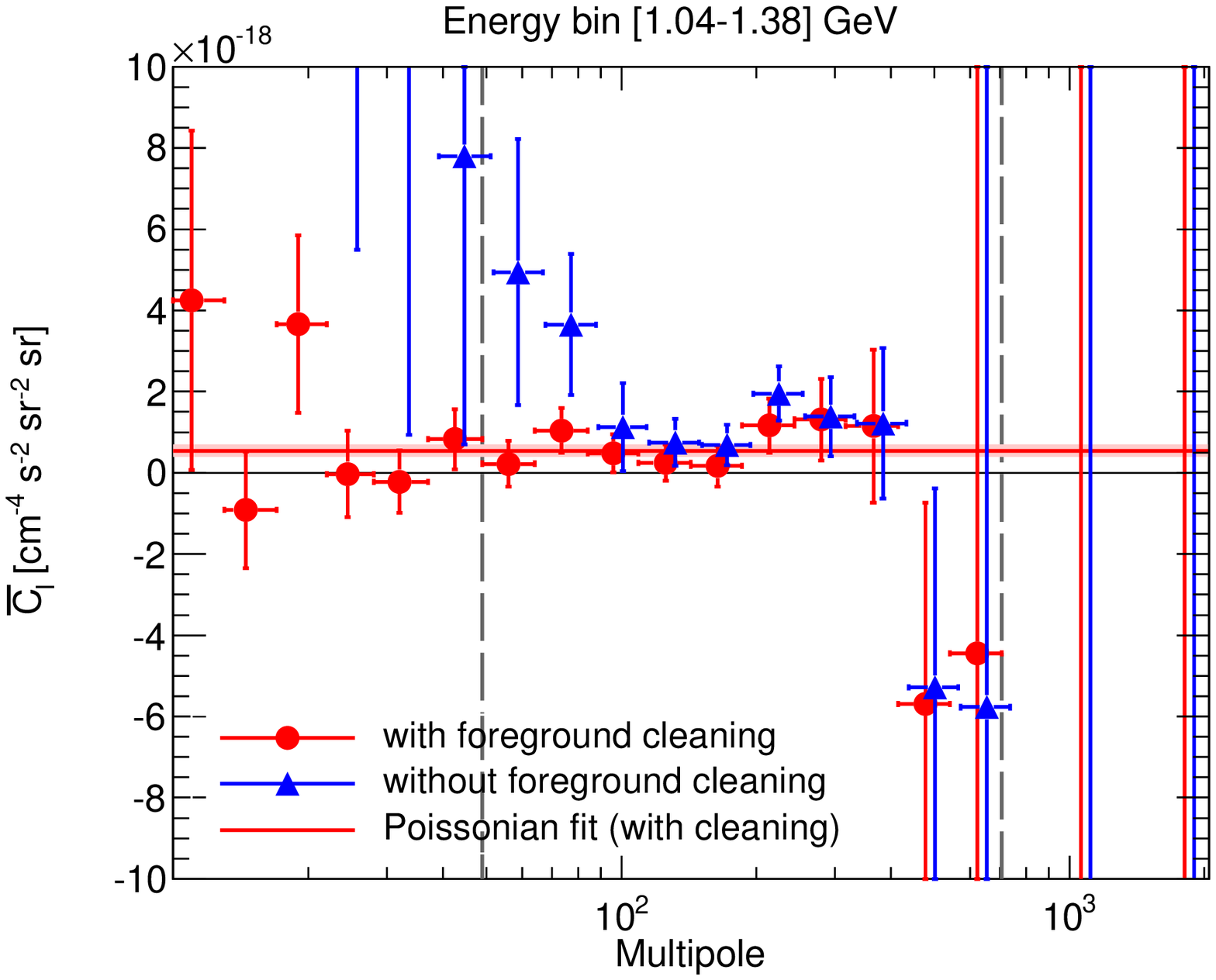}
\includegraphics[width=0.49\textwidth]{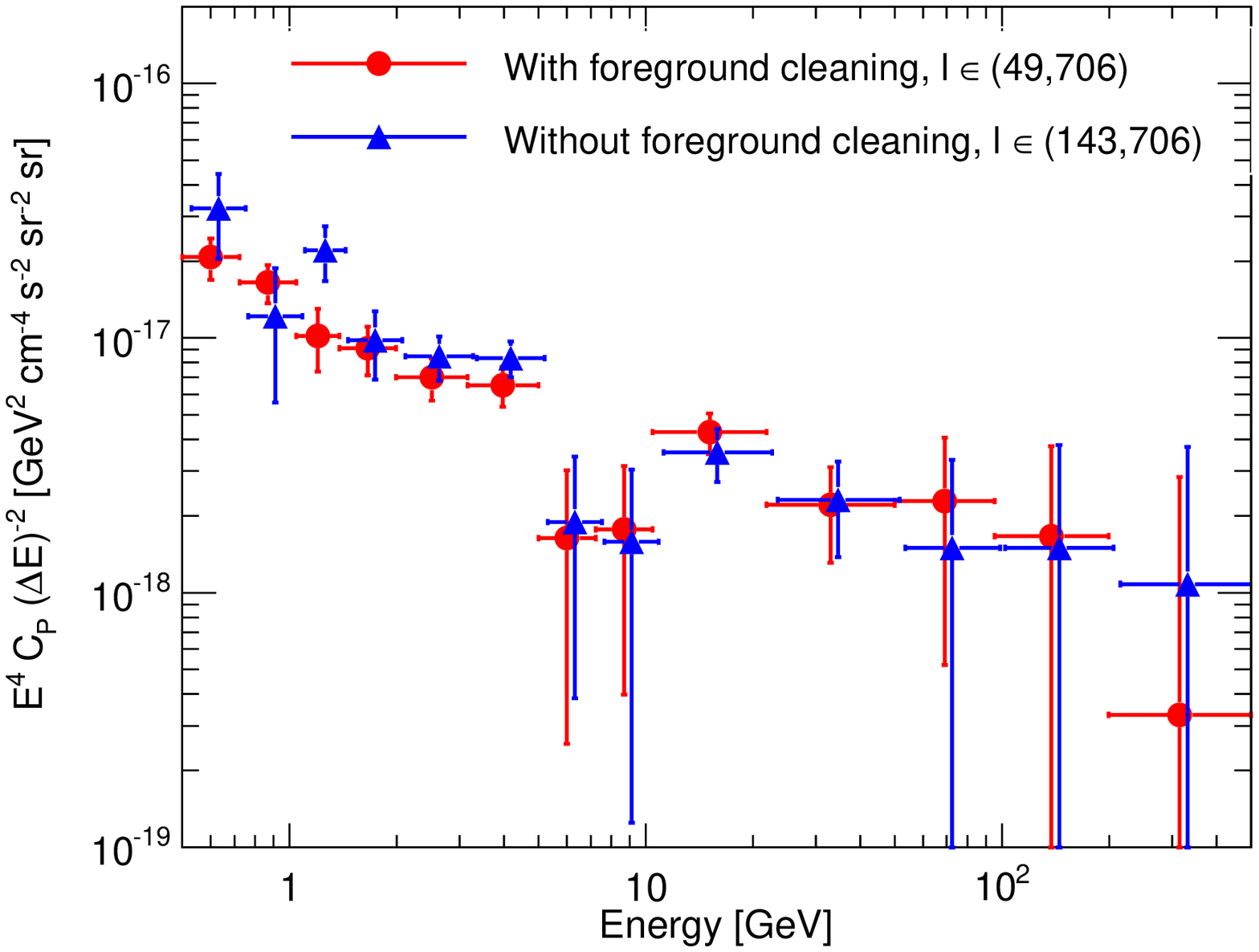}
\caption{\label{fig:cleannoclean} {\it Left:} Auto-APS in the energy bin between 1.04 and 1.38 GeV, comparing the data with (red circles) and without (blue triangles) foreground cleaning. The solid red line indicates the best-fit $C_{\rm P}$ for the case with foreground cleaning, and the pink band its 68\% CL error. The two dashed vertical lines mark our signal region in multipole. {\it Right:} Poissonian auto-APS as a function of the energy for the case with foreground cleaning and a signal region between $\ell=49$ and 706 (red circles) and for the case without foreground cleaning and a signal region between $\ell=143$ and 706 (blue triangles). Default data selection and 3FGL mask are used.}
\end{figure*}

\begin{figure*}
\includegraphics[width=0.49\textwidth]{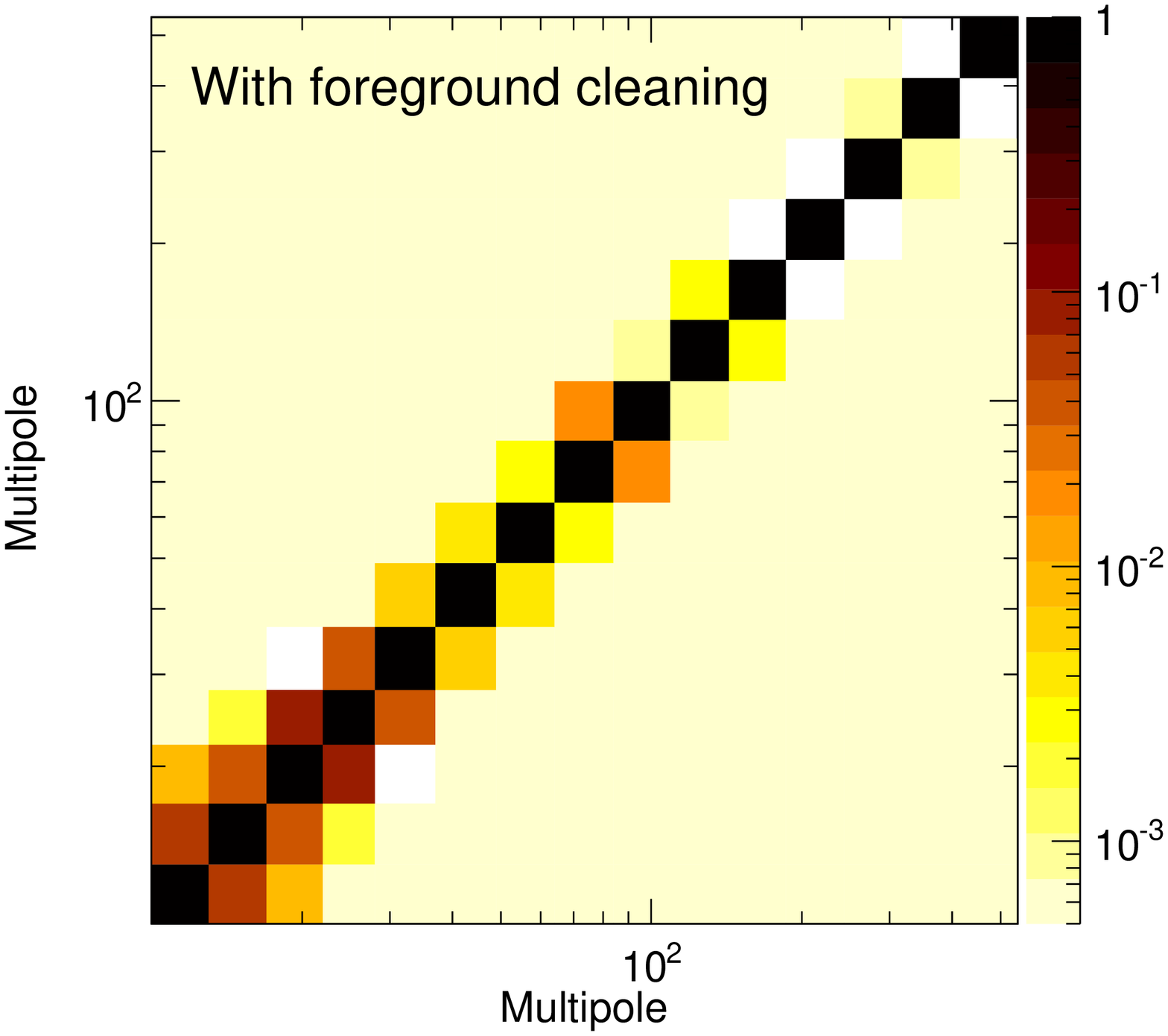}
\includegraphics[width=0.49\textwidth]{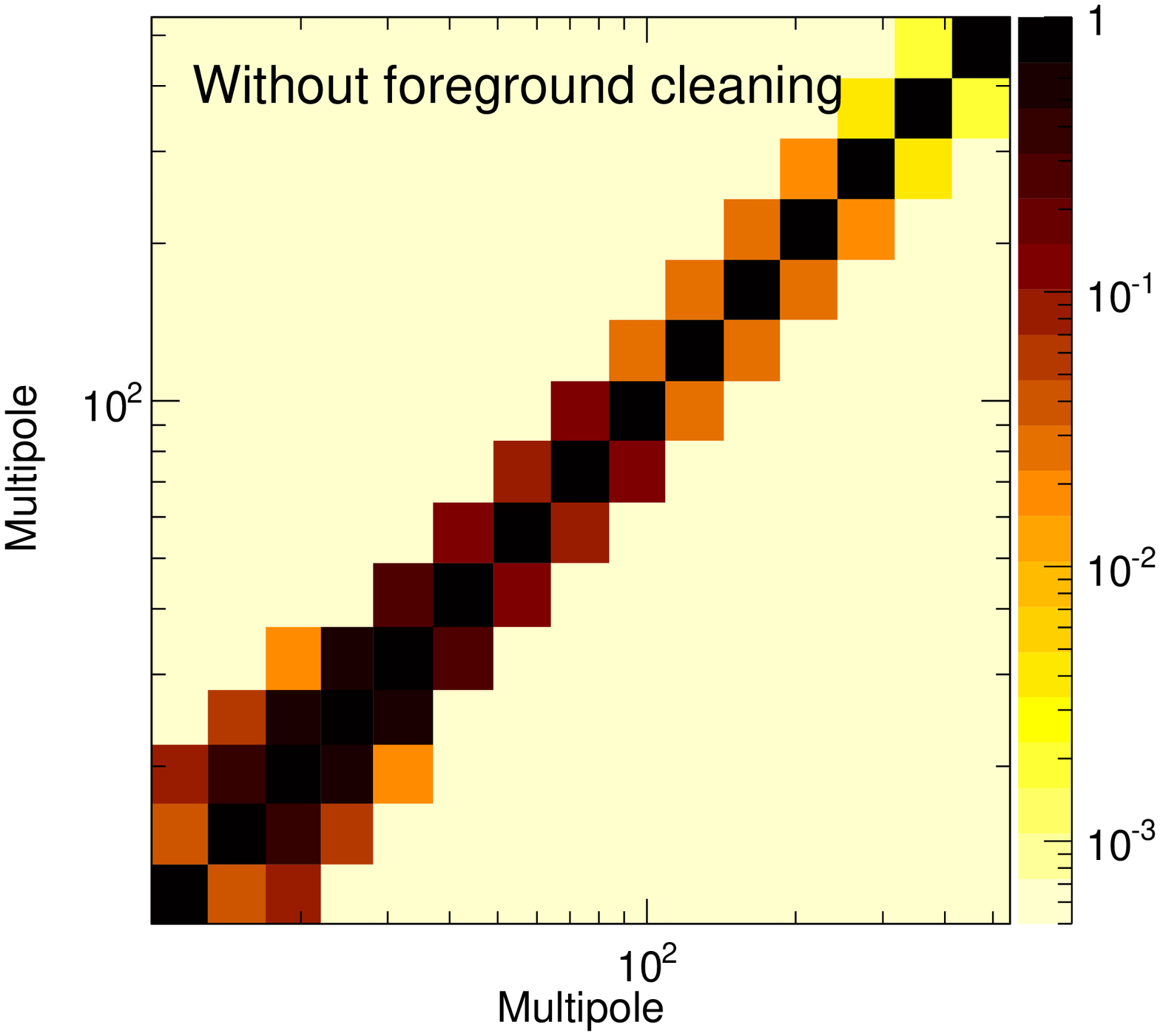}
\caption{\label{fig:covariance_foreground} Normalized covariance matrix ($\sigma_{i,j}/ \sqrt{\sigma_{i,i} \sigma_{j,j}}$) of the binned $\overline{C_{\ell}}$ shown in the left panel of Fig.~\ref{fig:cleannoclean}, i.e., for the energy bin between 1.04 and 1.38 GeV, default data selection and default mask covering 3FGL sources. The left panel shows the case with foreground cleaning, while the right panel is for the uncleaned case. The comparison between the two panels indicates that large covariances are present in the case without foreground cleaning up to multipoles $\ell \sim 100$.}
\end{figure*}

\subsubsection{Foreground cleaning}
The Galactic diffuse emission is bright, especially at low energies, and 
generally displays an approximate symmetry around the disk of the Galaxy, 
leading to excess power at low multipoles, corresponding to large angular 
scales. The measured auto-APS calculated both with and without performing 
foreground cleaning is shown in the left panel of Fig.~\ref{fig:cleannoclean} 
(red dots and blue triangles, respectively) for a selected energy bin at low 
energy. The default mask and data set are used. The effect of foreground 
cleaning is dramatic at low multipoles, significantly reducing the measured 
$\overline{C_\ell}$ below $\ell\sim$50. On the other hand, our analysis only 
considers multipoles larger than $\ell=49$, where the effect of foreground 
cleaning is smaller, although still important enough to be non-negligible. 
Above $\ell\sim$150, however, it is clear its impact becomes subdominant, and 
the APS could be measured even without performed any cleaning. This is 
confirmed by the right panel of Fig.~\ref{fig:cleannoclean}, where the best-fit 
Poissonian auto-APS for the case with foreground cleaning and our signal 
region, i.e. $\ell$ between 49 and 706 (red circles, the same as in 
Fig.\ref{fig:autoCPvsE}), is compared to the best-fit $C_{\rm P}$ for the case 
without foreground cleaning but performing the fit only between $\ell=143$ 
and 706 (blue triangles), i.e. neglecting the first four bins in multipole 
inside our signal region. Errors at low energies are larger for the uncleaned
case than for the cleaned one. This is due to the fact that, at low energies, 
only the few $\overline{C_\ell}$ with $\ell\alt300$ play a role in the 
determination of the best-fit $C_{\rm P}$, since at larger $\ell$ the beam 
suppression is too strong. Therefore, cutting the signal region at $\ell=143$
means that the best-fit $C_{\rm P}$ is determined only by very few data. 
Nonetheless, the two cases are found to be in good agreement within their 
uncertainties at all energies. From this we can conclude that the foreground 
cleaning is effective even down to $\ell \sim 50$, therefore validating 
our choice for the signal region in multipole.

Also, from the left panel of Fig.~\ref{fig:cleannoclean}, it is clear that the
binned APS $\overline{C_\ell}$ without foreground cleaning is characterized by 
much larger error bars than with cleaning, at least for $\ell \alt 150$.
The reason for this can be understood by looking at the covariance matrix of 
the binned auto-APS: in Fig.~\ref{fig:cleannoclean} the errors on 
$\overline{C_{\ell}}$ are simply the square root of the diagonal elements of 
the covariance matrix, while the full covariance matrix is plotted in 
Fig.~\ref{fig:covariance_foreground}, for the data between 1.04 and 1.38 GeV, 
with (left panel) and without foreground cleaning (right panel). Each pixel in 
the panels corresponds to a pair of bins in multipole and its color provides 
the covariance between those two bins. We do not directly plot the covariance 
matrix but, instead, each element $\sigma_{i,j}$ is divided by the square root 
of the product of the corresponding diagonal elements 
$\sqrt{\sigma_{i,i} \sigma_{j,j}}$. The main difference between the two 
panels is at low multipoles, where the case without foreground cleaning is 
characterized by a large covariance among different bins. This large 
covariance causes the diagonal terms at $\ell \lesssim 30$ to correlate with
diagonal terms at higher multipoles. But multipoles $\ell \lesssim 30$ are 
characterized by larger $\overline{C_\ell}$ (and, thus, also larger errors) for 
the uncleaned data set than for the cleaned one. This translates into large 
error bars also around around $\ell \sim$50-100, for the case without 
foreground cleaning.

\begin{figure*}
\includegraphics[width=0.49\textwidth]{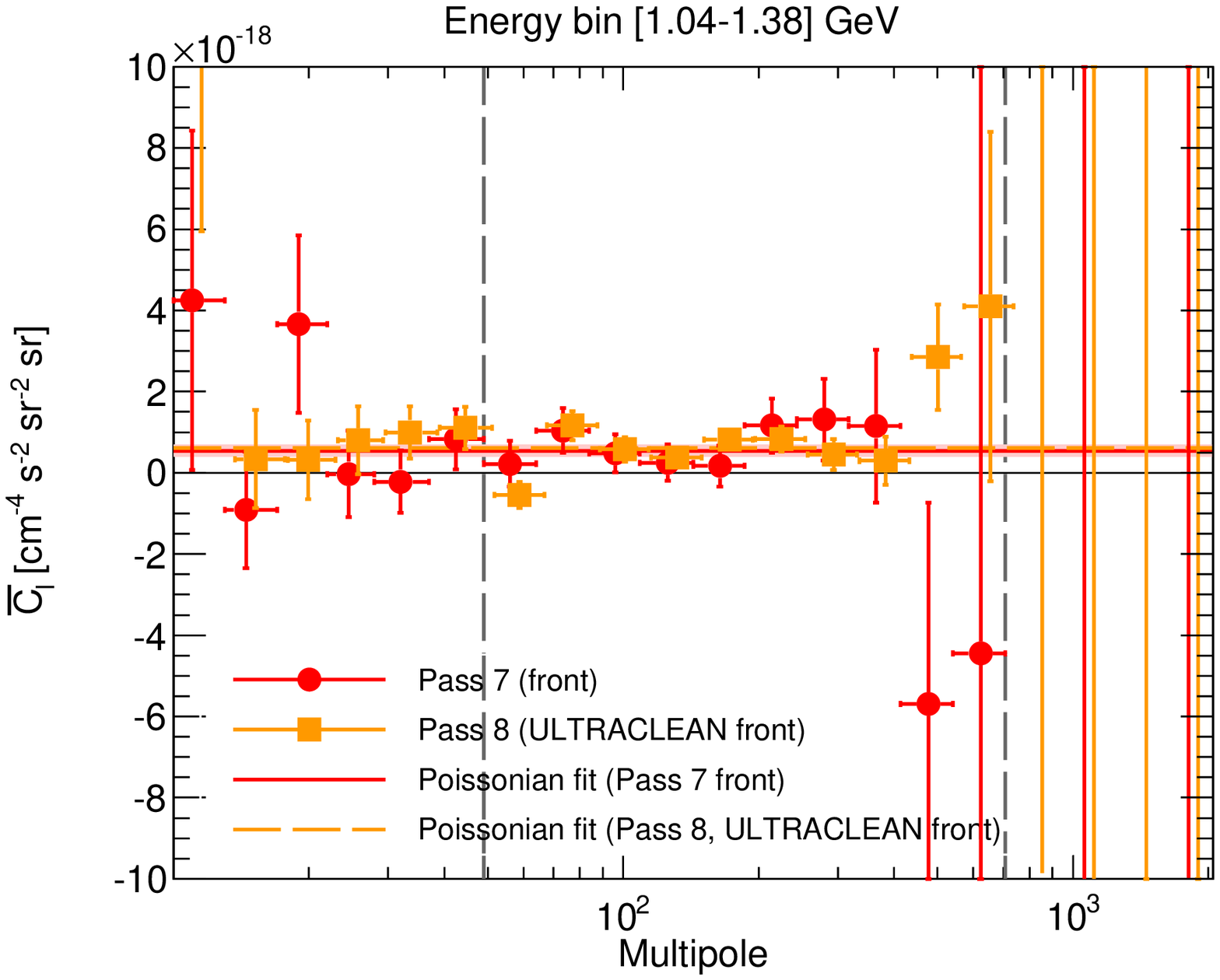}
\includegraphics[width=0.49\textwidth]{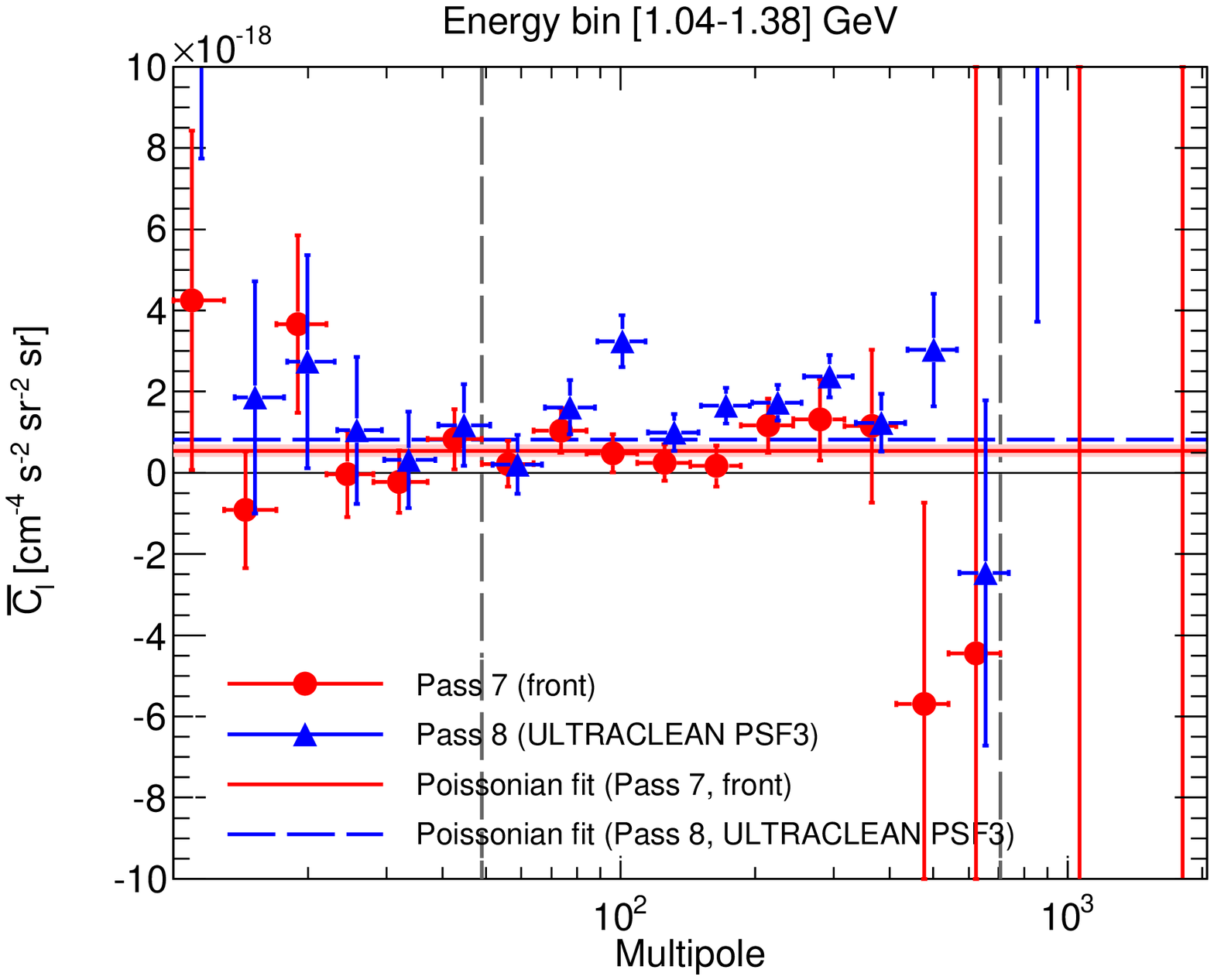}
\caption{\label{fig:p7p8} {\it Left:} Comparison of the auto-APS measurement in the energy bin between 1.04 and 1.38 GeV between the default Pass 7 data set (red circles) and the Pass 8 front event selection (orange squares). {\it Right:} same as the left panel but the comparison is between the Pass 7 data (red circles) and the Pass 8 PSF3 event selection (blue triangles). The solid red line marks the best-fit $C_{\rm P}$ for the default Pass 7 data set, with the pink band indicating its 68\% CL error. The dashed orange (blue) line gives the best-fit $C_{\rm P}$ for Pass 8 front (PSF3) in the left (right) panel. The vertical grey dashed lines mark the signal region between $\ell=49$ and 706. The default mask covering 3FGL sources is applied.}
\end{figure*}

Therefore, the introduction of the foreground cleaning reduces the intensity 
of the signal at $\ell \lesssim 50$ and it considerably removes the coupling 
between multipoles, leading to smaller and weakly correlated estimated errors. 
It also justifies the use of Eqs.~\ref{eqn:chi2_CP} and \ref{eq:logLmethod} for 
the determination of the Poissonian APS, since they are valid only under the 
hypothesis that covariances are negligible\footnote{Note that in 
Ref.~\cite{Ackermann:2012uf} the covariance between multipole bins was not 
discussed.}.

\begin{figure}
\includegraphics[width=0.49\textwidth]{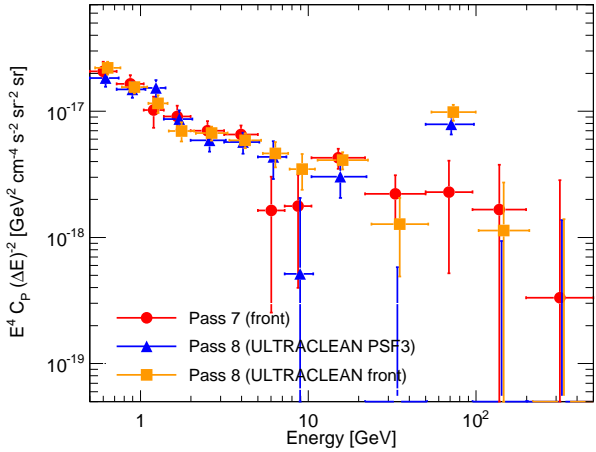}
\caption{\label{fig:CP_p7_p8} Poissonian auto-APS as a function of energy for the Pass 7 (red circles), Pass 8 front (orange squares) and Pass 8 PSF3 (blue triangles) data sets. The default mask covering 3FGL sources is applied.}
\end{figure}

\subsubsection{Data selection}
Next we consider the impact of our choice of data set. As described in
Sec.~\ref{sec:data}, our analysis is based on P7REP\_ULTRACLEAN\_V15 front 
events. For comparison, we now show the results for two different event 
selections using Pass 8 data, i.e., the most recent revision of the 
event-level {\it Fermi} LAT reconstruction analysis \cite{Atwood:2013rka}. In 
particular, we will use the P8R2\_ULTRACLEANVETO\_V6 class, designed to reduce 
the cosmic-ray contamination significantly. We consider separately two event 
selections, i.e. only Pass 8 front-converting events and the so-called PSF3 
events. PSF3 is a new selection available with Pass 8 data and it is 
characterized by an improved angular resolution. The effective area for the 
PSF3 events is roughly a factor of two smaller than that for the front events, 
since PSF3 represents the quartile of events with the best angular resolution, 
while the front events constitute approximately half the total events gathered 
by the LAT. The same analysis pipeline applied to the Pass 7 data is employed 
to the Pass 8 data, including foreground cleaning with the same Galactic 
diffuse model (refitted to the Pass 8 events outside the mask).

In Fig.~\ref{fig:p7p8} we compare the measured auto-APS in one energy bin 
between the default data set (i.e., Pass 7, denoted by red circles) and the 
two Pass 8 selections: Pass 8 front-converting events in the left panel 
(orange squares) and Pass 8 PSF3 in the right panel (blue triangles). The 
auto-APS is shown over the multipole range between $\ell=10$ and 2000. The two 
Pass 8 data sets are overall in good agreement with the default Pass 7 data 
set in the multipole range used for analysis, marked by the two grey vertical 
dashed lines in the figure. 

\begin{figure*}
\includegraphics[width=0.49\textwidth]{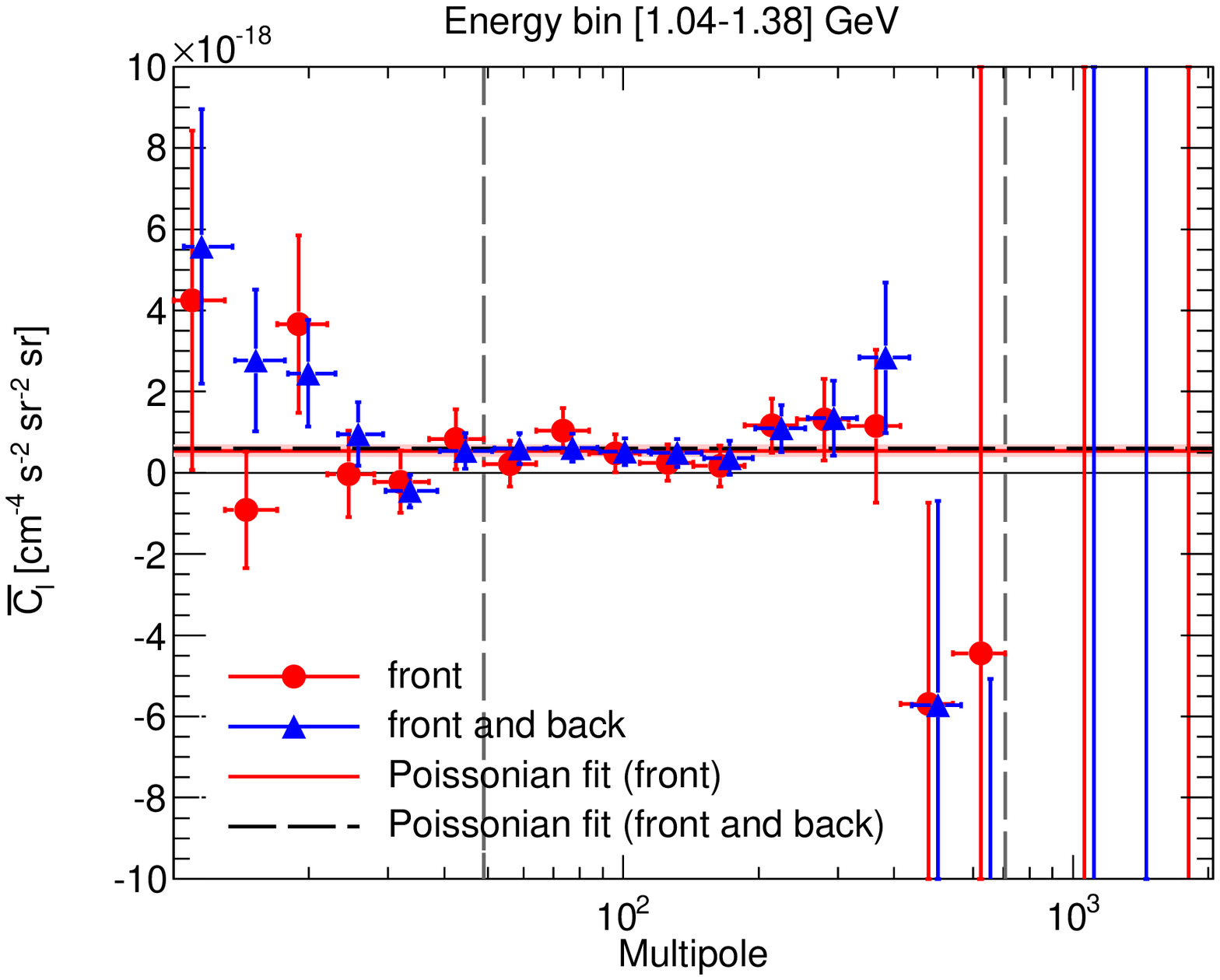}
\includegraphics[width=0.49\textwidth]{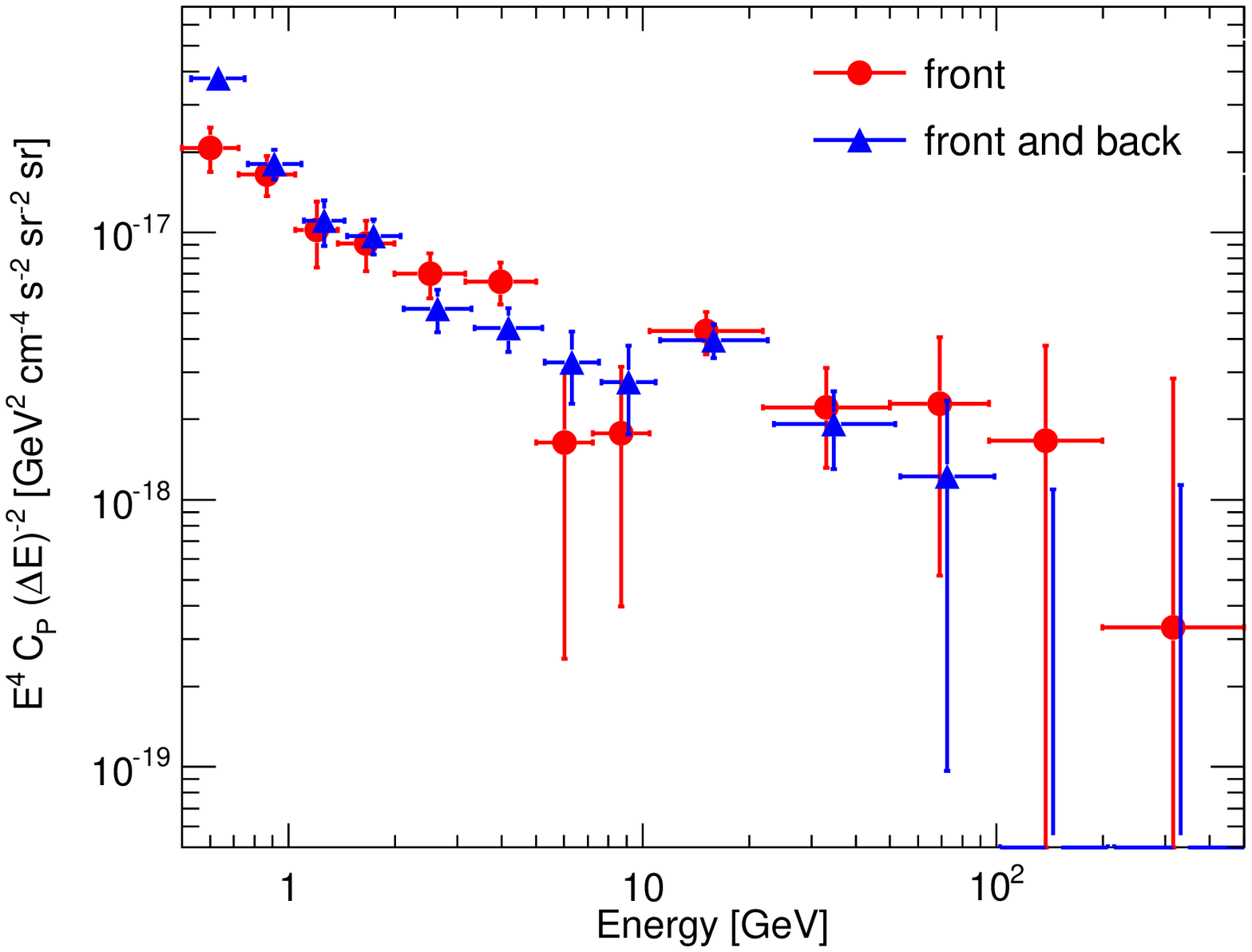}
\caption{\label{fig:frontback} {\it Left:} Comparison of the auto-APS in the energy bin between 1.04 and 1.38 GeV between the default data set which uses only front events (red circles) and the front+back data set (blue triangles). The solid red line marks the best-fit $C_{\rm P}$ for the default front-converting data, with the pink band indicating its 68\% CL error. The dashed blue line gives the best-fit $C_{\rm P}$ for the front+back data set. The vertical grey dashed lines mark the signal region between $\ell=49$ and 706. {\it Right:} Poissonian auto-APS as a function of energy for the front events (red circles) and the front+back ones (blue triangles). The default mask covering 3FGL sources is applied for both panels.}
\end{figure*}

In Fig.~\ref{fig:CP_p7_p8} we show the anisotropy energy spectra for the
three data sets discussed above. Their Poissonian auto-APS agree well within 
the measurement uncertainties in the various energy bins. The sharp drop in 
$C_{\rm P}$ around $\sim$7~GeV apparent in the Pass 7 data is less significant 
in the Pass 8 PSF3 data and absent in the Pass 8 front data, suggesting that 
the feature in the Pass 7 data may be the result of a statistical 
fluctuation. Also, with Pass 8, the auto-APS around 70 GeV has a larger value
than with Pass 7, although the difference is only at the 2$\sigma$ level and, 
thus, not very significant. We stress that this is only a qualitative 
comparison and a more thorough analysis of the Pass 8 data should be 
performed. With Pass 8, the measurement of the auto-APS and cross-APS is 
expected to improve in several ways, e.g. taking advantange of the new PSF 
classes (from PSF0 to PSF3), especially at low energies where the measurement 
uncertainties in the Pass 7 data are dominated by the suppression induced
by the beam window functions and (potentially) by the leaking from bright 
sources outside the mask (see Sec.\ref{sec:more_masks}). Also, new data
selections are available with Pass 8, characterized by different balances 
between effective area and cosmic-ray contamination. In fact, the improvement 
expected from using Pass 8 PSF3 or Pass 8 front data can already been seen in 
the reduction of the error bars for the blue triangles and orange squared in 
Fig.~\ref{fig:CP_p7_p8}, with respect to the red circles, especially at 
around 100 GeV. A detailed study with Pass 8 is beyond the scope of the 
present analysis and is left for future work.

We further investigate the impact of event selection by comparing the results 
obtained from the Pass 7 data using front data only (i.e., our default choice)
to the results obtained using both the front and back data. Including 
back-converting events in the analysis has the advantage of increasing the 
statistics by enlarging the effective area by a factor of $\sim$2. However, 
the average PSF for the front+back data set is poorer than for the front 
events alone, leading to a larger suppression due to the beam window function 
and to a stronger leakage outside the mask from bright point-like sources. In 
this comparison it should be kept in mind that the data sets are not 
independent since, by definition, the front+back data set contains all the
front-converting events. Also, it is important to note that due to the poorer 
PSF of the front+back data set, our source-masking scheme may not be 
sufficiently effective for that data set, particularly at low energies where 
the PSF is broadest (see also the discussion in Sec.~\ref{sec:more_masks}).

The left panel of Fig.~\ref{fig:frontback} shows the auto-APS 
$\overline{C_\ell}$ in a specific energy bin. Red circles refer to the Pass 7 
front data set and the blue triangles to the Pass 7 front+back one. The right 
panel indicates the Poissonian auto-APS as a function of energy, with the same 
color code. The measured $\overline{C_{\ell}}$ is in good agreement between the 
two data sets at all multipoles in our signal region. The same is true for the 
Poissonian $C_{\rm P}$, except in the lowest energy bin, where the 
front+back data yields a significantly higher $C_{\rm P}$. This discrepancy is 
consistent with the possibility that, for the front+back data set, the mask 
employed here (covering all sources in the 3FGL) is not big enough to get rid 
of the emission of the sources at low energies. Note that, also in this case, 
the significance of the dip at $\sim$7 GeV is strongly reduced.

\begin{figure*}
\includegraphics[width=0.49\textwidth]{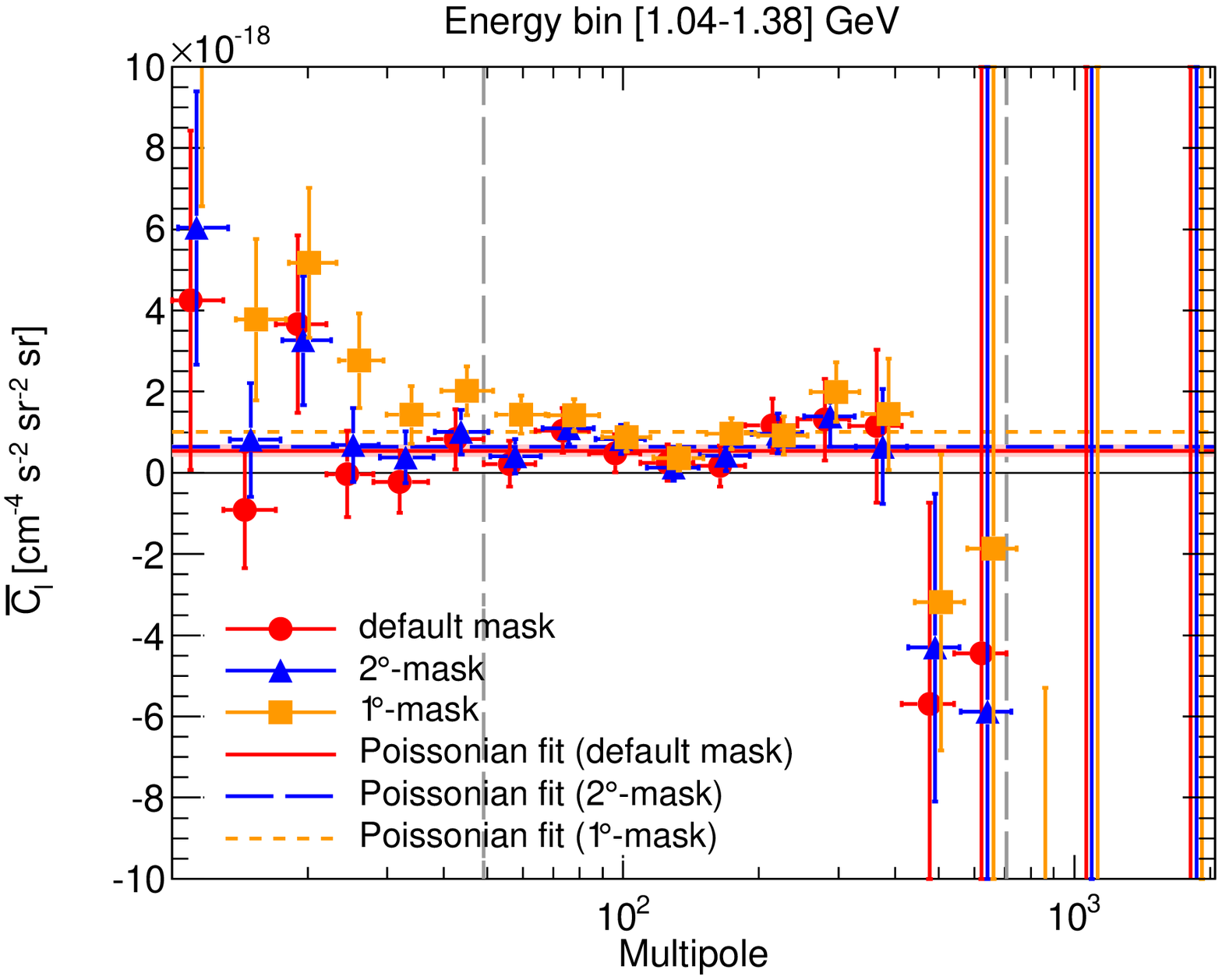}
\includegraphics[width=0.49\textwidth]{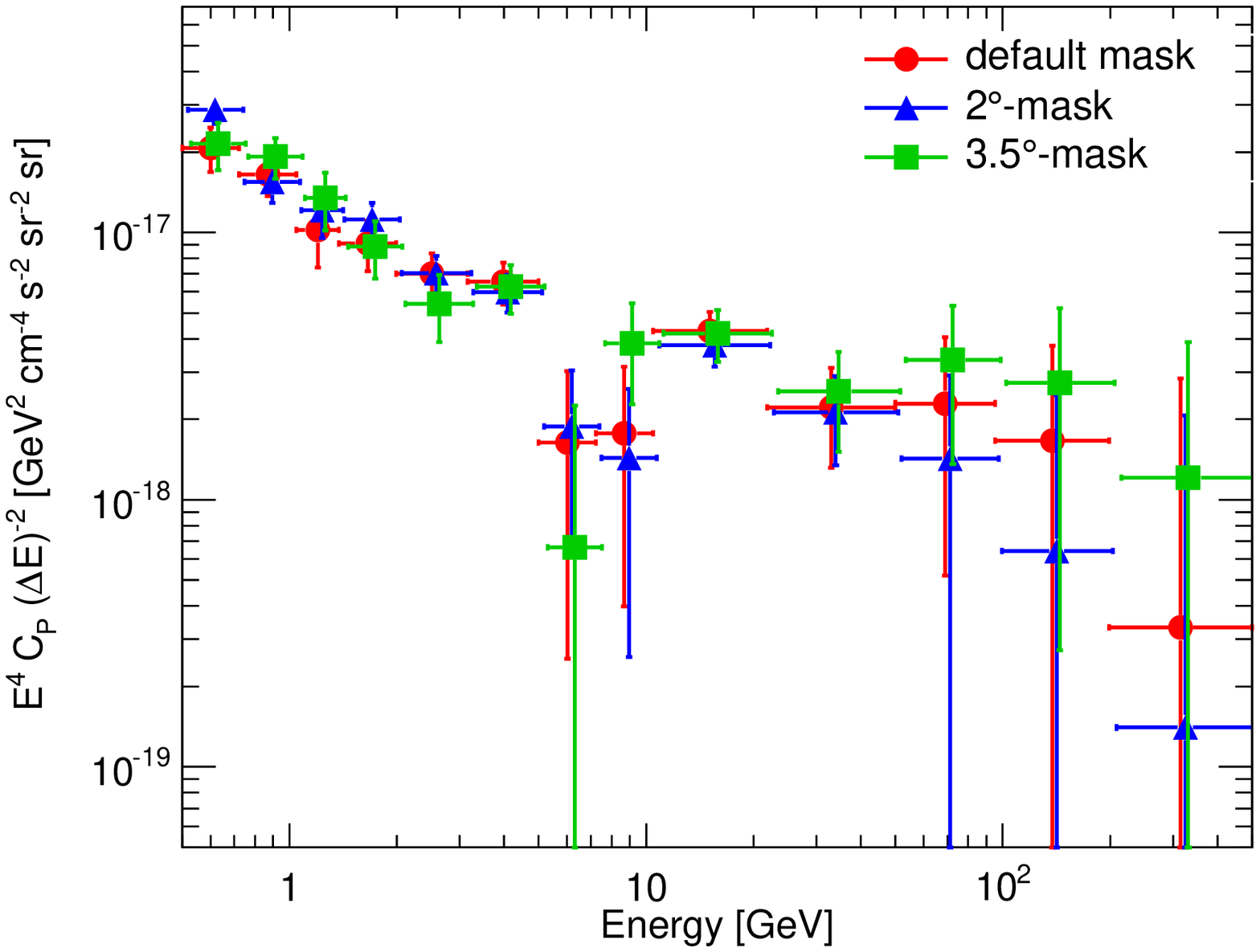}
\caption{\label{fig:1deg2deg} {\it Left:} Comparison of the auto-APS in the energy bin from 1.04 to 1.38 GeV among the case with the default mask covering the sources in 3FGL (red circles), the case with the $2^{\circ}$-mask (blue triangles) and the one with the $1^\circ$-mask (orange squares). The solid red line marks the best-fit $C_{\rm P}$ for the default mask, with the pink band indicating its 68\% CL error. The long-dashed blue line gives the best-fit $C_{\rm P}$ when the $2^\circ$-mask is employed and the short-dashed orange one for the case with the $1^\circ$-mask. The vertical grey dashed lines mark the signal region between $\ell=49$ and 706. {\it Right:} Poissonian auto-APS as a function of energy for the case with the default mask (red circles), the $2^\circ$-mask (blue triangles) and the 3.5$^\circ$-mask (green squares).}
\end{figure*}

\subsubsection{Mask around resolved sources}
\label{sec:more_masks}
We now investigate the effect of any possible leakage of emission outside the 
mask around the resolved sources. We recall that our default mask excludes 
(in addition to a latitude cut of $|b|<30^\circ$) a disk with a radius of 
3.5$^\circ$ around the 500 brightest sources in the 3FGL catalog, a disk with a
radius of 2.5$^\circ$ around the following 500 sources, one with a radius of
1.5$^\circ$ for the following 1000 sources and, finally, a disk with a region
with a 1.0$^\circ$-radius around all the remaining ones. This is what we refer 
to as our {\it default} mask when covering the sources in 3FGL. However, in 
order to validate our choice, we consider four additional masks. They are 
defined as follows:
\begin{itemize}
\item {\it 4$^\circ$-mask} excludes a disk with a radius of 4$^\circ$ around the
500 brightest sources in the 3FGL catalog, a disk with a radius of 3$^\circ$
for the following 500 sources, one with a radius of 1.5$^\circ$ for the
next 1000 sources and one with a radius of 1$^\circ$ for the remaining ones;
\item {\it 3.5$^\circ$-mask} excludes a disk with a radius of 3.5$^\circ$ around 
the 500 brightest sources in the 3FGL catalog, a disk with a radius of 
2.5$^\circ$ for the following 500 sources, one with a radius of 2.0$^\circ$ for 
the next 1000 sources and a disk with a radius of 1.5$^\circ$ for the remaining 
ones; \\
\item {\it 2$^\circ$-mask} covers a disk with a radius of 2.0$^\circ$ around the 
500 brightest sources and a disk with a radius of 1.0$^\circ$ around the 
remaining sources; \\
\item {\it 1$^\circ$-mask} excludes a disk with a radius of 1.0$^\circ$ around 
each source.
\end{itemize}

Our default mask is located between the 2$^\circ$-mask and the 3.5$^\circ$-mask, 
in terms of masked area. 
The specific details of the masks considered are not the result
of an a-priori analysis and, thus, they are somewhat subjective. However, our 
goal is to identify a reasonable mask that is as small as possible without 
suffering from leakage from point-like sources. As proved in the following, our 
default mask provides a suitable choice.

Above few GeV, where the PSF is narrower, we expect the $1^\circ$-mask to be 
sufficient to exclude the emission of the sources detected in 3FGL. However, 
at low energies some leakage may appear. Results are summarized in 
Fig.~\ref{fig:1deg2deg}. The left panel shows the measured auto-APS in the e
nergy bin between 1.04 and 1.38 Gev, for the 1.0$^\circ$-mask (orange squares), 
for the 2.0$^\circ$-mask (blue triangles) and for the default one (red circles). 
It is clear that there is a significant contamination due to power leakage 
outside the 1.0$^\circ$-mask, especially at $\ell<50$, but up to $\ell\sim$80. 
The other two more aggresive masks give consistent results in this energy bin. 
In the right panel, we plot the anisotropy energy spectrum for the 
$2^\circ$-masks (blue triangle), for the default one (red circle) and for the 
3.5$^\circ$-mask (green squares). While, at high energies, the three cases 
yield consistent results, the $2^\circ$-mask shows still an excess of power in 
the first energy bin. On the other hand, results for the 3.5$^\circ$-mask are 
consistent with our default mask. The anisotropy energy spectrum for the 
4$^\circ$-mask (not shown in Fig.~\ref{fig:1deg2deg} for clarity) is also 
consistent with the default case. This validates our choice of the latter as 
our fiducial mask when dealing with 3FGL sources\footnote{We also test an 
additional mask that covers exactly the same region of sky as our default mask 
for 3FGL sources but it also masks the region around Loop-I and the Galactic 
Lobes. The best-fit $C_{\rm P}$ with this more aggressive cut are compatible 
with the default Poissonian $C_{\rm P}$ in Fig.~\ref{fig:autoCPvsE}, within 
their statistical errors.}.

A similar validation is performed on the mask covering the sources in 2FGL.
We find that cutting a 1$^\circ$-disk around all 2FGL sources leads to some
power excess at low energies. However, extending the mask by covering a disk 
with a radius of $2^\circ$ for all sources is enough to get rid of the leakage 
and there is no need of more aggressive masks as for the case of 3FGL sources.
This is probably due to the fact that, when masking 2FGL sources, the measured 
power spectra are intrinsically larger than when masking sources in 3FGL 
(see Sec.~\ref{sec:results_analysis}). Thus, the contimation from leakage 
has {\it relatively} a minor impact.

\subsection{Effect of the gamma-ray emission from the Sun}
Steady gamma-ray emission from the Sun was detected in the \emph{Fermi} LAT 
data in Ref.~\cite{Abdo:2011xn} from 0.1 to 10 GeV. Later, 
Ref.~\cite{Ng:2015gya} extended the detection up to 100 GeV, also establishing 
that the flux varies with time and anti-correlates with Solar activity. Gamma 
rays are produced from the interaction of cosmic rays with the Solar atmosphere 
\cite{Seckel:1991ffa}, as well as from Inverse-Compton (IC) scattering of 
cosmic-ray electrons and positrons with Solar photons \cite{Orlando:2006zs,
Moskalenko:2006ta,Orlando:2013pza}.

The emission is quite difficult to see by eye because, even if quite 
significant, it is spread over the path followed by the Sun in the sky, i.e.
the ecliptic. However, it may still induce some feature in the auto- and 
cross-APS. We test this possibility by masking the region of $1.5^\circ$ 
above and below the ecliptic. The auto- and cross-APS obtained after having
introduced this additional mask are compatible with our default case within
their uncertainty. Thus, we conclude that the effect of the Sun on the 
measured anisotropies is negligible\footnote{The high-energy emission of the 
Moon peaked at about 200 MeV, with a similar intensity than the Sun 
\cite{Cerutti:2016gts}. However, at higher energies, it has an energy spectrum 
that is steeper than that of the Sun. Therefore, above 500 MeV, the effect of 
the Moon on the APS measurement is expected to be negligible.}. 

\subsection{Comparison with previous measurement}
\label{sec:old_measurement}
We conclude this section by comparing our new measurement to the previous 
(indeed, the first) anisotropy measurement from Ref.~\cite{Ackermann:2012uf}.  
Our current analysis includes many improvements with respect to the original
one, both from the perspective of the data set (as we now use Pass 7 
Reprocessed events and IRFs, compared to the Pass 6 events used in 
Ref.~\cite{Ackermann:2012uf}) and in terms of analysis method, including an
improved calculation of the noise term $C_{\rm N}$, the deconvolution of the 
mask (performed now with {\sc PolSpice}) and a MC-validated procedure to 
bin the auto-APS in multipoles and to estimate its error. The improved data 
set also allows us to measure the auto-APS with better precision over a larger 
multipole range covering the window between $\ell=49$ and 706, while the 
analysis in Ref.~\cite{Ackermann:2012uf} was restricted to $\ell=155-504$. We
also extend the energy range, spanning the interval between 500 MeV and 500
GeV, compared to the original 1--50 GeV range. Moreover, we use an improved 
diffuse model for foreground cleaning, compared to what was available at the 
time of Ref.~\cite{Ackermann:2012uf}.

\begin{figure}
\includegraphics[width=0.49\textwidth]{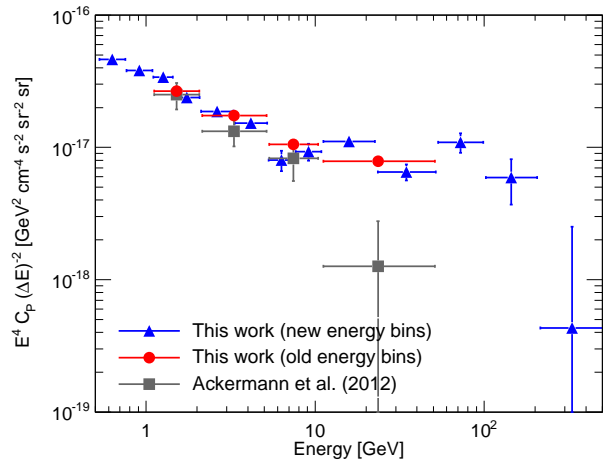}
\caption{\label{fig:oldnewcp} Poissonian auto-APS as a function of energy for the default data set in this analysis, with 13 energy bins (blue triangles) and with the 4 energy bins used in Ref.~\cite{Ackermann:2012uf} (red circles). Note that data are obtained using the mask that covers a $2^\circ$-circle around each source in 2FGL. The grey squares denote the measurement from Ref.\cite{Ackermann:2012uf} using the same mask.}
\end{figure}

In Fig.~\ref{fig:oldnewcp} we compare the anisotropy energy spectrum reported 
in Ref.~\cite{Ackermann:2012uf} for the mask covering the sources in 2FGL 
(grey squares) to our new measurement calculated for the same mask but with
our new default data set. We report our results for the 13 energy bins used 
in this work (blue triangles) and we also compute the auto-APS in the same 4 
energy bins used in Ref.~\cite{Ackermann:2012uf} (red circles). 
While there is a slight trend toward a higher $C_{\rm P}$ in our current 
measurement compared to the original one, we find good consistency with 
Ref.~\cite{Ackermann:2012uf}. The only exception is the highest energy bin of 
the original analysis, which is lower than the current measurement and 
inconsistent at about 3$\sigma$. Many factors may lead to the small systematic 
increase of the new $C_{\rm P}$ in the first 3 bins and to the larger difference
in the last energy bin. However, we attribute this trend primarily to the 
way the data are binned in multipole and to the way the Poissonian fit 
$C_{\rm P}$ is determined. As discussed in Sec.~\ref{sec:clvalidation}, in this
analysis we follow a different procedure with respect to the original analysis
in Ref.~\cite{Ackermann:2012uf}, after having verified that the latter can 
lead to a downward bias of both the $\overline{C_\ell}$ and the best-fit 
$C_{\rm P}$.

We end by noting that a concern about the auto-APS in 
Ref.~\cite{Ackermann:2012uf} being somewhat underestimated was raised already 
in Ref.~\cite{Chang:2013ada}. However, in that case it was claimed that the 
correct anisotropy should have been a factor 5-6 larger than the measured one 
for each energy. In the light of the present analysis, this is true only for 
the highest energy bin, while for the others the difference is only of 
20-30\%, and not significant within error bars.

\section{Interpretation in terms of source populations}
\label{sec:interpretation}
In this section we provide a phenomenological interpretation of our 
measurement in terms of different populations of unresolved sources. The main 
observables that we consider are the results of the Poissonian fits to the 
auto- and cross-APS, i.e., the anisotropy energy spectrum.

\begin{figure*}
\includegraphics[width=0.49\textwidth]{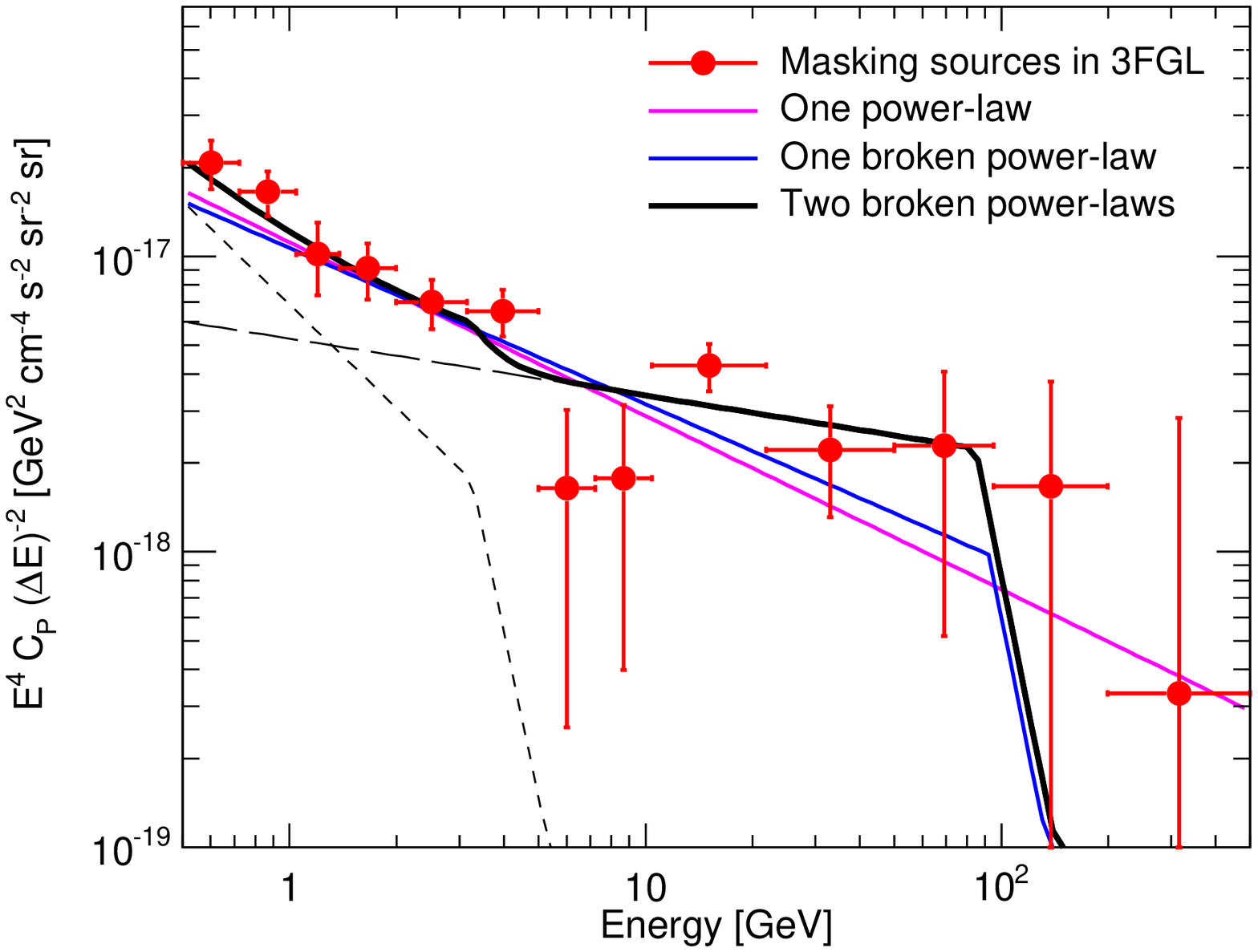}
\includegraphics[width=0.49\textwidth]{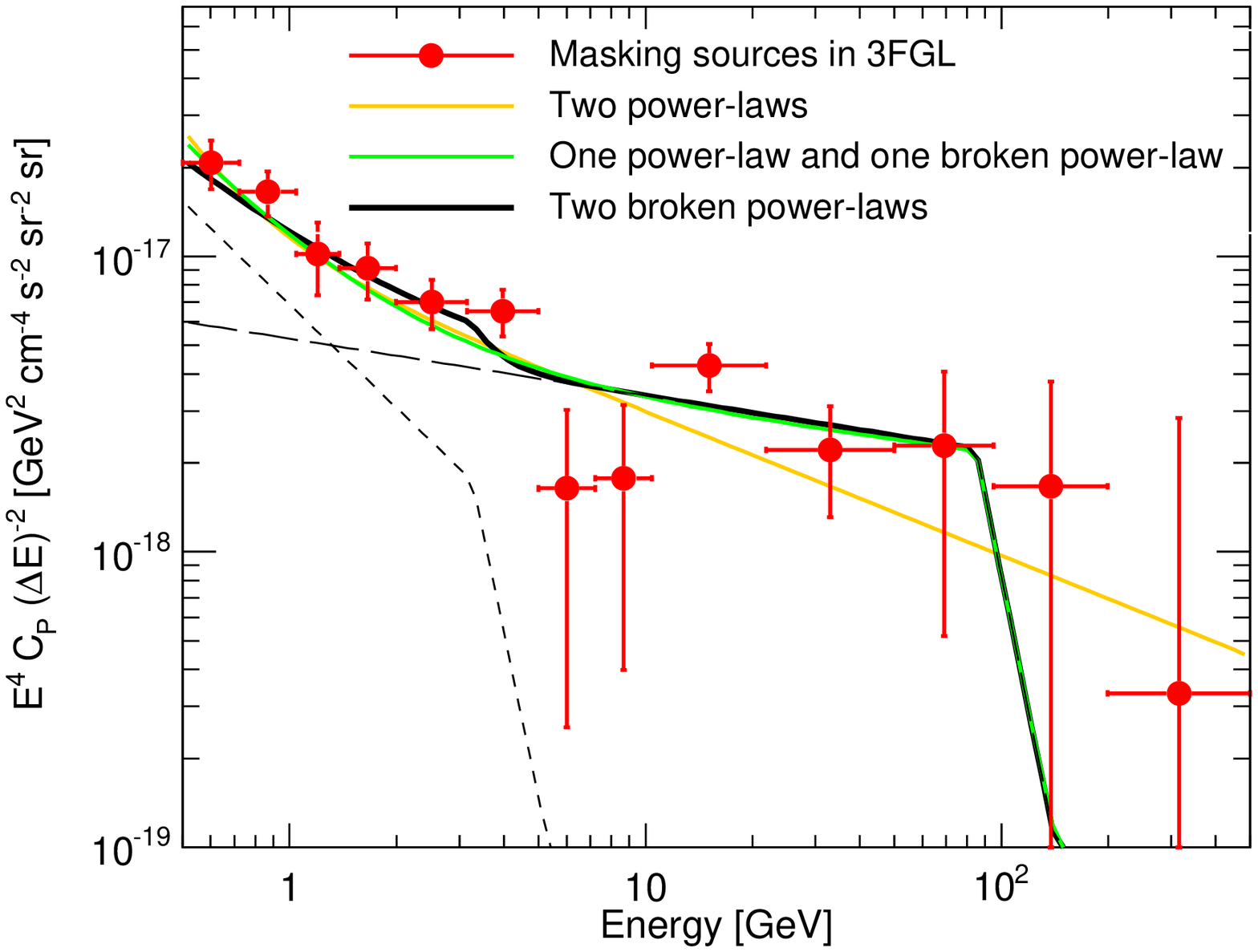}
\caption{\label{fig:CP_auto_fit} Anisotropy energy spectrum for the default data set masking the sources in 3FGL (red circles). The different lines correspond to the best-fit models to the measured auto- and cross-APS with one or two populations of unresolved sources. The solid magenta line and the solid blue one (left panel) are for one population emitting as a power law or as a broken power law, respectively. The solid yellow line (right panel) is for two populations with power-law energy spectra. The solid green line (right panel) shows the best-fit in the case of one population emitting as a power law and another as a broken power law. Finally, the thicker solid black line (present in both panels) represents the case of two populations emitting as broken power laws. This is the scenario that best fits the data. In this case, the contribution of the two components are shown as short-dashed and long-dashed black lines.}
\end{figure*}

\begin{table*}
\caption{\label{tab:popfit} Best-fit values for the parameters defining the populations assumed to describe the measured auto- and cross-APS. See the text for the definition of the parameters. The normalizations ($A$, $A_1$ and $A_2$) are measured in $\mbox{cm}^{-2} \mbox{s}^{-1} \mbox{sr}^{-1}$ and the energy breaks ($E_b$, $E_{b,1}$ and $E_{b,2}$) are measured in GeV. Errors are given at 68\% CL. The table also indicates the number of degrees of freedom $N_{\rm dof}$ (i.e., the number of fitted data points minus the number of free parameters), the $\chi^2$ of the best-fit solution, the $\chi^2$ of the best-fit point per degree of freedom and the corresponding $p$-value.}
\begin{ruledtabular}
\begin{tabular}{c|c|c|c|c|c|c|c||c|c|c|c}
\multicolumn{8}{c||}{} & $N_{\rm dof}$ & $\chi^2$ & $\chi^2 / N_{\rm dof}$ & $p$-value \\
\hline
\multicolumn{8}{c||}{\bf One power law} & \multicolumn{4}{c}{} \\
\hline
$\log_{10}(A)$ & $\alpha$ & & & & & & & & & & \\
\hline
$-8.48_{-0.01}^{+0.01}$ & $2.29_{-0.01}^{+0.02}$ & & & & & & & 89 & 135.31 & 1.52 & 0.001 \\
\hline

\hline
\multicolumn{8}{c||}{\bf One broken power law} & \multicolumn{4}{c}{} \\
\hline
$\log_{10}(A)$ & $\alpha$ & $\beta$ & $E_{b}$ & & & & & & & & \\
\hline
$-8.49_{-0.01}^{+0.01}$ & $2.26_{-0.02}^{+0.02}$ & $>3.74$ & $92.20_{-16.66}^{+16.02}$ & & & & & 87 & 118.57 & 1.36 & 0.010 \\
 & & at 68\% CL & & & & & & & & & \\
\hline

\hline
\multicolumn{8}{c||}{\bf Two power laws} & \multicolumn{4}{c}{} \\
\hline
$\log_{10}(A_1)$ & $\alpha_1$ & $\log_{10}(A_2)$ & $\alpha_2$ & & & & & & & & \\
\hline
$-8.52_{-0.04}^{+0.03}$ & $2.24_{-0.05}^{+0.03}$ & $-8.81_{-0.22}^{+0.14}$ & $3.27_{-0.45}^{+0.78}$ & & & & & 87 & 127.60 & 1.47 & 0.003 \\
\hline

\hline
\multicolumn{8}{c||}{\bf Two broken power laws} & \multicolumn{4}{c}{} \\
\hline
$\log_{10}(A_1)$ & $\alpha_1$ & $\beta_1$ & $E_{b,1}$ & $\log_{10}(A_2)$ & $\alpha_2$ & $\beta_2$ & $E_{b,2}$ & & & & \\
\hline
$-8.58_{-0.05}^{+0.04}$ & $2.58_{-0.12}^{+0.18}$ & $>3.49$ & $3.26_{-0.64}^{+1.05}$ & $-8.64_{-0.05}^{+0.04}$ & $2.10_{-0.05}^{+0.05}$ & $>3.86$ & $84.65_{-15.71}^{+10.28}$ & 83 & 91.58 & 1.10 & 0.240 \\
 & & at 68\% CL & & & & at 68\% CL & & & & & \\
\hline

\hline
\multicolumn{8}{c||}{\bf One power law and one broken power law} & \multicolumn{4}{c}{} \\
\hline
$\log_{10}(A_1)$ & $\alpha_1$ & $\log_{10}(A_2)$ & $\alpha_2$ & $\beta_2$ & $E_{b,2}$ & & & & & & \\
\hline
$-8.56_{-0.09}^{+0.06}$ & $2.71_{0.18}^{0.26}$ & $-8.68_{-0.13}^{+0.10}$ & $2.08_{-0.45}^{+0.88}$ & $>$3.89 & $84.79_{16.13}^{10.60}$ & & & 85 & 98.86 & 1.16 & 0.140 \\
 & & & & at 68\% CL & & & & & & & \\
\end{tabular}
\end{ruledtabular}
\end{table*}

We assume the sources responsible for the signal to be point-like and
unclustered \cite{Ando:2006mt,Cuoco:2012yf,DiMauro:2014wha,Cuoco:2015rfa}, and 
that they give rise only to a Poissonian auto- and cross-APS. We also assume 
each population to be characterized by a common intensity energy spectrum 
$F_a(E)$. The index $a$ runs over the number of source populations contributing 
to the signal. The contribution of the $a$-th source population to the 
auto-APS will be proportional to $F_a(E_i)^2$, while it will be proportional to 
$F_a(E_i)F_a(E_j)$ in the case of the cross-APS. Our choice of interpreting the 
auto- and cross-APS data in a phenomenological way is motivated by the desire 
to be model-independent. Alternative interpretations in terms of 
physically-motivated models of astrophysical gamma-ray emitters are on-going. 
We start by considering one single source population with a power-law 
spectrum, i.e. 
\begin{equation}
F(E) = A \, \left(\frac{E}{E_0} \right)^{-\alpha},
\end{equation}
with $E_0=1$ GeV. We fit both the best-fit Poissonian auto- and cross-APS 
taken from Tab.~\ref{tab:cp_3FGL}, i.e. for the mask around 3FGL 
sources. This scan and the following ones are performed with {\sc MultiNest} 
3.9 \cite{Feroz:2007kg,Feroz:2008xx,Feroz:2013hea} with 20000 live-points and 
a tolerance of $10^{-4}$ in order to provide a good sampling of the likelihood. 
The prior probability is chosen to be flat for all free parameters, between 
-15.0 and -5.0 for $\log_{10}(A)$ (and all the normalizations, measured in 
$\mbox{cm}^{-2} \mbox{s}^{-1} \mbox{sr}^{-1}$), between 0.0 and 5.0 for the 
slopes and between 5.0 GeV and 500.0 GeV for the energy breaks (see later). 
The best-fit solution is reported in Tab.~\ref{tab:popfit} and is represented 
by a solid magenta line in the left panel of Fig.~\ref{fig:CP_auto_fit}. The 
best fit has a $\chi^2$ per degree of freedom which is 1.52, corresponding to 
a $p$-value of 0.001.

Alternatively, we also consider a broken power law parametrized as follows:
\begin{equation}
F(E) = \left\{
\begin{array}{cl}
A \, (E/E_0)^{-\alpha} & \mbox{if } E \geq E_b \\
A \, (E_0/E_b)^{+\alpha-\beta} (E/E_0)^{-\beta} & \mbox{otherwise}
\end{array} \right. .
\label{eqn:broken_PL}
\end{equation}
In this case, the best-fit is reported in Tab.~\ref{tab:popfit} and shown as 
a solid blue line in the left panel of Fig.~\ref{fig:CP_auto_fit}. Its 
$\chi^2$ per degree of freedom is 1.36 with a $p$-value of 0.01.

Then, we allow for the possibility of two independent populations, starting
with the case of two power laws. The best-fit values are reported in 
Tab.~\ref{tab:popfit} and the model is represented by the solid yellow line
in the right panel of Fig.~\ref{fig:CP_auto_fit}: one population explains the 
data points below a few GeV and another one reproduces the data at higher 
energies. The best fit has a $\chi^2$ per degree of freedom of 1.47, 
corresponding to a $p$-value of 0.003.

We also consider the possibility of two broken power laws. With a $\chi^2$ per
degree of freedom of 1.10, it represents the best description to the data.
The model is shown as a solid black line in both panels of 
Fig.~\ref{fig:CP_auto_fit}: one broken power law reproduces the data at low 
energies (short-dashed black line) and the other one at higher energies 
(long-dashed black line). The best-fit solution for the cross-APS is shown in 
Figs.~\ref{fig:crosscorrcp1} and \ref{fig:crosscorrcp2} in 
Appendix~\ref{sec:appendix}.

Finally, we also test the hypothesis of one population emitting as a power 
law and one as a broken power law. This interpretation is characterized by
a $\chi^2$ per degree of freedom of 1.16 (with a $p$-value of 0.14) and it is 
represented in the right panel of Fig.~\ref{fig:CP_auto_fit} by a solid green 
line. The fit is slightly worse than the case with two broken power laws, 
especially around 3-4 GeV.

The difference between the $\chi^2$ of the best-fit solution for a model with 
one population and the same quantity for the model with two populations 
can be used as a TS to determine whether we can exclude the 
one-source-population scenario. From the values of the $\chi^2$ in 
Tab.~\ref{tab:popfit}, the exclusion is at 95\% CL in all cases\footnote{In
the comparison between one and two populations of sources, if the number of 
additional degrees of freedom is 2 (as for the case in which the sources in
the second population emit as power laws), then the 95\% CL exclusion 
corresponds to a $\Delta\chi^2$ of 5.99. On the other hand, if the second
population emits as a broken power law, the number of additional degrees of
freedom is 4 and the 95\% CL limit is obtained for a $\Delta\chi^2$ of 9.49.}. 
We can test how the different interpretations perform also in a Bayesian 
framework. Indeed, we can define the Bayes Factor $B$ as the ratio of the 
so-called ``evidence'' for two competing models (given the data) and it can 
be used to discriminate between them. In particular, with a $\ln B=0.5$, there 
is not a preference between the interpretation with one or two power laws 
(according to the Jeffrey's scale \cite{Trotta:2008qt}), while, with a 
$\ln B = 3.1$, there is a weak preference for the two-broken-power-laws 
solution over the one-broken-power-law one.

\section{Simulating the gamma-ray emission induced by Dark Matter}
\label{sec:simulations}
From this section onwards we focus our attention on the DM-induced gamma-ray 
emission: we first summarize how we simulate this component, and then we 
analyze our mock gamma-ray sky maps by computing their auto- and cross-APS. 
This will constitute our prediction for the APS associated with DM that will 
be compared to the measured auto- and cross-APS presented in the previous 
sections.

The simulated DM signal needs to account for all DM structures (halos and 
subhalos) around us, including the emission generated in the halo of our own 
Milky Way (MW). We divide the DM auto- and cross-APS into different components 
that are discussed separately in the following subsections (from 
Sec.~\ref{sec:EG-MSII} to Sec.~\ref{sec:GAL-AQ}). We follow closely the 
semi-analytical procedure developed in Ref.~\cite{Fornasa:2012gu}, i.e., we 
directly employ catalogs of DM (sub)halos from $N$-body simulations and 
complement them with well-motivated recipes to account for the emission of DM 
halos and subhalos below the mass resolution of the simulations. As in 
Ref.~\cite{Fornasa:2012gu}, we make use of the Millennium-II and Aquarius 
simulations, from the Virgo Consortium \cite{BoylanKolchin:2009nc,
Navarro:2008kc,Springel:2008cc} to simulate the Galactic and extragalactic 
components, respectively.

We take particular care in estimating the systematic uncertainties associated
with the DM auto- and cross-APS. In particular, each time we introduce a 
quantity that is not well determined, we consider a reasonable range of 
variability for it and determine its impact on the final DM signal.

We separately consider gamma-ray emission produced by annihilations or decays
of DM particles, organizing our predictions in the form of {\sc HEALPix}
maps with {\ttfamily Nside}=512. This corresponds to 3145728 pixels and
an angular size of approximately $0.115^\circ$. The order is lower than the one
used in the data analysis (see Sec.~\ref{sec:analysis}). However, note that 
we will only compare our predictions for the DM signal to the measured 
spectra below $\ell=706$, i.e. for angular scales larger than 0.25$^\circ$.

The gamma-ray flux (in units of cm$^{-2}$s$^{-1}$) produced by DM annihilations 
in the $i$-th energy bin and coming from the pixel centered towards direction 
$\mathbf{n}_j$ can be written as follows:
\begin{eqnarray}
\Phi(E_i,\mathbf{n}_j) & = & 
\frac{\langle \sigma_{\rm ann}v \rangle}{8\pi m_\chi^2} 
\int_{\Delta\Omega_j} d\Omega_{\mathbf{n}} \, \int_{0.0}^{2.15} dz \, 
\frac{c(1+z)^3}{H(z)} \rho^2_\chi(z,\mathbf{n}_j) \nonumber \\
& & \int_{E_i}^{E_{i+1}} dE_\gamma \frac{dN^{\rm ann}_\gamma(E_\gamma(1+z))}{dE} 
\nonumber \\
& & \exp[-\tau_{\rm EBL}(E_\gamma(1+z))],
\label{eqn:annihilation}
\end{eqnarray}
where the integration $d\Omega_{\mathbf{n}}$ extends over the pixel centered on
$\mathbf{n}_j$. For redshifts higher than $\sim$2, the evolution of the DM 
density field, combined with the larger comoving volume probed attenuates the 
signal to a negligible level. The interaction with the Extragalactic Background 
Light (EBL) additionally reduces the emisson from large 
redshifts\footnote{Refs.~\cite{Profumo:2009uf,Fornasa:2012gu,Zavala:2009zr} 
show that more than 90\% of the emission is produced below $z=2$.}. The EBL 
attenuation is modeled in Eq.~\ref{eqn:annihilation} by the factor 
$\exp(-\tau_{\rm EBL}(E_\gamma(1+z)))$, which is taken from 
Ref.~\cite{Dominguez:2010bv}. The thermal average of the cross section times
the relative velocity and the mass of the DM particles are expressed by 
$\langle \sigma_{\rm ann}v \rangle$ and $m_\chi$, while $c$ and $H(z)$ are the 
speed of light and the Hubble parameter. The function 
$\rho_{\chi}(z,\mathbf{n}_j)$ denotes the DM density at redshift $z$ towards 
the direction $\mathbf{n}_j$. The photon yield $dN^{\rm ann}_\gamma/dE$ determines 
the number of photons produced per annihilation. Different mechanisms of 
gamma-ray production contribute to $dN^{\rm ann}_\gamma/dE$. We specify which 
contribution is included when we discuss the different components of the 
total DM signal.

In the case of decaying DM, the expected gamma-ray emission is written as
follows:
\begin{eqnarray}
\Phi(E_i,\mathbf{n}_j) & = & \frac{1}{4\pi \, m_\chi \tau} 
\int_{\Delta\Omega_j} d\Omega_{\mathbf{n}} \, \int_{0.0}^{2.15} dz \, \frac{c}{H(z)} 
\rho_\chi(z,\mathbf{n}_j) \nonumber \\
& & \int_{E_i}^{E_{i+1}} dE_\gamma \frac{dN^{\rm decay}_\gamma(E_\gamma(1+z))}{dE} 
\nonumber \\
& & \exp[-\tau_{\rm EBL}(E_\gamma(1+z))].
\label{eqn:decay}
\end{eqnarray}
Contrary to Eq.~\ref{eqn:annihilation}, Eq.~\ref{eqn:decay} depends linearly 
on the DM density and it features the DM decay lifetime $\tau$, instead of
$\langle \sigma_{\rm ann}v \rangle$.

\subsection{Extragalactic resolved main halos and subhalos (EG-MSII)}
\label{sec:EG-MSII}
We label halos and subhalos as ``resolved''  if they are present in the 
Millennium-II catalog \cite{BoylanKolchin:2009nc} with a mass larger than 
$6.89 \times 10^8 M_\odot/h$. We employ the same procedure used in
Ref.~\cite{Fornasa:2012gu} to fill the region below $z=2.15$ with copies of 
the original Millennium-II simulation box (see Refs.~\cite{Zavala:2009zr} and
\cite{Fornasa:2012gu} for further details). This provides a possible 
realization of the distribution of resolved extragalactic DM halos and 
subhalos along the past light-cone. The sky map of their emission (i.e., what 
we call ``EG-MSII'' in the following) is obtained by determining, for each 
pixel in the map, which DM structures fall inside the angular area of the 
pixel (completely or partially, according to their size) and by summing 
together their gamma-ray flux. In the case of annihilating DM, the 
annihilation rate of a DM halo or subhalo is computed from $V_{\rm max}$ and 
$r_{\rm max}$ (i.e., the maximal circular velocity and the distance from the 
center of the halo where this occurs) and by assuming that all DM structures 
are characterized by a Navarro-Frenk-White (NFW) density profile 
\cite{Navarro:1996gj}. A different choice of density profile would affect the 
overall intensity of the DM-induced emission (by a factor as large as 10, 
between extreme cases such as the Moore \cite{Moore:1999gc} and Burkert 
\cite{Burkert:1995yz} profiles \cite{Profumo:2009uf,Zavala:2009zr}) but it 
would not affect the shape of the auto- and cross-APS since only a relatively 
small number of the halos in EG-MSII appear as extended, i.e., covering more 
than one pixel in our sky map. In the case of decaying DM, the decay rate of a 
halo depends only on its mass, which is independent of the choice of the
density profile.

The Millennium-II and Aquarius $N$-body simulations were performed assuming 
cosmological parameters favored by WMAP 1. Adopting the most recent values in 
agreement with the Planck mission \cite{Ade:2015xua} could modify the 
clustering and abundance of DM structures in the simulations. However, it was 
shown that the increased matter density $\Omega_m$ and the decreased linear 
fluctuation amplitude $\sigma_8$ (with respect to WMAP 1) have compensating 
effects \cite{Guo:2012fy} and, therefore, we neglect the dependence of our 
results on the cosmological parameters (see also Ref.~\cite{Ludlow:2013vxa}). 

As in Refs.~\cite{Zavala:2009zr,Fornasa:2012gu}, the way the copies of the 
Millennium-II simulation box are positioned around the observer is a random 
process. Changing their orientation modifies the distribution of resolved DM 
halos and subhalos, affecting the shape of the auto- and cross-APS for 
EG-MSII. Ref.~\cite{Fornasa:2012gu} showed that this is just a 10\% effect 
that can be neglected in comparison with other sources of uncertainty that 
will be mentioned later.

For EG-MSII, the photon yield $dN^{\rm ann}_\gamma/dE$ includes the primary 
gamma-ray emission (taken from Ref.~\cite{Cirelli:2010xx}), i.e. hadronization
of particles produced in the annihilation, final state radiation and internal
bremsstrahlung. We also consider secondary emission, namely the photons 
up-scattered by the IC of DM-induced electrons onto the cosmic microwave 
background (see Ref.~\cite{Fornasa:2012gu} for details). In the case of 
decaying DM, the photon yield is determined as 
$dN^{\rm decay}_i(E)/dE = dN^{\rm ann}_i(2E)/dE$, where $i$ stands for either 
photons or electrons. 

\subsection{Extragalactic unresolved main halos (EG-UNRESMain)}
\label{sec:EG-UNRESMain}
The emission of unresolved main halos (i.e. with a mass smaller than 
$6.89 \times 10^8 M_\odot/h$) all the way down to the mass of the smallest 
self-bound halos $M_{\rm min}$, is referred to as ``EG-UNRESMain''. $M_{\rm min}$ 
depends on the nature of the DM particle and on its interactions with normal 
matter but, at least within the context of supersymmetric WIMPs, values 
between $10^{-12} M_\odot/h$ and $1 \mbox{ } M_\odot/h$ are reasonable, while 
$M_{\rm min}=10^{-6} M_\odot/h$ has become a popular benchmark 
\cite{Profumo:2006bv,Bringmann:2009vf}.

In order to estimate EG-UNRESMain, we assume that unresolved main halos share 
the same clustering properties of main halos with a mass between 
$1.39 \times 10^8 M_\odot/h$ and $6.89 \times 10^8 M_\odot/h$. These main halos 
are just below our threshold of resolved DM structures. They are barely 
resolved in the Millennium-II simulations (with a number of particles between
20 and 100) and they populate a regime in mass where the linear halo bias 
reaches a plateau \cite{BoylanKolchin:2009nc}. Assuming that this remains true 
even below $1.39 \times 10^8 M_\odot/h$, it is reasonable to think that 
unresolved main halos will have a similar linear halo bias as the main halos 
with a mass between $1.39 \times 10^8 M_\odot/h$ and $6.89 \times 10^8 M_\odot/h$. 
Therefore, their emission can be accounted for simply by artificially 
enhancing the emission of the main halos between $1.39 \times 10^8 M_\odot/h$ 
and $6.89 \times 10^8 M_\odot/h$. Such an enhancement is implemented as follows:
\begin{equation}
\int_{M_{\rm min}}^{1.39 \times 10^8 M_\odot/h} dM \frac{dn_h(M)}{dM} L_h^i(M),
\label{eqn:unresolved_main_halos}
\end{equation}
where $dn_h/dM$ is the main-halo mass function and $L_h^i(M)$ is the gamma-ray
flux produced by a single main halo with mass $M$. The index $i$ stands for 
``ann'' or ``decay'', accordingly. The mass function is assumed to follow a 
power law in mass, down to $M_{\rm min}$. Its normalization and slope are fixed 
by fitting the abundance of main halos above the mass resolution of 
Millennium-II, separately in the different snapshots of the simulation, in 
order to reproduce the redshift dependence of $dn_h/dM$.

Accounting for the contribution of main halos below the mass resolution of
Millennium-II by enhancing the emission of the main halos with a mass between
$1.39 \times 10^8 M_\odot/h$ and $6.89 \times 10^8 M_\odot/h$ is equivalent to
assuming that the two populations of DM halos share the same spatial 
distribution. Following the formalism introduced in Ref.~\cite{Ando:2005xg} 
this is also equivalent to assuming that they are characterized by the same 
2-halo term.

In the case of annihilating DM, the computation of $L_h^{\rm ann}$ in 
Eq.~\ref{eqn:unresolved_main_halos} is very sensivite to the concentration of 
the halo $c(M,z)$. Contrary to Ref.~\cite{Fornasa:2012gu}, we consider only 
the concentration model described in Ref.~\cite{Sanchez-Conde:2013yxa}: this 
model allows for $c(M,z)$ to flatten as $M$ decreases. Consequently, it 
agrees with the results of the recent $N$-body simulations in 
Refs.~\cite{Anderhalden:2013wd,Ishiyama:2014uoa} and is a more accurate model
than the ones from Ref.~\cite{Springel:2008zz,Gao:2011rf}. At $z=0$ and for 
$M_{\rm min}=10^{-6} M_\odot/h$, Eq.~\ref{eqn:unresolved_main_halos} is 24 times 
larger than the emission of all main DM halos with a mass between 
$1.39 \times 10^8 M_\odot/h$ and $6.89 \times 10^8 M_\odot/h$. For 
$M_{\rm min}=10^{-12} M_\odot/h$ ($M_{\rm min}=1 \mbox { } M_\odot/h$), the number is 
28 (15).

For decaying DM, $L_h^{\rm decay}(M,z)=M$ and the enhancement (with respect to
the emission of DM halos with a mass between $1.39 \times 10^8 M_\odot/h$ and 
$6.89 \times 10^8 M_\odot/h$) is 6.7, 6.5 and 5.8 for $M_{\rm min}=10^{-12}$, 
$10^{-6}$ and 1 $M_\odot/h$, respectively.

\subsection{Extragalactic unresolved subhalos}
\label{sec:unresolved_subhalos}
In order to account for the emission of the subhalos of unresolved halos, we
modify Eq.~\ref{eqn:unresolved_main_halos} as follows:
\begin{equation}
\int_{M_{\rm min}}^{1.39 \times 10^8 M_\odot/h} dM \frac{dn_h(M)}{dM} L_h^i(M) B^i(M,z).
\label{eqn:unresolved_subhalos}
\end{equation}
The additional term $B^i(M,z)$ is the so-called boost factor, describing how
much the emission of main halos increases when the contribution of their 
subhalos is included. $B^{\rm decay}$ is equal to 1 for all DM halos and 
redshifts, while the value of $B^{\rm ann}(M,z)$ is quite uncertain. We consider 
two scenarios that we believe bracket the current uncertainty on 
$B^{\rm ann}(M,z)$:
\begin{itemize}
\item LOW scenario: this prescription is the same as in 
Ref.~\cite{Fornasa:2012gu} and it is motivated by the parametrization 
(performed in Ref.~\cite{Kamionkowski:2010mi} and extended in 
Ref.~\cite{SanchezConde:2011ap}) of the probability $P(\rho,r)$ of finding a 
value of the DM density between $\rho$ and $\rho+d\rho$ in the data of the 
Via Lactea II $N$-body simulation. For this scenario and at $z=0$, the 
overall enhancement in Eq.~\ref{eqn:unresolved_subhalos} (with respect to the 
emission of DM main halos with a mass between $1.39 \times 10^8 M_\odot/h$ and 
$6.89 \times 10^8 M_\odot/h$) is 160, 88 and 23 for $M_{\rm min}=10^{-12}$, 
$10^{-6}$ and 1 $M_\odot/h$, respectively. \\

\item HIGH scenario: this is the same as the LOW recipe below $10^8 M_\odot/h$
while it predicts a boost factor 5 times larger above that mass. Indeed, 
recent results favor boost factors that are larger than the ones of the LOW 
framework. Ref.~\cite{Bartels:2015uba} developed a semi-analytic model that 
accounts for the mass accretion rate of subhalos in larger host halos. The 
model also describes the effect of tidal stripping and dynamical friction 
experienced by subhalos. Including these effects increases the boost factor 
by a factor of 2-5, relative to the LOW scenario. A similar increase is 
expected when one accounts for the fact that the concentration of subhalos 
changes according to the distance of the subhalo from the center of the host 
halo \cite{Moline:2016pbm}. Finally, Refs.~\cite{Zavala:2013bha,
Zavala:2013lia} developed a new statistical method to describe the behaviour 
of DM particles in collapsed structures, based on the modelling of the 
so-called Particle Phase-Space Average Density (P$^2$SAD). 
Ref.~\cite{Zavala:2015ura} demonstrated that, when computed in the case of DM 
subhalos, the P$^2$SAD is universal over subhalos of halos with a mass that 
goes from that of dwarf galaxies to that of galaxy clusters. Employing a 
reasonable parametrisation of the P$^2$SAD, Ref.~\cite{Zavala:2015ura} found 
boost factors that are as much as a factor of 5 larger than the LOW case, at 
least for massive DM halos.

The boost factor predicted in Ref.~\cite{Zavala:2015ura} for DM halos with a 
mass below $10^8 M_\odot/h$ is, however, moderate (see their 
Fig.~5\footnote{Note that the boost factor in Ref.~\cite{Zavala:2015ura} is 
defined in a different way than in Ref.~\cite{Fornasa:2012gu}. Whenever we 
take some information from Ref.~\cite{Zavala:2015ura}, we translate it into 
the same definition used in Ref.~\cite{Fornasa:2012gu}.}). Thus, when we 
compute the emission of unresolved main halos from $10^{-6} M_\odot/h$ to 
$10^8 M_\odot/h$ as in Eq.~\ref{eqn:unresolved_subhalos}, and we include the 
boost factor of Ref.~\cite{Zavala:2015ura}, we find a very similar result to 
that of the LOW scenario. However, predictions are different for massive DM 
halos: for objects with a mass larger than $10^8 M_\odot/h$ a boost factor 5 
times larger than for the LOW case is viable. We assume that such an increment 
is the same for all masses above $10^8 M_\odot/h$ and we do not concern 
ourselves with which mechanism (or combinations of mechanisms) is responsible 
for it among the ones mentioned above, since those studies agree on an 
increase of this magnitude. For the case with $M_{\rm min}=10^{-12} M_\odot/h$ 
($M_{\rm min}=1 \mbox{ } M_\odot/h$), the HIGH boost factor is defined to be a 
factor 8.4 (2.1) larger than the LOW one (see Fig. 5 of 
Ref.~\cite{Zavala:2013lia}\footnote{Fig. 5 of Ref.~\cite{Zavala:2013lia} 
refers to the boost factor of MW-like DM halo. Assuming that similar results 
apply for all DM halos with a mass larger than $10^8 M_\odot/h$ is therefore an 
approximation.}).
\end{itemize}

As in the case of resolved structures (EG-MSII), the emission of unresolved 
halos and subhalos is computed including the primary gamma-ray emission and 
that resulting from the IC scattering off the cosmic microwave background. 
The total extragalactic signal is defined as the sum of EG-MSII and 
EG-UNRESMain, boosted for the emission of unresolved subhalos. We refer to the 
total emission as ``EG-LOW'' and ``EG-HIGH'', depending on the subhalo boost 
factor scheme employed.

\subsection{The smooth halo of the Milky Way (GAL-MWsmooth)}
\label{sec:GAL-MWsmooth}
As in Ref.~\cite{Fornasa:2012gu}, the emission of the smooth halo of the MW
(called ``GAL-MWsmooth'') is modeled by assuming that the MW halo follows an
Einasto profile \cite{Einasto:1965}:
\begin{equation}
\log \left( \frac{\rho}{\rho_s} \right) = - \frac{2}{\alpha} 
\left[ \left( \frac{r}{r_s} \right)^\alpha - 1 \right].
\end{equation}
The parameters in the above equation that provide the best fit to the data of 
the highest resolution halo, Aq-A-1, in the Aquarius simulation 
\cite{Navarro:2008kc,Springel:2008cc} are: 
$\rho_s = 7.46 \times 10^{15} h^2 M_\odot/\mbox{Mpc}^3$, $r_s=11.05 \mbox{ kpc}/h$ 
and $\alpha=0.170$. This corresponds to a total MW halo mass of 
$1.34 \times 10^{12} M_\odot/h$, defined as the amount of DM contained in a 
sphere with an average density of 200 times the critical density of the 
Universe. Observationally, the mass of the MW DM halo remains uncertain: 
Fig. 1 of Ref.~\cite{Wang:2015ala} shows how different methods (including, 
e.g., MW mass modeling, dynamics of different tracers and the study of the 
orbits of Andromeda and the MW) suggest values that go from 
$5.0 \times 10^{11} M_\odot$ to $2.0 \times 10^{12} M_\odot$. Halo Aq-A-1 
described above is on the higher end of this range. In order to account for 
the uncertainty on the MW mass, we assume that $\rho_s$ can vary from its 
nominal value of $7.46 \times 10^{15} h^2 M_\odot/\mbox{Mpc}^3$ down to 
$1.87 \times 10^{10} h^2 M_\odot/\mbox{Mpc}^3$. The latter corresponds to a MW 
mass that is 1/4 of the value of Aq-A-1\footnote{Since the observer is located 
approximately at a distance of $R_0$=8.5 kpc from the center of the MW, the 
best-fit Einasto profile to Aq-A-1 corresponds to a local DM density of 0.45 
GeV/cm$^3$. An uncertainty of a factor 4 on $\rho_s$ would generate a 
variability of the same size on the local DM density.}. Note that the 
intensity of the DM-induced gamma-ray emission is proportional to $\rho_s^2$ 
and to $\rho_s$ for an annihilating and decaying DM candidate, respectively. On 
the other hand, the auto- and cross-APS of GAL-MWsmooth are proportional to 
$\rho_s^4$ for annihilation-induced gamma rays and to $\rho_s^2$ for decaying 
DM. Thus, the uncertainty on the MW halo mass is a major systematic for the 
predicted signal we are interested in.

\begin{figure*}
\includegraphics[width=0.49\textwidth]{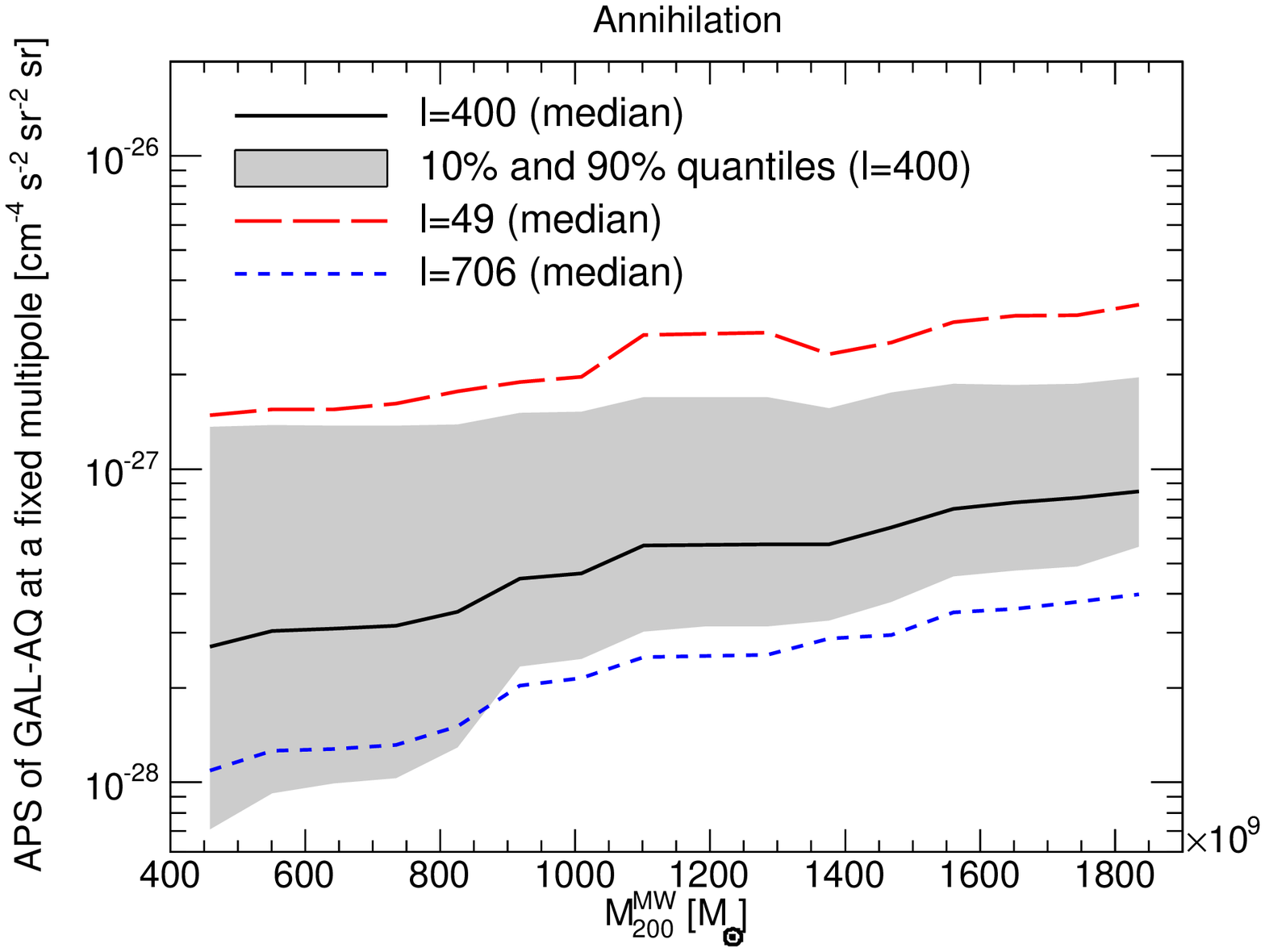}
\includegraphics[width=0.49\textwidth]{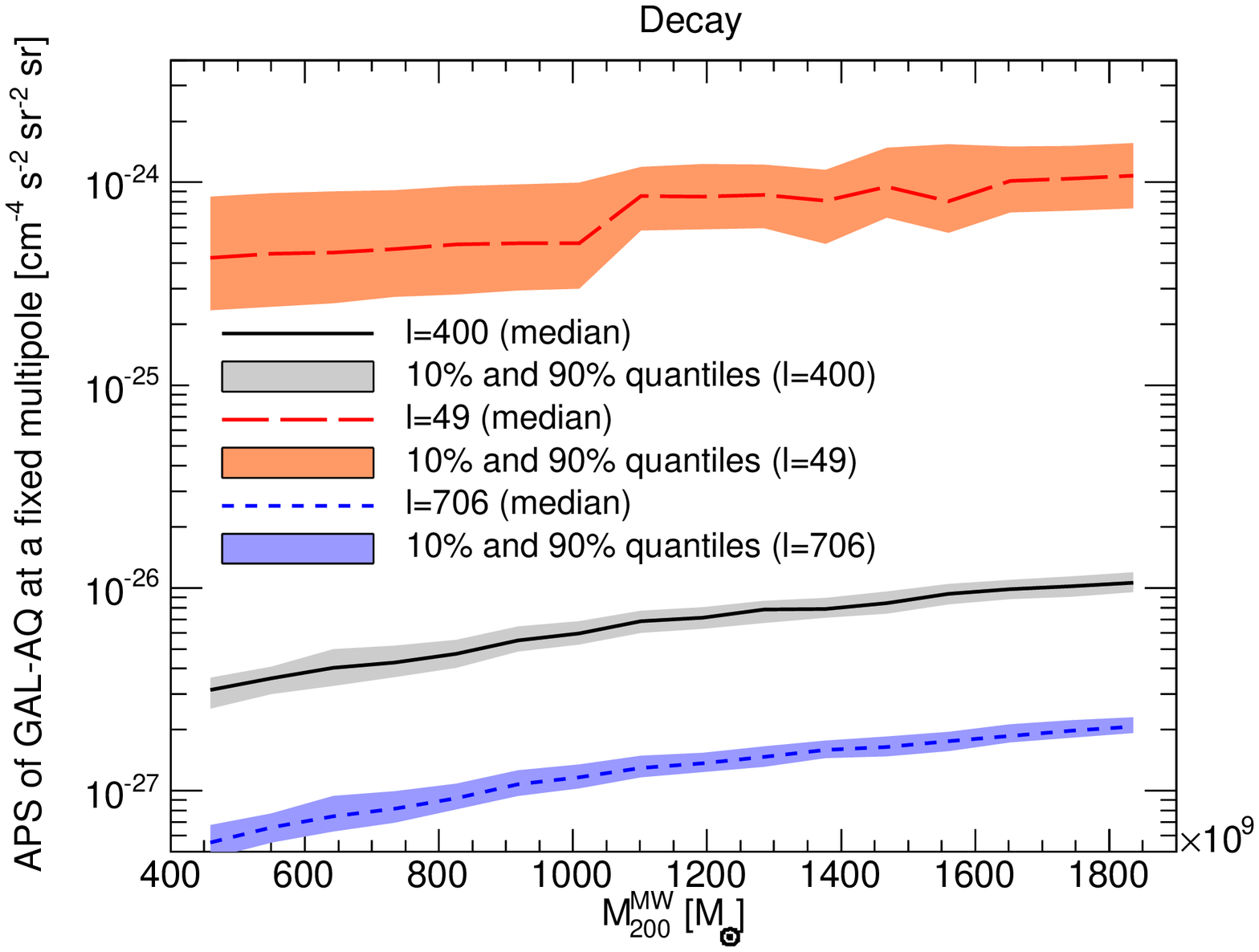}
\caption{Dependence of the APS of the GAL-AQ component on the MW mass. {\it Left:} The lines show the auto-APS at a fixed multipole ($\ell=49$ for the long-dashed red line, $\ell=400$ for the solid black one and $\ell=706$ for the short-dashed blue one) as a function of the mass of the DM halo of the MW, in our simulation of GAL-AQ described in the text. The auto-APS is computed between 0.5 and 0.72 GeV, for $m_\chi=2.203$ TeV with a thermal annihilation cross section $\langle \sigma_{\rm ann}v \rangle = 3 \times 10^{-26} \mbox{cm}^3\mbox{s}^{-1}$ and for annihilations into $b\bar{b}$. For each value of the MW mass, 100 realizations of GAL-AQ are computed for different positions of the observer. The lines refer to the median of the distribution of the corresponding auto-APS, while the grey band denotes the variability between the 10\% and the 90\% quantiles of the distribution. The band is present only for the case with $\ell=400$ for clarity. {\it Right}: The same as in the left panel but for decaying DM. The auto-APS is computed in the same energy bin, for the same $m_\chi$ and a decay lifetime of $2 \times 10^{27}$s.}
\label{fig:MW_mass}
\end{figure*}

Ref.~\cite{Navarro:2008kc} also showed that a NFW profile provides a 
reasonable fit to the Aq-A-1 data. We do not consider this alternative here, 
since the difference compared to the Einasto profile described above would be 
evident only below $\sim 30^\circ$ from the center of the MW, i.e. in a region 
located inside the mask considered in the data analysis (see 
Sec.~\ref{sec:masks} and Refs.~\cite{Fornasa:2012gu} and 
\cite{Bertone:2008xr}).

In the case of GAL-MWsmooth, the photon yield is computed including primary
emission and secondary emission from IC. The latter is computed using a full
model for the interstellar radiation field of the MW, as described in 
Ref.~\cite{Fornasa:2012gu}, and not just the IC scattering off the cosmic 
microwave background. We also include the hadronic emission, produced in the 
interactions of DM-induced protons with the interstellar medium (see 
Ref.~\cite{Fornasa:2012gu}).

\subsection{The subhalos of the Milky Way (GAL-AQ)}
\label{sec:GAL-AQ}
The last component to be considered is called GAL-AQ and it accounts for the
emission of the subhalos of the DM halo of the MW. We derive this component
from the subhalo catalog produced in the Aquarius $N$-body simulation. 
Structures with a mass larger than $1.71 \times 10^5 M_\odot$ are treated as 
``resolved''. We place the observer at a distance of $R_0=8.5$ kpc from the 
center of the MW and we compute the sky map of the emission of resolved 
Galactic subhalos by identifying which subhalos fall within each pixel of the 
map and summing up their gamma-ray flux. We neglect the contribution of 
unresolved subhalos (i.e., with a mass smaller than $1.71 \times 10^5 M_\odot$), 
as they do not contribute to the auto- and cross-APS as argued in 
Refs.\cite{Fornasa:2012gu} and \cite{Ando:2009fp}\footnote{However, unresolved 
Galactic subhalos are expected to contribute to the intensity of the 
DM-induced emission. This should be kept in mind when, in 
Fig.~\ref{fig:intensity}, we compute our predictions for the DM-induced 
gamma-ray intensity.}.

As in Ref.~\cite{Fornasa:2012gu}, only the primary gamma-ray emission is
considered when computing GAL-AQ.

Depending on the exact position of the observer on the sphere with radius $R_0$ 
and centered on the Galactic Center, the distribution of resolved subhalos 
changes, and so does the intensity of GAL-AQ and its auto- and cross-APS. We 
estimate this variability by producing 100 realizations of GAL-AQ, changing,
each time, the position of the observer on the sphere. We compute the auto-
and cross-APS for each realization\footnote{Here and for all the sky maps 
simulating the DM-induced emission, the auto- and cross-APS are computed on
the masked gamma-ray sky with the {\ttfamily anafast} routine of {\sc HEALPix} 
after having subtracted the monopole and dipole contributions.} and note that, 
for annihilating DM, the 10\% quantile of the distribution of the auto-APS 
(at $\ell=400$) among the 100 realizations is a factor $\sim$1.5 below the 
median, while the 90\% quantile is a factor $\sim$2.3 above. See the grey 
band in the left panel of Fig.~\ref{fig:MW_mass}. These numbers are 2.1 and 
5.6 (1.4 and 2.2) at $\ell=49$ ($\ell=706$). We suspect that the distribution 
gets more peaked (i.e. less variable) at large multipoles because it becomes 
more sensitive to the inner structure of DM subhalos (which is constant among 
the realizations), instead of their distribution in the sky. In the case of 
decaying DM, the 10\% and 90\% quantiles are always less than a factor 1.5 
away from the median (see the bands in the right panel of 
Fig.~\ref{fig:MW_mass}). The variability induced by changing the position of 
the observer is an important component in the total uncertainty of our DM 
predictions and it will be considered in the following sections.

When discussing GAL-MWsmooth (Sec.~\ref{sec:GAL-MWsmooth}), we considered the
effect of allowing the MW mass to decrease by a factor 4 with respect to
the nominal value of Aq-A-1. This has an impact also on GAL-AQ as the number 
of subhalos in a DM structure is found to be proportional to the mass of the 
host halo \cite{BoylanKolchin:2009an,Klypin:2010qw}. If we define $k$ as the 
fraction by which we decrease the MW mass, we consider 16 values of $k$, from 
0.0 to 0.25. For each $k$, we randomly remove a fraction $k$ of the subhalos 
in the Aquarius catalog, to simulate a lighter MW DM halo. For each value of 
$k$, we produce 100 realizations of GAL-AQ for different positions of the
observer. We compute the auto- and cross-APS for each of the realizations. In 
Fig.~\ref{fig:MW_mass}, the lines show the median of the distribution of the 
auto-APS of GAL-AQ (for a fixed multipole) as a function of $k$ (and, thus, as 
a function of the MW mass). The left panel is for annihilating DM and the 
right one for decaying DM. The solid black line is for the auto-APS at 
$\ell=400$, while the long-dashed red (short-dashed blue) one is for $\ell=49$ 
($\ell=706$). The coloured band (when present) denotes the scatter between the 
10\% and the 90\% quantiles in the distribution among the 100 realizations. 
For annihilating DM (left panel), the band becomes larger as the map is 
populated by less and less subhalos and, therefore, it depends more and more 
on their distribution. The variability induced by our partial knowledge of the 
MW mass is another important source of uncertainty that will be considered in 
the following sections.

For some values of the DM mass, annihilation cross section and decay lifetime, 
the gamma-ray flux of some DM subhalos in GAL-AQ may exceed the \emph{Fermi} 
LAT source sensitivity threshold. These DM subhalos would appear as resolved 
sources in the sky and they would be included in the 3FGL catalog. Since the 
auto- and cross-APS are measured masking the sources in 3FGL, DM subhalos that 
are bright enough to be detected should be neglected when simulating GAL-AQ. 
Being very bright, they may be responsible for a significant fraction of the 
auto- and cross-APS of GAL-AQ. Thus, neglecting them may affect significantly 
our predictions for GAL-AQ, as noticed in 
Ref.~\cite{Lange:2014ura}\footnote{One should also check that none of the DM 
halos or subhalos in EG-MSII are bright enough to be detected individually. We 
do not perform such a test because, even if some DM structures were to be 
removed, this would hardly affect the prediction for the auto- and cross-APS 
of EG-LOW and EG-HIGH.}. In order to test this, we define the so-called 
particle physics factors $\Phi_{\rm PP}^{\rm ann}$ and $\Phi_{\rm PP}^{\rm decay}$, 
which gathers all the terms in Eqs.~\ref{eqn:annihilation} and \ref{eqn:decay} 
that do not depend on the DM distribution. More precisely:
\begin{equation}
\Phi_{\rm PP}^{\rm ann} = \frac{\langle \sigma_{\rm ann} v \rangle}{2 m_\chi^2} 
\int_{E_{\rm thr}}^{m_\chi} E \frac{dN^{\rm ann}_\gamma}{dE} dE
\end{equation}
and
\begin{equation}
\Phi_{\rm PP}^{\rm decay} = \frac{1}{m_\chi \tau} 
\int_{E_{\rm thr}}^{m_\chi/2} E \frac{dN^{\rm decay}_\gamma}{dE} dE,
\end{equation}
where we choose a reference energy $E_{\rm thr}$ of 0.1 GeV. We consider a 
reasonable range for the particle physics factors that goes from $10^{-30}$ to
$10^{-25} \mbox{cm}^{3}\mbox{s}^{-1} \mbox{GeV}^{-1}$ for $\Phi_{\rm PP}^{\rm ann}$ and 
from $10^{-30}$ and $10^{-24} \mbox{s}^{-1}$ for 
$\Phi_{\rm PP}^{\rm decay}$\footnote{For a DM mass of 200 GeV and annihilations 
(decays) into $b\bar{b}$, the range mentioned above corresponds to a variation
between $3.0 \times 10^{-28} \mbox{cm}^{3} \mbox{s}^{-1}$ and 
$3.0 \times 10^{-22} \mbox{cm}^{3} \mbox{s}^{-1}$ for 
$\langle \sigma_{\rm ann}v \rangle$ (between $1.6 \times 10^{23}$ s and 
$1.6 \times 10^{29}$ s for $\tau$).}. This is divided in 50 logarithmic bins 
and, for each bin, we build 100 realizations of GAL-AQ, varying the position 
of the observer. For each particle physics factor and for each realization, we 
identify the subhalos (if any) with an energy flux above 0.1 GeV that is 
larger than the sensitivity flux in 3FGL at $|b|>10^\circ$, i.e. 
$3 \times 10^{-12} \mbox{erg cm}^{-2} \mbox{s}^{-1}$ \cite{Ackermann:2015yfk}. We 
consider the energy flux and not the number flux, since the \emph{Fermi} LAT
sensitivity, expressed in terms of the energy flux, is more independent of the 
shape of the gamma-ray energy spectrum than when it is expressed by the number
flux. Also, we use the sensitivity obtained for point sources, noting that the 
majority of the emission in a DM clump comes from a region within its scale 
radius $r_s$. In a typical realization of GAL-AQ, almost all resolved DM 
subhalos have a $r_s$ that corresponds to an angular size smaller than 1 
degree. This is a reasonable value for the angular resolution of \emph{Fermi} 
LAT at the energies of interest here. Thus, DM subhalos in GAL-AQ are rarely 
extended and the use of the point-source sensitivity is well motivated.

\begin{figure}
\begin{center}
\includegraphics[width=0.49\textwidth]{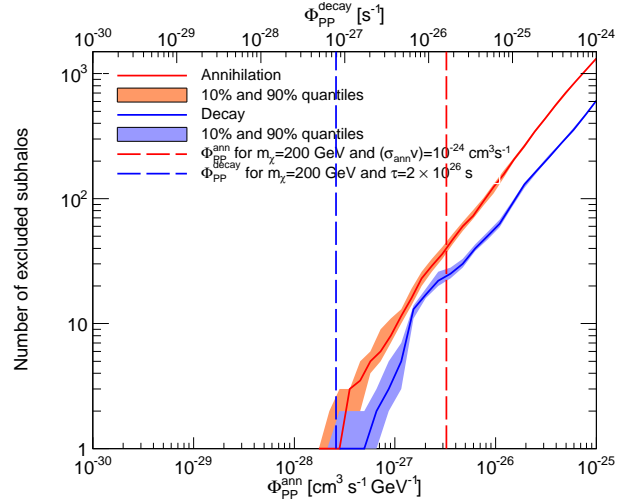}
\caption{The solid lines show the number of the DM subhalos in GAL-AQ with an energy flux above 0.1 GeV that is larger than $3 \times 10^{-12} \mbox{erg cm}^{-2} \mbox{s}^{-1}$, i.e., the point-source sensitivity of \emph{Fermi} LAT in the 3FGL catalog \cite{Ackermann:2015yfk}. The solid lines denote the median over 100 independent realizations differing by the position of the observer and they are plotted as a function of the particle physics factor, in the case of an annihilating DM candidate (red line and bottom axis) and for a decaying one (blue line and top axis). In both cases, annihilations/decays into $b\bar{b}$ are considered. The colored bands indicate the 10\% and 90\% quantiles among the 100 realizations. For reference, the $\Phi_{\rm PP}^{\rm ann}$ for $m_\chi=200$ GeV, $\langle \sigma_{\rm ann} v \rangle = 10^{-24} \mbox{cm}^3 \mbox{s}^{-1}$ and annihilation into $b\bar{b}$ is marked by the dashed red line. Finally, the dashed blue line corresponds to the $\Phi_{\rm PP}^{\rm decay}$ for $m_\chi=200$ GeV, $\tau=2 \times 10^{26} \mbox{s}$ and decaying into $b\bar{b}$.}
\label{fig:number_halos}
\end{center}
\end{figure}

In Fig.~\ref{fig:number_halos}, the solid lines show the median (over the 100 
realizations) of the number of subhalos that have been excluded because they
are too bright, as a function of the annihilation (red line, bottom axis) and 
decay (blue line, top axis) particle physics factor. At the upper end of the 
range considered for $\Phi_{\rm PP}$, this correction affects between 500 and 
2000 DM subhalos for annihilating and decaying DM, respectively. These numbers
correspond to approximately 1-2\% of the total number of subhalos considered 
in the Aquarius catalog\footnote{Note that only 1010 unidentified sources are 
present in 3FGL \cite{Acero:2015gva}. Therefore, a particle physics factor 
that yields more than 1010 DM subhalos with a flux larger than the 3FGL source 
sensitivity should be excluded. As we will see in the following sections, the 
region in the parameter space of DM that is not excluded by the measured 
auto- and cross-APS does not correspond to those extreme values of particle 
physics factor.}. The colored bands indicate the variability associated with 
the 10\% and 90\% quantiles among the 100 realizations. The dashed vertical 
lines are included as a reference and they correspond to the particle physics 
factor for an annihilating DM candidate with a mass of 200 GeV and 
$\langle \sigma_{\rm ann} v \rangle = 10^{-24} \mbox{cm}^3 \mbox{s}^{-1}$ (dashed 
red line) and for a decaying DM candidate with the same mass and 
$\tau=2 \times 10^{26} \mbox{s}$ (dashed blue line). In both cases 
annihilations/decays into $b\bar{b}$ are considered and the values chosen for 
$\langle \sigma_{\rm ann}v \rangle$ and $\tau$ correspond approximately to their
exclusion limits (for $m_\chi=200$ GeV and for the REF benchmark scenario, see 
later) as they will be computed in the following sections. This tells us that, 
for a given DM mass, the allowed region in the parameter space of decaying 
DM would have almost no DM subhalos that are too bright. On the other hand, 
the impact of bright subhalos may be important in the case of annihilating 
DM and this effect will be considered when deriving the exclusion limits on 
$\langle \sigma_{\rm ann}v \rangle$.

\begin{figure*}
\begin{center}
\includegraphics[width=0.49\textwidth]{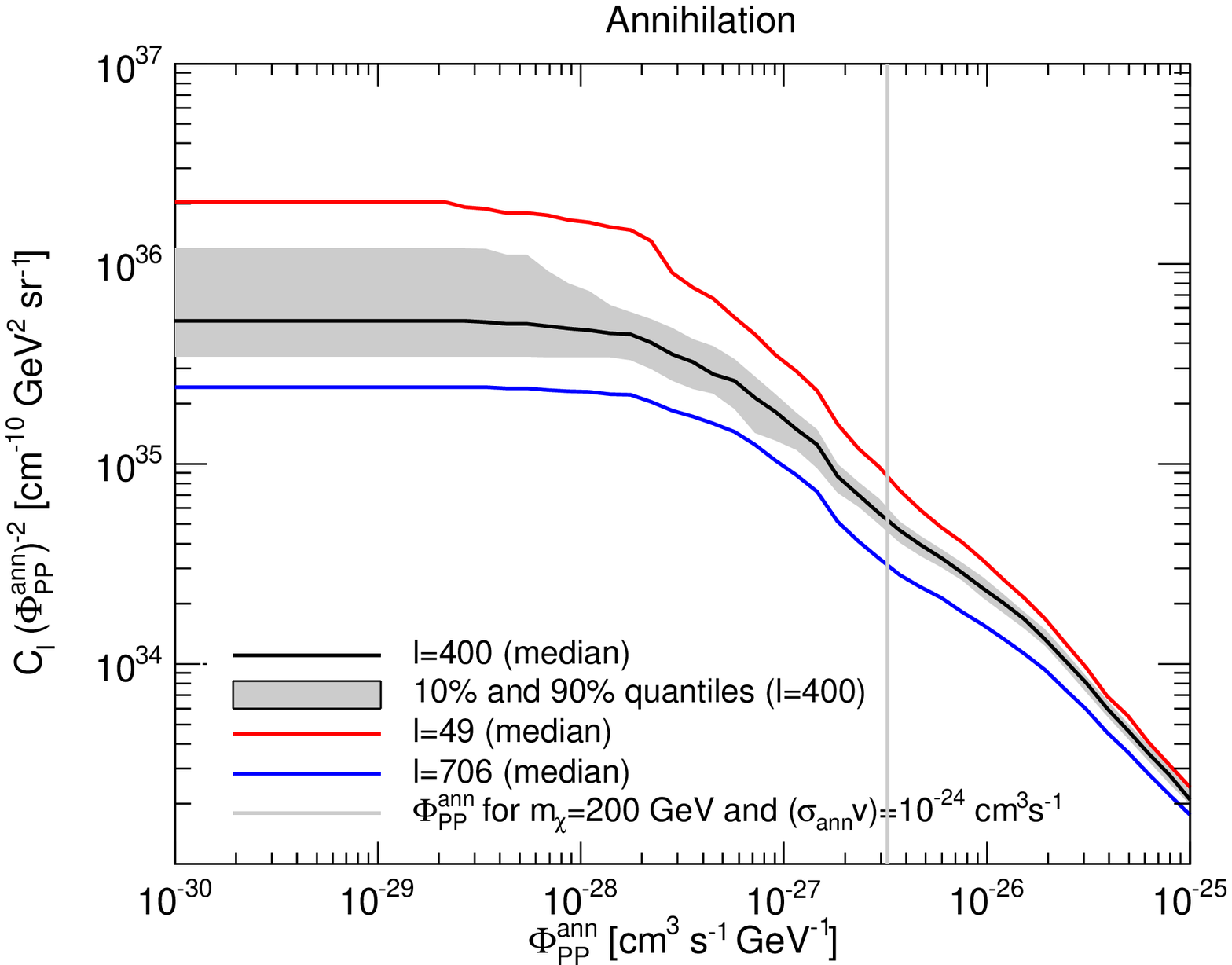}
\includegraphics[width=0.49\textwidth]{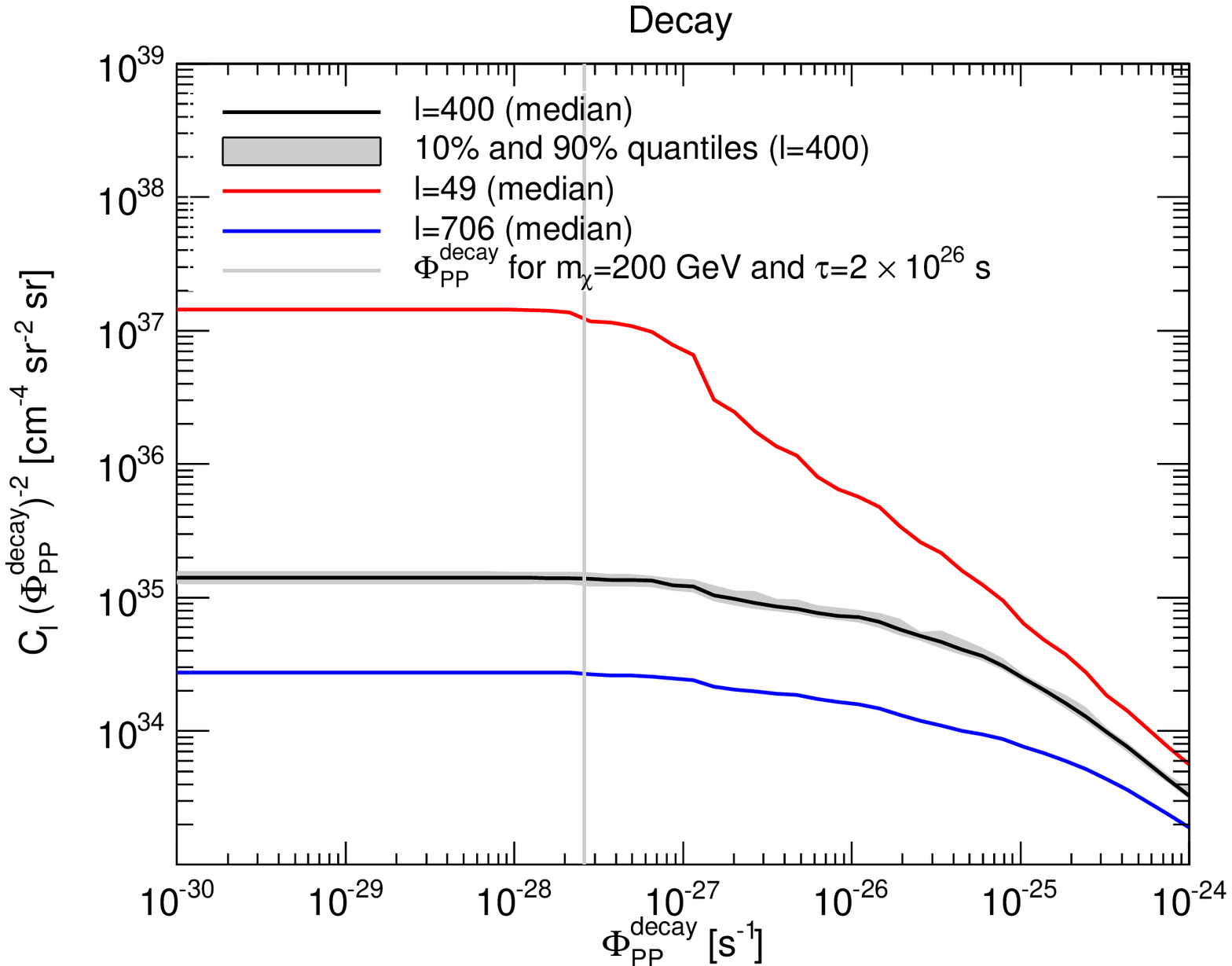}
\caption{Dependence of the APS of the GAL-AQ component on the Particle Physics factor defined in the text. {\it Left:} The solid lines show the DM-induced APS (computed above 0.1 GeV, for a fixed multipole and multiplied by $(\Phi_{\rm PP}^{\rm ann})^{-2}$) as a function of $\Phi_{\rm PP}^{\rm ann}$, neglecting the DM subhalos that would be detected individually according to the \emph{Fermi} LAT sensitivity threshold in the 3FGL catalog \cite{Ackermann:2015yfk}. The black, red and blue lines are for $\ell=400$, $\ell=49$ and $\ell=706$. They indicate the median over the 100 realizations with different positions for the observer, while the grey band (only for the case with $\ell=400$) shows the variabilty between the 10\% and 90\% quantiles. For reference, the value of $\Phi_{\rm PP}^{\rm ann}$ for $m_\chi=200$ GeV, $\langle \sigma_{\rm ann} v \rangle = 10^{-24} \mbox{cm}^3 \mbox{s}^{-1}$ and annihilation into $b\bar{b}$ is marked by the grey vertical line. {\it Right:} The same as in the left panel, but for a decaying DM candidate. The vertical grey line is the particle physics factor of a DM candidate with $m_\chi=200$ GeV, $\tau=2 \times 10^{26} \mbox{s}$ and decaying into $b\bar{b}$. Note that the default mask covering 3FGL sources is employed when computing the auto-APS.}
\label{fig:APS_xi} 
\end{center}
\end{figure*}

In Fig.~\ref{fig:APS_xi} we see the effect on the auto-APS of neglecting the 
DM subhalos in Aquarius that are too bright. The left (right) panel shows the 
auto-APS at a specific multipole, for an annihilating (decaying) DM candidate 
as a function of $\Phi_{\rm PP}^{\rm ann}$ ($\Phi_{\rm PP}^{\rm decay}$). The auto-APS 
is multiplied by $(\Phi_{\rm PP}^{\rm ann})^{-2}$ and by 
$(\Phi_{\rm PP}^{\rm decay})^{-2}$, respectively, so that deviations with respect to 
a horizontal line indicate how much the auto-APS is suppressed due to the 
excluded subhalos. Note that the default mask covering the sources in 3FGL is 
used when computing the auto-APS. The solid black line is for $\ell=400$, 
while the red and blue ones are for $\ell=49$ and $\ell=706$. They indicate 
the median over the 100 realizations, while the grey band (sometimes difficult 
to see because it is too narrow) represents the variablity between the 10\% 
and 90\% quantiles. As we anticipated in Fig.~\ref{fig:number_halos}, the 
effect of neglecting bright subhalos starts to be important around 
$10^{-28} \mbox{cm}^3 \mbox{s}^{-1} \mbox{GeV}^{-1}$ in the left panel and around 
$10^{-27} \mbox{s}^{-1}$ in the right panel. The same values of the Particle 
Physics factor marked by the vertical lines in Fig.~\ref{fig:number_halos} are 
plotted in Fig.~\ref{fig:APS_xi} by the solid grey lines. 

The effect of DM subhalos that are too bright is accounted for by defining the 
following quantity:
\begin{equation}
\mathcal{\kappa}(\Phi_{\rm PP},\ell) = 
\frac{C_{\ell}(\Phi_{\rm PP})}{C_{\ell}(\Phi_{\rm PP}^{\rm min})},
\end{equation}
where $\Phi_{\rm PP}^{\rm min}=10^{-30} \mbox{cm}^{3} \mbox{s}^{-1} \mbox{GeV}^{-1}$ 
for annihilating DM and $\Phi_{\rm PP}^{\rm min}=10^{-30} \mbox{s}^{-1}$ for decaying
DM. $\mathcal{\kappa}$ is computed by using the median over the 100 
realizations. We will employ it as a correction factor to account for the
bright DM subhalos that should be masked and it will be multiplied by the APS 
of GAL-AQ with {\it all} the DM subhalos.

\subsection{Results}
\label{sec:benchmarks}
In this section we define some benchmark cases that we will use in the 
following to discuss our main results:
\begin{itemize}
\item REF: this is our reference case and it is constructed by summing 
EG-MSII and EG-LOW, with $M_{\rm min}=10^{-6} M_\odot$. We also include 
GAL-MWsmooth (for the nominal value of the MW DM halo, taken from Aq-A-1, 
of $1.34 \times 10^{12} M_\odot/h$) and the median of GAL-AQ over the 100 
realizations produced for the nominal MW DM halo mass; \\
\item MAX: we build this case by maximizing all the uncertainties considered 
(and discussed in the previous sections). Thus, we take it as a good estimate
of the largest signal that can be associated with DM (for a given value of the
particle physics factors and of $M_{\rm min}$). The MAX benchmark is defined by 
summing EG-MSII, EG-HIGH (for $M_{\rm min}=10^{-6} M_\odot$), GAL-MWsmooth (for a 
nominal mass of the MW DM halo) and the 90\% quantile among the 100 
realizations of GAL-AQ relative to a $1.34 \times 10^{12} M_\odot/h$ MW; \\
\item MIN: contrary to MAX, this benchmark is obtained by tuning all the 
uncertainties considered above to their minimal configuration. In particular,
we sum EG-MSII, EG-LOW (for $M_{\rm min}=10^{-6}M_\odot$), GAL-MWsmooth (for a MW 
mass that is 1/4 of the nominal value of Aq-A-1) and the 10\% quantile of the 
100 realizations of GAL-AQ for a MW mass that is 1/4 of the value of Aq-A-1.
\end{itemize}

In order to discuss the effect of changing $M_{\rm min}$, we also compute 
the MIN and MAX benchmarks for $M_{\rm min}=10^{-12} M_\odot$ and 
$M_{\rm min}=1 \mbox{ } M_\odot$.

\begin{figure*}
\begin{center}
\includegraphics[width=0.49\textwidth]{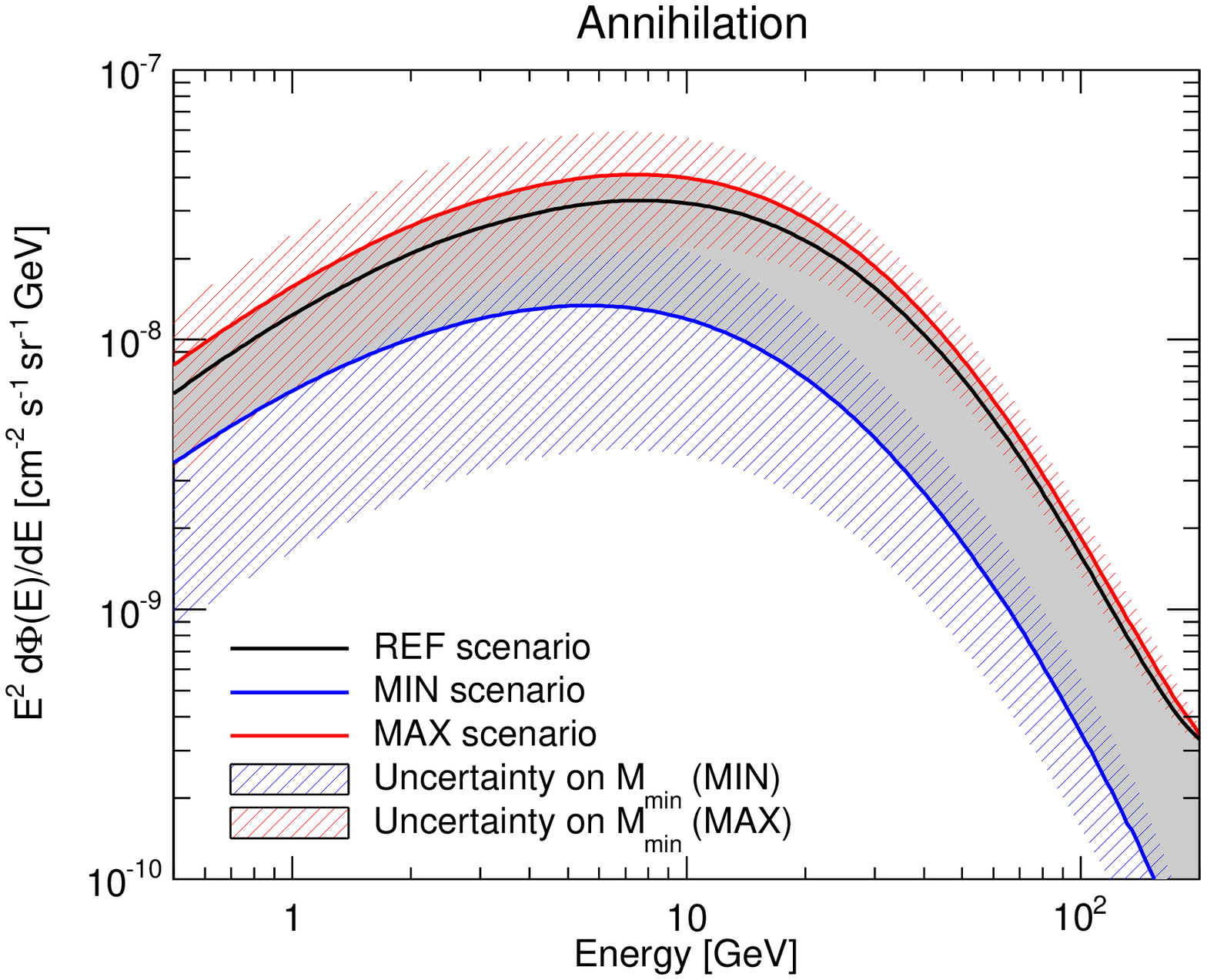}
\includegraphics[width=0.49\textwidth]{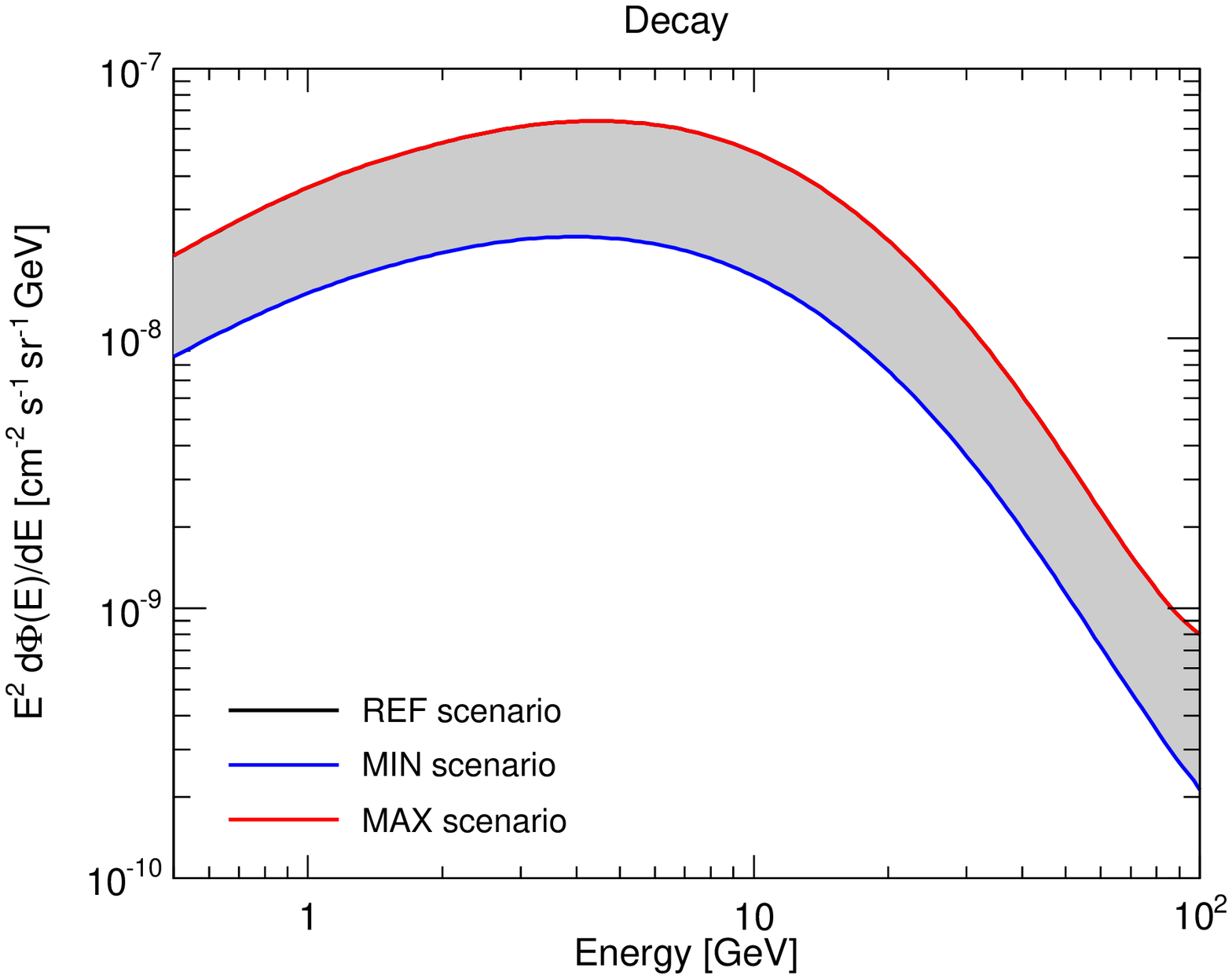}
\caption{{\it Left:} The predicted energy spectrum of the gamma-ray emission induced by the annihilation of a DM particle with a mass of 212 GeV, $\langle \sigma_{\rm ann}v \rangle = 3 \times 10^{-26} \mbox{cm}^3 \mbox{s}^{-1}$ and annihilation into $b\bar{b}$. The black line stands for the REF scenario, while the red and blue ones are for the MAX and MIN cases. Thus, the grey band determines the variability between the MIN and MAX scenarios. In all cases $M_{\rm min}=10^{-6}M_\odot$ (see text for details). The red and blue shaded areas indicate how the MAX and MIN benchmarks change if we let $M_{\rm min}$ vary between $10^{-6}$ and 1 $M_\odot$. {\it Right}: The same as in the left panel but for a decaying DM particle with a mass of 212 GeV and a decaying lifetime of $2 \times 10^{27}$ s. The red and black lines overlap.}
\label{fig:intensity}
\end{center}
\end{figure*}

Fig.~\ref{fig:intensity} shows our predictions for the intensity of the
DM-induced emission, averaged over the whole sky. The left panel is for 
annihilating DM with a mass of 212 GeV and 
$\langle \sigma_{\rm ann}v \rangle = 3 \times 10^{-26}\mbox{cm}^3 \mbox{s}^{-1}$, 
while the right panel is for a decaying candidate with the same mass and 
$\tau=2 \times 10^{27}$ s. Annihilations and decays produce gamma rays via 
$b\bar{b}$. The solid black line is for the REF scenario, while the red and 
blue ones are for the MIN and MAX benchmark (for $M_{\rm min}=10^{-6} M_\odot$). 
Thus, the grey band between the red and blue lines indicates how much our 
predictions change when accounting for the uncertainties mentioned above. 

For annihilating DM, in the case of the REF benchmark, the emission is 
contributed, almost equally, by EG-LOW and GAL-MWsmooth. Thus, the difference 
between REF and MAX comes entirely from the different subhalo boost employed 
to describe unresolved extragalactic DM structures (see 
Sec.~\ref{sec:unresolved_subhalos}). The boost factor is larger at higher
redshifts and, therefore, at energies close to $m_\chi$, where the emission is 
dominated by nearby sources, the red line approaches the black one. On the 
other hand, in the LOW scenario the contribution of GAL-MWsmooth is suppressed 
and, therefore, the intensity of the LOW benchmark is almost halved compared 
to REF.

Subhalo boosts do not affect the predictions for decaying DM and, therefore,
the REF and the MAX benchmarks in the right panel overlap\footnote{The 
contribution of GAL-AQ is subdominant.}. The lower intensity of the LOW case 
is, as before, due to the suppression of GAL-MWsmooth.

In the left panel, the blue and red shaded areas indicate how our predictions 
for MIN and MAX change when allowing $M_{\rm min}$ to vary in the range 
mentioned above. These uncertainty bands get larger for smaller energies, as 
the signal becomes sensitive to the emission at higher redshifts. Changing the 
minimal DM halo mass has a very minor effect on decaying DM and, thus, the 
shaded bands are not present.

The predictions of Fig.~\ref{fig:intensity} can be compared with Fig.~12 of
Ref.~\cite{Fornasa:2012gu}: the main difference is the fact that our brightest 
configuration (i.e. the MAX scenario) predicts almost one order of magnitude 
less gamma-ray flux than in Ref.~\cite{Fornasa:2012gu}, given the new 
definition of the HIGH subhalo boost factor. On the other hand, our 
predictions are compatible with the results of Ref.~\cite{Ackermann:2015tah}.

\begin{figure*}
\begin{center}
\includegraphics[width=0.49\textwidth]{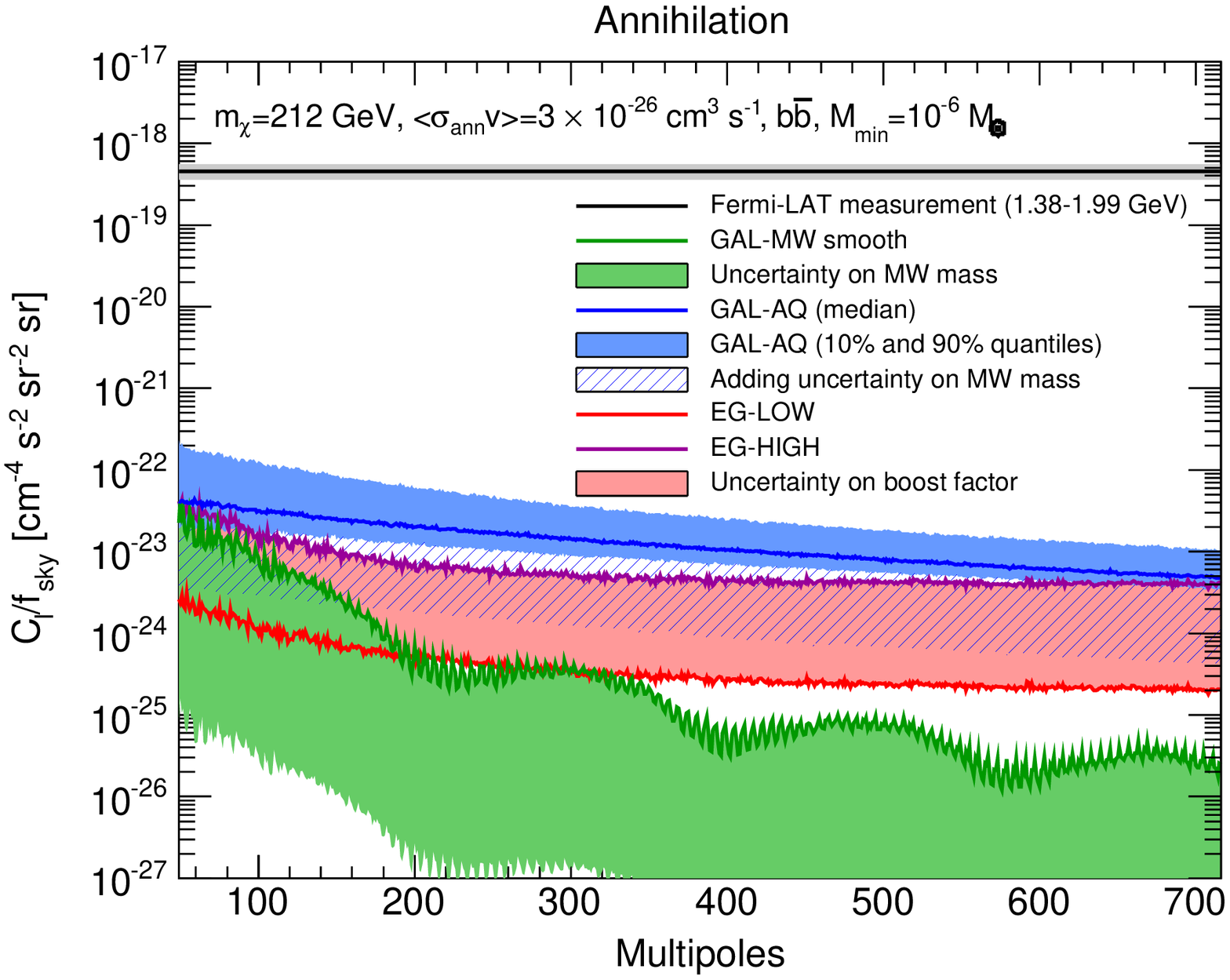}
\includegraphics[width=0.49\textwidth]{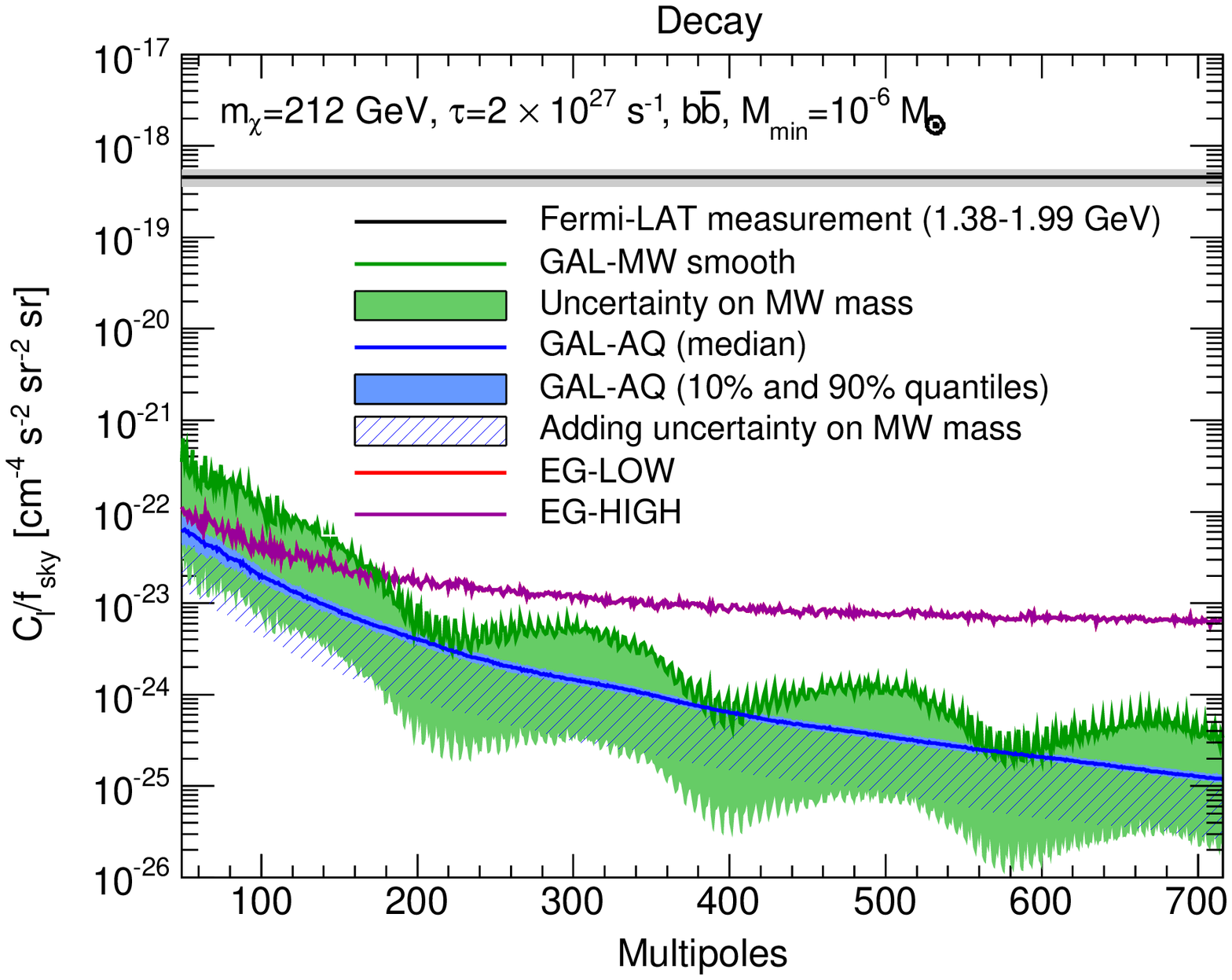}
\caption{{\it Left:} Auto-APS in the energy bin between 1.38 and 1.99 GeV, for a DM candidate with a mass of 212 GeV, $\langle \sigma_{\rm ann}v \rangle = 3 \times 10^{-26} \mbox{cm}^3 \mbox{s}^{-1}$ and annihilation into $b\bar{b}$. The auto-APS is divided by $f_{\rm sky}$ to correct for the presence of the mask described in Sec.~\ref{sec:masks}. The black solid line and the grey band indicate the Poissonian auto-APS measured in this energy bin and for the mask around 3FGL sources (see Sec.~\ref{sec:results_analysis}). The solid blue line is the median of the auto-APS for GAL-AQ over the 100 realizations with different positions for the observer and the blue band shows the variability between the 10\% and 90\% quantiles. The uncertainty band on GAL-AQ extends downwards (shaded blue area) if we account for an uncertainty of a factor 4 in the value of the mass of MW DM halo. The red and purple lines show the auto-APS for EG-LOW and EG-HIGH, for $M_{\rm min}=10^{-6}M_\odot$. The green line stands for the GAL-MWsmooth component and the green band accounts for an uncertainty of a factor 4 in the mass of MW DM halo. The wiggles in this component are due to the mask applied to cover the Galactic plane (see text for details). {\it Right}: The same as in the left panel but for a decaying DM particle with a mass of 212 GeV and a decaying lifetime of $2 \times 10^{27}$ s. The red and and purple lines overlap.}
\label{fig:APS}
\end{center}
\end{figure*}

In Fig.~\ref{fig:APS} we show the predicted auto-APS in the energy bin 
between 1.38 and 1.99 GeV, for the same DM candidates considered in 
Fig.~\ref{fig:intensity}. Note that these correspond to particle physics 
factors that are quite low and therefore none of the DM subhalos in GAL-AQ 
would be resolved by \emph{Fermi} LAT. We can then neglect the 
$\mathcal{\kappa}$ correction discussed in Sec.~\ref{sec:GAL-AQ}. The left
panel is for annihilating DM and the right one for decaying DM. The predicted 
intensity APS is shown separately for the different components discussed 
above. We also include the Poissonian auto-APS measured by \emph{Fermi} LAT in 
the same energy bin (solid black line) and its estimated error (grey band), 
for the mask around 3FGL sources. To compare the measurement with the 
predicted DM signal, the latter needs to be corrected for the presence of the 
mask. In the analysis of the \emph{Fermi} LAT data, this is done automatically 
by {\sc PolSpice} (see Sec.~\ref{sec:masks}), while we correct our predictions 
by dividing the auto- and cross-APS obtained from the masked sky by 
$f_{\rm sky}$, i.e., the fraction of the sky outside the default mask defined 
in Sec.~\ref{sec:masks}. Such a recipe is based on the assumption that the 
masked region is characterised by the same clustering properties of the 
unmasked one. We test this hypothesis by computing the auto-APS of our 
simulated sky maps with and without the mask. For the extragalactic signal we 
find that, indeed, dividing the masked auto-APS by $f_{\rm sky}$, we reproduce 
the unmasked one. On the other hand, for GAL-AQ, the masked auto-APS is 0.43 
times smaller than the unmasked one (approximately at all multipoles). This 
factor is larger than $f_{\rm sky}$, which suggests that the distribution of DM 
subhalos outside the mask is slightly more isotropic than the distribution 
inside the mask. This is to be expected since DM subhalos are more clustered 
towards the center of the host halo. Finally, the auto-APS of GAL-MWsmooth is 
significantly different if we include the mask: the intensity of the auto-APS 
decreases as we are covering the region which produces the majority of the 
emission. Also, the morphology of the mask induces some spurious fluctuations 
on the auto-APS. A more sophisticated algorithm should be employed to correct 
for these features. Alternatively, the wiggles would be reduced by smoothing 
the edges of the mask. However, we note that, in the signal region defined in 
Sec.~\ref{sec:analysis}, the GAL-MWsmooth component is not responsible for 
the majority of the signal and, therefore, using the reconstructed auto-APS 
of GAL-MWsmooth would not considerably change our results. Therefore, for both 
the GAL-AQ and the GAL-MWsmooth, we simply apply the $1/f_{\rm sky}$ correction.

In the left panel of Fig.~\ref{fig:APS} we note that the dominant contribution
is GAL-AQ: the solid blue line is the median over the 100 realizations with 
different observers (in the case of a MW DM halo with the same mass as Aq-A-1), 
while the filled blue band shows the variabilty between the 10\% and 90\% 
quantiles. If we also include the possibility that the MW DM halo may be up to
4 times lighter (see Sec.~\ref{sec:GAL-AQ} and Fig.~\ref{fig:MW_mass}), the 
uncertainty band extends downwards to include the shaded blue band. Over the 
signal region, GAL-AQ is not constant and it decreases by approximately a 
factor of 10. The extragalactic signal is plotted in red and purple, for a LOW 
and HIGH subhalo boost, respectively. This uncertainty gives rise to the pink 
band that covers approximately one order of magnitude. The extragalactic 
component becomes nearly constant for $\ell \gtrsim 300$ but, over the whole 
signal region, it decreases by a factor of 10. Finally, the GAL-MWsmooth is 
plotted in green and the green band indicates how much the signal decreases
when the mass of MW DM halo is allowed to decrease by up to a factor of 4 with 
respect to the value of Aq-A-1.

If we had considered $M_{\rm min}=10^{-12}M_\odot/h$ instead, the intensity of 
EG-LOW and EG-HIGH would have been approximately 4 times larger, while it 
would have decreased by a factor 50 if we had considered 
$M_{\rm min}=1 \mbox{ } M_\odot/h$. However, since the EG-LOW and EG-HIGH are not
dominant components, the effect of changing $M_{\rm min}$ on the total DM signal 
will not be as large.

In the right panel we follow the same color coding: the main difference with
respect to the case of annihilating DM is the fact that the extragalactic 
contribution dominates the signal for most of the measured signal region. 
There is no uncertainty associated with the boost factor and, therefore, the 
red and purple lines coincide. As in the left panel, the auto-APS is nearly 
constant for $\ell \gtrsim 300$ and it decreases by a factor $\sim$50 overall. 
Another important difference, with respect to the case of DM annihilation, is 
the fact that the auto-APS of GAL-AQ is much steeper, decreasing by a factor 
$\sim$600 from $\ell=49$ to $\ell=716$.

Independently of how the different components are summed together\footnote{In
principle, one should include the cross-correlation between the different 
components considered. We do not expect any correlation between extragalactic
and Galactic emission. The cross-correlation between GAL-MWsmooth and GAL-AQ
was computed in Ref.~\cite{Fornasa:2009qh} and it is at least one order of
magnitude below the auto-correlation of GAL-AQ. We neglect it here.}, producing
the different REF, MIN and MAX scenarios described above, the total signal 
associated with DM is {\it not} Poissonian but decreases at smaller angular
scales. This will be crucial when comparing our predictions to the 
\emph{Fermi} LAT data.

\section{Using the auto- and cross-APS to constrain Dark Matter annihilation and decay}
\label{sec:limits}
In this section we compare the predictions for the DM-induced auto- and 
cross-APS obtained in Sec.~\ref{sec:simulations} with the updated 
\emph{Fermi} LAT measurement of the IGRB auto- and cross-APS presented in 
Sec.~\ref{sec:results_analysis}. Such a comparison will allow us to determine 
whether the data are consistent with a DM interpretation or how we can use 
them to constrain the nature of DM. We follow two complementary approaches that 
will be described separately in the following subsections. Neither method
finds a significant detection of DM in the auto- and cross-APS data and, 
therefore, the measurement is used to derive exclusion limits on the 
intensity of the DM-induced gamma-ray emission, as a function of $m_\chi$.

\subsection{Conservative exclusion limits}
\label{sec:conservative}
This first strategy is motivated by the desire to be conservative. In 
particular, the DM-induced APS, for any energy bin or combination of bins, 
must not exceed the measured data. For a certain benchmark case (among REF, 
MIN and MAX) and for a certain value of $M_{\rm min}$, imposing that constraint 
will translate into upper limits on the intensity of the DM-induced signal or, 
by fixing $m_\chi$ and the annihilation/decay channel, into upper limits on 
$\langle \sigma_{\rm ann}v \rangle$ for annihilating DM and into lower limits 
on $\tau$ for decaying DM. We refer to these exclusion limits as 
``conservative''.

\begin{figure*}
\begin{center}
\includegraphics[width=0.49\textwidth]{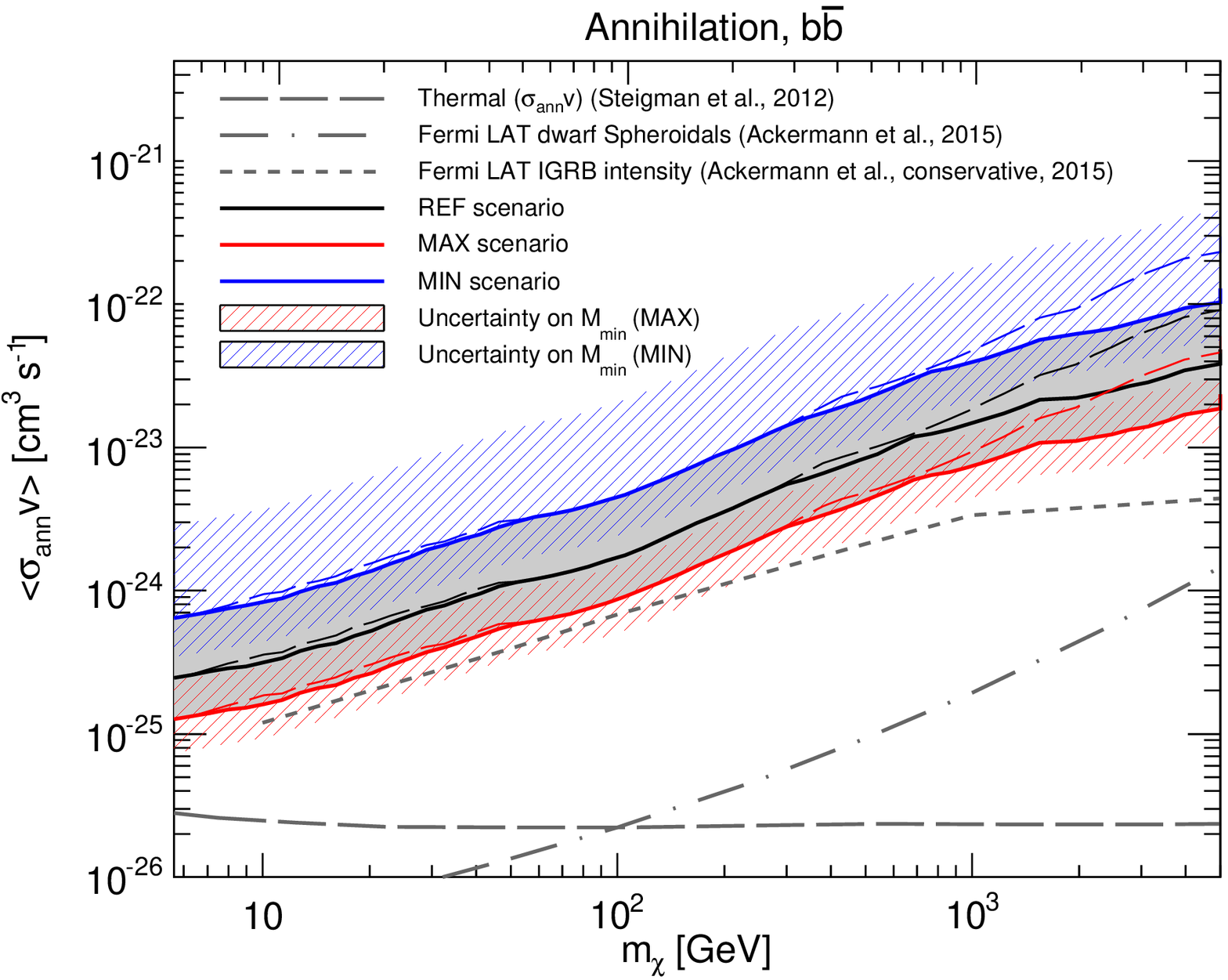}
\includegraphics[width=0.49\textwidth]{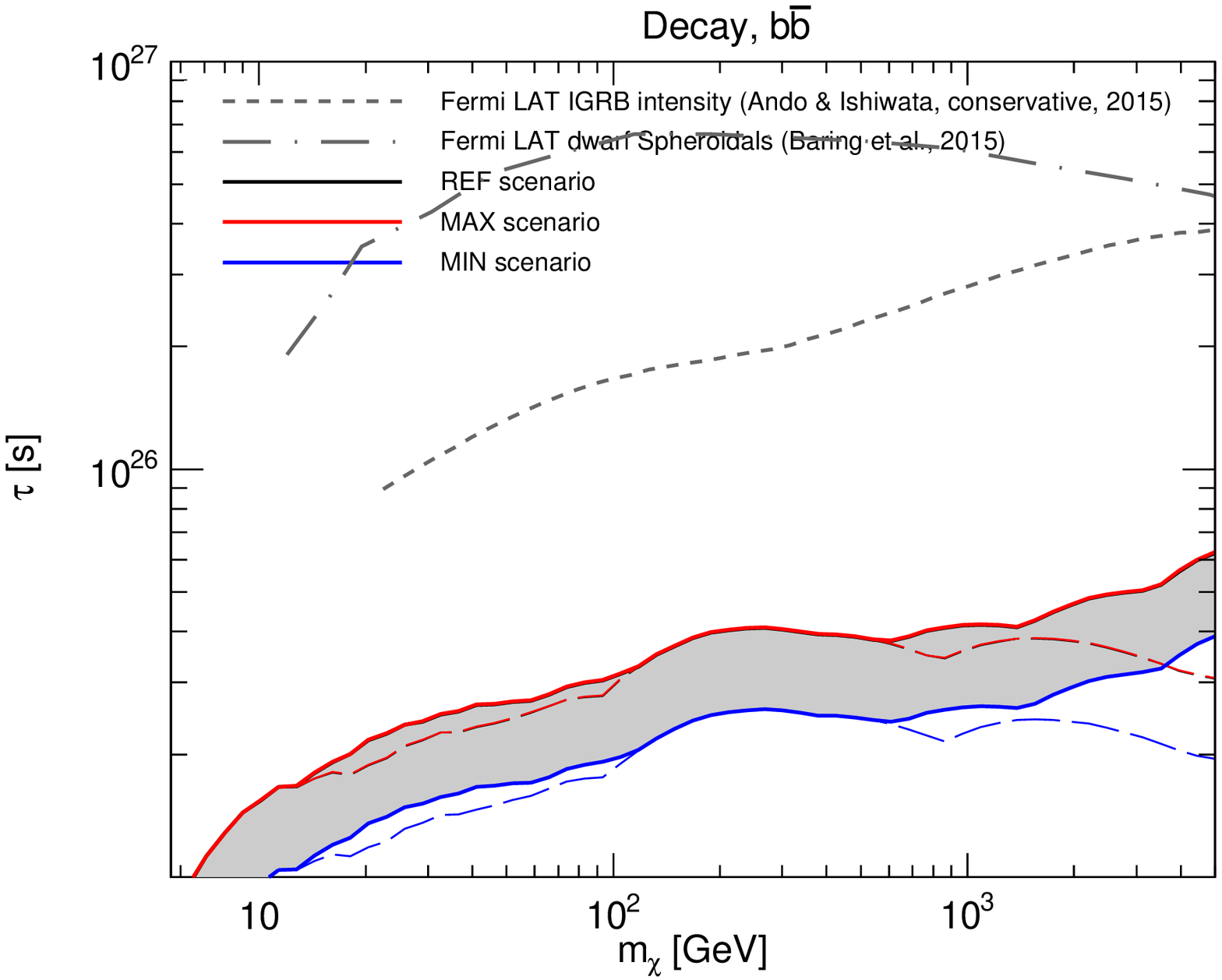}
\caption{Conservative exclusion limits on annihilating and decaying DM from the new APS measurement. {\it Left:} The solid lines show the upper limits on $\langle \sigma_{\rm ann} v \rangle$ derived from the auto- and cross-APS measured in Sec.~\ref{sec:analysis}, as a function of $m_\chi$, for $M_{\rm min}=10^{-6} M_\odot$ and annihilations into $b\bar{b}$. The limits follow the conservative approach described in the text. The black line is for the REF scenario, while the red and blue ones are for MAX and MIN, respectively. The grey band between the MIN and MAX scenario represents our estimated total astrophysical uncertainty for $M_{\rm min}=10^{-6} M_\odot$, accounting for all the sources of uncertainty mentioned in Sec.~\ref{sec:simulations}. The red and blue shaded bands describe the effect of changing $M_{\rm min}$ between $10^{-12} M_{\odot}$ and 1 $M_{\odot}$, for the MAX and MIN scenario. In the case of the black, red and blue dashed lines, the upper limits are derived only considering the measured auto-APS and neglecting the cross-APS. For comparison, the long-dashed grey line marks the annihilation cross section for thermal relics from Ref.~\cite{Steigman:2012nb} and the dash-dotted grey line the upper limit obtained in Ref.~\cite{Ackermann:2015zua} from the combined analysis of 15 dwarf spheroidal galaxies. Finally, the short-dashed grey line shows the conservative upper limit derived in Ref.~\cite{Ackermann:2015tah} from the intensity of the IGRB. {\it Right}: The same as in the left panel but for the lower limits on $\tau$ for decaying DM. The short-dashed grey line represents the lower limit obtained in Fig. 6 of Ref.~\cite{Ando:2015qda} from the IGRB intensity, while the dash-dotted grey one is obtained from the combined analysis of 15 dwarf spheroidal galaxies in Ref.~\cite{Baring:2015sza}.}
\label{fig:exclusion_limits_conservative}
\end{center}
\end{figure*}

We consider 60 values of $m_\chi$ between 5 GeV and 5 TeV. For each value of
$m_\chi$, we compute the DM-induced auto- and cross-APS in the 13 energy bins
defined in Sec.~\ref{sec:data}, for the 3 benchmarks described in 
Sec.~\ref{sec:benchmarks}, for 3 annihilation/decay channels (i.e., $b\bar{b}$, 
$\tau^+\tau^-$ and $\mu^+\mu^-$)\footnote{See Ref.~\cite{Fornasa:2012gu} on 
how to compute the emission for multiple annihilation/decay channels without
having to recompute, for each case, the mock sky maps from the $N$-body 
simulations.} and for 3 values of $M_{\rm min}$ (i.e., $10^{-12}$, $10^{-6}$ and 1 
$M_\odot$). The APS associated with DM for energy bins $i$ and $j$ is averaged 
over the signal region in multipole and we require it to be smaller than the 
Poissonian APS measured for that pair of energy bins plus 1.64 times its error: 
$\langle C_{\ell,{\rm DM}}^{i,j} \rangle < C_{\rm P}^{i,j} + 1.64 \, 
\sigma_{C_{\rm P}^{i,j}}$. Assuming that the measured $C_{\rm P}$ has a Gaussian 
probability distribution with a central value of $C_{\rm P}$ and a standard 
deviation of $\sigma_{C_{\rm P}}$, values further away than 1.64 times 
$\sigma_{C_{\rm P}}$ from the central value correspond to a cumulative probability 
distribution larger than 0.95. Thus, excluding them provides a 95\% CL 
exclusion bound. For each $m_\chi$, we take the most stringent limit among all 
the combinations of energy bins.

Fig.~\ref{fig:exclusion_limits_conservative} shows the upper limits on the 
$\langle \sigma_{\rm ann}v \rangle$ (left panel) and the lower limits on $\tau$ 
(right panel), as a function of $m_\chi$, for annihilations/decays into 
$b\bar{b}$ and for the different benchmark scenarios considered above. The 
black line is for REF while the blue and red ones are for MIN and MAX. Thus, 
the grey band represents our total systematic astrophysical uncertainty (for 
$M_{\rm min}=10^{-6}M_\odot$) and it is as large as approximately a factor of 5 or 
2, for annihilating and decaying DM, respectively. The limits have some 
wiggles because, depending on the DM mass, the emission peaks at different 
energies, and different combinations of energy bins are responsible for the 
exclusion limit. Solid lines are obtained considering all the possible 
combinations of energy bins, while for the black, blue and red dashed ones 
only the auto-APS is employed. The figure shows that, at large DM mass and 
both for annihilating and decaying DM, the exclusion limits are driven by the 
cross-APS and not by the auto-APS, approximately for $m_\chi>200$ GeV for 
annihilating DM and for $m_\chi>700$ GeV for decaying DM.

In the left panel, the red and blue shaded areas account for the effect of 
changing $M_{\rm min}$ between $10^{-12} M_\odot$ and 1 $M_\odot$ and they are 
computed only for the MIN and MAX scenarios. The effect is more important
for the MIN case since the emission from extragalactic DM structures 
contributes more to the total signal in this case (see Fig.~\ref{fig:APS}). If 
we include the variability on $M_{\rm min}$ in our budget for the systematic 
uncertainty, the systematic error grows to a factor of 40. Compared to the 
conservative upper limits on $\langle \sigma_{\rm ann}v \rangle$ derived by the 
intensity of the IGRB in Ref.~\cite{Ackermann:2015tah}, our uncertainty is 
approximately a factor of 2 larger. The long-dashed grey line is the thermal 
annihilation cross section computed in Ref.~\cite{Steigman:2012nb}. The line 
marks the beginning of the region where, for WIMP DM, one can find 
annihilation cross sections that correspond to a relic DM abundance in 
agreement with the Planck data \cite{Ade:2015xua}. Unfortunately, our 
conservative upper limits do not probe this region, as they are, at least, a 
factor 3 away from it. Also, the REF upper limit is approximately two orders 
of magnitude higher than the upper limit derived from the observation of 15 
dwarf spheroidal galaxies performed by \emph{Fermi} LAT 
\cite{Ackermann:2015zua} and included here as a grey dash-dotted line. 
Finally, it is a factor 2 higher than the conservative limits that can be 
derived from the intensity of the IGRB (short-dashed grey line) 
\cite{Ackermann:2015tah}, at least below 100 GeV. Above this value, the IGRB 
intensity leads to an even more stringent exclusion.

In the right panel, it can be seen that the lower limits on $\tau$ derived 
here from the auto- and cross-APS are a factor of 5 less stringent than the 
lower limits obtained in Fig.~6 of Ref.~\cite{Ando:2015qda} from the IGRB 
intenstity, in their conservative scenario. The REF scenario is also at least
a factor of 5 from the dash-dotted grey line showing the lower limits obtained
from the combined analysis of 15 dwarf spheroidal galaxies in 
Ref.~\cite{Baring:2015sza}.

Figs.~\ref{fig:tau_conservative} and \ref{fig:mu_conservative} in
Appendix~\ref{sec:other_channel} show the exclusion limits on 
$\langle \sigma_{\rm ann}v \rangle$ and $\tau$ in the case of the $\tau$- and 
$\mu$- channels.

\subsection{Fit to the data and realistic exclusion limits}
\label{sec:fit}
In this section we describe our analysis of the auto- and cross-APS using a 
2-component model that includes a Poissonian term and a DM-induced one which, 
as we noticed in Fig.~\ref{fig:APS}, deviates from a Poissonian behaviour. The 
Poissonian component is interpreted as the APS of unresolved astrophysical 
sources, even if we do not try to predict its amplitude in terms of a specific 
model. This 2-component model will be used to fit the \emph{Fermi} LAT APS 
as a function of multipole.

The fit minimizes the $\chi^2$ defined as:
\begin{equation}
\chi^2 = \sum_{i,j,\ell} 
\frac{[ \overline{C_{\ell}}^{i,j} - C_{\ell,{\rm DM}}^{i,j} - C_{\rm P,astro}^{i,j} ]^2}
{[ \overline{\sigma_{\ell}}^{i,j} ]^2},
\label{eqn:lnlike}
\end{equation}
where the $i,j$ indexes in the sum extend over all the 91 independent 
combinations of energy bins and the $\ell$ index runs over the 10 bins in 
multipoles contained in the signal region. $\overline{C_\ell}^{i,j}$ indicates 
the APS measured in the $(i,j)$ combination of energy bins and in the $\ell$ 
multipole bin, while $C_{\ell,{\rm DM}}^{i,j}$ and $C_{\rm P,astro}^{i,j}$ are the DM 
and Poissonian components of our model in the same combination of energy bins 
and in the same multipole bin. Finally, $\overline{\sigma_{\ell}}^{i,j}$ is the 
experimental error associated to $\overline{C_\ell}^{i,j}$ and provided by 
{\sc PolSpice}. The DM APS $C_{\ell,{\rm DM}}^{i,j}$ are computed for the same 60 
values of $m_\chi$ as in the previous section, the same 3 annihilation/decay 
channels, 3 benchmark scenarios and 3 values of $M_{\rm min}$. The only 
remaining parameter needed to calculate $C_{\ell,{\rm DM}}^{i,j}$ is either 
$\langle \sigma_{\rm ann}v \rangle$ or $\tau$: they will be fixed to a specific 
value every time we compute $\chi^2$. On the other hand, the 91 independent 
values of $C_{\rm P,astro}^{i,j}$ in Eq.~\ref{eqn:lnlike} are left free in the fit. 
Putting the DM term to zero in Eq.~\ref{eqn:lnlike} defines our null 
hypothesis. In that case, the fit to the \emph{Fermi} LAT data leads us to the 
$C_{\rm P}$ estimators discussed in Sec.~\ref{sec:analysis}, whose auto-APS is 
plotted in Fig.~\ref{fig:autoCPvsE}. Including the DM component, for a fixed 
$m_\chi$, annihilation/decay channel, benchmark scenario and $M_{\rm min}$, we 
repeat the minimization of $\chi^2$ in Eq.~\ref{eqn:lnlike}, for different 
values of $\langle \sigma_{\rm ann}v \rangle$ and $\tau$. 

\begin{figure}
\begin{center}
\includegraphics[width=0.49\textwidth]{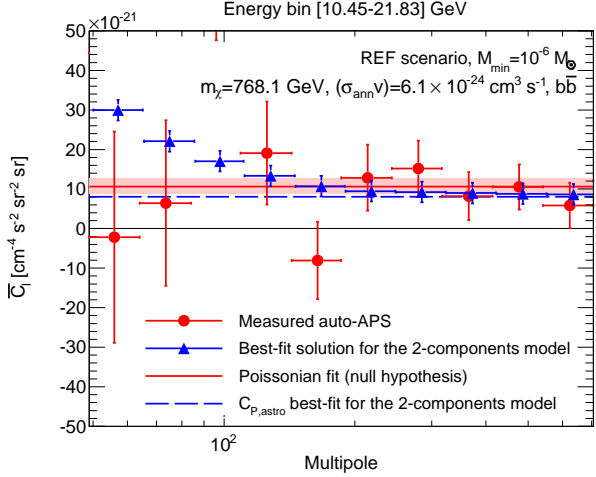}
\caption{\label{fig:fit} Example of a fit to the binned APS $\overline{C_\ell}$ in our particular energy bins, in terms of the 2-component model described in the text. The red circles show the measured auto-APS in the energy bin between 10.4 and 21.8 GeV, as a function of multipole. The solid red line is the Poissonian best-fit APS in the null hypothesis and the pink band denotes its estimated 68\% CL error. The dashed blue line denotes the best-fit of the Poissonian component when DM is included in the fit, for a DM mass of 768.1 GeV and a $\langle \sigma_{\rm ann}v \rangle$ of $6.12 \times 10^{-24} \mbox{cm}^3 \mbox{s}^{-1}$, annihilation into $b\bar{b}$ and a REF scenario with $M_{\rm min}=10^{-6} M_\odot$. The best-fit signal (Poissonian plus DM component) is plotted by means of the blue triangles.}
\end{center}
\end{figure}

We show an example in Fig.~\ref{fig:fit}, for the case of the REF scenario for 
a DM candidate with a mass of 768.1 GeV, 
$\langle \sigma_{\rm ann}v \rangle = 6.12 \times 10^{-24} \mbox{cm}^3 \mbox{s}^{-1}$
annihilating into $b\bar{b}$. The value of the annihilation cross section 
corresponds approximately to the exclusion upper limit for that value of DM 
mass, as will be computed later. The red circles show the measured auto-APS as 
a function of $\ell$ in the signal region for one reference energy bin, i.e., 
the one between 10.4 and 21.8 GeV (when masking 3FGL sources). The solid red 
line with the pink band denotes the best-fit $C_{\rm P}$ in that energy bin for 
the null hypothesis (i.e., without DM), while the dashed blue line is the 
best-fit Poissonian component $C_{\rm P,astro}$ when the fit is done with the 
2-component model (i.e. including DM). The dashed line is lower than the solid 
one, since at these energies part of the signal is explained by DM and, 
therefore, there is less need of a Poissonian component. Energy bins not 
localized near the peak of the DM emission are only slightly affected by the 
inclusion of the DM term in the fit. The best-fit configuration for the 
two-components model is plotted by blue triangles: the inclusion of the DM 
term makes it multipole-dependent so that it decreases by a factor of $\sim$3 
over the signal region.

\begin{figure*}
\begin{center}
\includegraphics[width=0.49\textwidth]{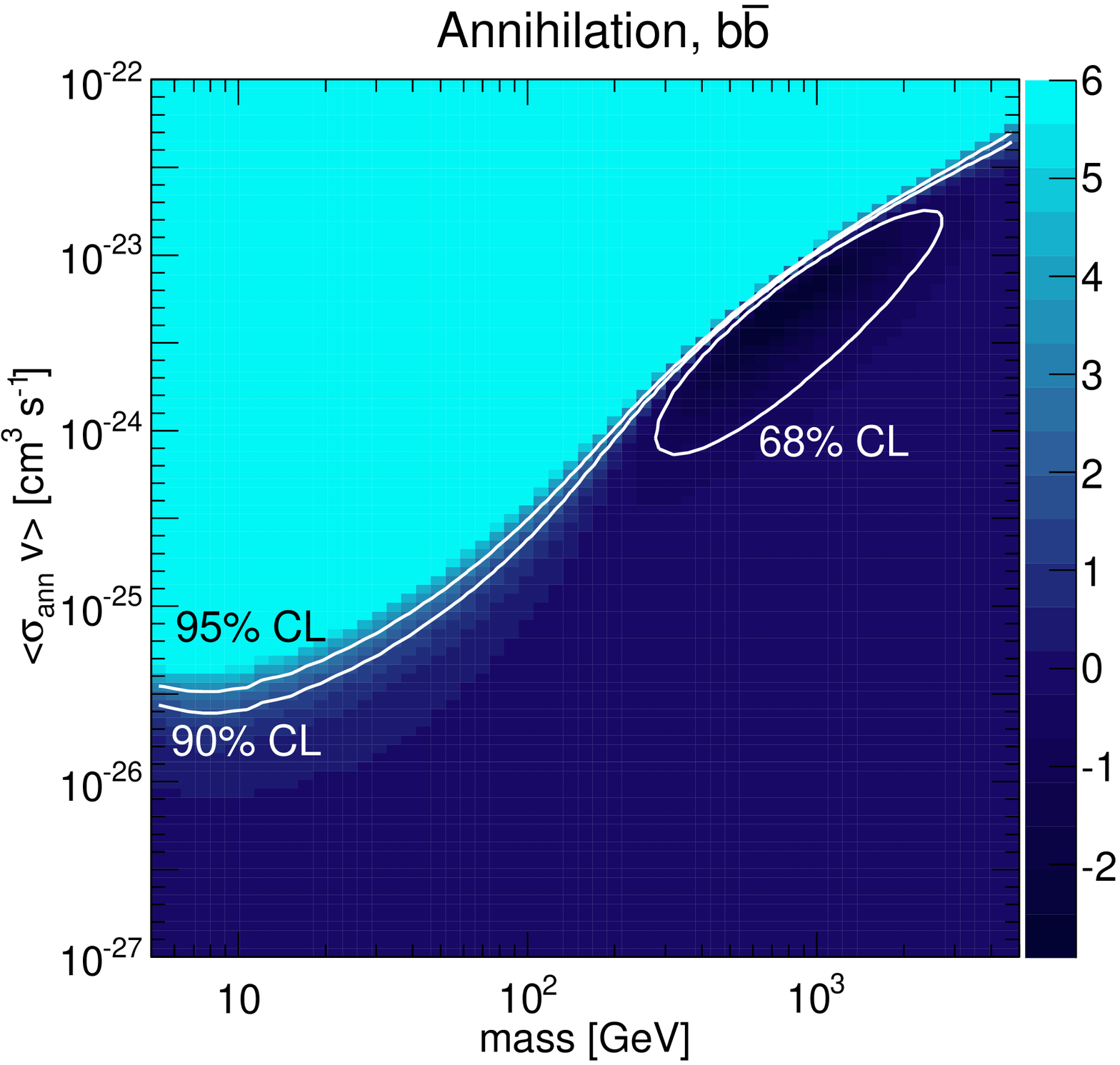}
\includegraphics[width=0.49\textwidth]{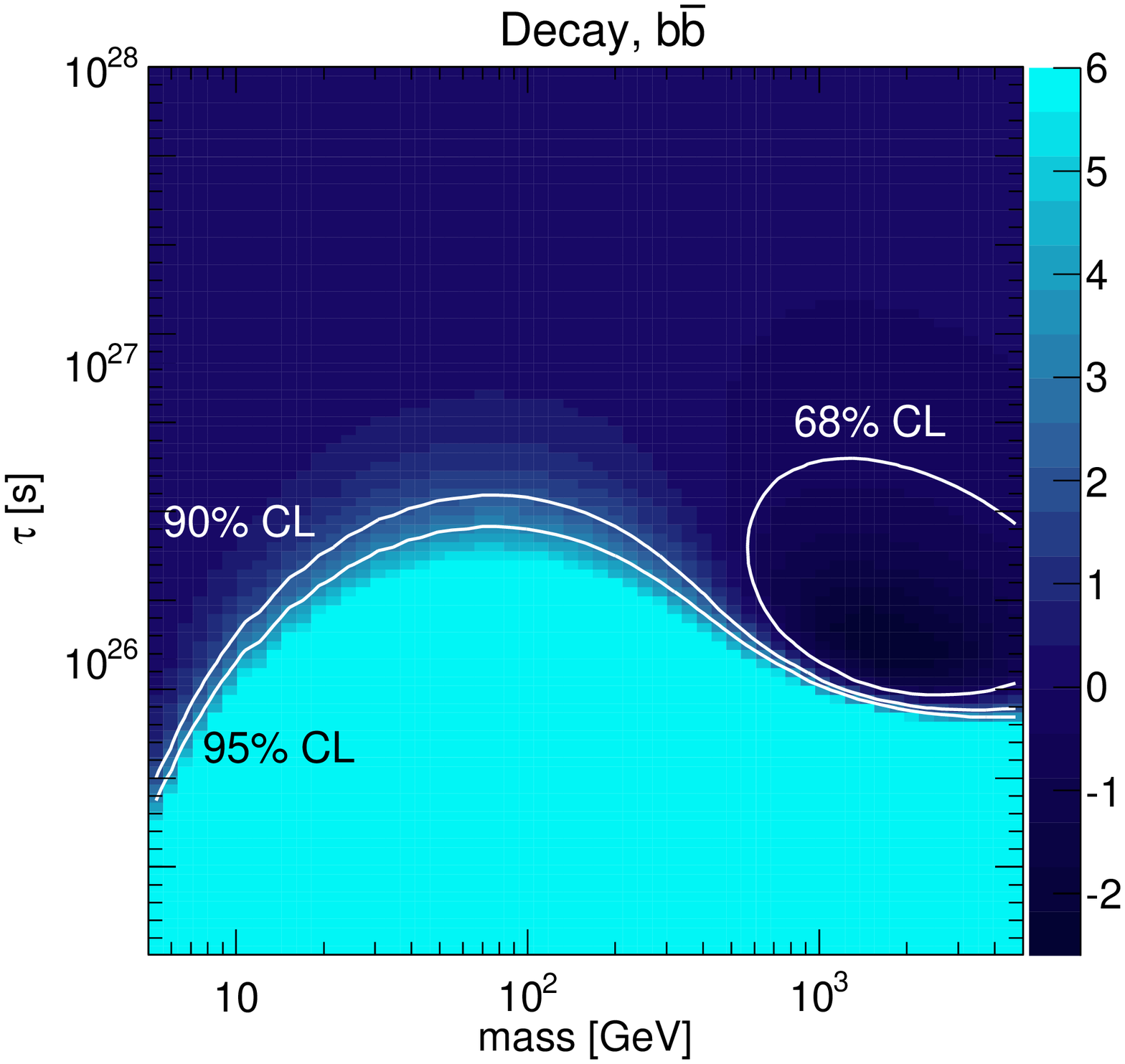}
\caption{$\Delta\chi^2$ between the best-fit solution for the 2-component scenario and the best fit in the null hypothesis. Results presented here refer to the REF scenario for annihilation/decay into $b\bar{b}$ and $M_{\rm min}=10^{-6} M_\odot$. {\it Left:} Each point in the $(m_\chi,\langle \sigma_{\rm ann} v \rangle)$ parameter space is colored according to its $\Delta\chi^2$, i.e. the difference between the $\chi^2$ of the best fit to the auto- and cross-APS in terms of the 2-component model and the $\chi^2$ of the best fit of the null hypothesis (i.e. no DM). The closed white contour marks the 68\% CL region. The 90\% and 95\% CL regions are below the white open curves labelled ``90\% CL'' and ``95\% CL'' respectively. {\it Right}: The same as the left panel but for decaying DM.}
\label{fig:contours}
\end{center}
\end{figure*}
 
We note that, including the DM component, it is possible to find a 
configuration that improves the $\chi^2$ of the best-fit point with respect 
to the null hypothesis, at least for DM masses above few hundreds of GeV. This 
is probably due to the fact that the measured auto-APS is slightly 
multipole-dependent. We can quantify the improvement in the fit provided by 
the DM component by building a 2-dimensional grid in 
$(m_\chi,\langle \sigma_{\rm ann}v \rangle)$ for annihilating DM and in 
$(m_\chi,\tau)$ for decaying DM and plotting the TS $\Delta\chi^2$, i.e. the 
difference between the $\chi^2$ of the best fit for the null hypothesis and 
the $\chi^2$ of the best fit in the case of the 2-component model. This is 
shown in Fig.~\ref{fig:contours}, where the left panel refers to annihilating 
DM and the right one to decaying DM (for the $b$ channel and a REF scenario 
with $M_{\rm min}=10^{-6}M_\odot$). In both panels, the closed area indicate the
region where the 2-component model is preferred over the null hypothesis at
a 68\% CL. The 90\% and 95\% CL regions are open and bounded by the 
corresponding white lines\footnote{The 68\% CL ares is obtained by identifying
the region where the best-fit solution for the 2-component model has a 
TS $\Delta\chi^2$ of 2.30 larger than the null hypothesis. The values are 4.61 
and 5.99 for the 90\% and 95\% CL regions.}. This tells us that including the 
DM component in the model provides a better fit to the auto- and cross-APS 
measured in Sec.~\ref{sec:analysis}, with a significance between 1 and 
1.6$\sigma$. This is too small to consider as significant. Thus, we conclude 
that the data do not significantly prefer the addition of a DM component and 
we use the measured auto- and cross-APS to derive constraints on the DM 
signal.

The contour plots for the $\tau$- and the $\mu$-channel can be seen in
Appendix~\ref{sec:other_channel}. In both cases, the 68\% CL region is the
only closed one.

\begin{figure*}
\begin{center}
\includegraphics[width=0.49\textwidth]{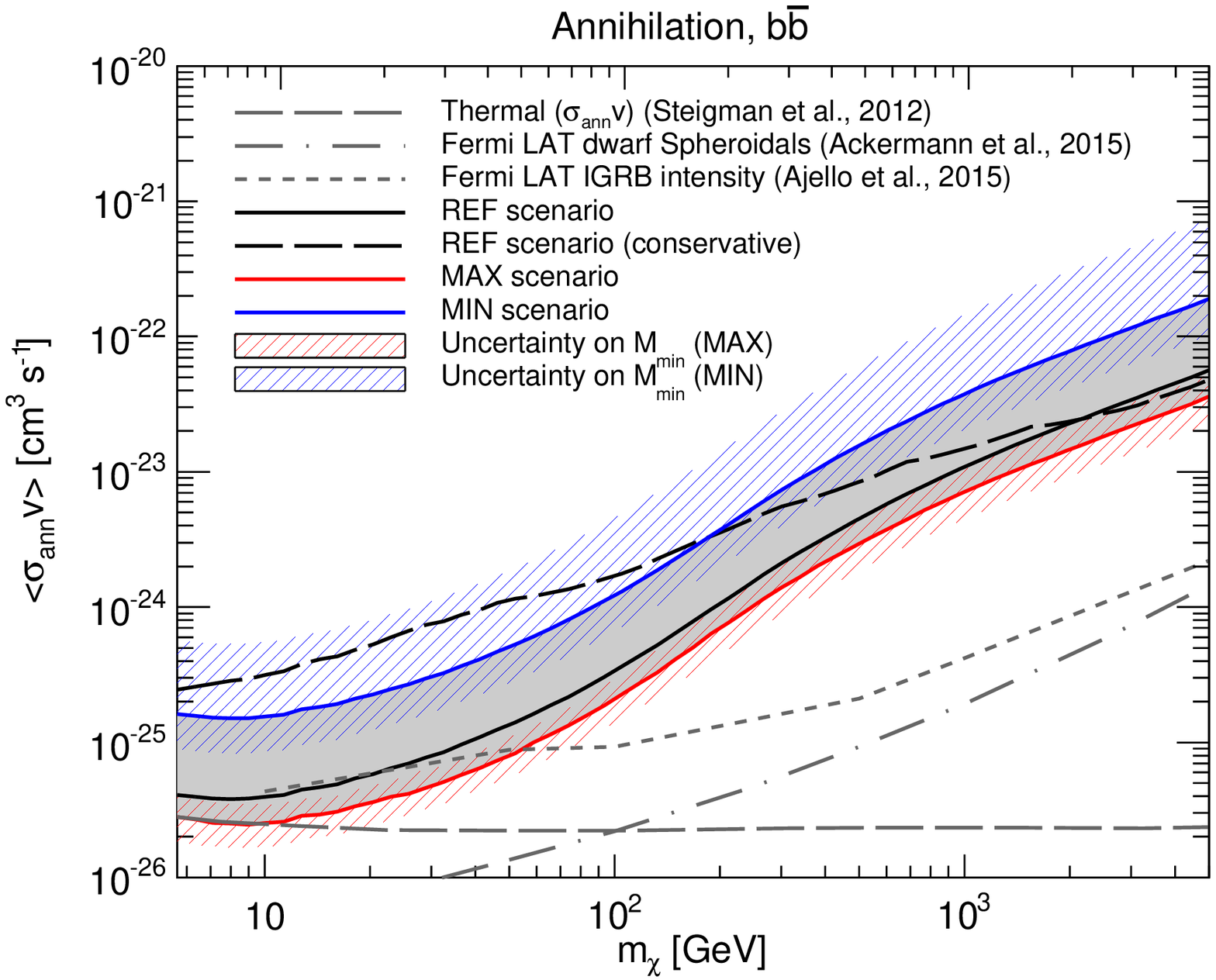}
\includegraphics[width=0.49\textwidth]{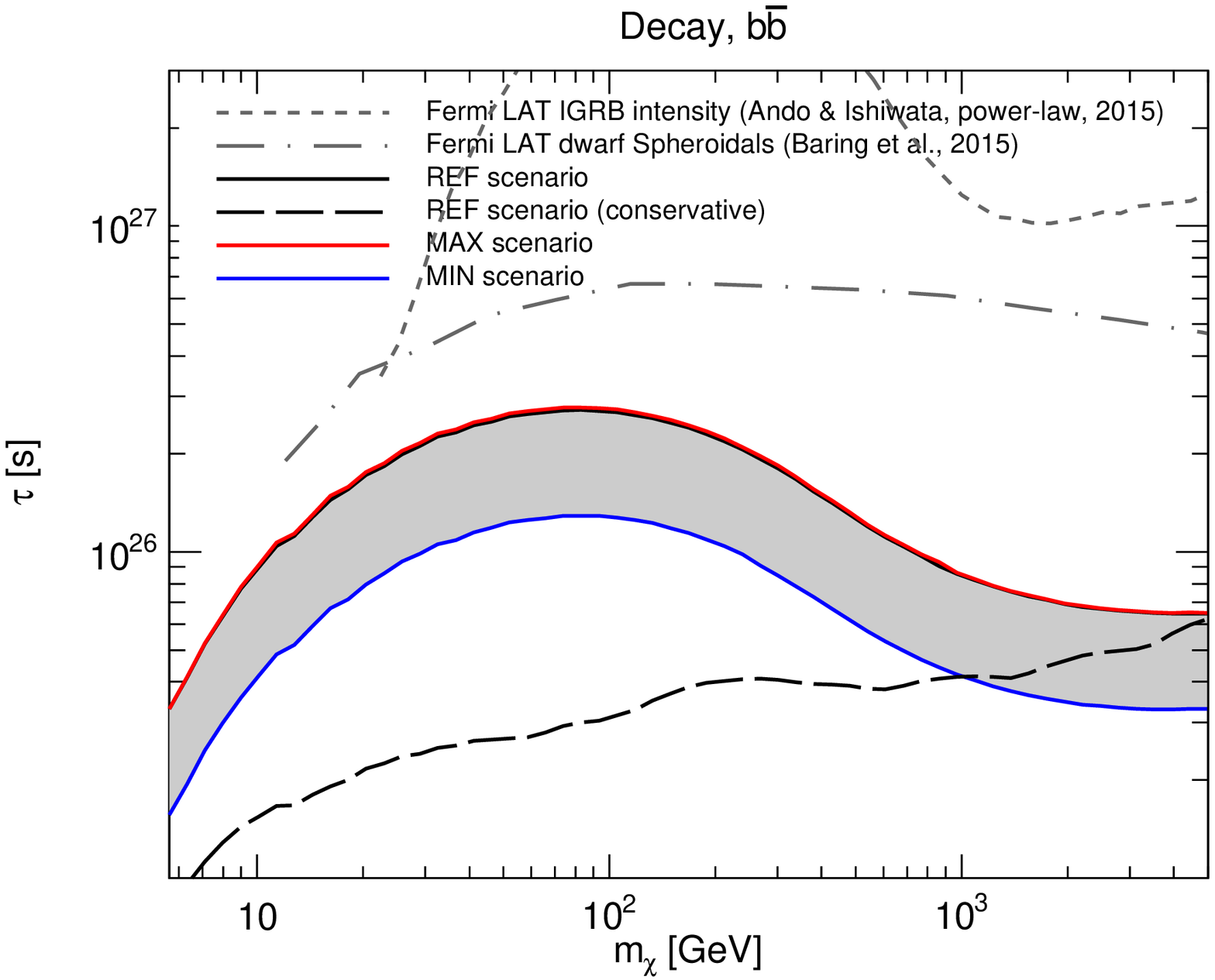}
\caption{Exclusion limits on annihilating and decaying DM from the fit to the binned $\overline{C_\ell}$ in terms of the 2-component model. {\it Left:} The solid lines show the upper limits that can be derived on $\langle \sigma_{\rm ann}v \rangle$ as a function of $m_\chi$ (for annihilations into $b\bar{b}$ quarks and $M_{\rm min}=10^{-6} M_\odot$) by fitting the \emph{Fermi} LAT data with a 2-component model that includes astrophysical sources and DM (see text for details). The black, blue and red lines correspond to the REF, MIN and MAX scenario. The blue and red shaded areas indicate how the MIN and MAX upper limits change when leaving $M_{\rm min}$ free to vary between $10^{-12} M_\odot$ and $1 \mbox { } M_\odot$. The black dashed line is the REF upper limit in the conservative case, from Fig.~\ref{fig:exclusion_limits_conservative}, while the long-dashed grey line is the thermal annihilation cross section from Ref.~\cite{Steigman:2012nb}. The dot-dashed grey line is the upper limits derived in Ref.~\cite{Ackermann:2015zua} from the combined analysis of 15 dwarf spheroidals, while the short-dashed grey line comes from the analysis of the IGRB intensity performed in Ref.~\cite{Ajello:2015mfa}. {\it Right}: The same as in the left panel but for the lower limits on $\tau$, in the case of decaying DM. The short-dashed grey line represents the lower limit obtained in Ref.~\cite{Ando:2015qda} from the IGRB intensity. The line is taken from Fig.~5 of Ref.~\cite{Ando:2015qda}, where the IGRB is interpreted in terms of a component with a power-law emission spectrum and a DM contribution. Finally, the dot-dashed grey line is the upper limit from the analysis of 15 dwarf spheroidal galaxies performed in Ref.~\cite{Baring:2015sza}.}
\label{fig:exclusion_limits_bestfit}
\end{center}
\end{figure*}

For each value of $m_\chi$, $M_{\rm min}$, annihilation/decay channel and 
benchmark scenario, the exclusion limits on $\langle \sigma_{\rm ann}v \rangle$ 
and $\tau$ are derived by scanning on $\langle \sigma_{\rm ann}v \rangle$ and 
$\tau$ until we find the values that correspond to a best fit with a TS 
$\Delta\chi^2$ of 3.84 with respect to the null hypothesis. Such a 
value is derived assuming that $\Delta\chi^2$ follows a $\chi^2$ probability 
distribution with one degree of freedom (i.e., 
$\langle \sigma_{\rm ann}v \rangle$ or $\tau$) and noting that values larger 
than 3.84 fall outside $5\%$ of the cumulative distribution probability. This 
recipe provides the 95\% CL exclusion limits on 
$\langle \sigma_{\rm ann}v \rangle$ and $\tau$ that are summarised in the left 
and right panels of Fig.~\ref{fig:exclusion_limits_bestfit}, respectively. 

In the left panel, as in Fig.~\ref{fig:exclusion_limits_conservative}, the 
black, blue and red solid lines correspond to the REF, MIN and MAX scenarios. 
The difference between MIN and MAX covers slightly more than a factor of 5. 
The blue and red shaded regions around the solid lines of the same color 
indicate how the upper limits change when we leave $M_{\rm min}$ free to vary. 
This extends the range of the total systematic uncertainty to approximately a 
factor of 20. For comparison, the black dashed line is the REF upper limit in 
its conservative version (from Fig.~\ref{fig:exclusion_limits_conservative}). 
Fitting the data with the 2-component model generates exclusion limits that 
are approximately a factor of 10 stronger, at least at low DM masses. As the 
mass increases, the method employed in this section starts to perform 
progressively worse and the solid black line gets closer to the dashed one. 
This is due to the fact that, for $m_\chi>150$ GeV, the data slightly prefer 
the interpretation with DM as opposed to the null hypothesis. The figure also 
includes the thermal cross section from Ref.~\cite{Steigman:2012nb} as a 
long-dashed grey line: our upper limit for the REF case is slightly above it, 
below 10 GeV. It is also more than a factor 10 weaker than the upper limit 
derived from the observation of 15 dwarf spheroidal galaxies in 
Ref.~\cite{Ackermann:2015zua}. Finally, the short-dashed grey line indicates 
the exclusion limits obtained in Ref.~\cite{Ajello:2015mfa} by studying the 
intensity of the IGRB with a 2-component model that, similarly to what is done 
here, includes both a generic model-independent astrophysical contribution and 
a DM one. Our REF limit is slightly stronger than the short-dashed grey line 
for $m_\chi<30$ GeV, suggesting that the study of the IGRB anisotropies could 
{\it in principle} be a more effective way of constraining DM than the IGRB 
intensity. However, for larger DM masses, our limit gets worse due, again, to 
the fact that the data slightly prefer an interpretation that includes DM.

The same color coding is used in the right panel for decaying DM. With no
dependence on $M_{\rm min}$, the band of the systematic uncertainty covers a 
factor 2, and the REF upper limit is even one order of magnitude above the
conservative one, at least at 60-70 GeV. For larger masses our limit worsens 
for the same reason as in the left panel. As in 
Fig.~\ref{fig:exclusion_limits_conservative}, the short-dashed grey line is 
the lower limit obtained from the analysis of the IGRB intensity from 
Ref.~\cite{Ando:2015qda}. The line refers to the case in which the IGRB is
modeled in terms of a component with a power-law energy spectrum and a DM
contribution. Above 20 GeV, where both lines are available, the analysis of the
IGRB intensity is always more powerful than the anisotropy study performed 
here. Finally, the dot-dashed grey line is the lower limit obtained from the
analysis of 15 dwarf spheroidal galaxies in Ref.~\cite{Baring:2015sza}. Our
REF scenario is always below this line, at least by a factor of 2.

In Appendix~\ref{sec:other_channel} we include the exclusion limits for the
$\tau$- and $\mu$-channels.

When fitting with the 2-component model, the 91 $C_{\rm P, astro}^{i,j}$ can vary
independently and they react to the presence of the DM component by 
reproducing the measured APS in those combinations of energy bins where the
DM component is subdominant. Therefore, it may difficult to interpret a
best-fit set of $C_{\rm P,astro}^{i,j}$ in terms of one or more populations of
actual astrophysical sources, e.g. unresolved blazars or star-forming galaxies.
A more physical approach can be obtained by considering the phenomenological
description presented in Sec.~\ref{sec:interpretation}. In this case, the
astrophysical component in the 2-component fit is described by means of the 
two-broken-power-law scenario (see Tab.~\ref{tab:popfit}). The latter depends
on 8 free parameters instead of 91. We employ this revised version of the 
2-component model to fit the binned auto- and cross-APS $\overline{C_{\ell}}$
in all the combinations of energy bins. The exclusion limits on 
$\langle \sigma_{\rm ann}v \rangle$ and $\tau$ are obtained by finding the 
configuration that yields a $\chi^2$ that is larger by 3.84 than the best-fit 
$\chi^2$ of the null hypothesis (i.e. no DM). The resulting upper limits are 
compatible with the ones showed in Fig.~\ref{fig:exclusion_limits_bestfit}
and, therefore, we decided not to show them.

\section{Discussion and conclusion}
\label{sec:conclusion}

In this paper we measure the auto-correlation and cross-correlation angular 
power spectrum (APS) of the diffuse gamma-ray emission detected by {\it Fermi} 
LAT at high Galactic latitudes in 81 months of observation. The measurement 
builds on a similar analysis based on 22 months of data and published in 
Ref.~\cite{Ackermann:2012uf}. With respect to the latter, this work takes 
advantage of the larger statistics, as well as of the improved event 
reconstruction achieved for Pass 7 Reprocessed events and instrument response 
functions. Other improvements, with respect to Ref.~\cite{Ackermann:2012uf}, 
consist of a revised method for binning the data in multipole and to compute 
the Poissonian auto- and cross-APS. We also correct the estimate of the photon 
noise and we employ a different method to account for the effect of the mask. 
Finally, we consider a more recent model of the diffuse Galactic foreground 
associated with the Milky Way (MW) disk.

The second part of the paper focuses on the auto- and cross-APS expected
from annihilation or decay of DM. We employ a hybrid approach to model the
distribution of DM, making use of catalogs of DM halos and subhalos from 
state-of-the-art $N$-body simulations and combining them with analytical 
recipes to account for DM structures below the mass resolution of the 
simulations. The methodology follows what was done in 
Ref.~\cite{Fornasa:2012gu}. Compared to the latter, this work discards the 
possibility of very large subhalo boost factors induced by na\"ive power-law 
extrapolations of the concentration parameter to low halo masses. We also 
account for the uncertainty associated with the mass of the MW, and we correct 
for the possibility of having very bright Galactic DM subhalos that would be 
individually resolved as gamma-ray sources. 

The main results of this papers are summarized in the following list.
\begin{itemize}
\item {\it Detection of auto- and cross-APS:} because of the 
instrumental improvements and of the refinements in the analysis mentioned
before, the measurement presented here probes a larger energy range (compared
to the original analysis in Ref.~\cite{Ackermann:2012uf}), between 0.5 and 
500 GeV, divided in 13 energy bins. We also compute, for the first time, the 
cross-APS between different energy bins. We detect significant auto-APS in 
almost all the energy bins below 21.83 GeV. Significant cross-APS is also 
measured in most combinations of energy bins (see Tabs.~\ref{tab:cp_3FGL} and 
\ref{tab:cp_2FGL}).

\item {\it  Independence on angular multipole:} our results cover a 
larger range in multipoles than the original analysis, i.e., from $\ell=49$ 
and 706. In this multipole range, the detected auto- and cross-APS are 
consistent with being Poissonian, i.e. constant in multipole. An alternative 
$\ell$-dependent model is also employed to fit the data but there is no 
significant preference for the $\ell$-dependent model over the Poissonian 
interpretation. If future data sets were able to detect a non-Poissonian 
behaviour, it would represent the first detection of scale dependence in 
gamma-ray anisotropies. Such a result would provide valuable insight into the 
nature of the Isotropic Gamma-Ray Background (IGRB), e.g. an upper limit on 
the contribution of sources like blazars or misaligned active galactic nuclei, 
which are associated with a Poissonian APS. It would also probe other possible 
sources like star-forming galaxies or Dark Matter (DM) structures, from which 
we expect a $\ell$-dependent auto- and cross-APS \cite{Ando:2009nk}.

\item {\it Detection of multiple source classes:} the anisotropy energy 
spectrum (i.e. the dependence of the auto- and cross-APS on the energy) is not 
featureless and it is best fitted by two populations of sources with 
broken-power-law energy spectra. The interpretation in terms of only one 
source population (whether emitting as a power law or broken power law) is 
excluded at 95\% CL. This suggests that the auto- and cross-APS result from a 
class of objects emitting mainly at low energies with a soft energy spectrum 
$\propto E^{-\alpha_1}$ with $\alpha_1 \sim$2.58, and a second population of 
harder objects with $\alpha_2\sim$2.10. The cross-over between the two source 
classes, according to our fit, happens at approximately 2 GeV. The harder 
spectral slope is compatible with that expected from BL Lacertae 
\cite{Acero:2015gva}, which are thought to dominate the IGRB at high energies. 
At lower energies, the spectral slope is similar to that of Flat-Spectrum 
Radio Quasars \cite{Ajello:2011zi} or of normal star-forming galaxies 
\cite{Ando:2009nk,Tamborra:2014xia} (see also Ref.~\cite{Brown:2016sbl}).

\item {\it Presence of an high-energy cut-off:} our best-fit 
interpretation shows a cut-off at around 85 GeV. This may be related to the 
absorption of the extragalactic background light (EBL), since a similar
feature is detected in the intensity energy spectrum of the IGRB in 
Ref.~\cite{Ackermann:2014usa}\footnote{The cut-off in the IGRB energy spectrum 
detected in Ref.~\cite{Ackermann:2014usa} is at slightly higher energies, 
namely around 200 GeV, depending on the model of the diffuse Galactic 
foreground employed. Notice, however, that the measurement in 
Ref.~\cite{Ackermann:2014usa} is performed masking the sources in the 2FGL,
while the value of $E_b$=85 GeV quoted above refers to the measurement after
masking the sources in the 3FGL. Thus, it is possible for the two cut-offs to 
be located at different energies.}. If we were able to confirm that the 
cut-off is associated with the EBL, this would be the first time that the 
absorption by the EBL is detected via anisotropies. One way to achieve such a 
confirmation would be to detect a significant cross-correlation between the 
same data set employed here and a catalog of tracers of the Large-Scale 
Structure of the Universe. The possibility of binning the catalog in redshift 
would allow us to perform a tomographic analysis and to select the emission 
coming from different comoving distances \cite{Ando:2013xwa,Ando:2014aoa}. 
Alternatively, the cut-off may be an intrinsic feature of the energy spectrum 
of the sources responsible for the auto- and cross-APS at high energies.

\item {\it Systematic uncertainties in the anisotropies induced by 
annihilating DM:} in the case of an annihilating DM candidate, an uncertainty 
of a factor 4 in the mass of the MW induces a variation of a factor $\sim$30 
in the auto- and cross-APS associated with Galactic subhalos. For a MW mass of 
the order of $10^{12} M_\odot$, Galactic subhalos dominate the expected 
anisotropic signal from DM. If the MW is less massive, i.e., few times 
$10^{11} M_\odot$, the extragalactic component starts to be important, at least 
for large subhalo boost models. For DM annihilations occuring in extragalactic 
DM halos and subhalos, the uncertainties on the subhalo boost factor (for a 
fixed $M_{\rm min}$) induce an uncertainty of a factor $\sim$20 on the expected 
auto- and cross-APS from this component. The gamma-ray emission produced by DM 
annihilations in the smooth halo of the MW generates a negligible anisotropic 
signal outside the adopted mask. The overall uncertainty on the predicted 
DM-induced APS (for a fixed $M_{\rm min}$) is of a factor of 20, similar to the 
one estimated in Ref.~\cite{Ackermann:2015tah} for the intensity of all-sky 
gamma-ray emission. Changing the value of $M_{\rm min}$ from $10^{-12}$ to 1 
$M_\odot$ approximately doubles the systematic uncertainty.

\item {\it Systematic uncertainties in the anisotropies induced by 
decaying DM:} In the case of decaying DM, the extragalactic signal dominates 
the expected auto- and cross-APS and the prediction is independent of the 
value of the subhalo boost factor. Decays in the smooth halo of the MW or in 
its subhalos are subdominant. The overall uncertainty (for a fixed value of
$M_{\rm min}$) is less than a factor of 2. Varying $M_{\rm min}$ over the range
mentioned before has a negligible effect in the case of decaying DM.

\item {\it Conservative exclusion limits on DM:} requiring that the 
DM-induced auto- and cross-APS does not exceed the measurement in any energy 
bin or combination of energy bins yields an upper limit on 
$\langle \sigma_{\rm ann}v \rangle$ that is at least a factor of 2 less 
stringent that the one obtained in Ref.~\cite{Ackermann:2015tah} from the 
analysis of the IGRB energy spectrum (for the REF scenario and the $b$ 
channel). In the case of annihilations into $b\bar{b}$, the constraint on the 
annihilation cross section reaches a value as low as 
$10^{-25} \mbox{cm}^3 \mbox{s}^{-1}$ for a DM mass of 5 GeV and, therefore, it 
is approximately two orders of magnitude less constraining than the one 
inferred from the observation of dwarf spheroidals galaxies. For decaying DM, 
the lower limit on $\tau$ is a factor of 5 weaker than the one from the IGRB 
intensity \cite{Ando:2015qda} and, at least, a factor of 5 weaker than the one 
from the analysis of dwarf spheroidal galaxies \cite{Baring:2015sza} (for the 
REF case and decays into $b\bar{b}$).

\item {\it Exclusion limits from the 2-component fit:} fitting the data 
with a 2-component model that includes DM provides more constraining exclusion 
limits. The resulting upper limit for annihilating DM (in the REF scenario for 
a $M_{\rm min}$ of $10^{-6} M_\odot/h$ and annihilations into $b\bar{b}$) is still 
a factor of 10 less constraining than the combined analysis of dwarf 
spheroidals from Ref.~\cite{Ackermann:2015zua}. However, below a DM mass of 30 
GeV, it is slightly better than what was derived in Ref.~\cite{Ajello:2015mfa} 
from the analysis of the IGRB intensity energy spectrum in terms of a 
2-component model. For decaying DM, the lower limits on $\tau$ are, at least, 
a factor of 2 less stringent than those obtained from the IGRB intensity 
energy spectrum \cite{Ando:2015qda} or from the combined analysis of dwarf 
spheroidals \cite{Baring:2015sza}.
\end{itemize}

The exclusion limits on DM, although they do not exclude new regions of the 
DM parameter space, are complementary to those computed from the intensity of 
the IGRB or from the observation of dwarf spheroidals and, therefore, they 
provide independent information. Also, they are expected to become more 
stringent as the measurement of the auto- and cross-APS will improve during 
the next years. Beside making use of the data collected after May 2015, future 
analyses will rely on Pass 8 data, benefiting from the new event classes and 
data selections available (see Sec.~\ref{sec:validations}). Also, future 
catalogs of sources, deeper than 3FGL, will explore faint sources that are now 
unresolved and will improve our modelling of those source classes. It will 
certainly be interesting to complement the measurement of gamma-ray 
anisotropies performed here with a similar observation at higher energies 
(which will be possible in the near future with the Cherenkov Telescope Array 
\cite{Acharya:2013sxa,Ripken:2012db}) or in the sub-GeV regime (with future 
satellites like ASTROGAM\footnote{http://astrogam.iaps.inaf.it/} and ComPair 
\cite{Moiseev:2015lva}). Finally, in a multi-messenger perspective, the study 
of gamma-ray anisotropies can be interfaced with similar analyses on the 
high-energy neutrinos recently discovered by IceCube \cite{Aartsen:2013bka,
Aartsen:2013jdh,Aartsen:2014gkd,Aartsen:2016xlq}. Since the  same sources that 
contribute to the IGRB (e.g. blazars, star-forming or radio galaxies) are 
expected to emit also neutrinos, the auto- and cross-APS measured in this work 
represents a useful indication for the {\it minimal} level of anisotropies 
that can be found in the distribution of neutrinos. A quantitative estimate 
of IceCube prospects to detect anisotropies can be found in 
Ref.~\cite{Aartsen:2014ivk}.

\begin{acknowledgments}
The \emph{Fermi} LAT Collaboration acknowledges generous ongoing support from 
a number of agencies and institutes that have supported both the development 
and the operation of the LAT as well as scientific data analysis. These 
include the National Aeronautics and Space Administration and the Department 
of Energy in the United States, the Commissariat \`a l'Energie Atomique and 
the Centre National de la Recherche Scientifique / Institut National de 
Physique Nucl\'eaire et de Physique des Particules in France, the Agenzia 
Spaziale Italiana and the Istituto Nazionale di Fisica Nucleare in Italy, the 
Ministry of Education, Culture, Sports, Science and Technology (MEXT), High 
Energy Accelerator Research Organization (KEK) and Japan Aerospace Exploration 
Agency (JAXA) in Japan, and the K.~A.~Wallenberg Foundation, the Swedish 
Research Council and the Swedish National Space Board in Sweden.
 
Additional support for science analysis during the operations phase is 
gratefully acknowledged from the Istituto Nazionale di Astrofisica in Italy 
and the Centre National d'\'Etudes Spatiales in France.

We thank Dr. Shin'ichiro Ando for useful discussion and for providing us with 
the lower limits on the decay lifetime from Ref.~\cite{Ando:2015qda}. We
also thanks Prof. John F. Beacom for the valuable discussion during the last
stages of the work.

MF gratefully acknowledges support from the Netherlands Organization for 
Scientific Research (NWO) through a Vidi grant (P.I.: Dr. Shin'ichiro Ando),
from the the Leverhulme Trust and the project MultiDark CSD2009-00064. The 
Dark Cosmology Centre is funded by the DNRF. JZ is supported by the EU under a 
Marie Curie International Incoming Fellowship, contract PIIF-GA-2013-62772. 
JG acknowledges support from NASA through Einstein Postdoctoral Fellowship 
grant PF1-120089 awarded by the Chandra X-ray Center, which is operated by the 
Smithsonian Astrophysical Observatory for NASA under contract NAS8-03060, and 
from a Marie Curie International Incoming Fellowship, contract 
PIIF-GA-2013-628997. GAGV is supported by CONICYT 
FONDECYT/POSTODOCTORADO/3160153, the Spanish MINECO's Consolider Ingenio 2010
Programme under grant MultiDark CSD2009-00064 and also partly by MINECO
under grant FPA2012-34694. MASC is a Wenner-Gren Fellow and acknowldeges the
support of the Wenner-Gren Foundantions to develop his research. EK thanks 
J.U. Lange for useful discussion on the bias in $C_{\rm P}$ due to binning 
presented in Sec.~\ref{sec:clvalidation}. FP acknowledges support from the 
Spanish MICINNs Consolider-Ingenio 2010 Programme under grant MultiDark 
CSD2009-00064, MINECO Centro de Excelencia Severo Ochoa Programme under grant 
SEV-2012-0249 and MINECO grant AYA2014-60641-C2-1-P. FZ acknowledges the 
support of the NWO through a Veni grant.
\end{acknowledgments}

\appendix
\section{Derivation of the photon noise $C_{\rm N}$}
\label{sec:photon_noise}
Let $n_i$ be the number of photons in a pixel $i$, $\bar n_i$ be its 
expectation value, and $\delta n_i\equiv n_i-\bar n_i$ be a fluctuation around 
the mean. The photon flux is given by $n_i/A_{i}$, where $A_i$ is the exposure 
in pixel $i$. Then, the photon flux per unit solid angle is given by 
$n_i/(A_{i}\Omega_{\rm pix})$ where $\Omega_{\rm pix}$ is the solid angle of 
pixel $i$.

The spherical harmonics coefficients $a_{\ell,m}$ are 
\begin{equation}
a_{\ell,m}=\int d\Omega_i \, \frac{\delta n_i}{A_{i}\Omega_{\rm pix}}
Y^*_{\ell,m}(\Omega_i) \,,
\end{equation}
where $\Omega_i$ denotes the direction of pixel $i$. The expectation value 
of the product between two coefficients is
\begin{equation}
\langle a_{\ell,m}a_{\ell',m'}^*\rangle = 
\int d\Omega_i \int d\Omega_j \, 
\frac{\langle\delta n_i\delta n_j\rangle}{A_{i}A_j\Omega_{\rm pix}^2} \, 
Y_{\ell m}^*(\Omega_i)Y_{\ell' m'}(\Omega_j) \,.
\end{equation}
If $\delta n_i$ is purely Poisson noise, 
$\langle \delta n_i \delta n_j \rangle / \Omega_{\rm pix}^2 = (\bar n_i/\Omega_{\rm pix}) \, \delta(\Omega_i-\Omega_j)$ where $\delta$ is the Dirac delta function. 
Thus:
\begin{equation}
\langle a_{\ell,m}a_{\ell',m'}^*\rangle = 
\int d\Omega_i \, \frac{\bar n_i}{A^2_{i}\Omega_{\rm pix}} \, 
Y_{\ell m}(\Omega_i)Y_{\ell' m'}^*(\Omega_i) \,.
\end{equation}
Now, we calculate the diagonal element, i.e., 
$C_\ell \equiv \sum_m \langle a_{\ell,m}a_{\ell,m}^* \rangle / (2\ell+1)$, 
obtaining:
\begin{equation}
C_\ell = \int \frac{d\Omega_i}{4\pi} \, \frac{\bar n_i}{A^2_{i}\Omega_{\rm pix}}
\,.
\end{equation}
The latter is independent on multipole $\ell$ and equivalent to the definition
of $C_{\rm N}^k$ in Eq.~\ref{eq:noise}.

\section{Auto-correlation angular spectra for all the energy bins}
\label{sec:appendixauto}
\begin{figure*}
\includegraphics[width=0.49\textwidth]{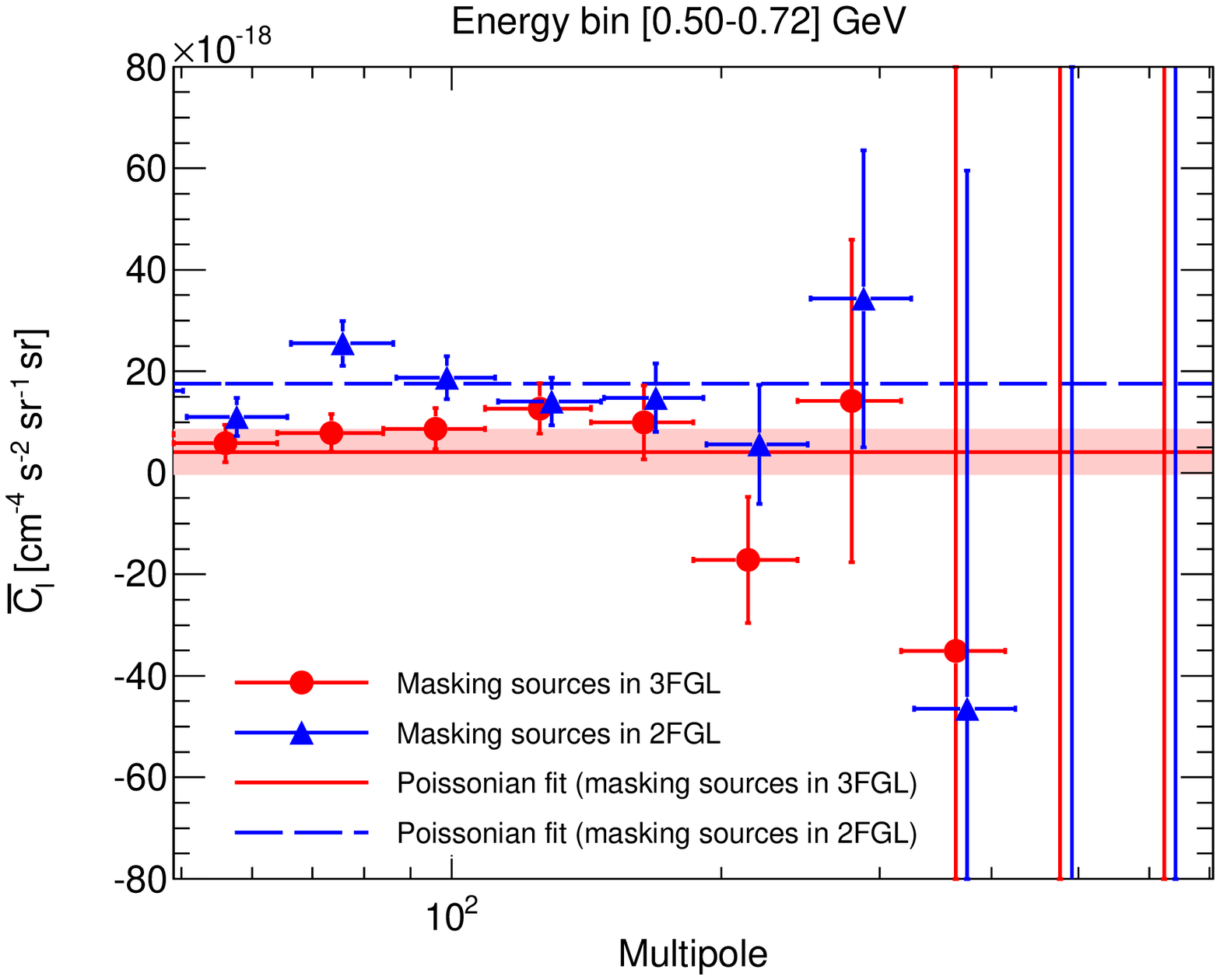}
\includegraphics[width=0.49\textwidth]{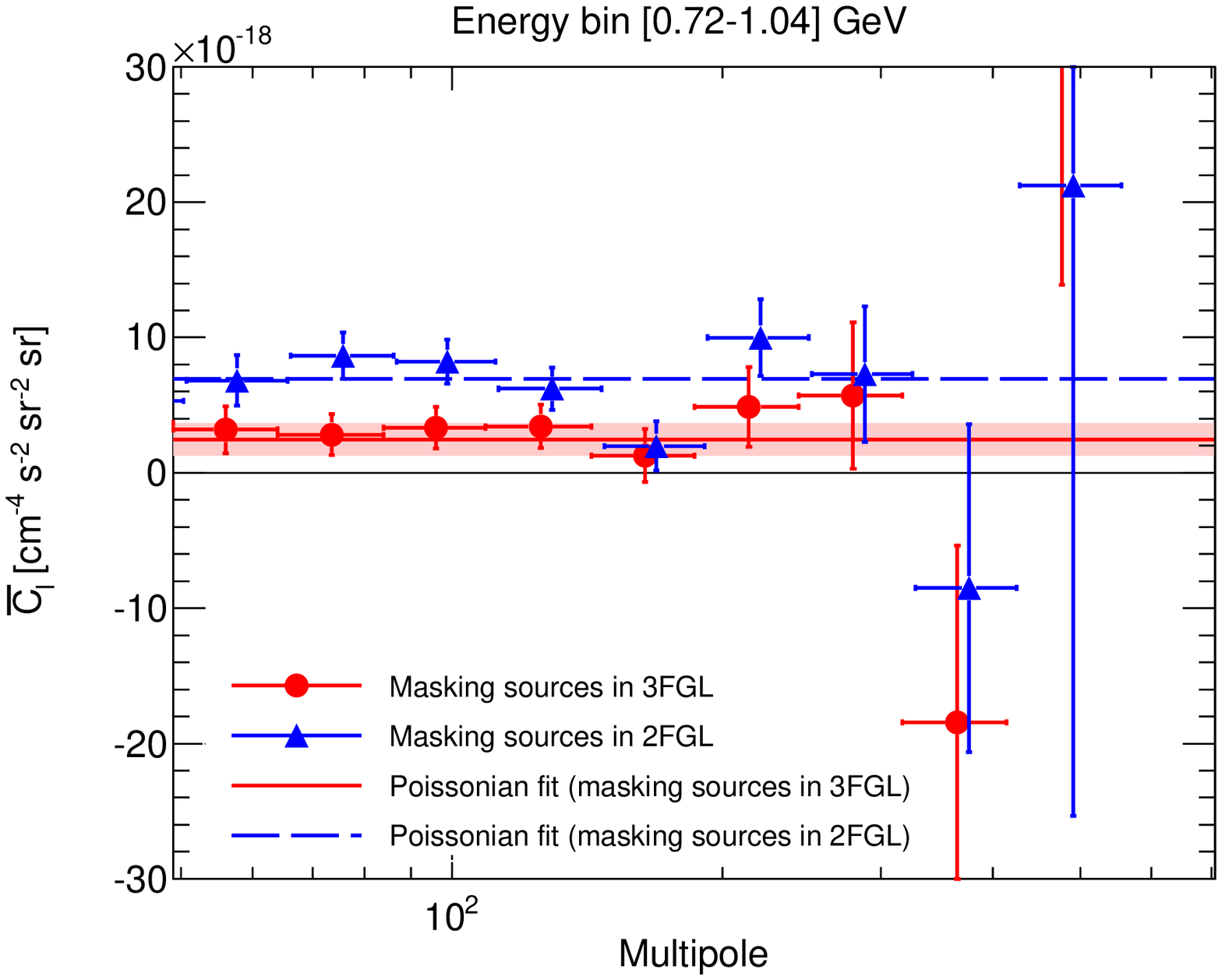}
\includegraphics[width=0.49\textwidth]{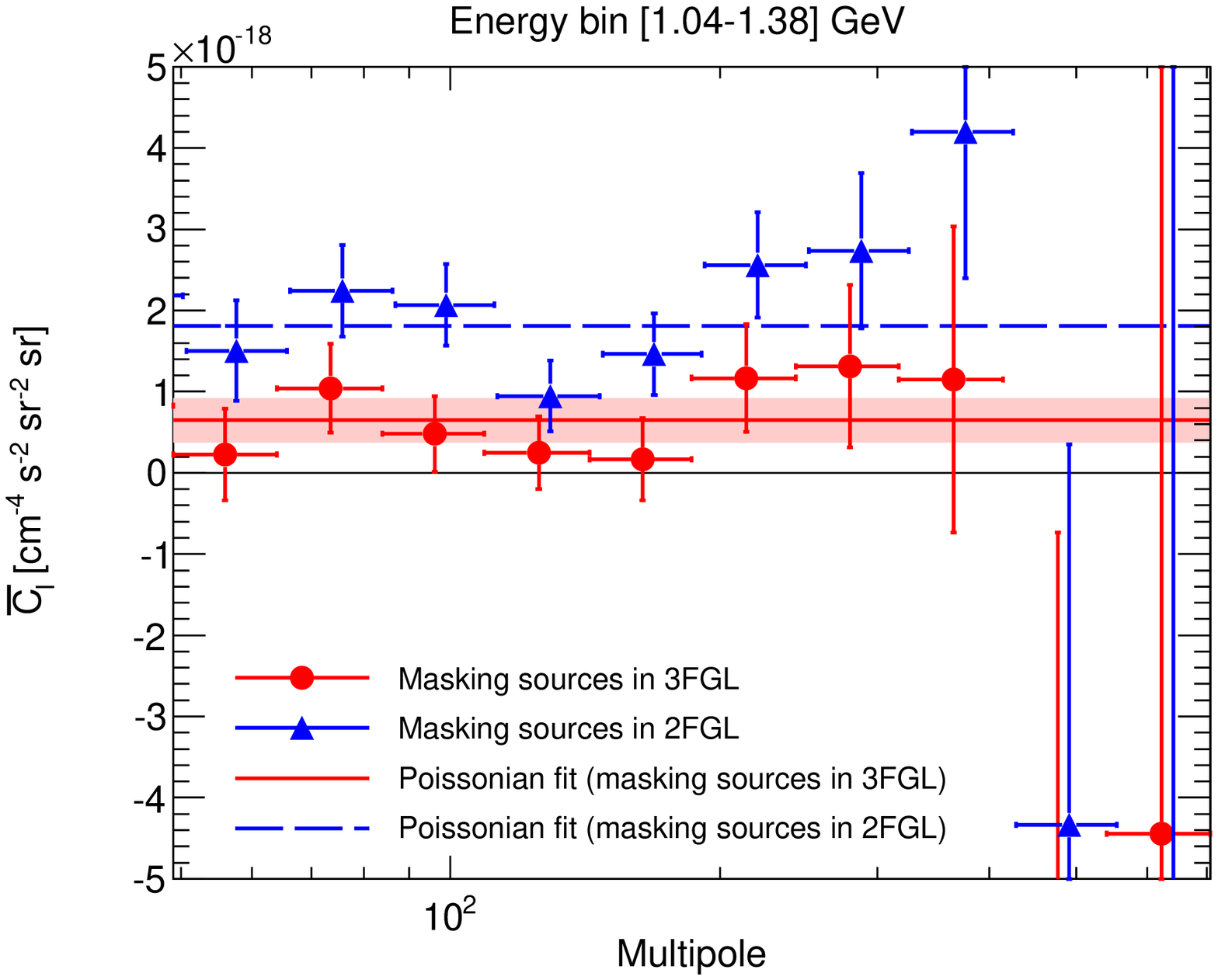}
\includegraphics[width=0.49\textwidth]{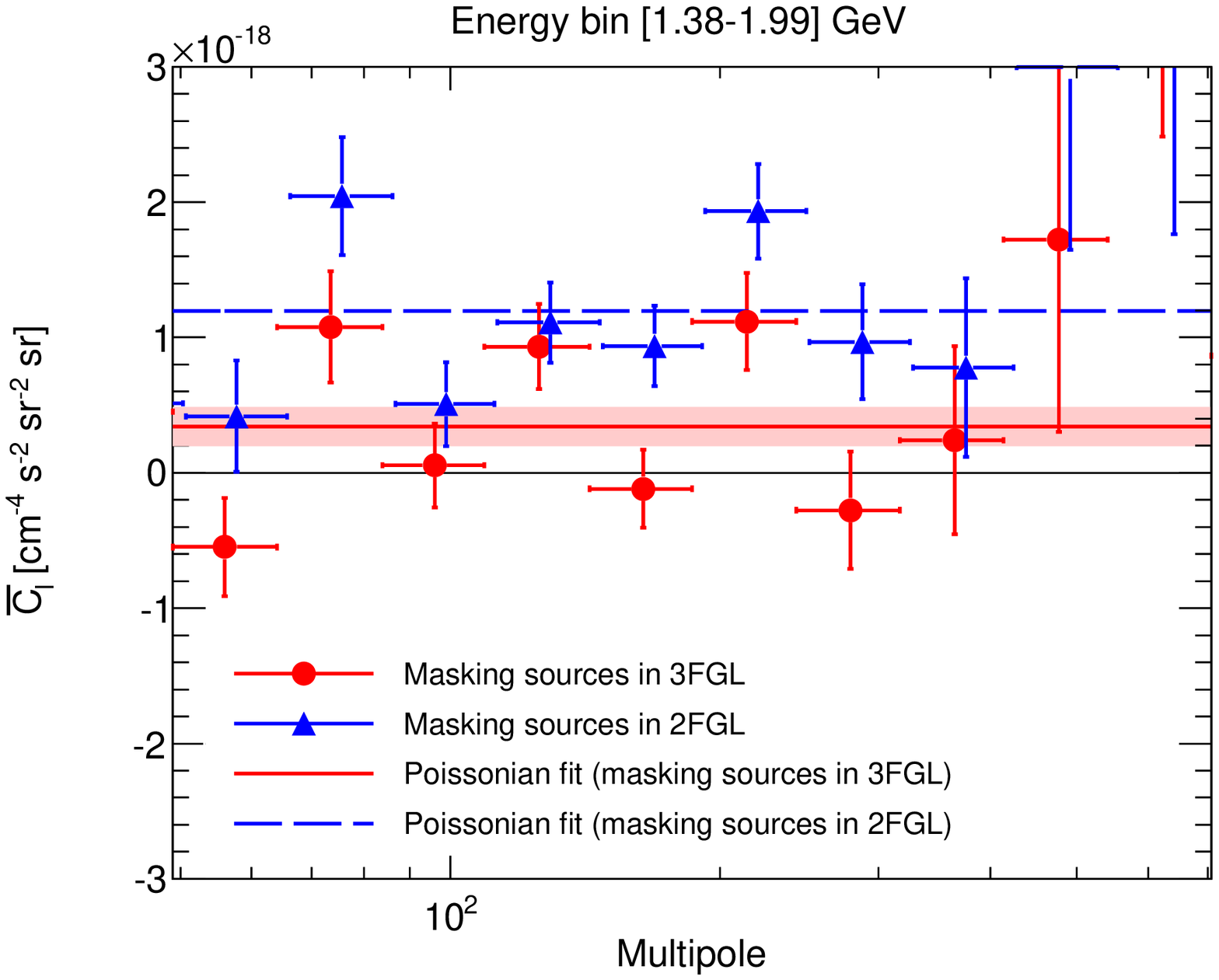}
\includegraphics[width=0.49\textwidth]{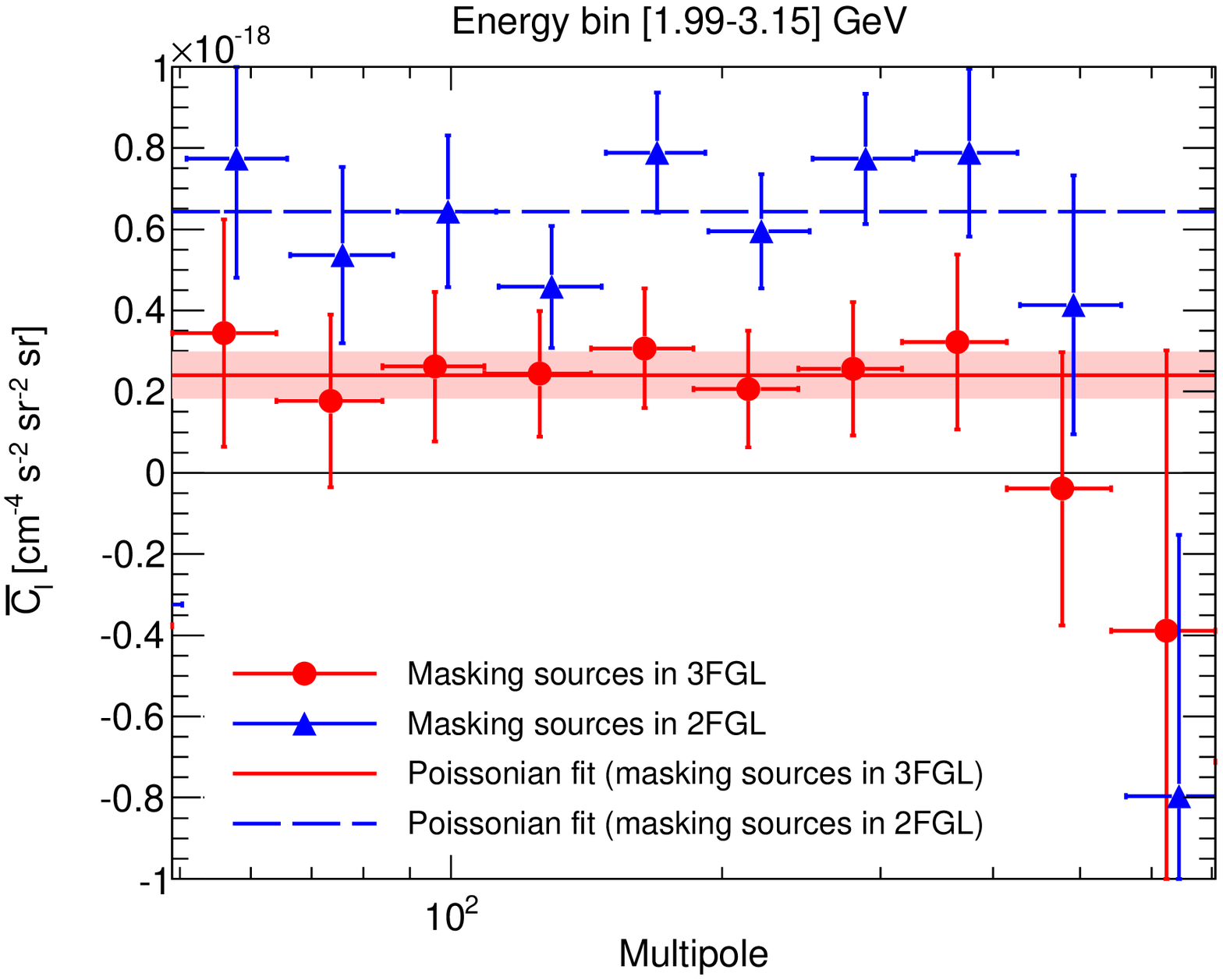}
\includegraphics[width=0.49\textwidth]{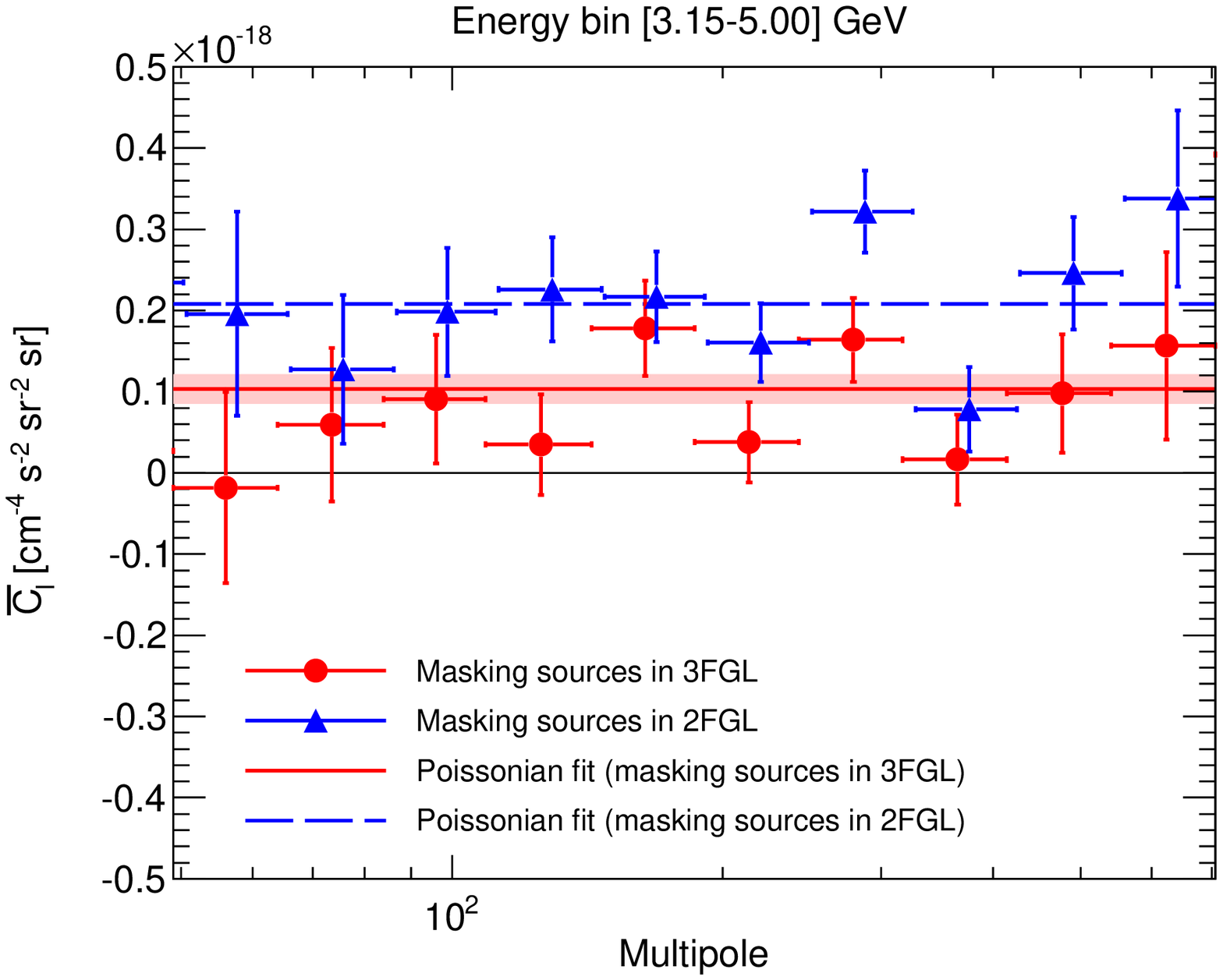}
\caption{\label{fig:cls1} Auto-APS of the IGRB for the first 6 energy bins and for the reference data set (P7REP\_ULTRACLEAN\_V15 front events) using the reference mask which excludes $|b|<30^\circ$ and 3FGL sources (red circles). The blue triangles show the same but masking the sources in 2FGL, instead. Data have been binned as described in Sec.~\ref{sec:clvalidation}. The solid red line shows the best-fit $C_{\rm P}$ for the red data points, with the pink band indicating its 68\% CL error. The dashed blue line corresponds to the best-fit $C_{\rm P}$ for the blue data points. The energy range is indicated on the top of each panel. Note that only the results in our signal region (i.e. between $\ell=49$ and 706) are plotted and that the scale of the $y$-axis can vary from panel to panel.}
\end{figure*}

\begin{figure*}
\includegraphics[width=0.49\textwidth]{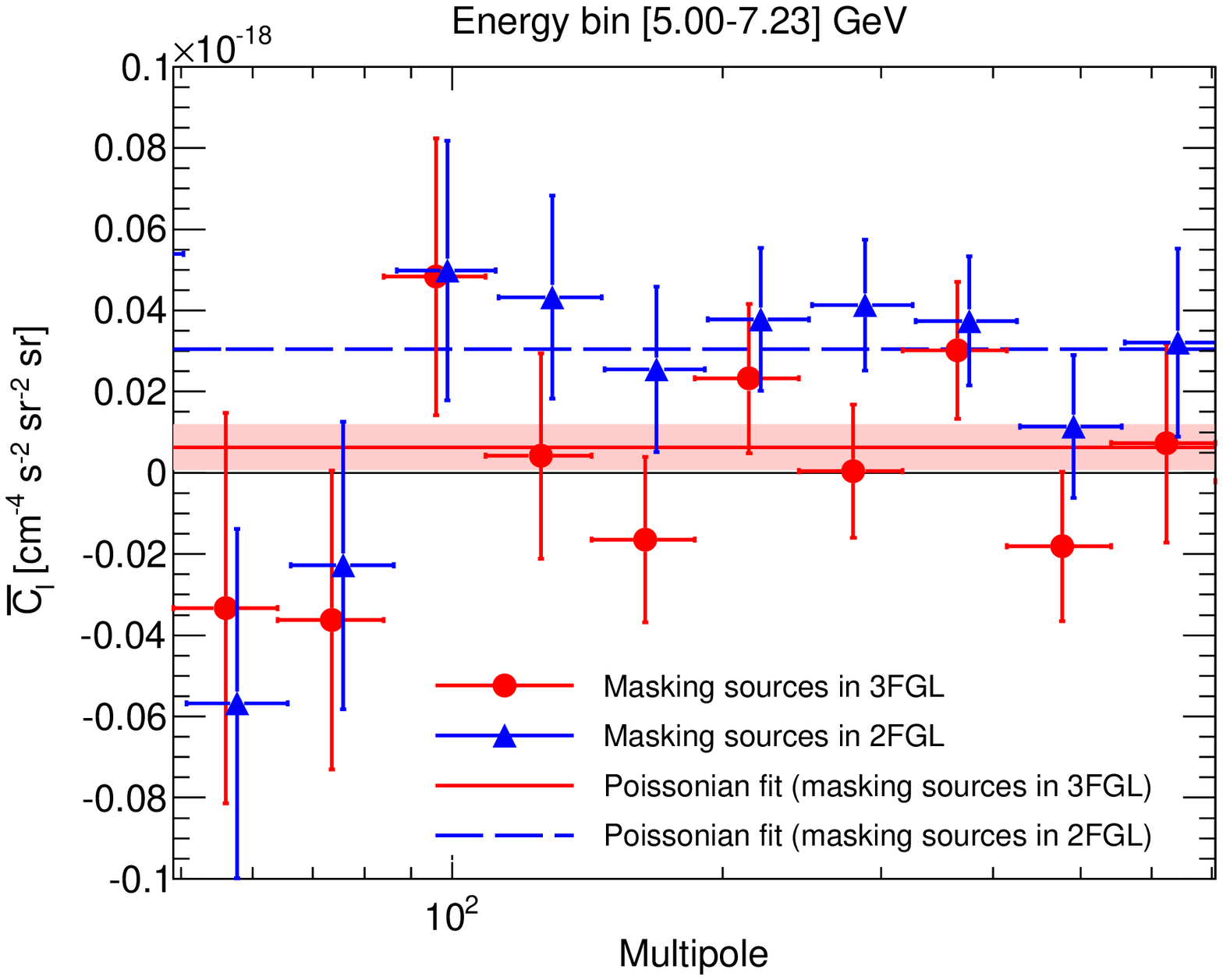}
\includegraphics[width=0.49\textwidth]{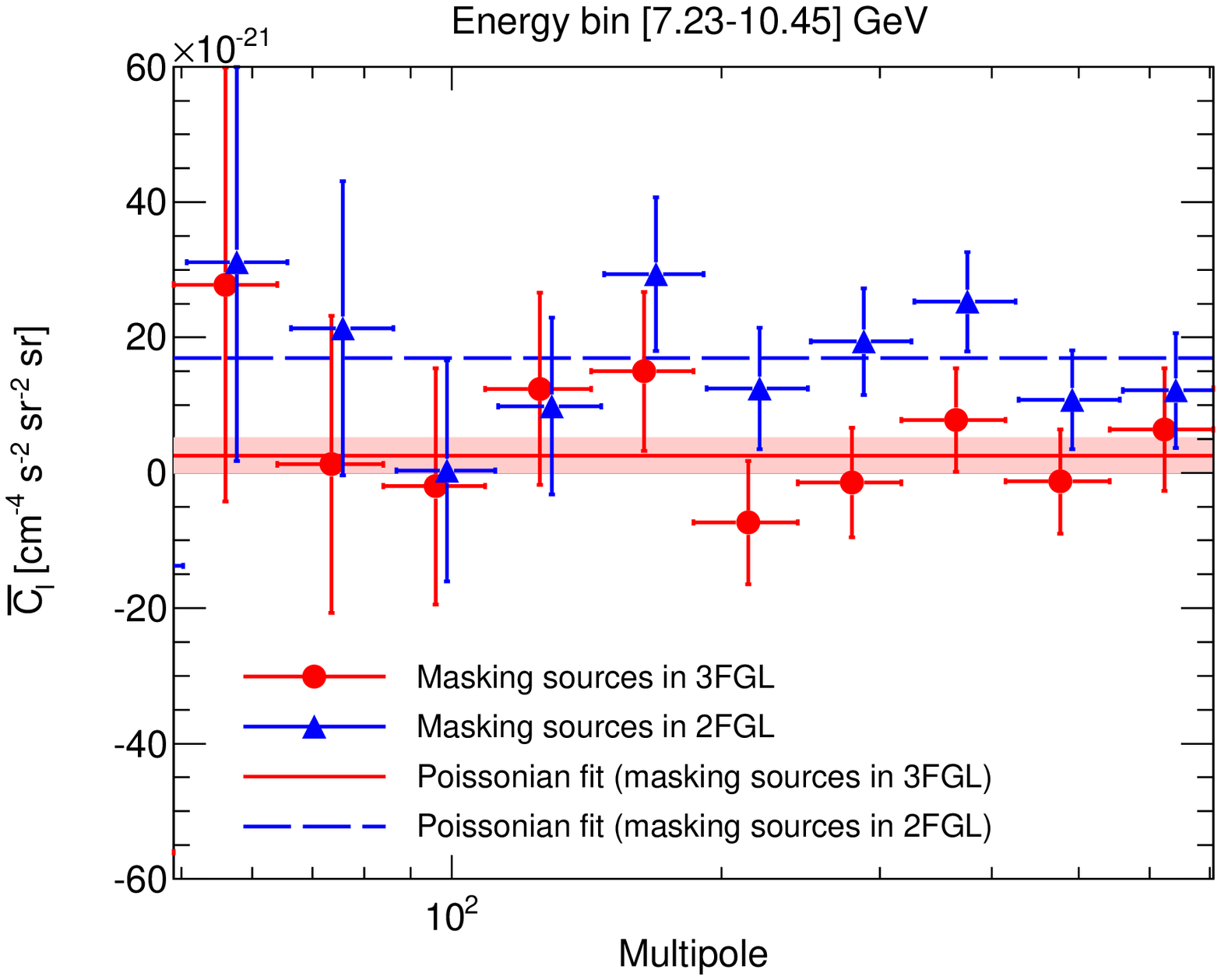}
\includegraphics[width=0.49\textwidth]{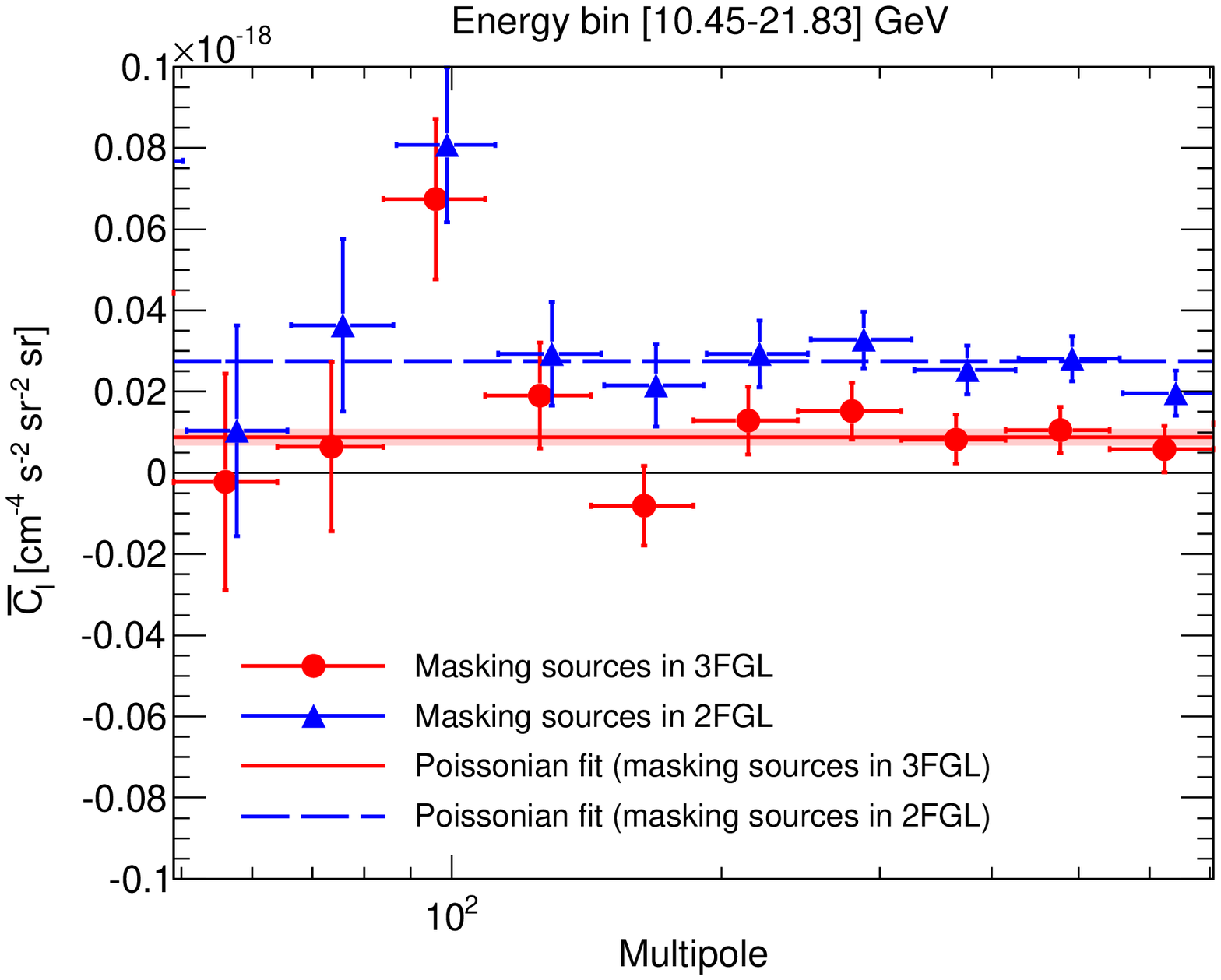}
\includegraphics[width=0.49\textwidth]{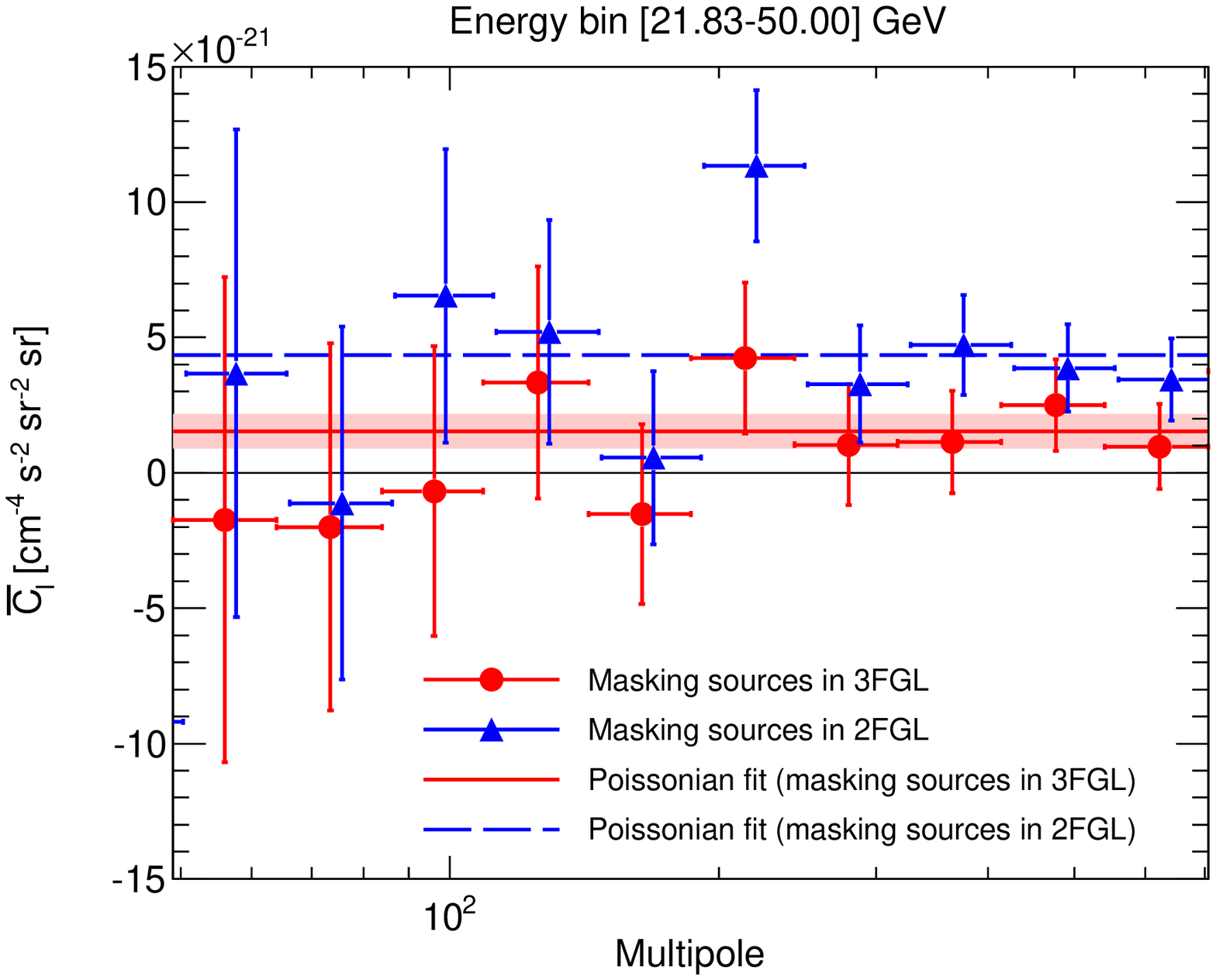}
\includegraphics[width=0.49\textwidth]{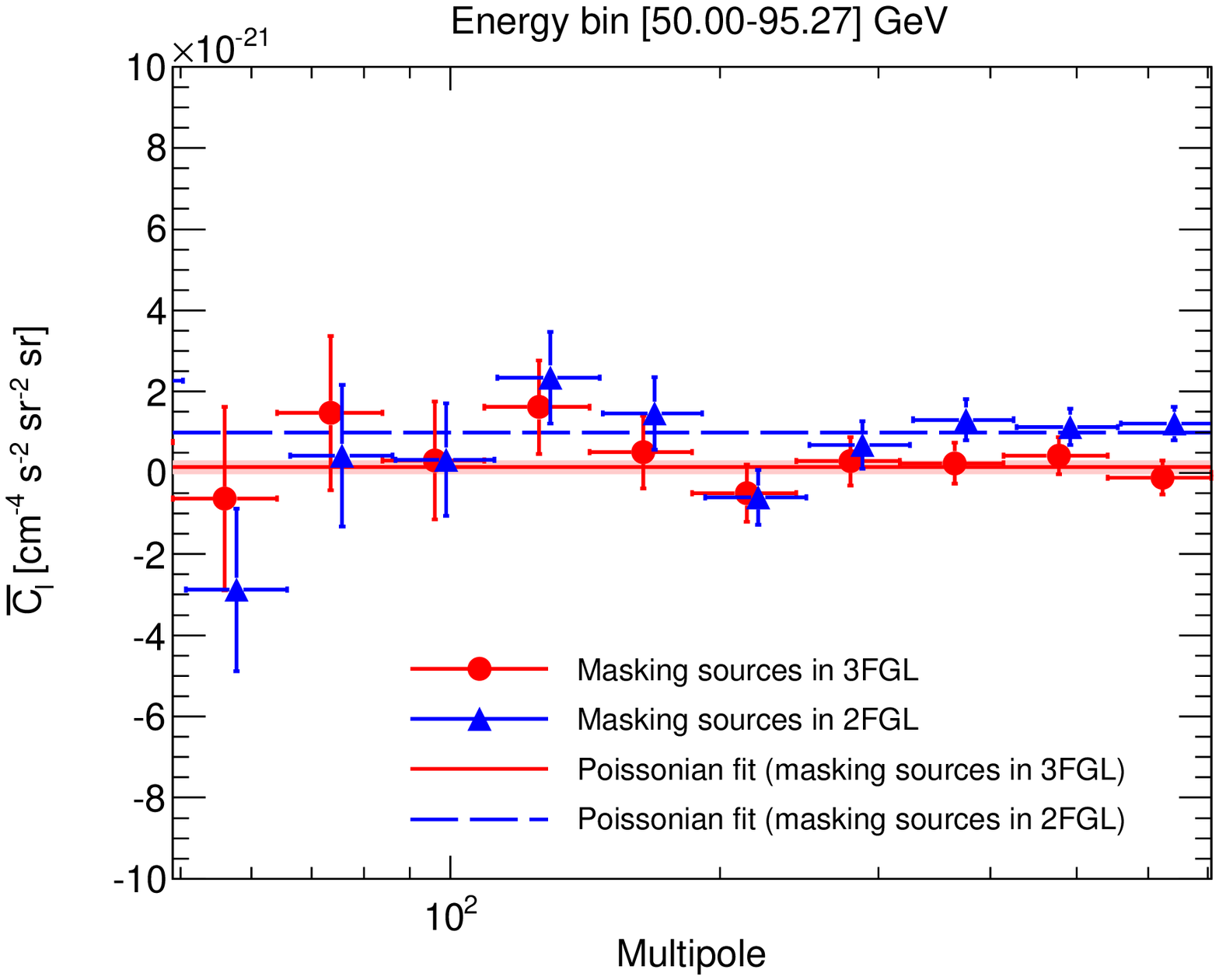}
\includegraphics[width=0.49\textwidth]{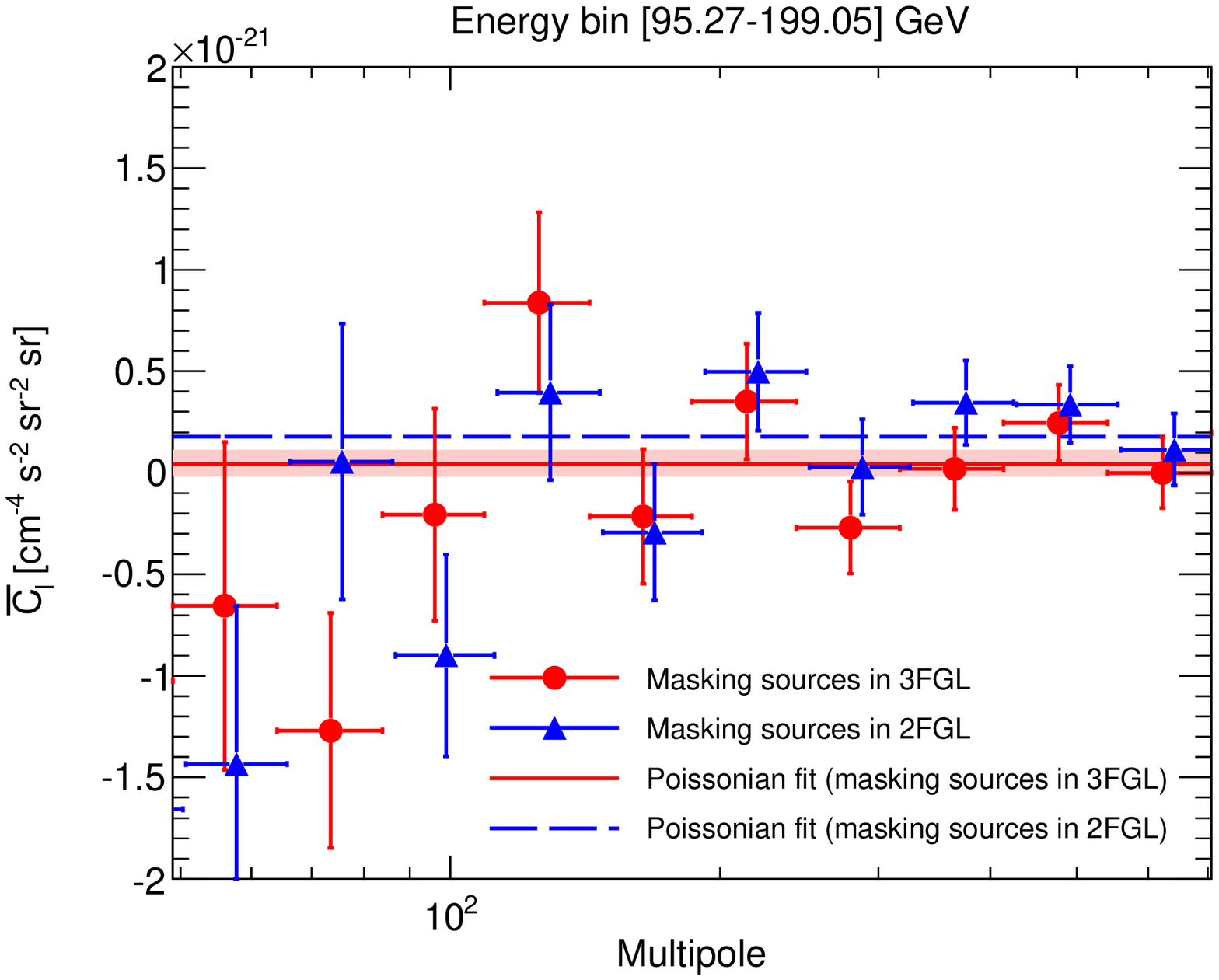}
\includegraphics[width=0.49\textwidth]{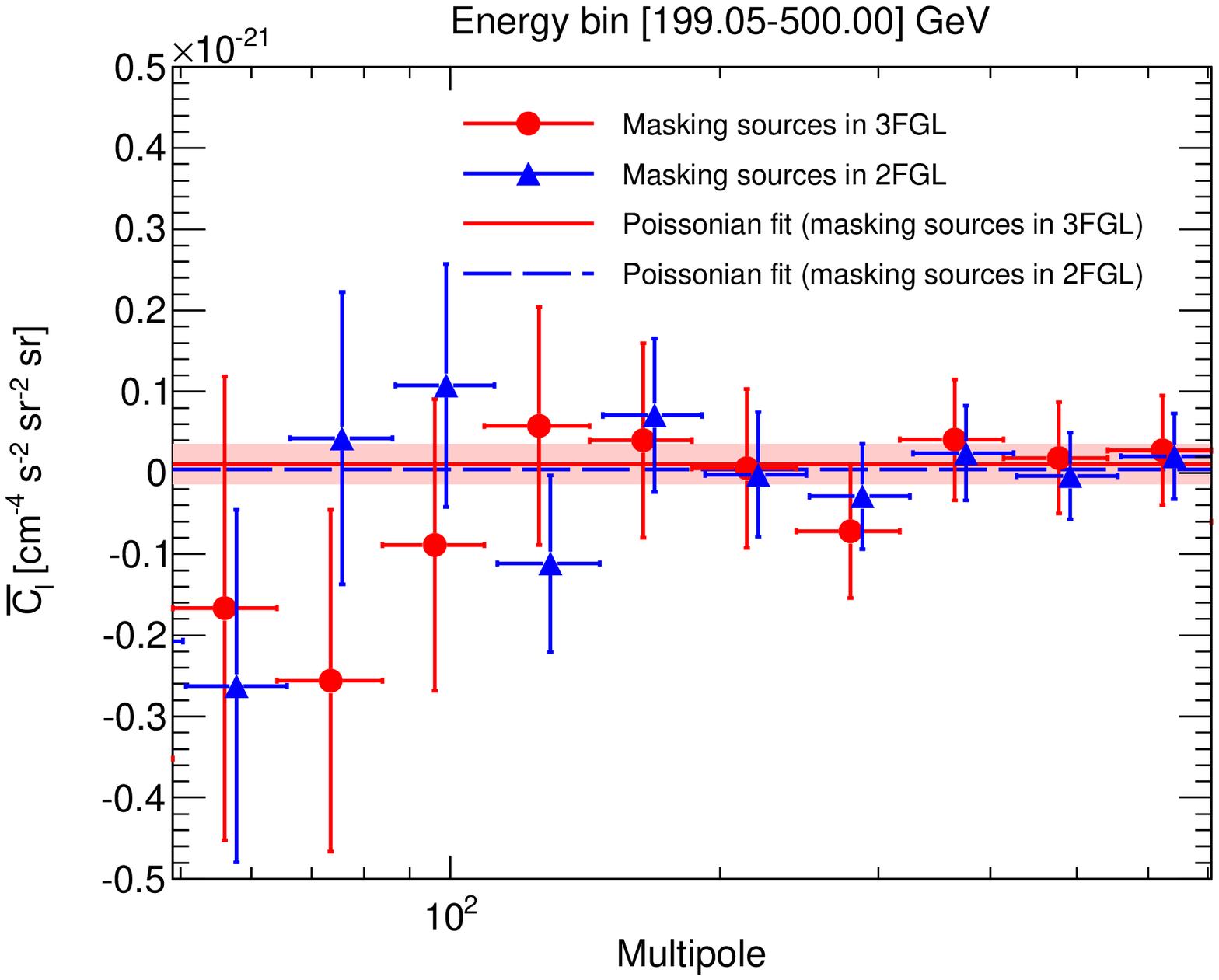}
\caption{\label{fig:cls2} Same as Fig.~\ref{fig:cls1} but for the last 7 energy bins.}
\end{figure*}

Figs.~\ref{fig:cls1} and \ref{fig:cls2} show the binned auto-APS 
$\overline{C_\ell}$ obtained as described in Sec.~\ref{sec:analysis}, for all
13 energy bins considered. The auto-APS is shown only within the signal 
region, i.e. between a multipole of 49 and 706. Red circles refer to the data 
set obtained with our default mask covering the sources in 3FGL and the solid 
red line marks the corresponding best-fit $C_{\rm P}$. The pink band denotes the 
68\% CL error on $C_{\rm P}$. The blue data points are for the same data set but 
using the default mask covering sources in 2FGL. The dashed blue line stands 
for the Poissonian best-fit to the blue triangles. Data are available at 
https://www-glast.stanford.edu/pub\_data/552. 

\section{Anisotropy energy spectrum for the cross-correlation angular power spectrum}
\label{sec:appendix}
\begin{figure*}
\includegraphics[width=0.49\textwidth]{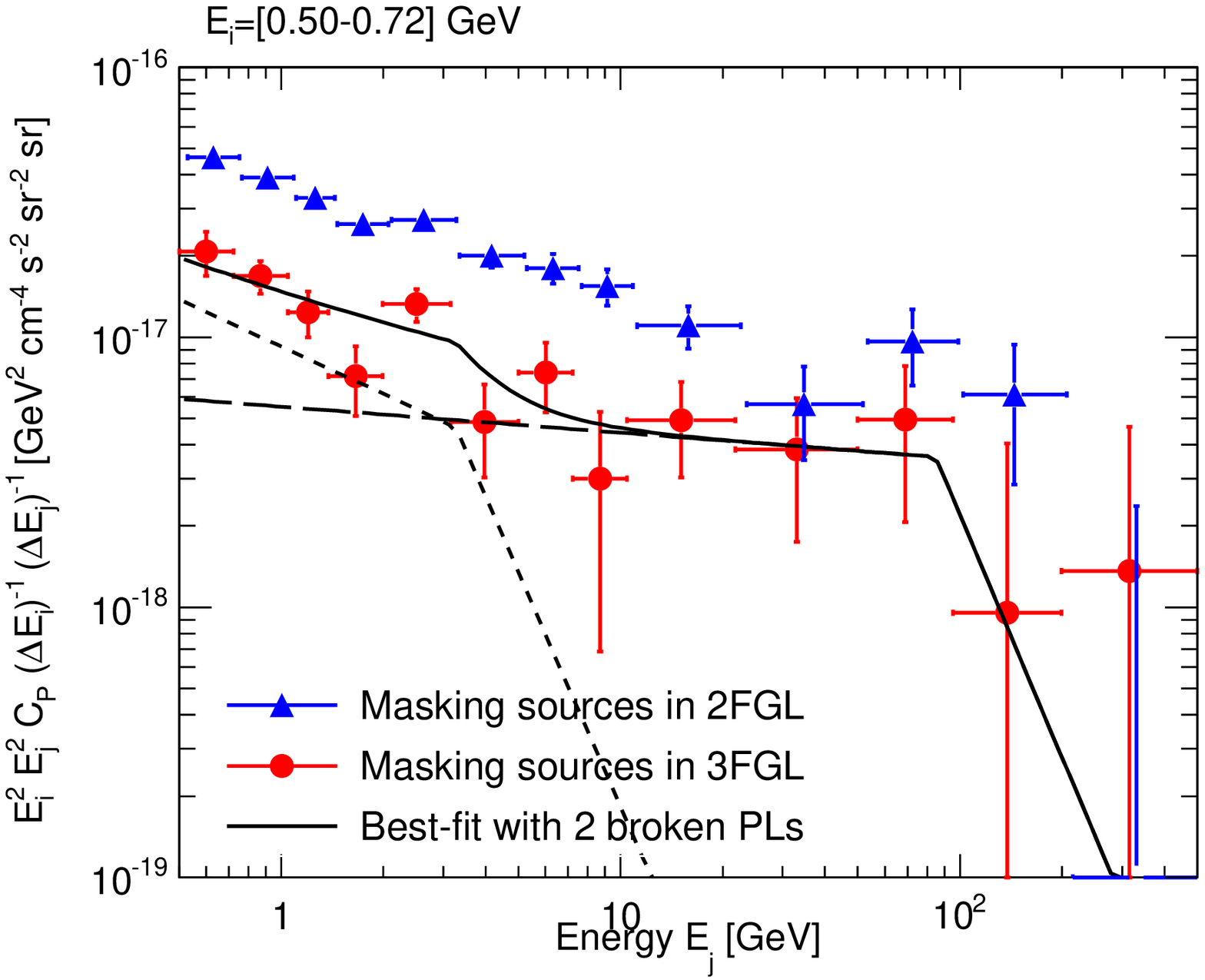}
\includegraphics[width=0.49\textwidth]{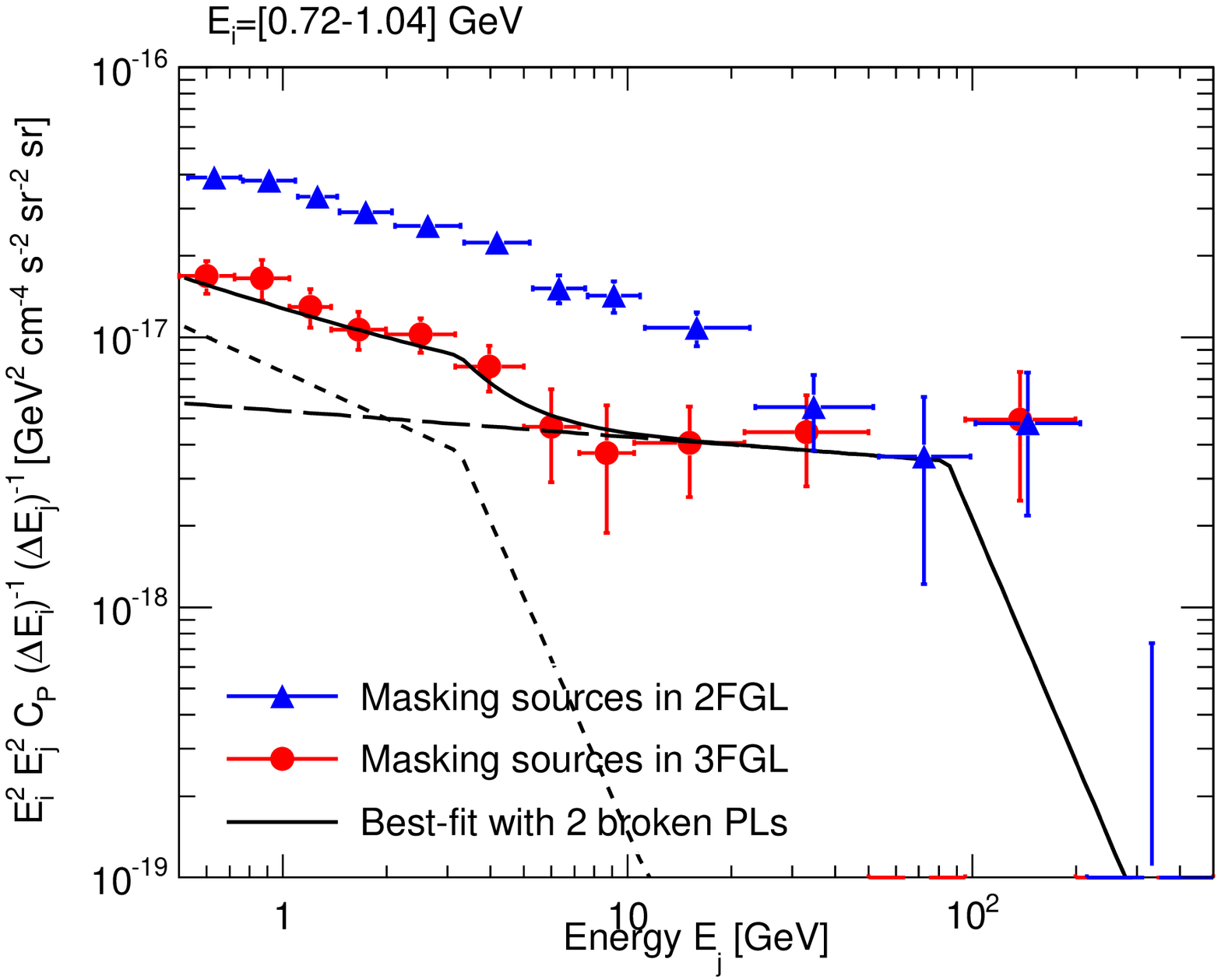}
\includegraphics[width=0.49\textwidth]{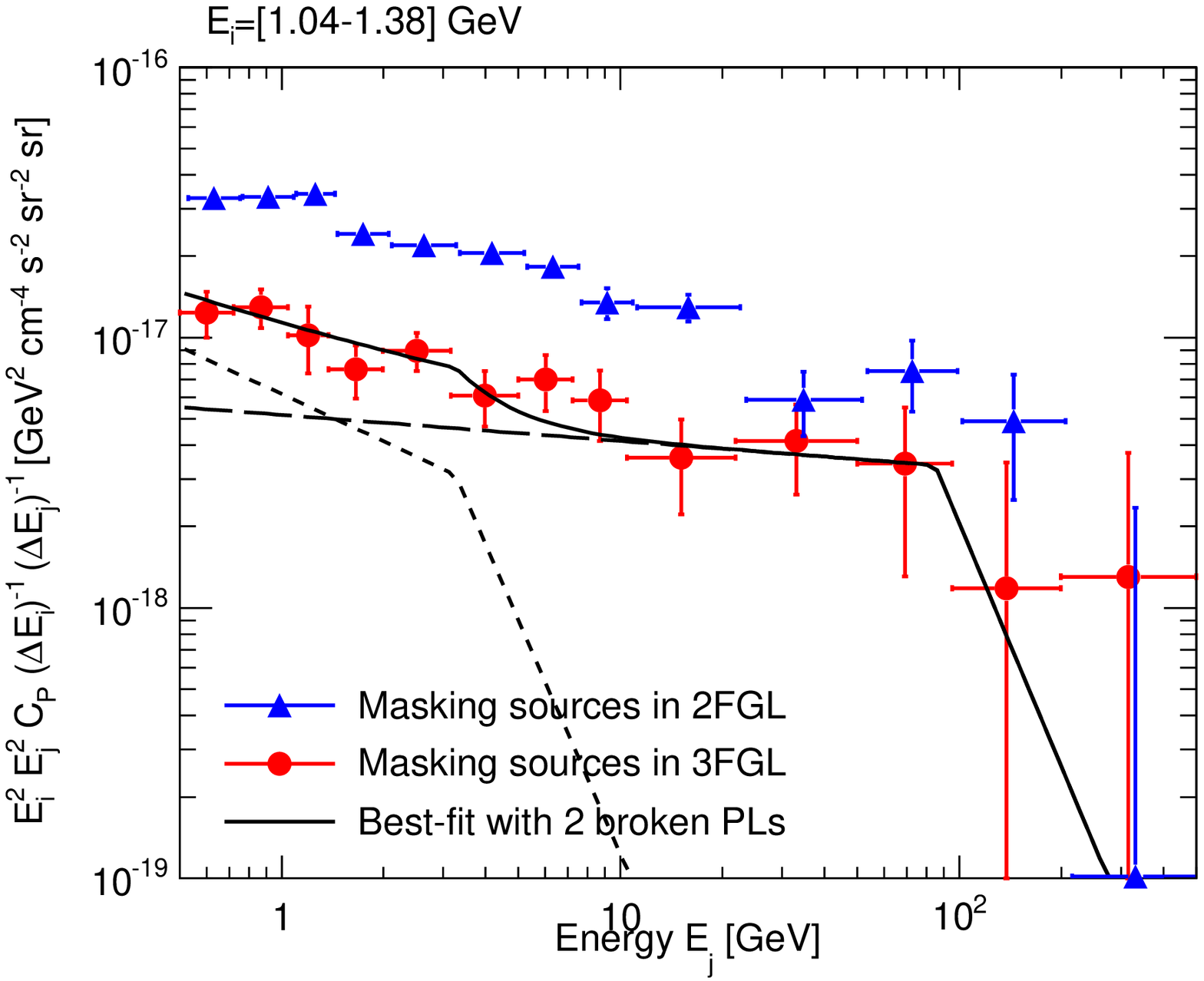}
\includegraphics[width=0.49\textwidth]{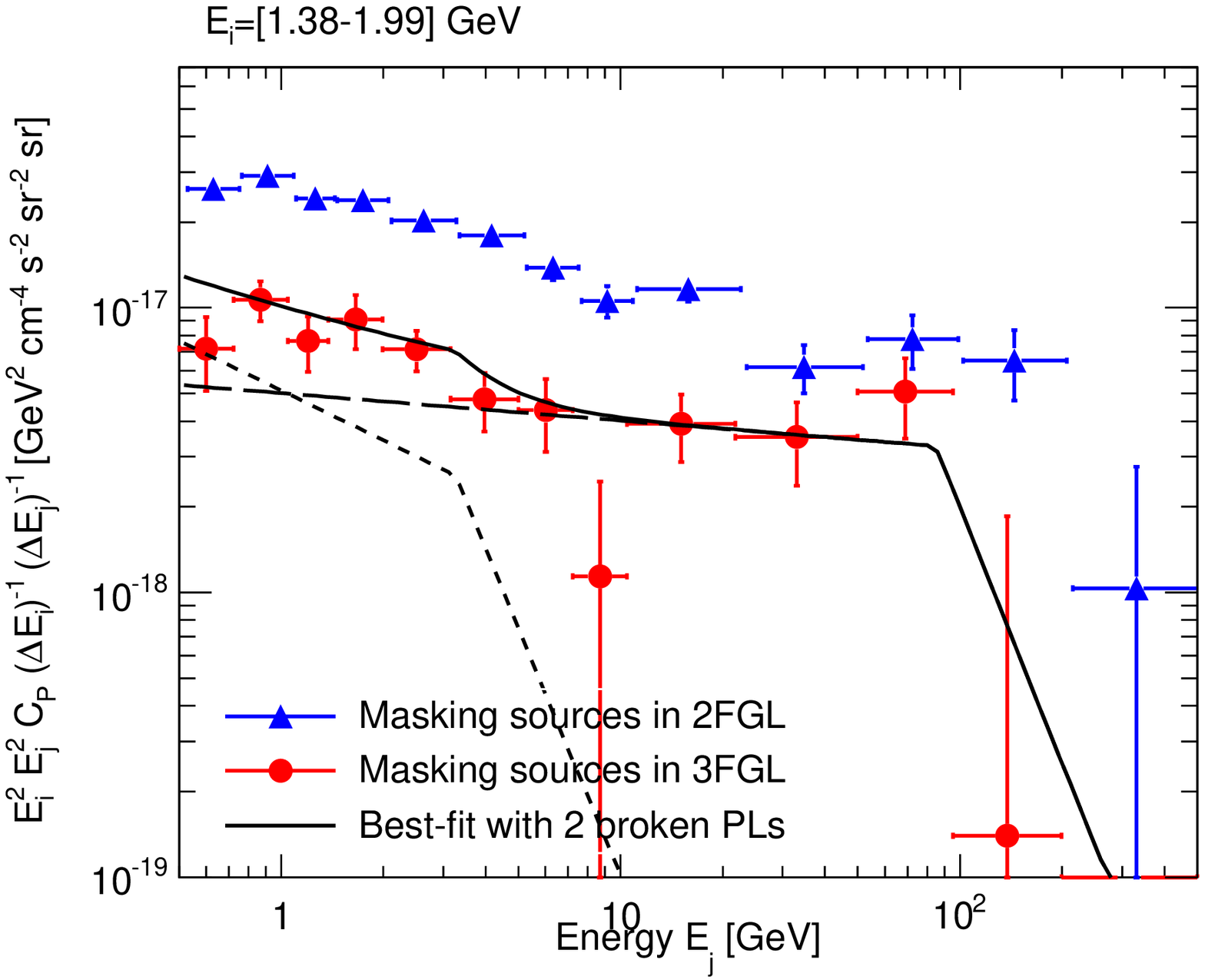}
\includegraphics[width=0.49\textwidth]{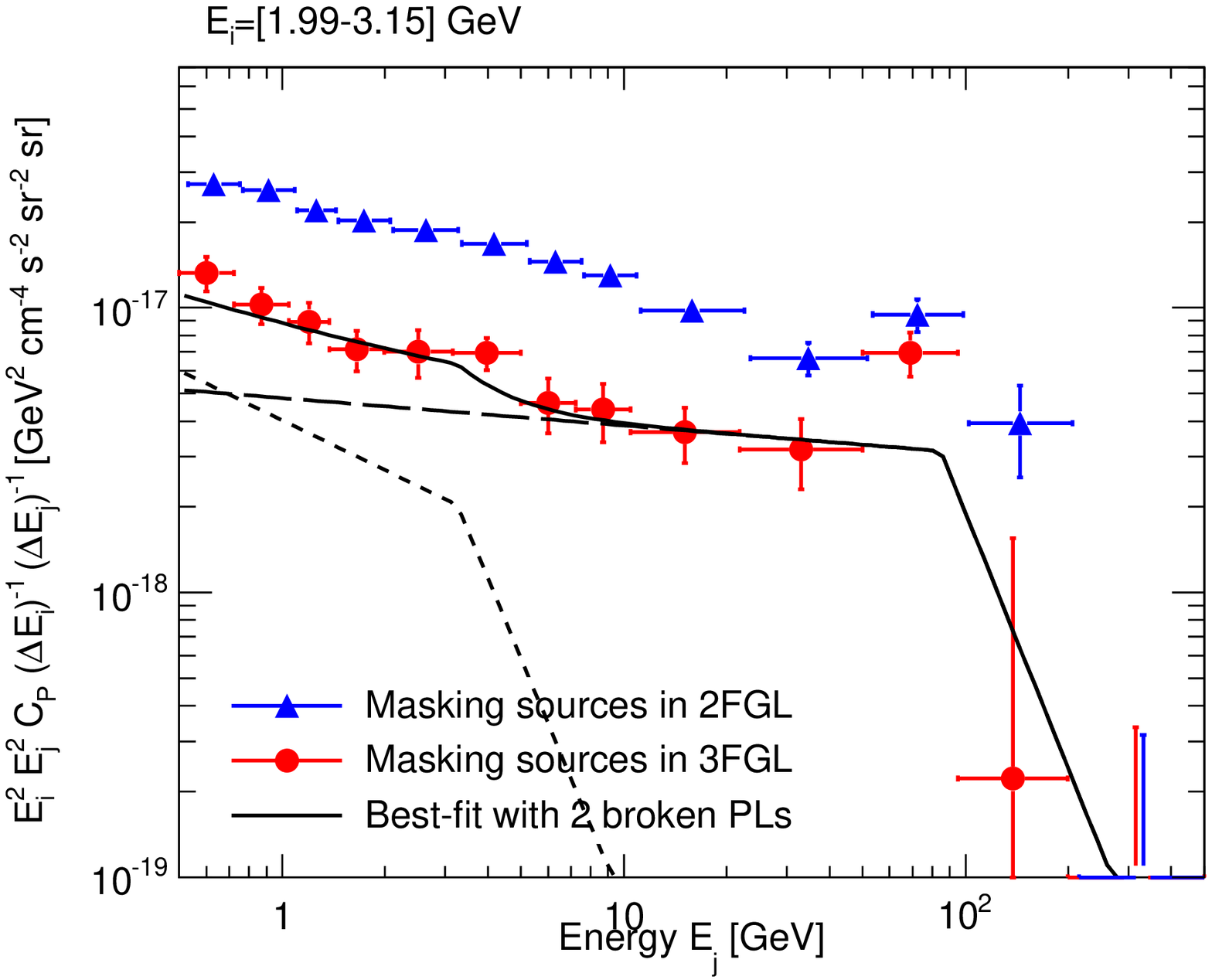}
\includegraphics[width=0.49\textwidth]{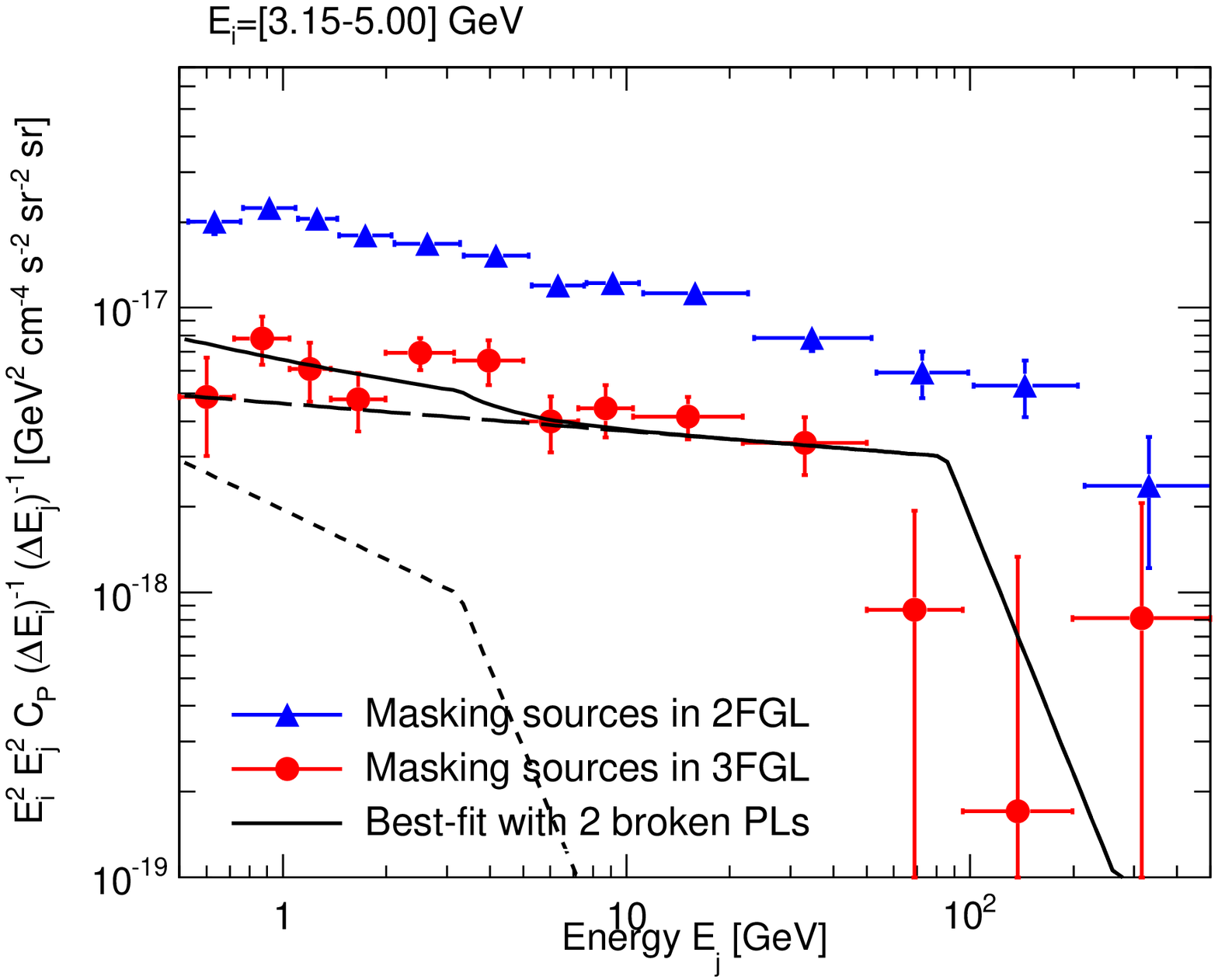}
\caption{\label{fig:crosscorrcp1} Depedence of the cross-APS on the energy. Each panel shows the best-fit Poissonian $C_{\rm P}$ for the cross-APS between the $i$-th and the $j$-th energy bins, as a function of $E_j$. Red circles are for the reference data set (P7REP\_ULTRACLEAN\_V15 front events) using the default mask masking 3FGL sources, while the blue triangles show the result for the same data set and for the default mask excluding 2FGL sources. The first 6 energy bins are shown in this figure and $E_i$ is indicated in the top of each panel. The solid black line is the best-fit solution when data are fitted assuming two independent populations of sources with broken-power-law energy spectra. The short-dashed and long-dashed black lines show the two populations independently.}
\end{figure*}

\begin{figure*}
\includegraphics[width=0.49\textwidth]{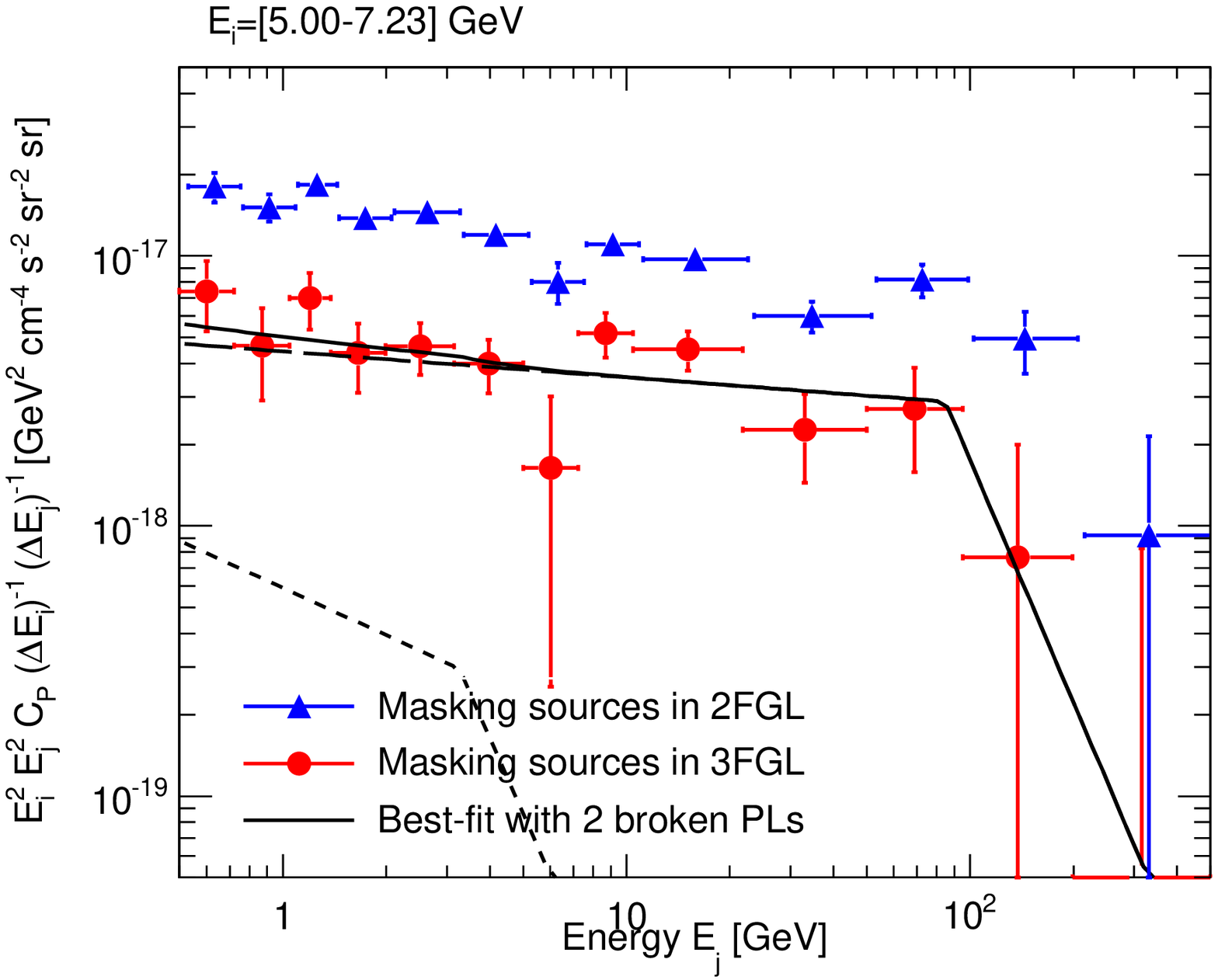}
\includegraphics[width=0.49\textwidth]{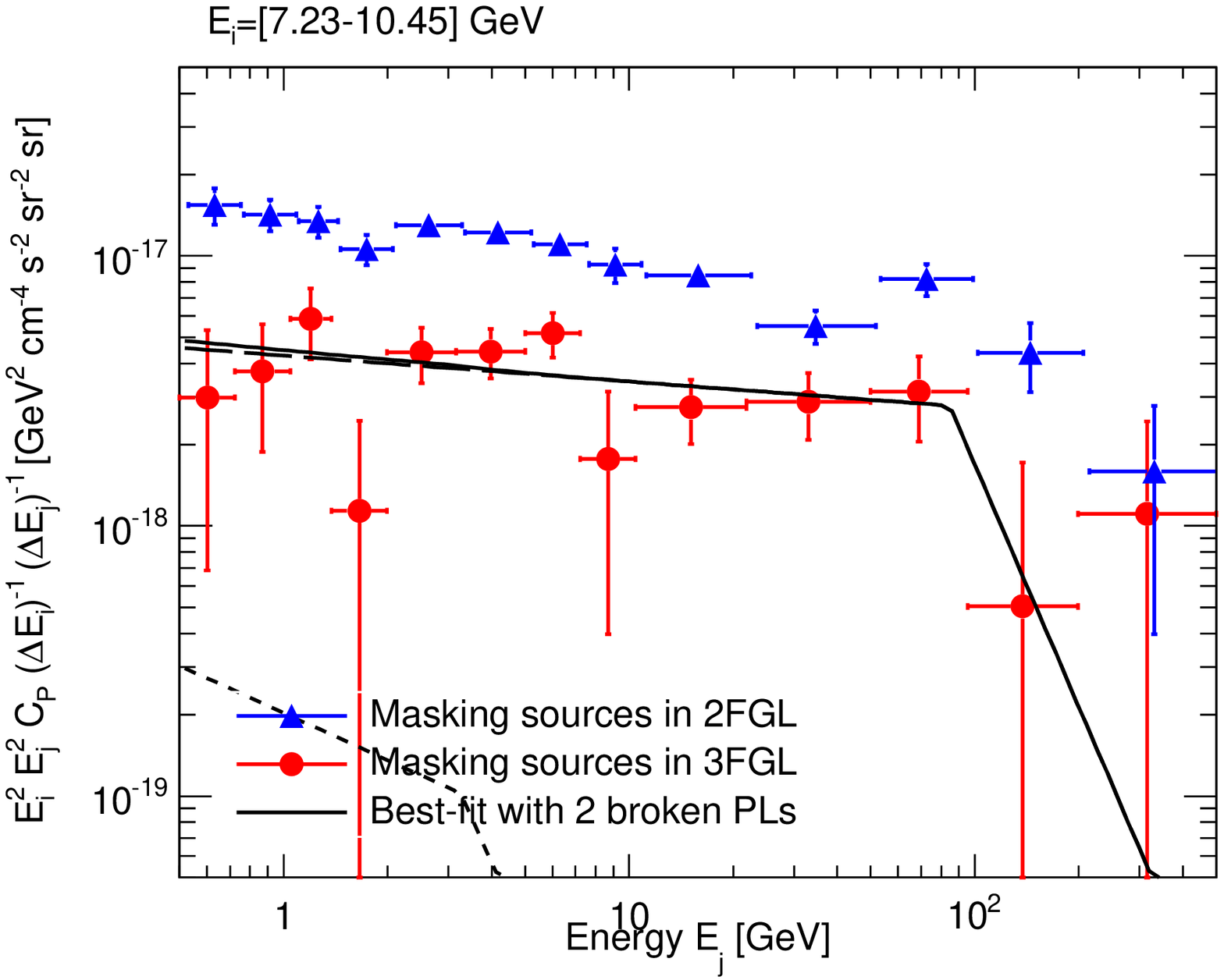}
\includegraphics[width=0.49\textwidth]{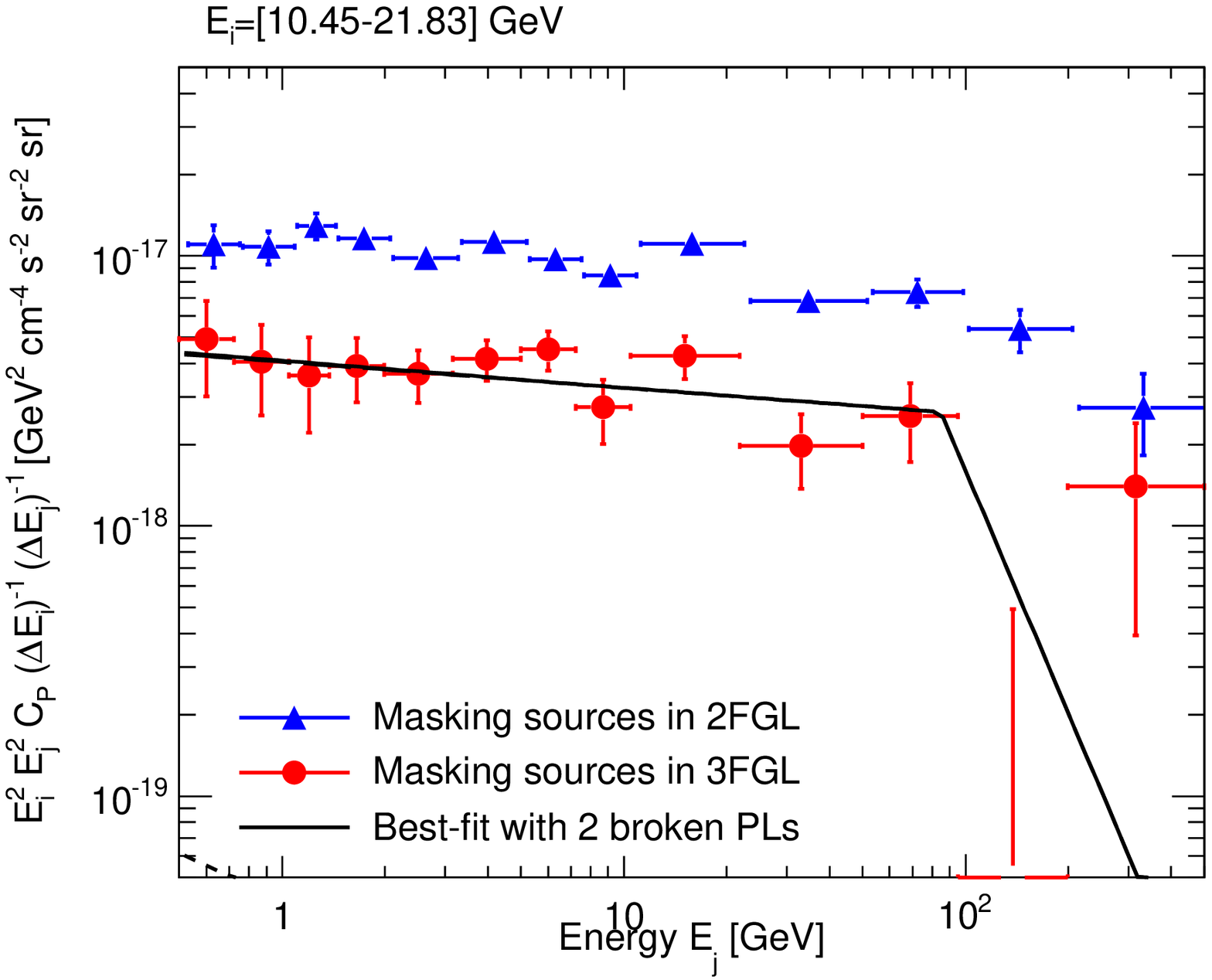}
\includegraphics[width=0.49\textwidth]{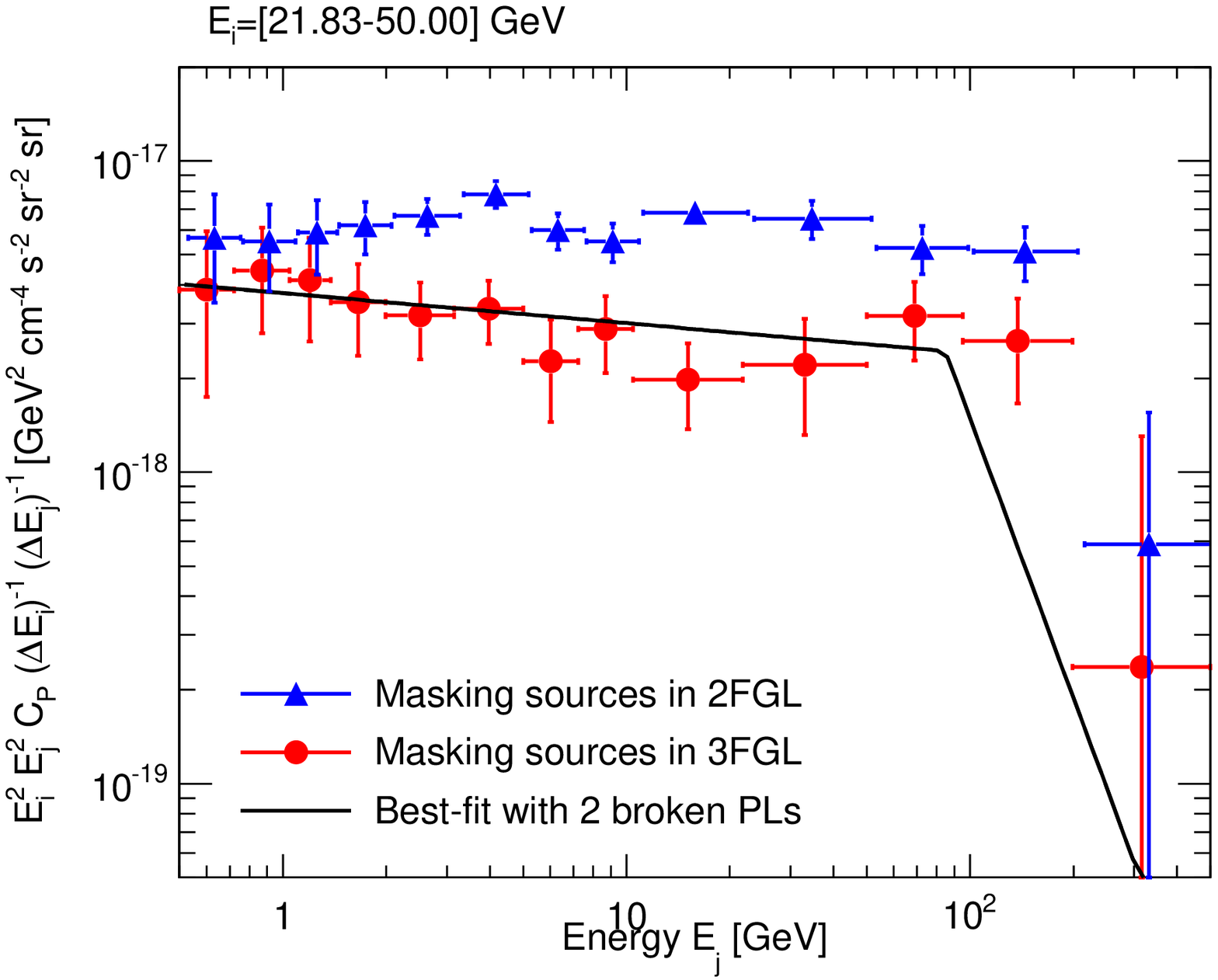}
\includegraphics[width=0.49\textwidth]{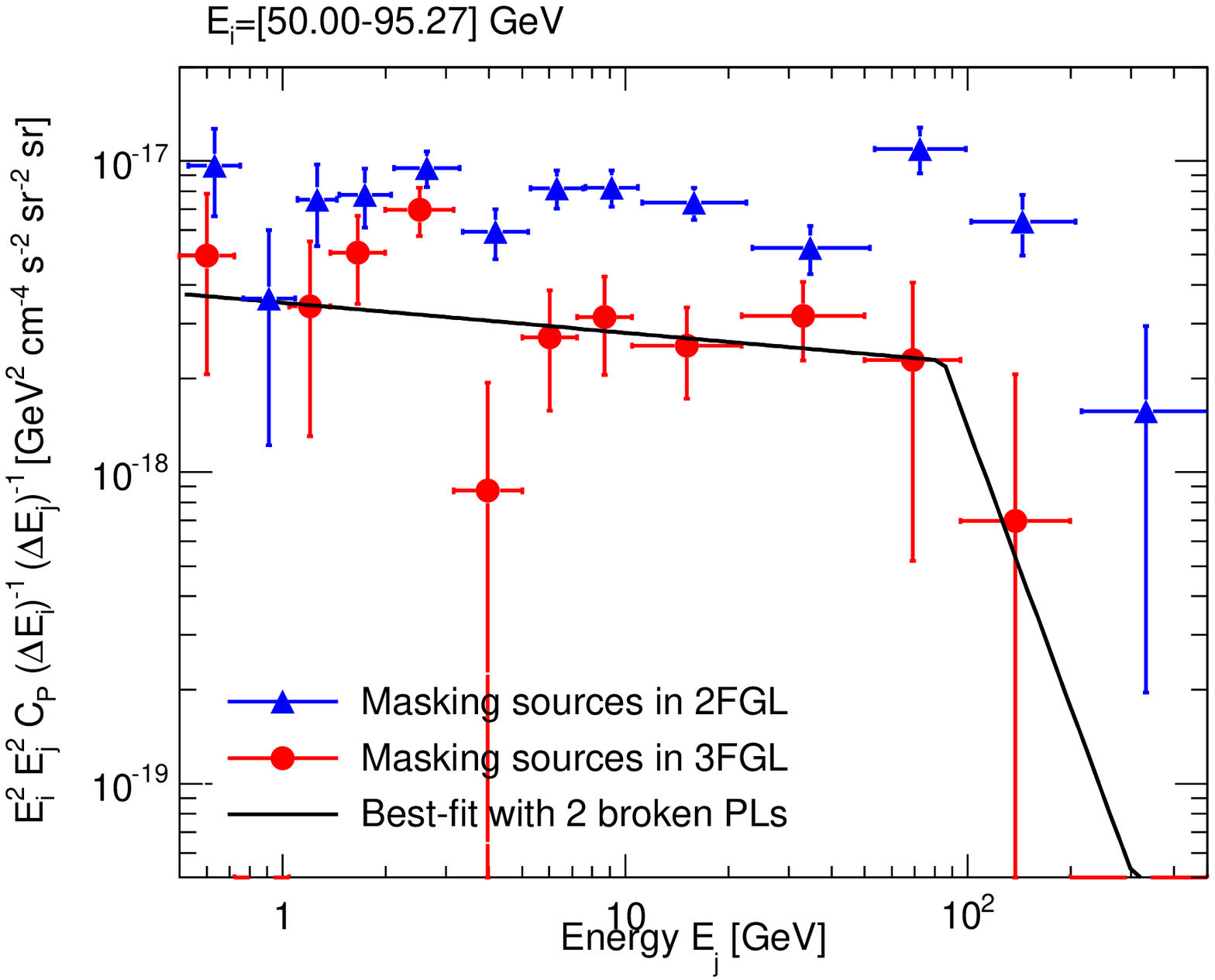}
\includegraphics[width=0.49\textwidth]{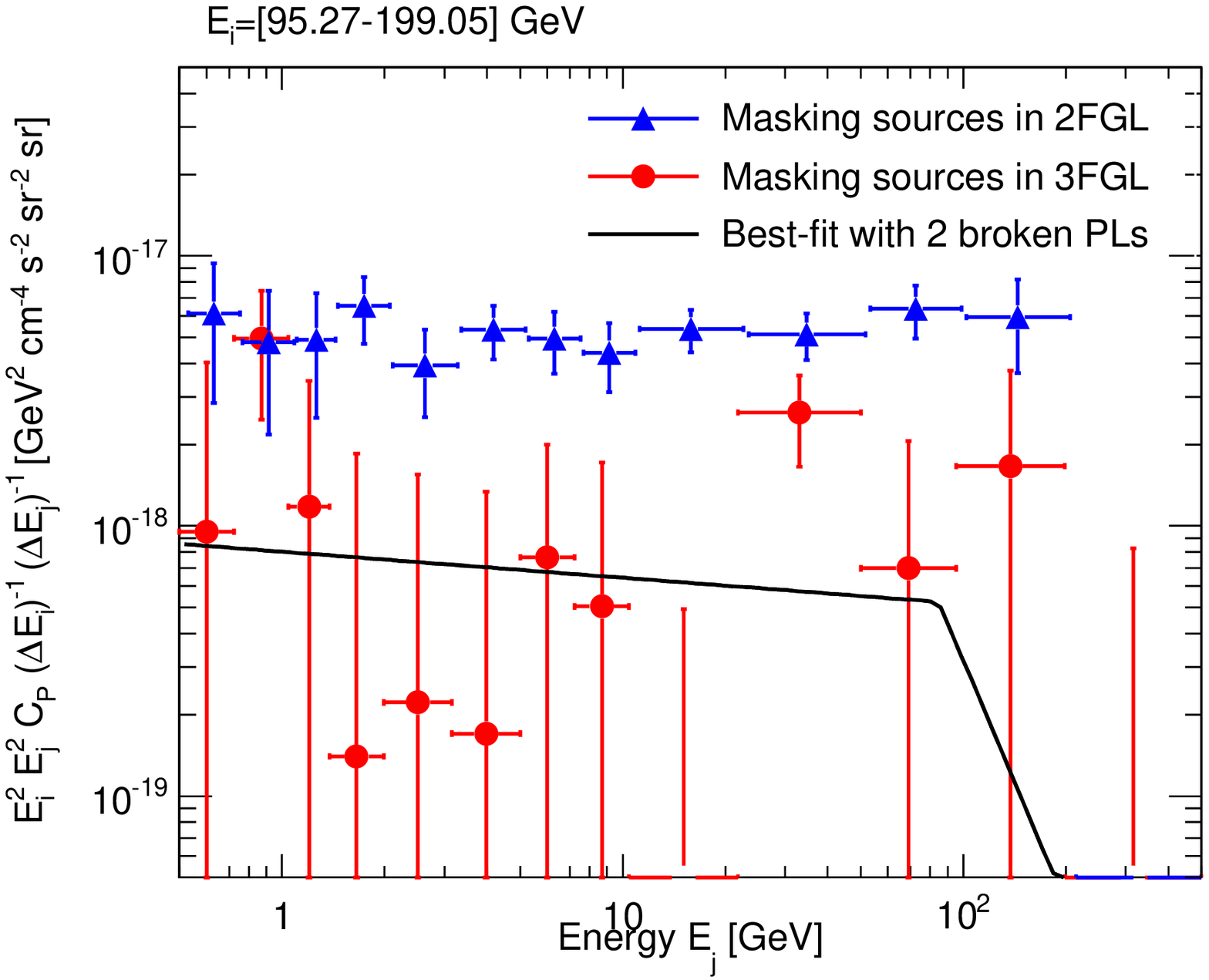}
\includegraphics[width=0.49\textwidth]{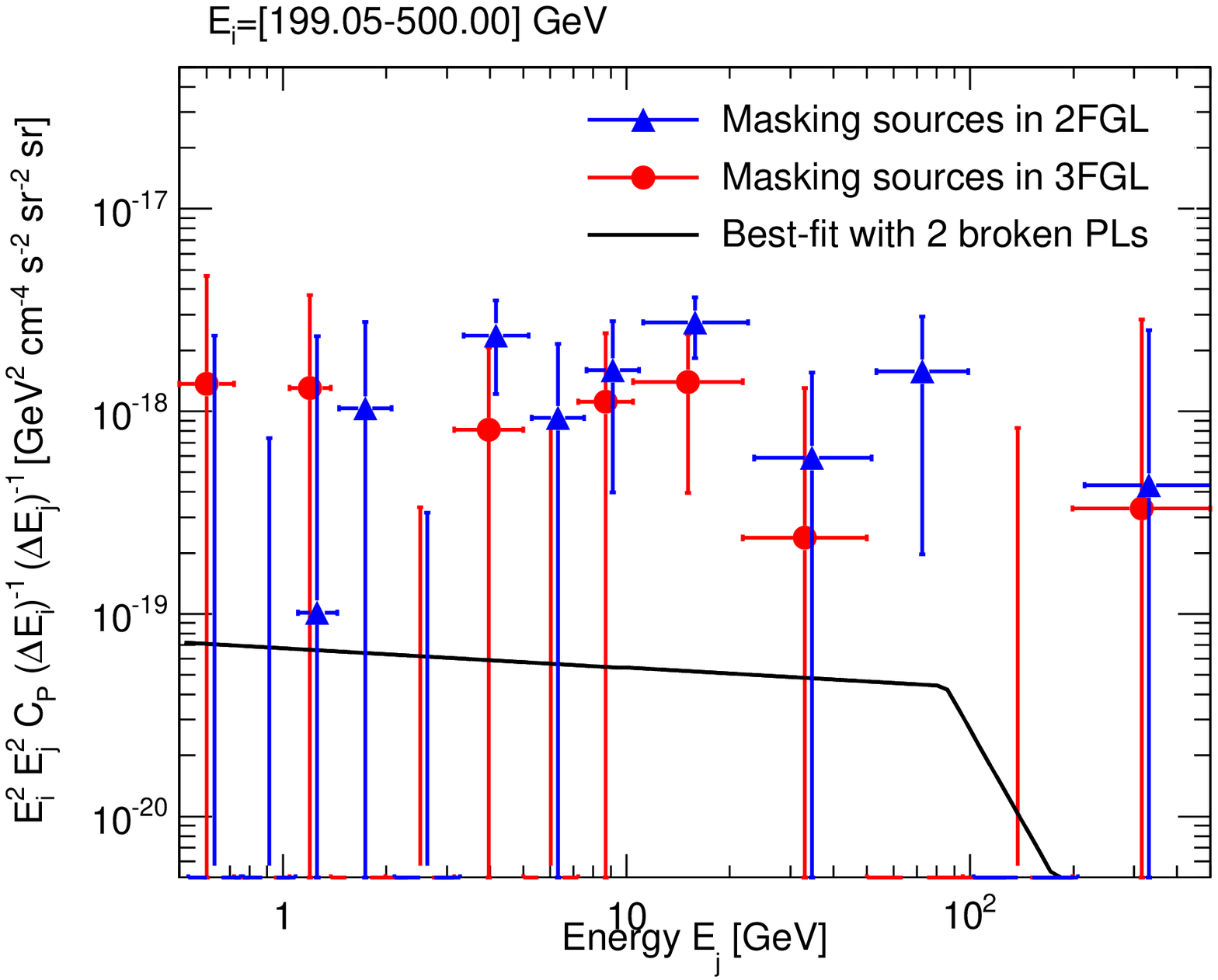}
\caption{\label{fig:crosscorrcp2} Same as Fig.~\ref{fig:crosscorrcp1}, for the last 7 energy bins.}
\end{figure*}

Figs.~\ref{fig:crosscorrcp1} and \ref{fig:crosscorrcp2} show the best-fit 
$C_{\rm P}$ for the cross-APS as a function of energy. Red circles refer to the 
mask covering 3FGL sources and blue triangles to the mask of 2FGL sources. The 
solid black line denotes the best-fit solution discussed in 
Sec.~\ref{sec:interpretation}, i.e., the one in terms of two populations of 
unresolved sources with broken-power-law energy spectra. The short-dashed and 
long-dashed black lines indicate the two source populations independently.
Data are available at https://www-glast.stanford.edu/pub\_data/552. 

\section{The dependence on energy of the cross-correlation coefficients}
\label{sec:cross_corr_coefficents}
\begin{figure*}
\includegraphics[width=0.49\textwidth]{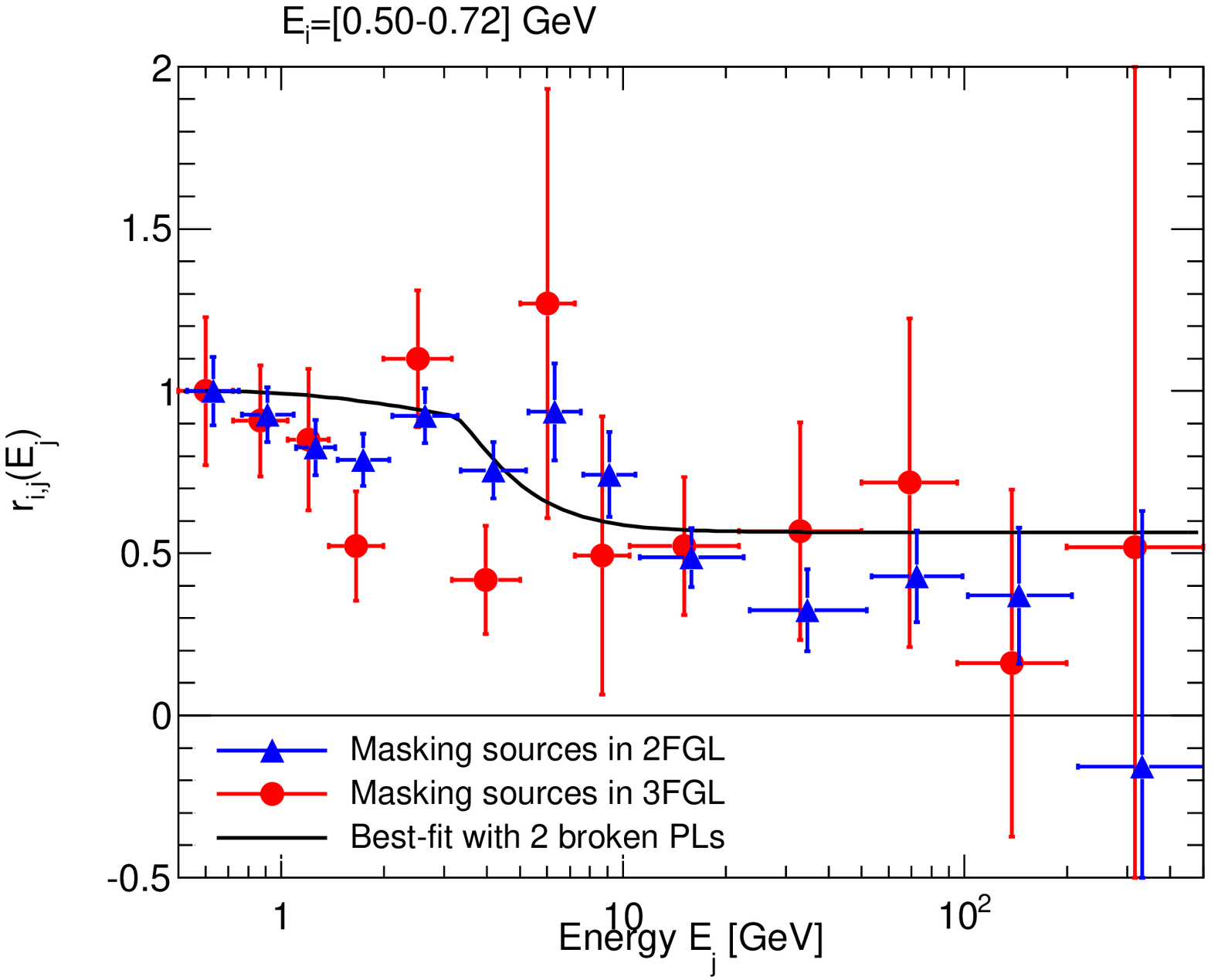}
\includegraphics[width=0.49\textwidth]{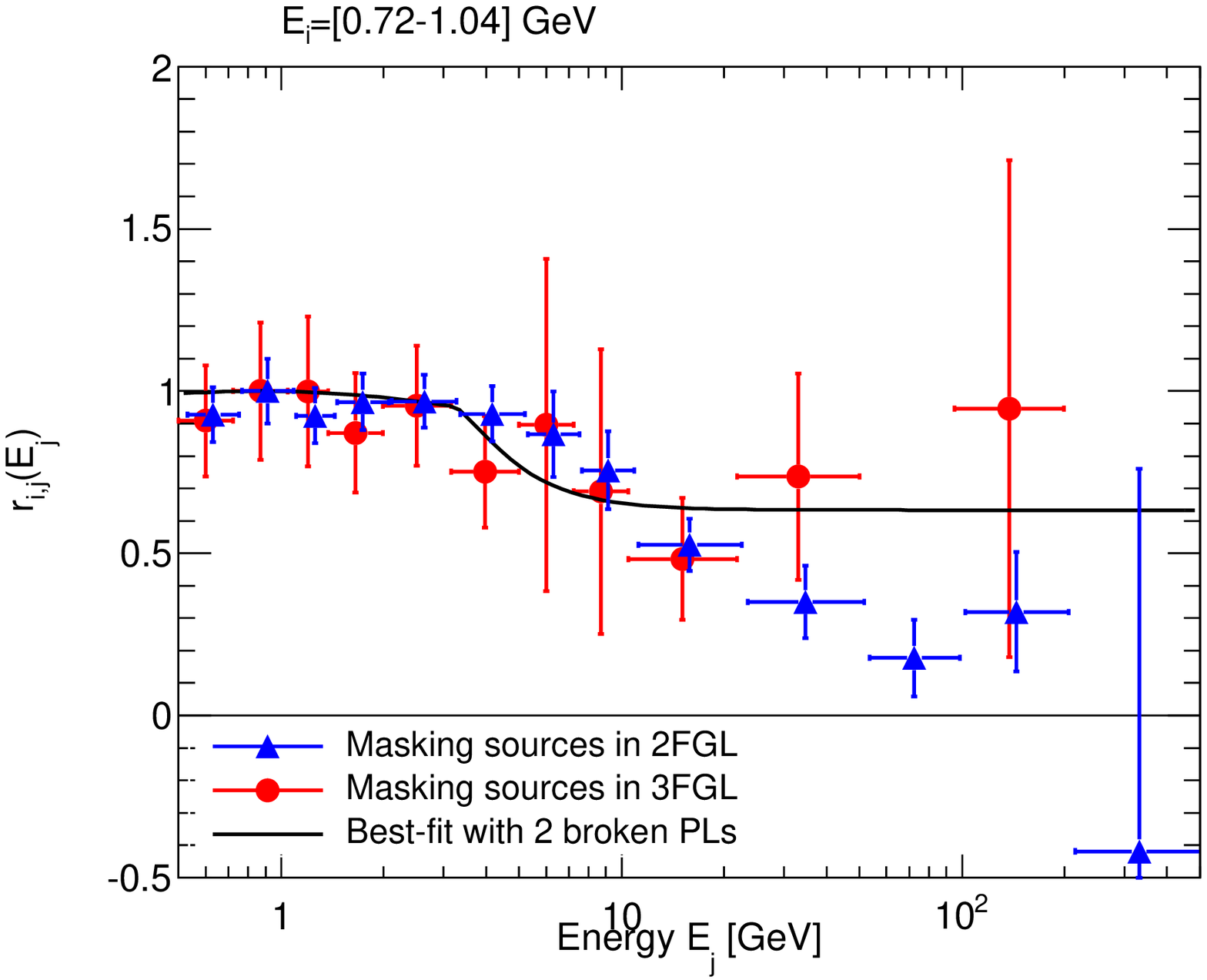}
\includegraphics[width=0.49\textwidth]{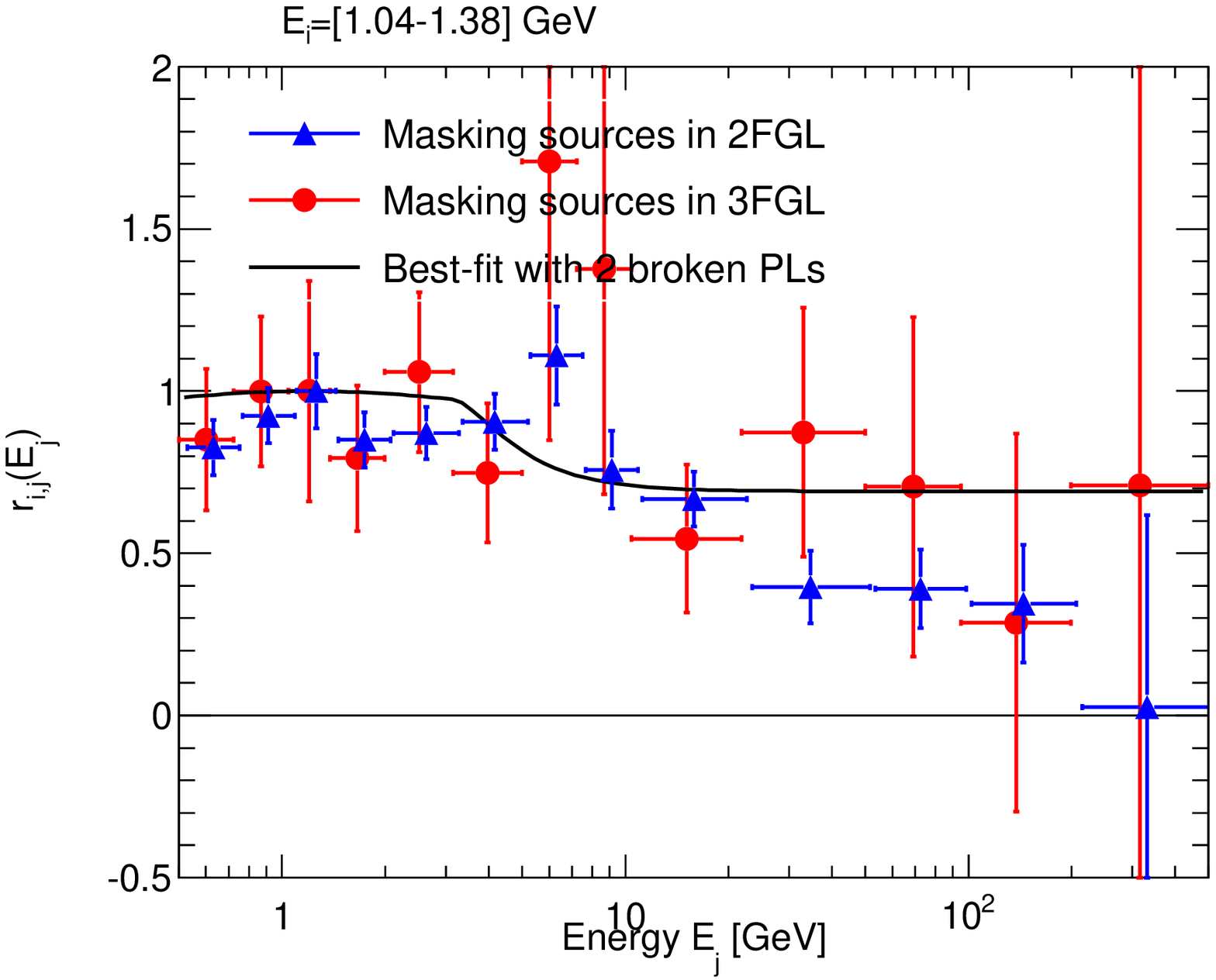}
\includegraphics[width=0.49\textwidth]{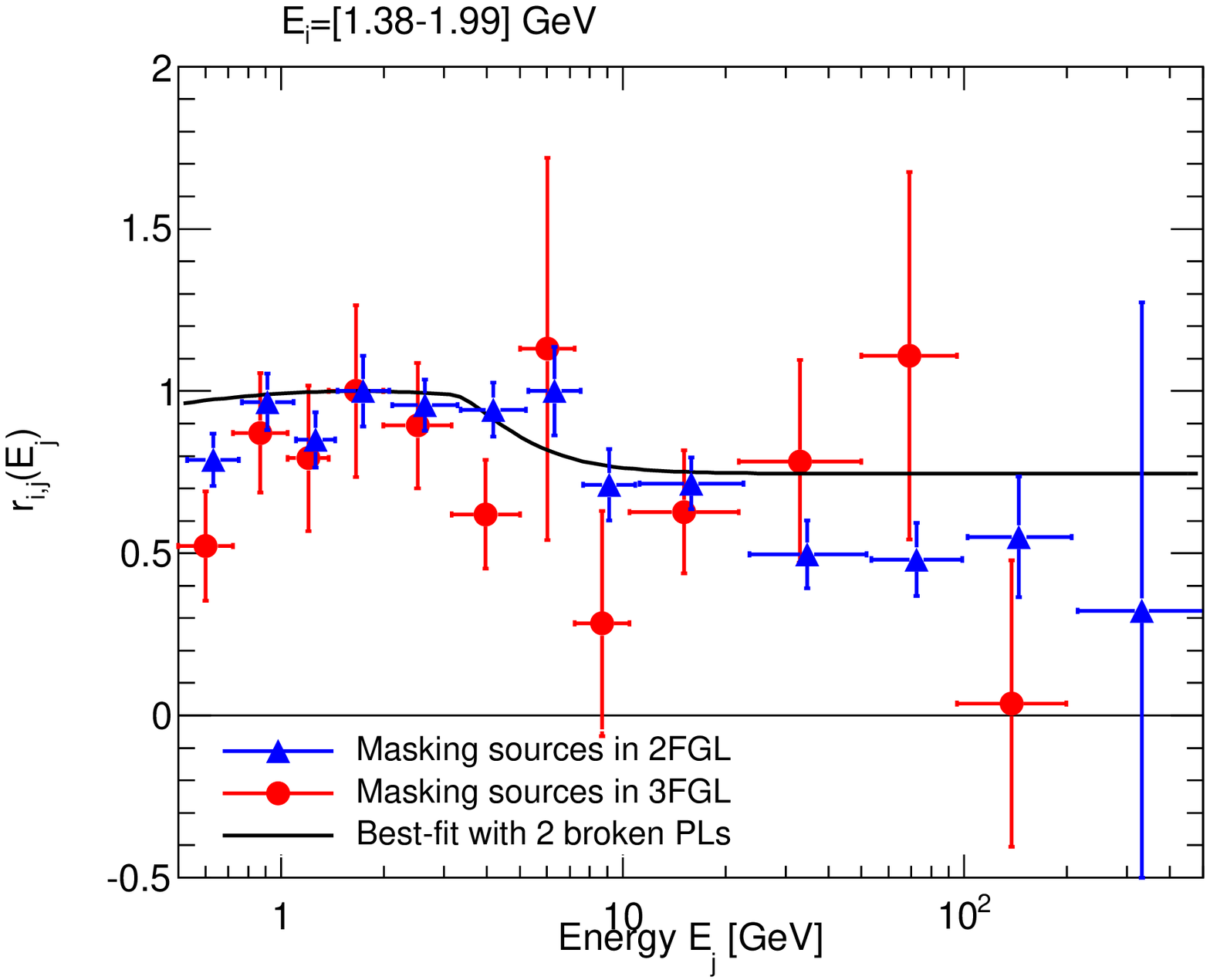}
\includegraphics[width=0.49\textwidth]{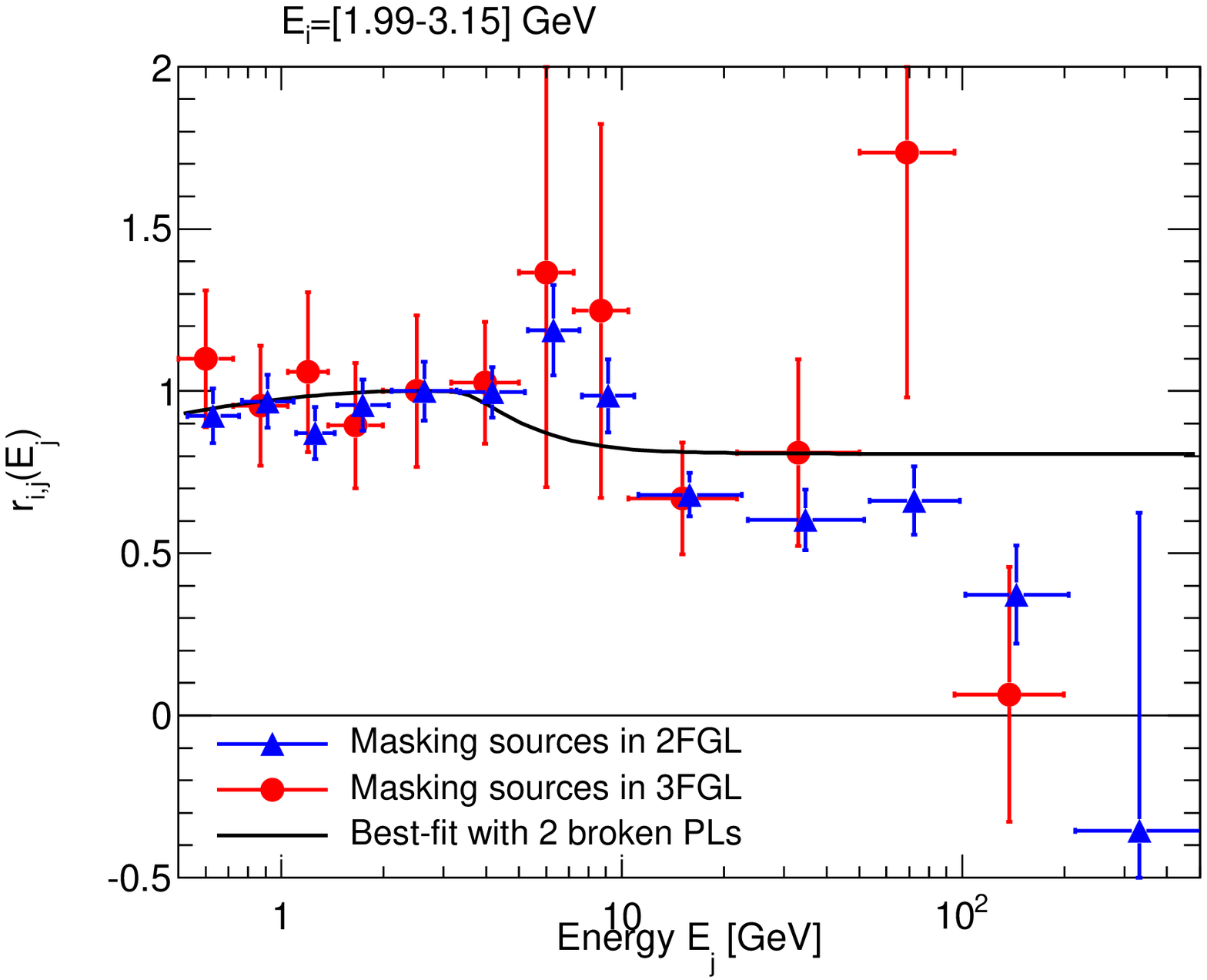}
\includegraphics[width=0.49\textwidth]{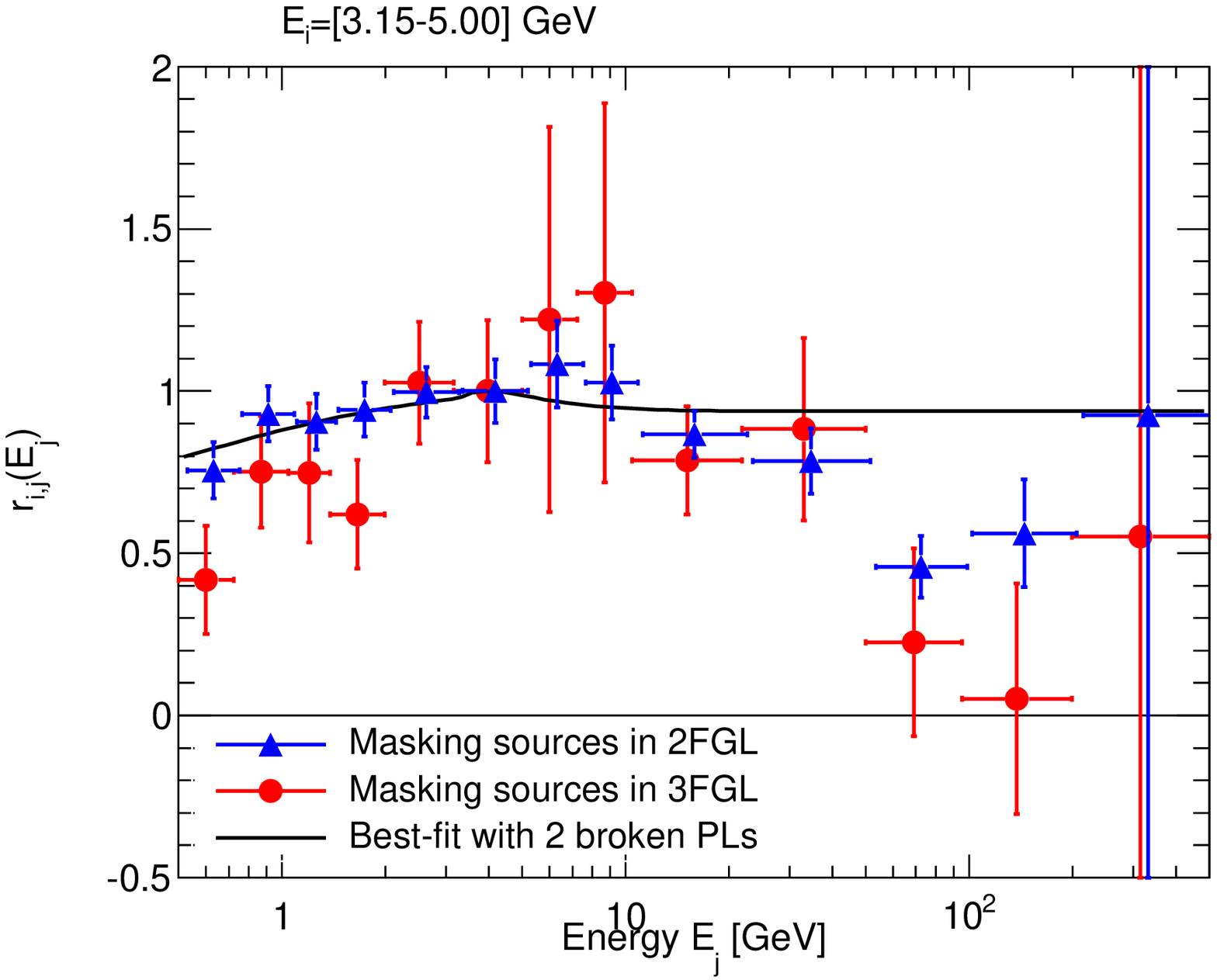}
\caption{\label{fig:cross_corr_coeff_1} Dependence of the cross-correlation coefficients on the energy. Each panel shows the cross-correlation coefficients $r_{i,j}$ defined in Sec.~\ref{sec:cross_correlation} between the $i$-th and the $j$-th energy bins, as a function of $E_j$. Red circles are for the reference data set (P7REP\_ULTRACLEAN\_V15 front events) using the default mask masking 3FGL sources, while the blue triangles show the result for the same data set and for the default mask excluding 2FGL sources. The first 6 energy bins are shown in this figure and $E_i$ is indicated in the top of each panel. The solid black line shows the $r_{i,j}$ corresponding to the best-fit solution when data are fitted masking 3FGL sources and assuming two independent populations of sources with broken-power-law energy spectra (see Sec.~\ref{sec:interpretation}).}
\end{figure*}

\begin{figure*}
\includegraphics[width=0.49\textwidth]{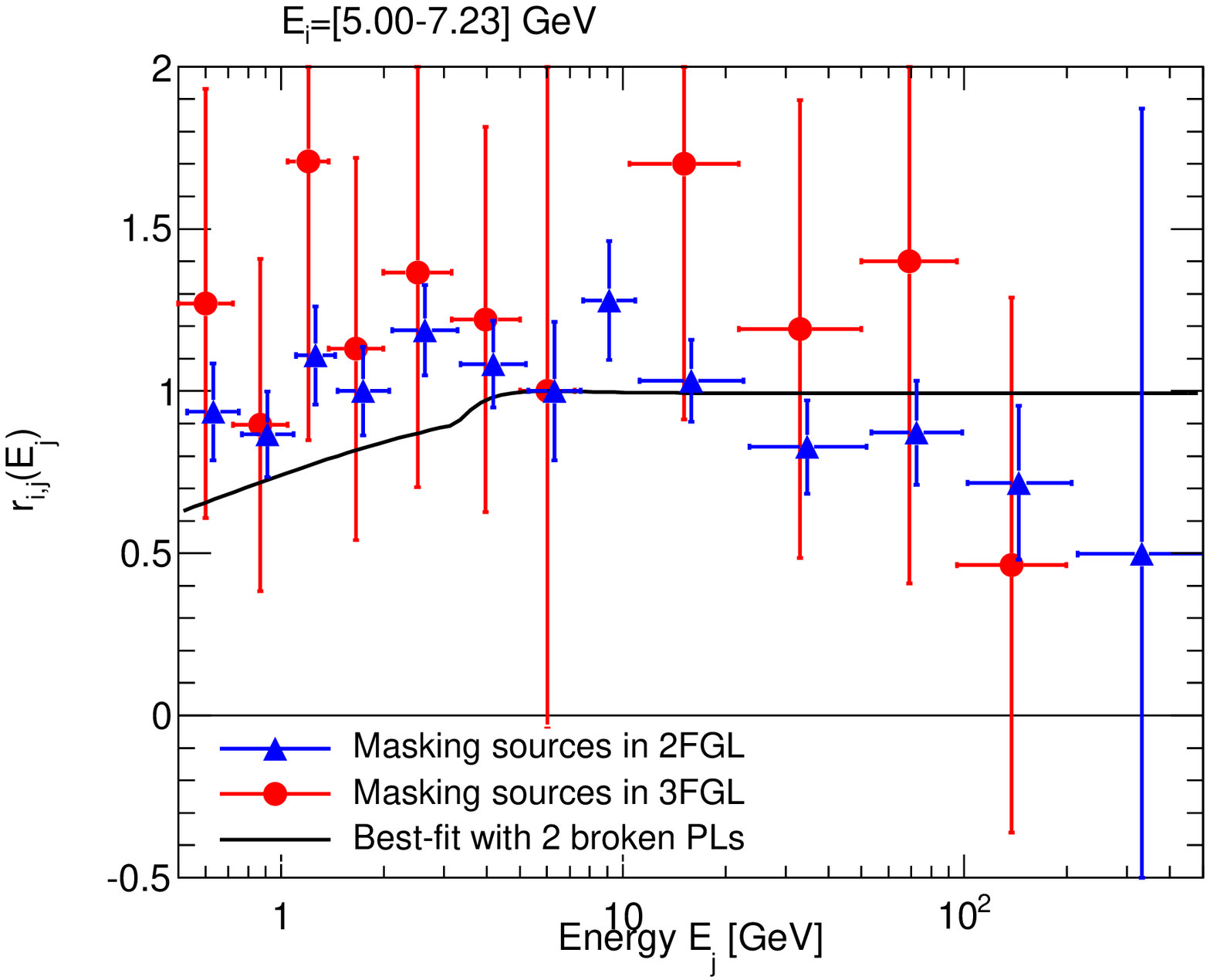}
\includegraphics[width=0.49\textwidth]{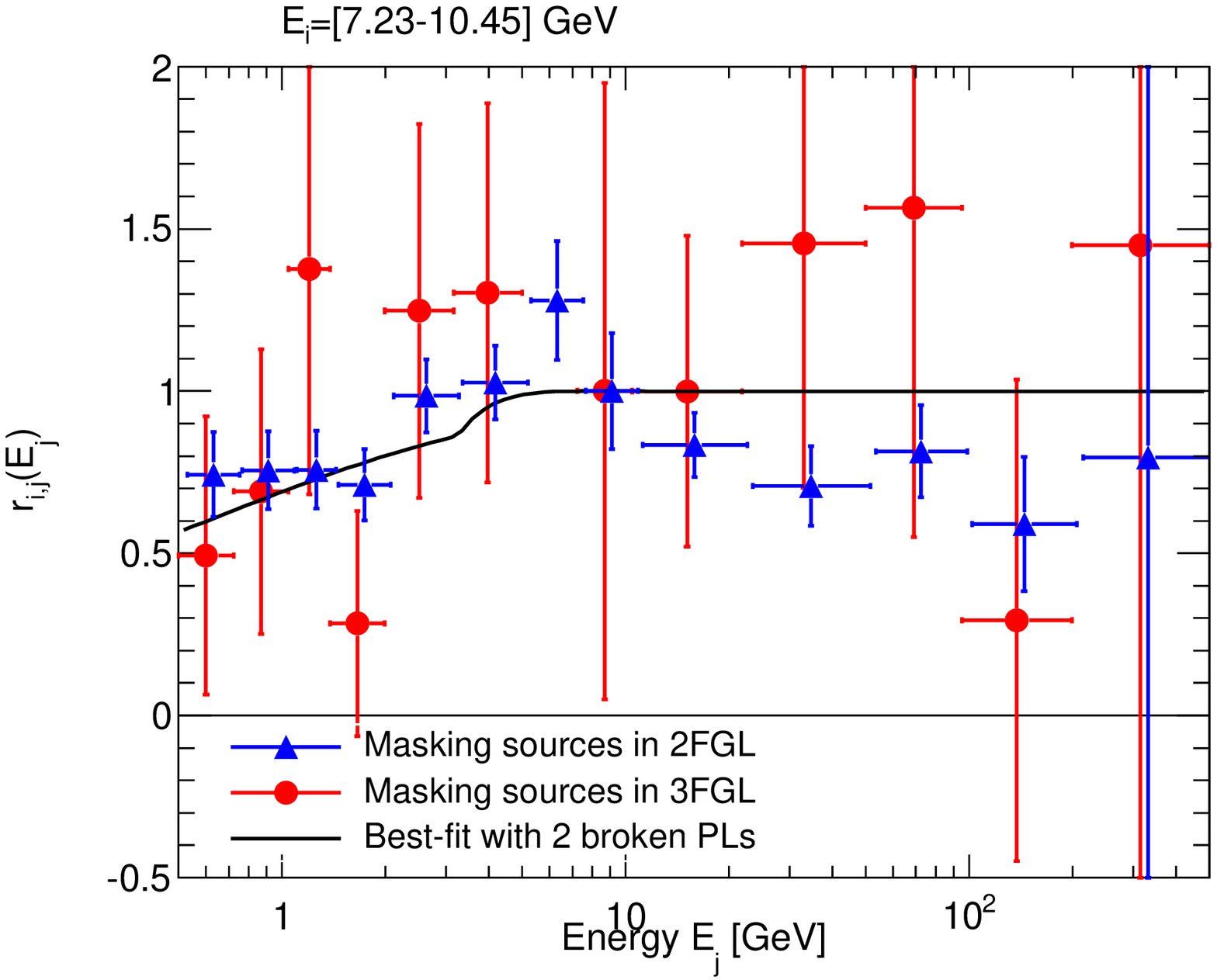}
\includegraphics[width=0.49\textwidth]{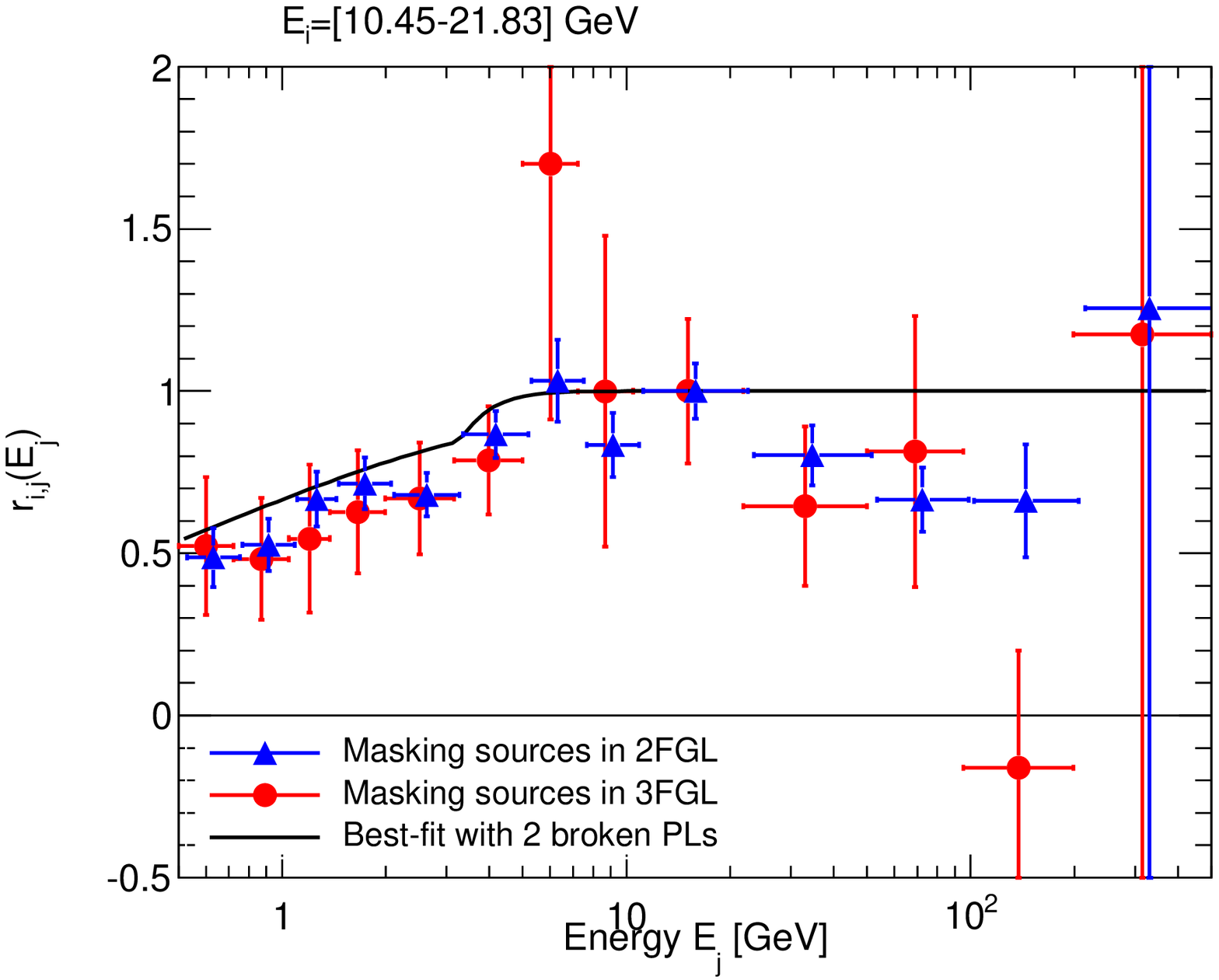}
\includegraphics[width=0.49\textwidth]{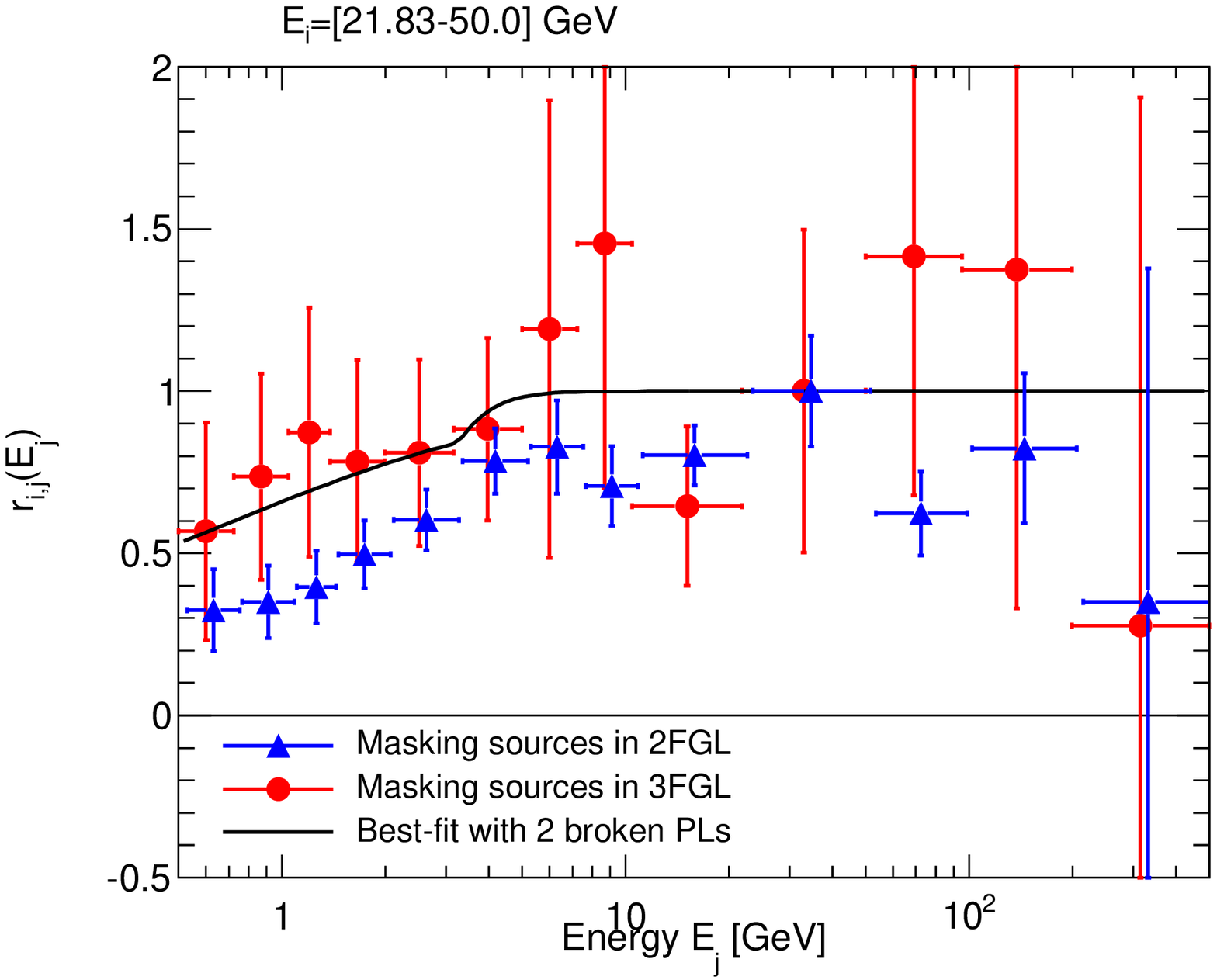}
\includegraphics[width=0.49\textwidth]{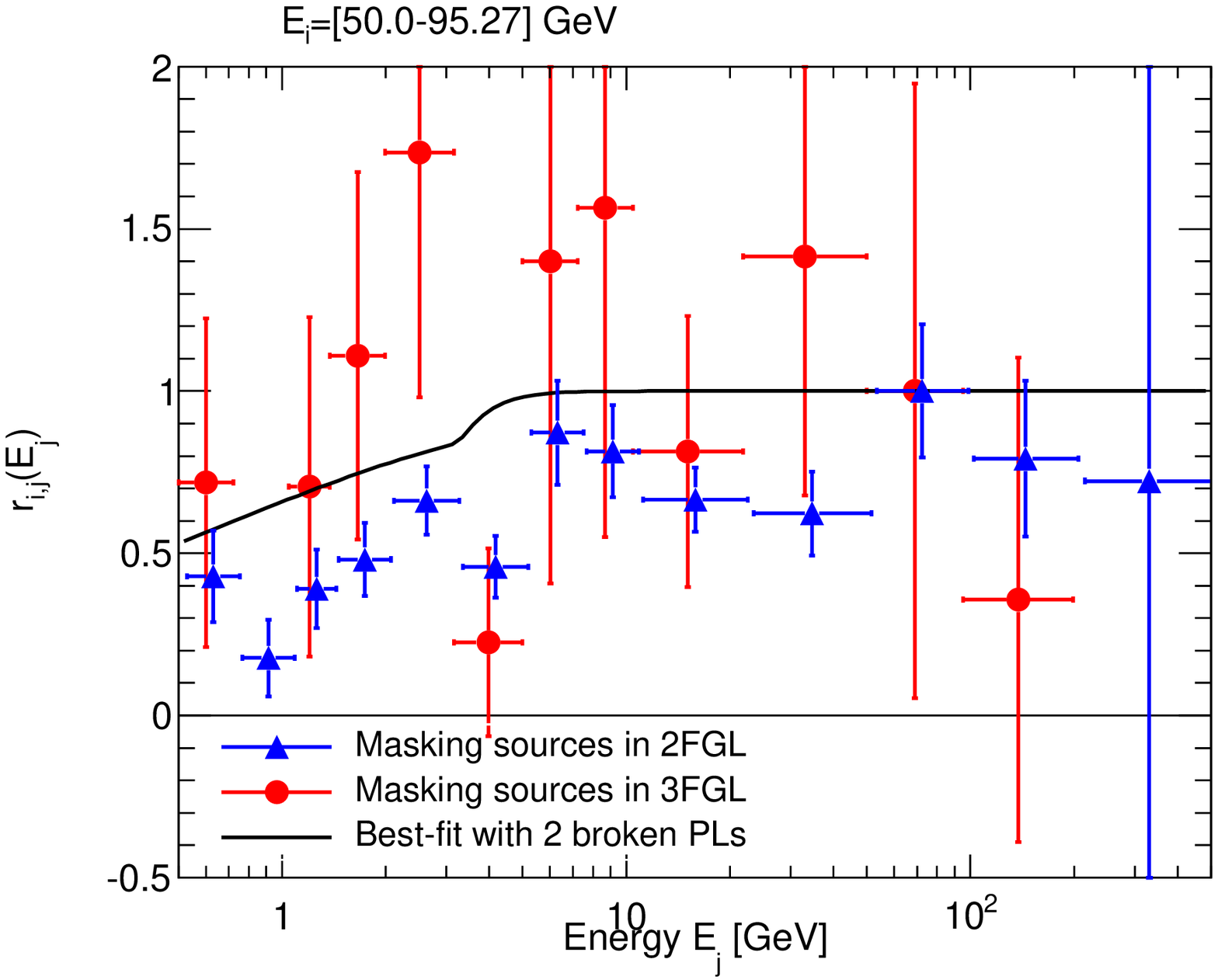}
\includegraphics[width=0.49\textwidth]{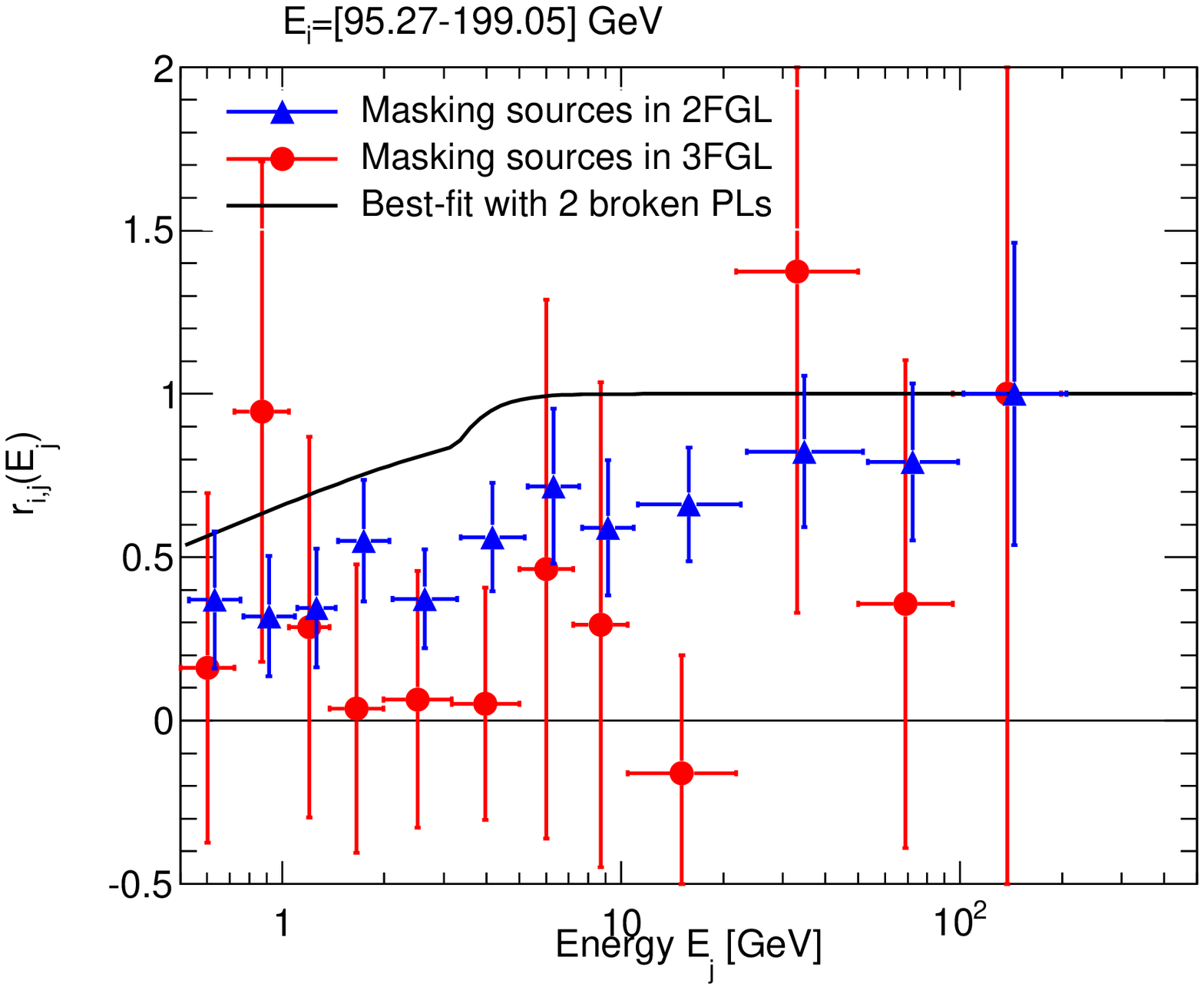}
\includegraphics[width=0.49\textwidth]{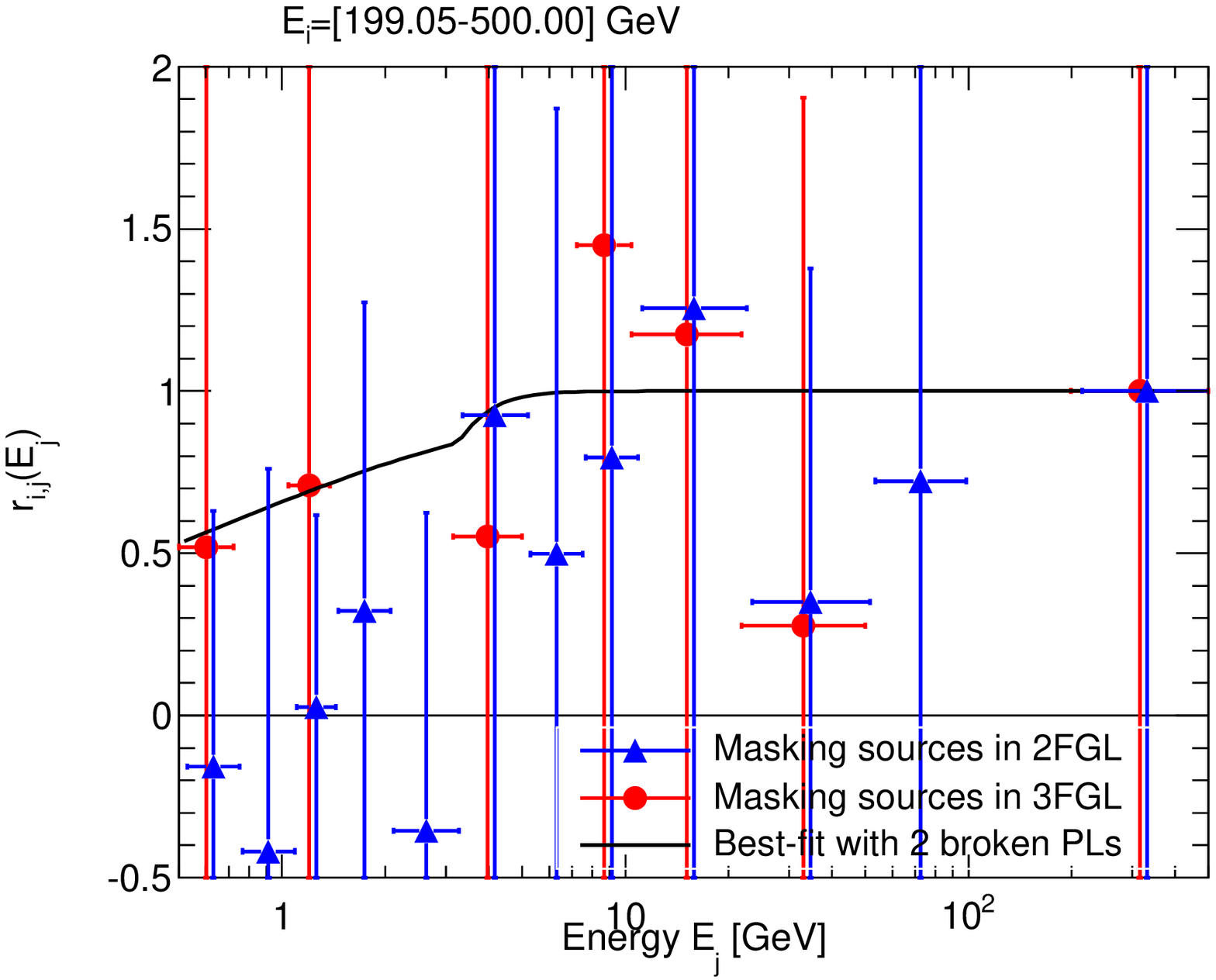}
\caption{\label{fig:cross_corr_coeff_2} Same as Fig.~\ref{fig:cross_corr_coeff_1}, for the last 7 energy bins.}
\end{figure*}

Figs.~\ref{fig:cross_corr_coeff_1} and \ref{fig:cross_corr_coeff_2} show the 
cross-correlation coefficients $r_{i,j}$ defined in 
Sec.~\ref{sec:cross_correlation} in terms of the best-fit auto- and cross-APS
$C_{\rm P}$. Each panel shows $r_{i,j}$ at a specific energy $E_i$, as a function
of $E_j$. Red circles refer to the mask covering 3FGL sources and blue 
triangles to the mask around 2FGL sources. The solid black line shows the 
cross-correlation coefficents corresponding to the best-fit solution discussed 
in Sec.~\ref{sec:interpretation} in the case of two populations of unresolved 
sources with broken-power-law energy spectra and masking 3FGL sources. The
fact that the blue triangles decrease with energy in the first panels, while 
they increase towards 1 in the last panels indicates the lack of correlation
between low and high energies. The same trend is noted for the red circles,
but with a lower significance.

\section{Exclusion limit on Dark Matter for the $\tau$ and $\mu$ channel}
\label{sec:other_channel}
\begin{figure*}
\begin{center}
\includegraphics[width=0.49\textwidth]{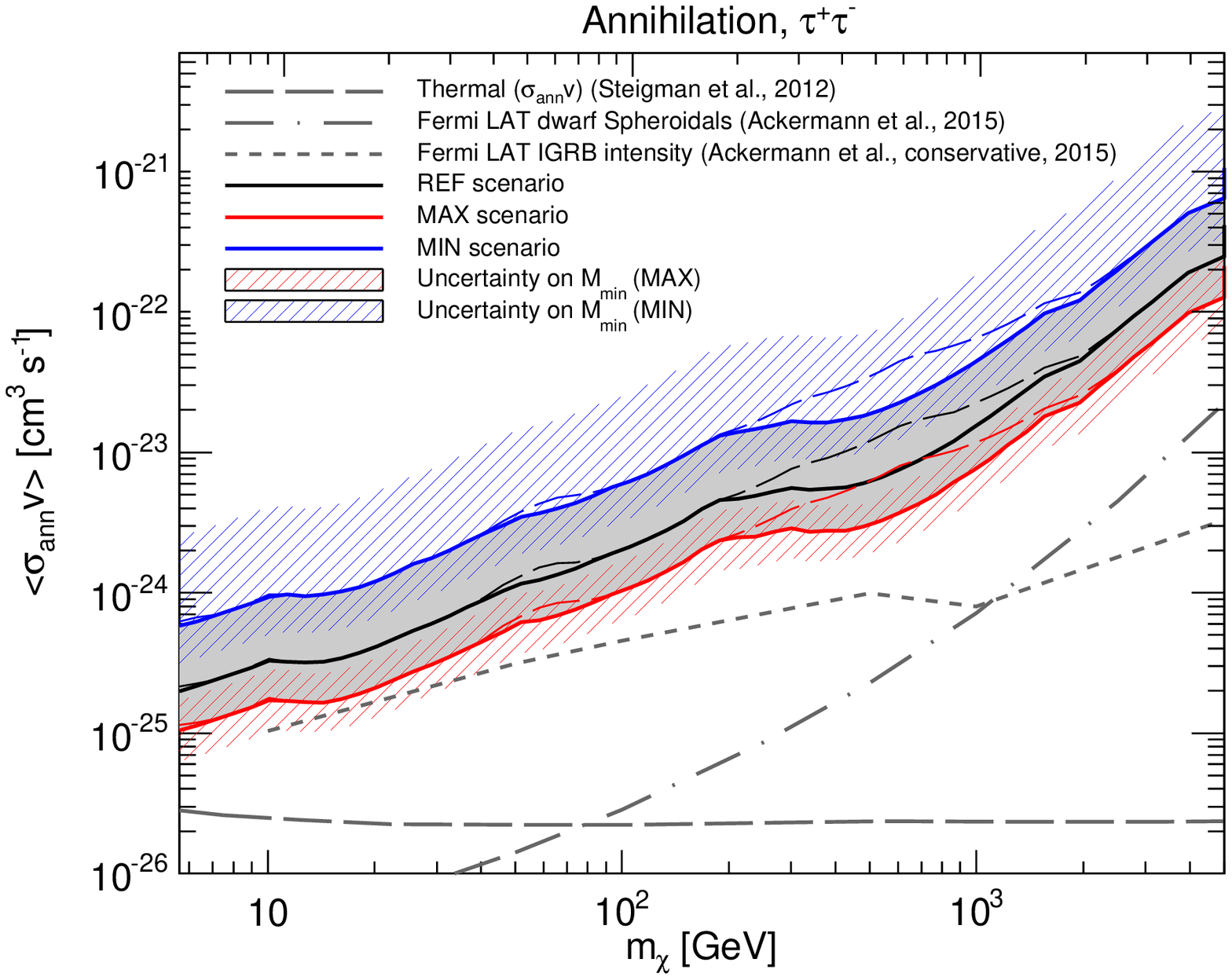}
\includegraphics[width=0.49\textwidth]{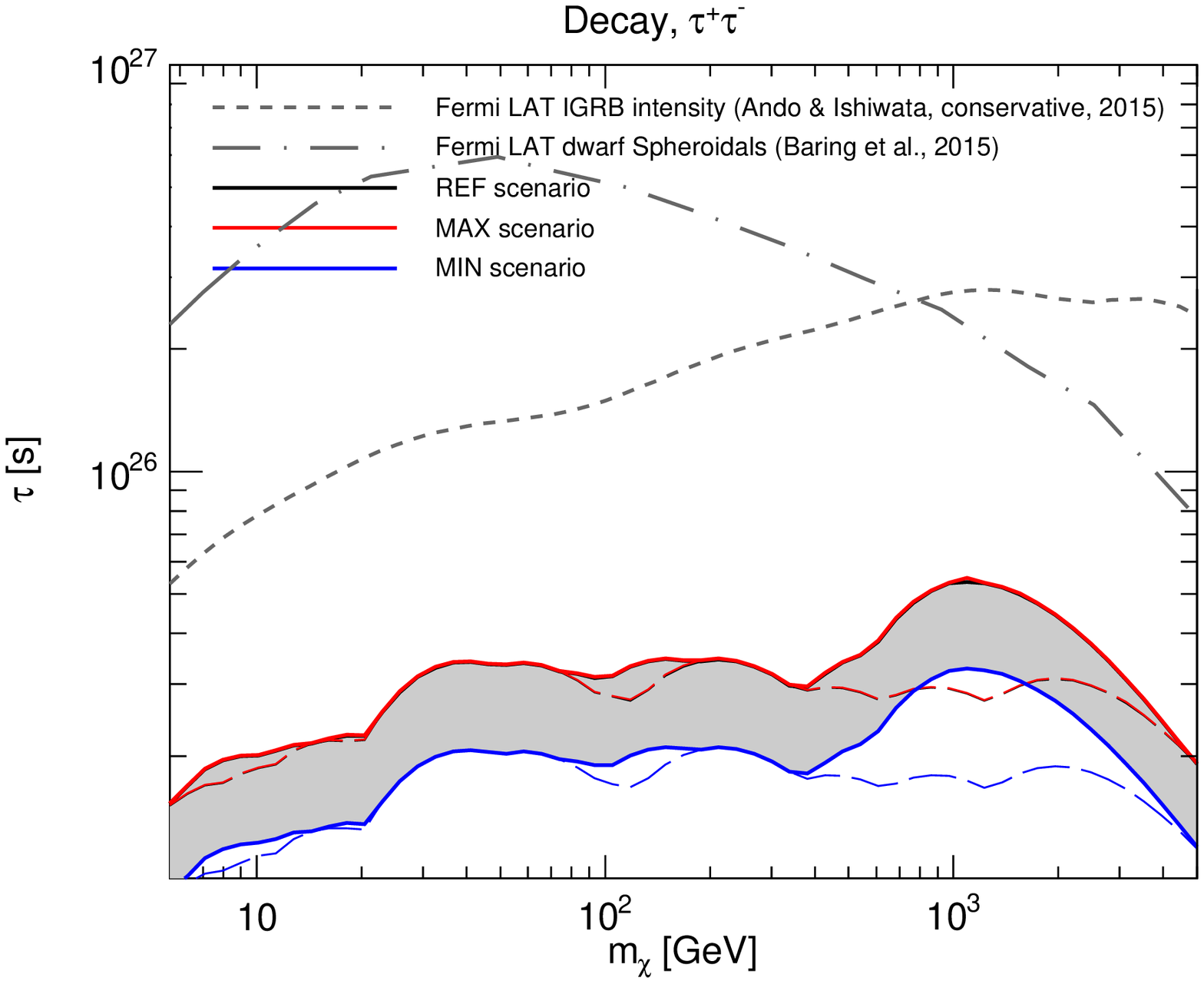}
\caption{\label{fig:tau_conservative} Conservative exclusion limits on annihilating and decaying DM from the new APS measurement, for the $\tau$ channel. {\it Left:} The solid lines show the upper limits on $\langle \sigma_{\rm ann} v \rangle$ derived from the auto- and cross-APS measured in Sec.~\ref{sec:analysis}, as a function of $m_\chi$, for $M_{\rm min}=10^{-6} M_\odot$ and annihilations into $\tau^+\tau^-$. The limits follow the conservative approach described in Sec.~\ref{sec:conservative}. The black line is for the REF scenario, while the red and blue ones are for MAX and MIN. The grey band between the MIN and MAX scenario represents our estimated total astrophysical uncertainty for $M_{\rm min}=10^{-6} M_\odot$, accounting for all the sources of uncertainty mentioned in Sec.~\ref{sec:simulations}. The red and blue shaded bands describe the effect of changing $M_{\rm min}$ between $10^{-12} M_{\rm min}$ and 1 $M_{\rm min}$, for the MAX and MIN scenario, respectively. In the case of the black, red and blue dashed lines, the upper limits are derived only considering the measured auto-APS and neglecting the cross-APS. For comparison, the long-dashed grey line marks the annihilation cross section for thermal relics from Ref.~\cite{Steigman:2012nb} and the dash-dotted grey line the upper limit obtained in Ref.~\cite{Ackermann:2015zua} from the combined analysis of 15 dwarf spheroidal galaxies. Finally, the short-dashed grey line shows the conservative upper limit derived in Ref.~\cite{Ackermann:2015tah} from the intensity of the IGRB. {\it Right}: The same as in the left panel but for the lower limits on $\tau$ for decaying DM. The short-dashed grey line represents the lower limit obtained in Fig. 6 of Ref.~\cite{Ando:2015qda} from the IGRB intensity, while the dash-dotted grey one is obtained from the combined analysis of 15 dwarf spheroidal galaxies in Ref.~\cite{Baring:2015sza}.}
\end{center}
\end{figure*}

Sec.~\ref{sec:limits} shows exclusion limits on the DM 
$\langle \sigma_{\rm ann} v \rangle$ and $\tau$ in the case of 
annihilations/decay into $b\bar{b}$. Here we calculate the same exclusion 
limits for two additional channels. Fig.~\ref{fig:tau_conservative} shows the 
upper limits on $\langle \sigma_{\rm ann}v \rangle$ (left panel) and on $\tau$ 
(right panel) as a function of the DM mass $m_\chi$, in the case of 
annihilations/decays into $\tau^+\tau^-$. The exclusion limits are obtained 
following the conservative approach described in Sec.~\ref{sec:conservative}. 
The solid black line refers to the REF scenario, while the solid red and solid
blue ones stand for the MAX and MIN benchmark. The solid red and solid black 
lines almost exactly overlap in the right panel. The dashed black, blue and 
red lines are obtained considering only the auto-APS measurement. The red and 
blue shaded band indicate the variability of the exclusion limits in the MAX 
and MIN scenario when $M_{\rm min}$ is left free to vary between 1 $M_\odot$ and 
$10^{-12} M_\odot$. The long-dashed grey line in the left panel shows the 
thermal annihilation cross section, as computed in 
Ref.~\cite{Steigman:2012nb}, while the dot-dashed grey line is the upper limit
obtained in Ref.~\cite{Ackermann:2015zua} from the analysis of 15 dwarf
spheroidal galaxies. Finally, the short-dashed grey line derives from the
analysis of the IGRB intensity performed in Ref.~\cite{Ackermann:2015tah}.
On the other hand, the short-dashed grey line in the right panel of 
Fig.~\ref{fig:tau_conservative} is obtained from the study of the IGRB 
intensity in Ref.~\cite{Ando:2015qda} and the dot-dashed grey line comes from
the observation of 15 dwarf spheroidal galaxies performed in 
Ref.~\cite{Baring:2015sza}.

\begin{figure*}
\begin{center}
\includegraphics[width=0.49\textwidth]{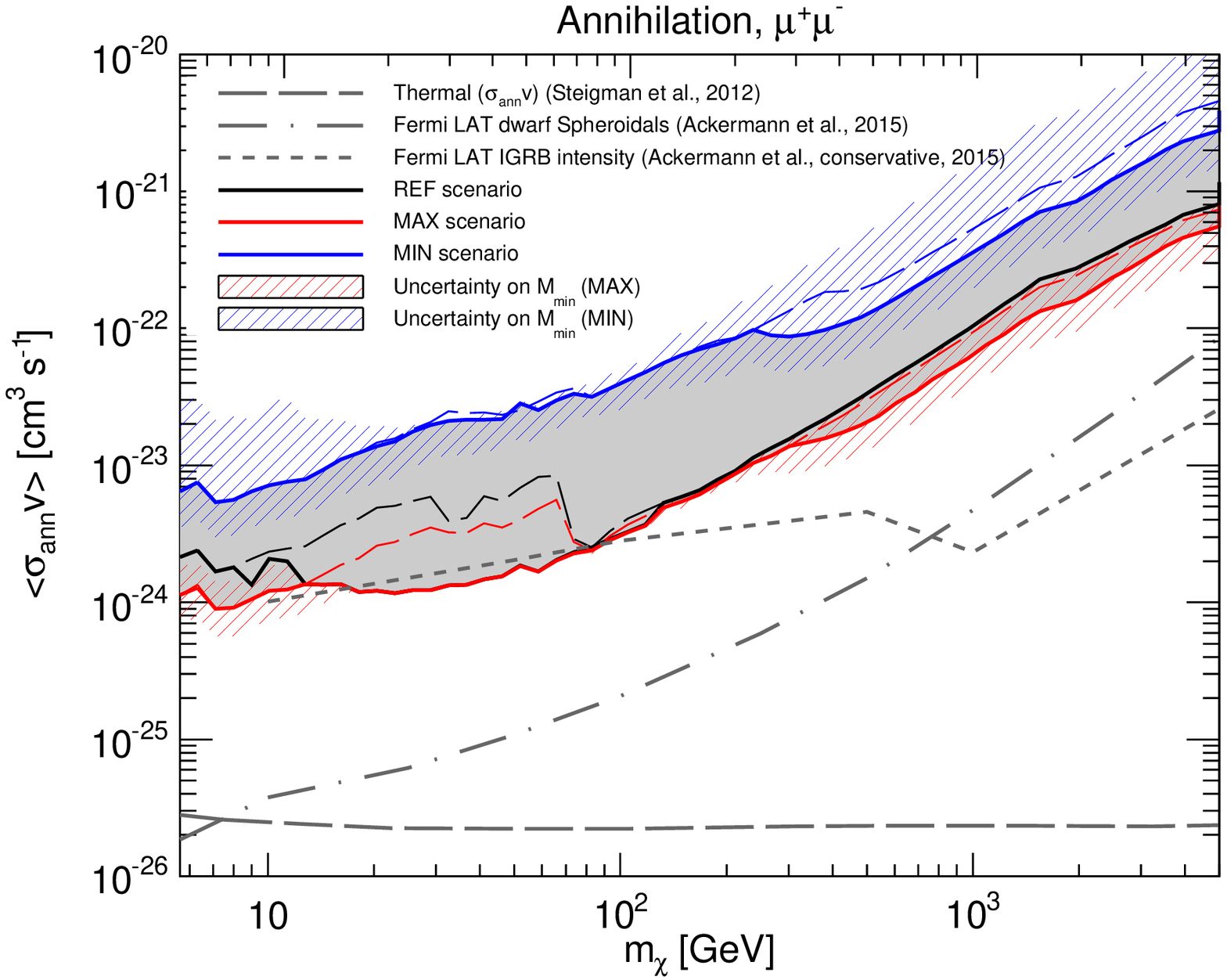}
\includegraphics[width=0.49\textwidth]{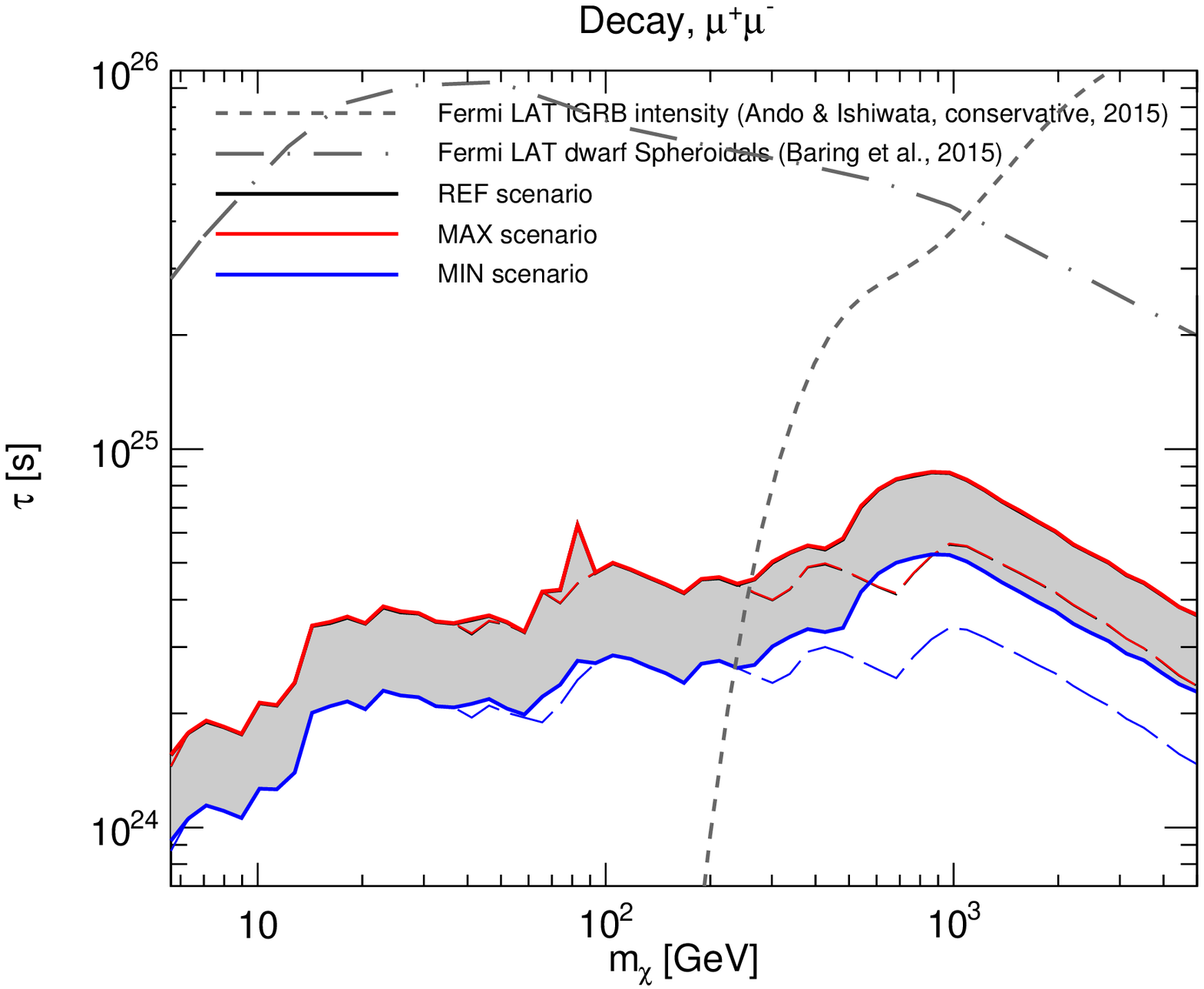}
\caption{\label{fig:mu_conservative} Same as Fig.~\ref{fig:tau_conservative} but for annihilations/decays into $\mu^+\mu^-$.}
\end{center}
\end{figure*}

Fig.~\ref{fig:mu_conservative} shows the same exclusion limits as in 
Fig~\ref{fig:tau_conservative} but for annihilations/decays into the 
$\mu^+\mu^-$. Between approximately 20 and 200 GeV, the DM-induced signal is
dominated by the IC emission associated with the smooth halo of the MW. That
is the reason why the solid black and red lines overlap, since the REF and 
MAX scenarios only differ in the computation of the boost factor for the
extragalactic component. For the same reason the blue and red shaded bands are 
reduced in width. Above 200 GeV, the IC emission for the extragalactic 
component starts to contribute more and the solid black and red lines deviate 
one from the other again.

\begin{figure*}
\begin{center}
\includegraphics[width=0.49\textwidth]{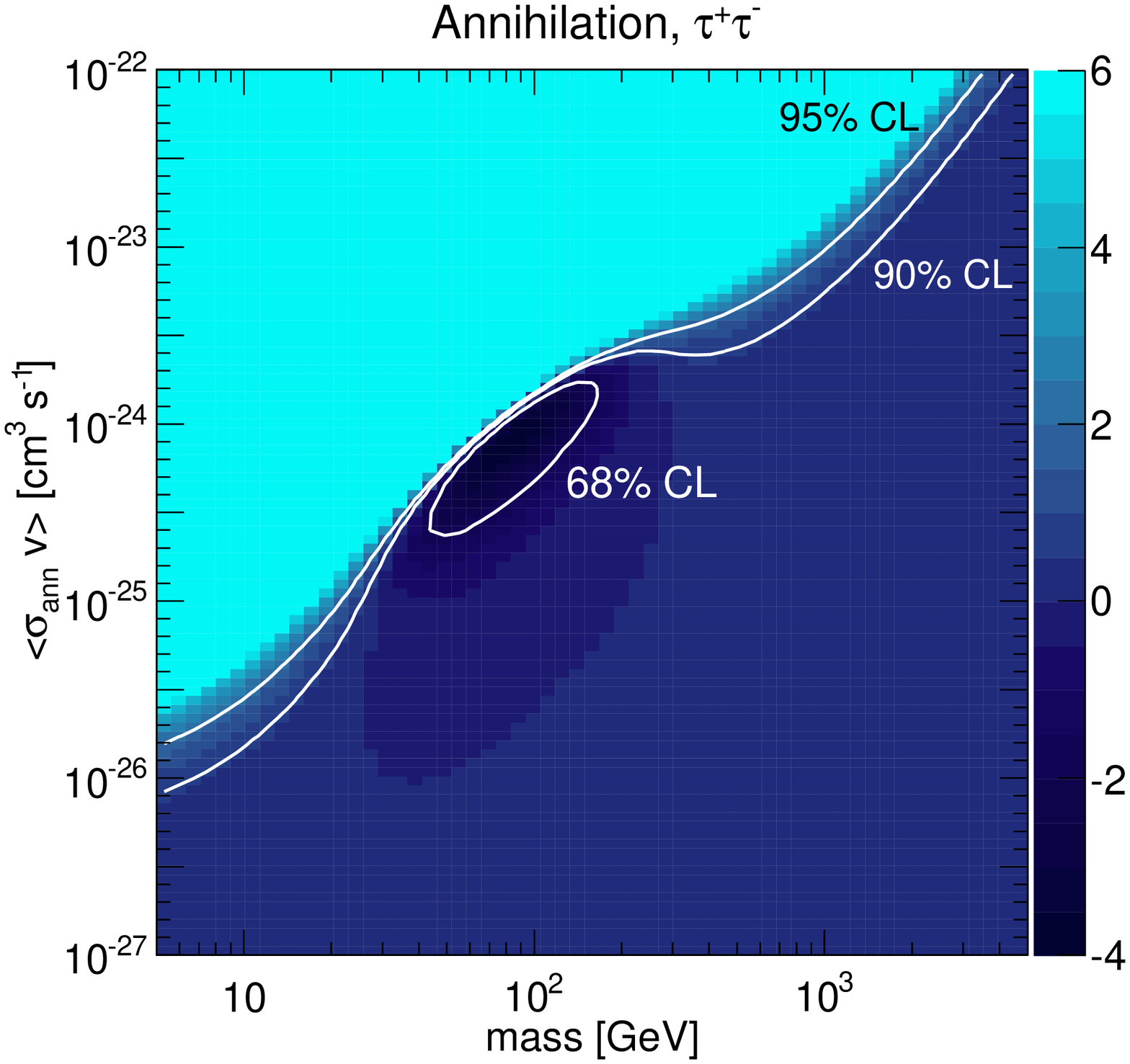}
\includegraphics[width=0.49\textwidth]{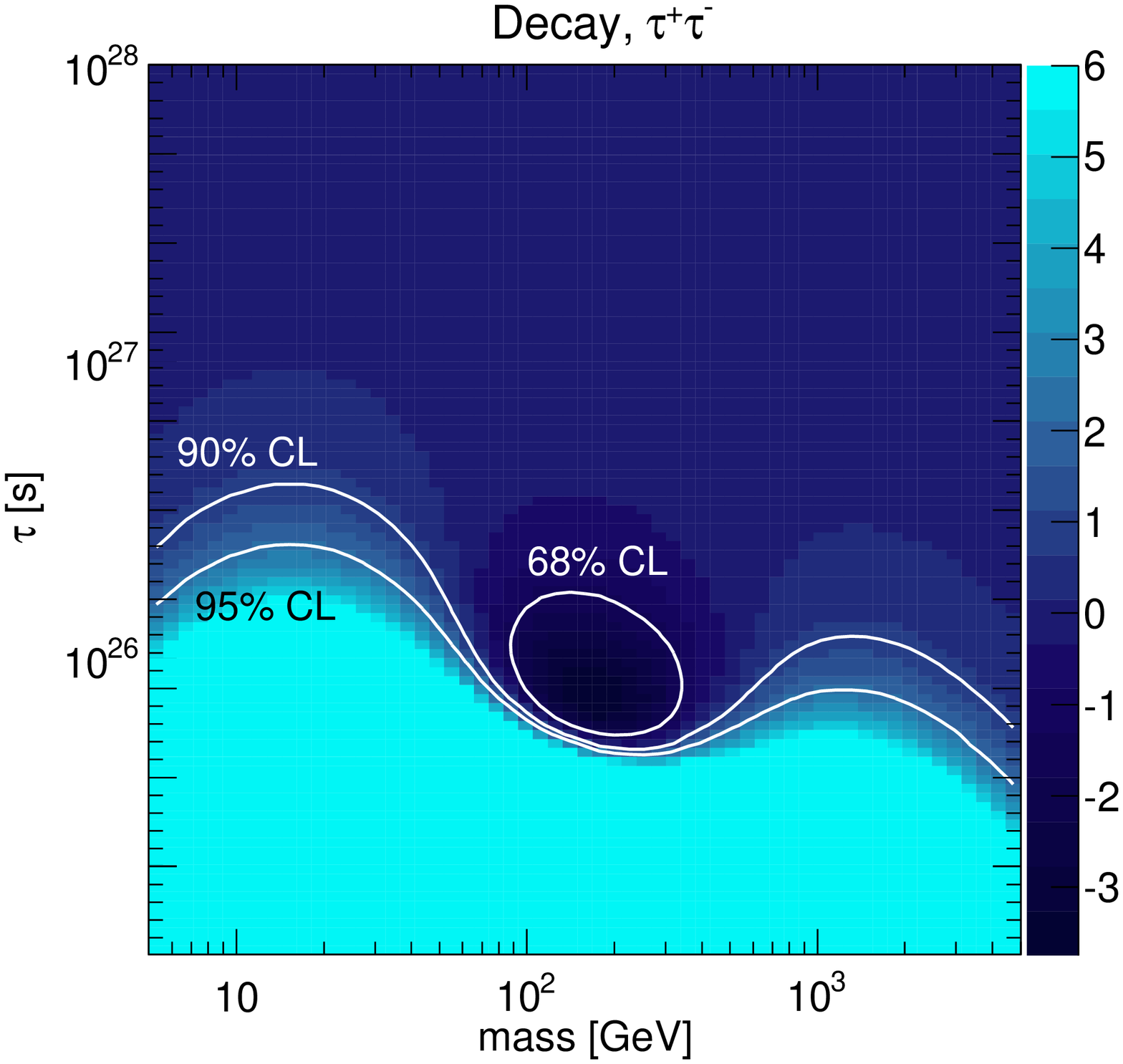}
\includegraphics[width=0.49\textwidth]{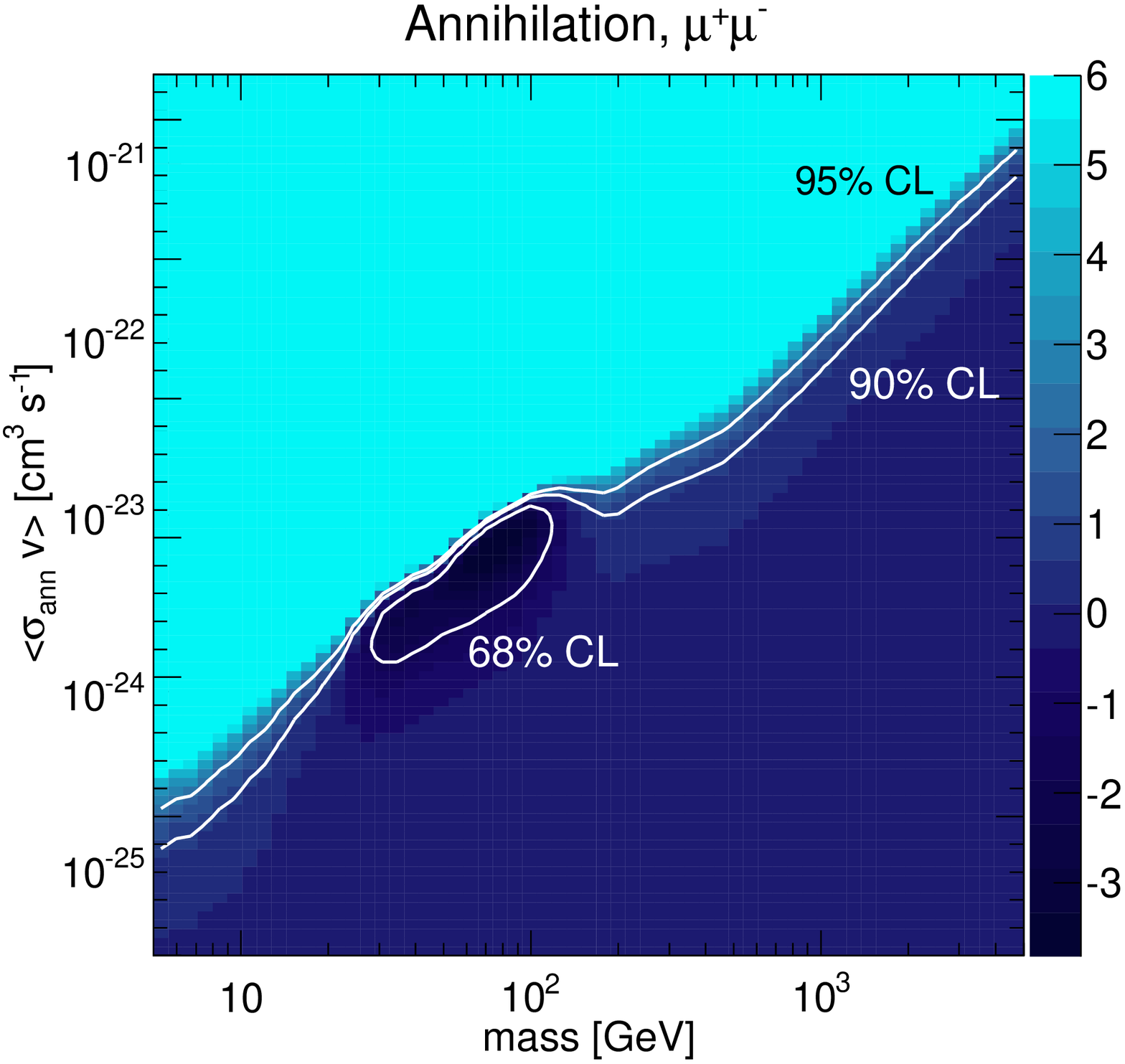}
\includegraphics[width=0.49\textwidth]{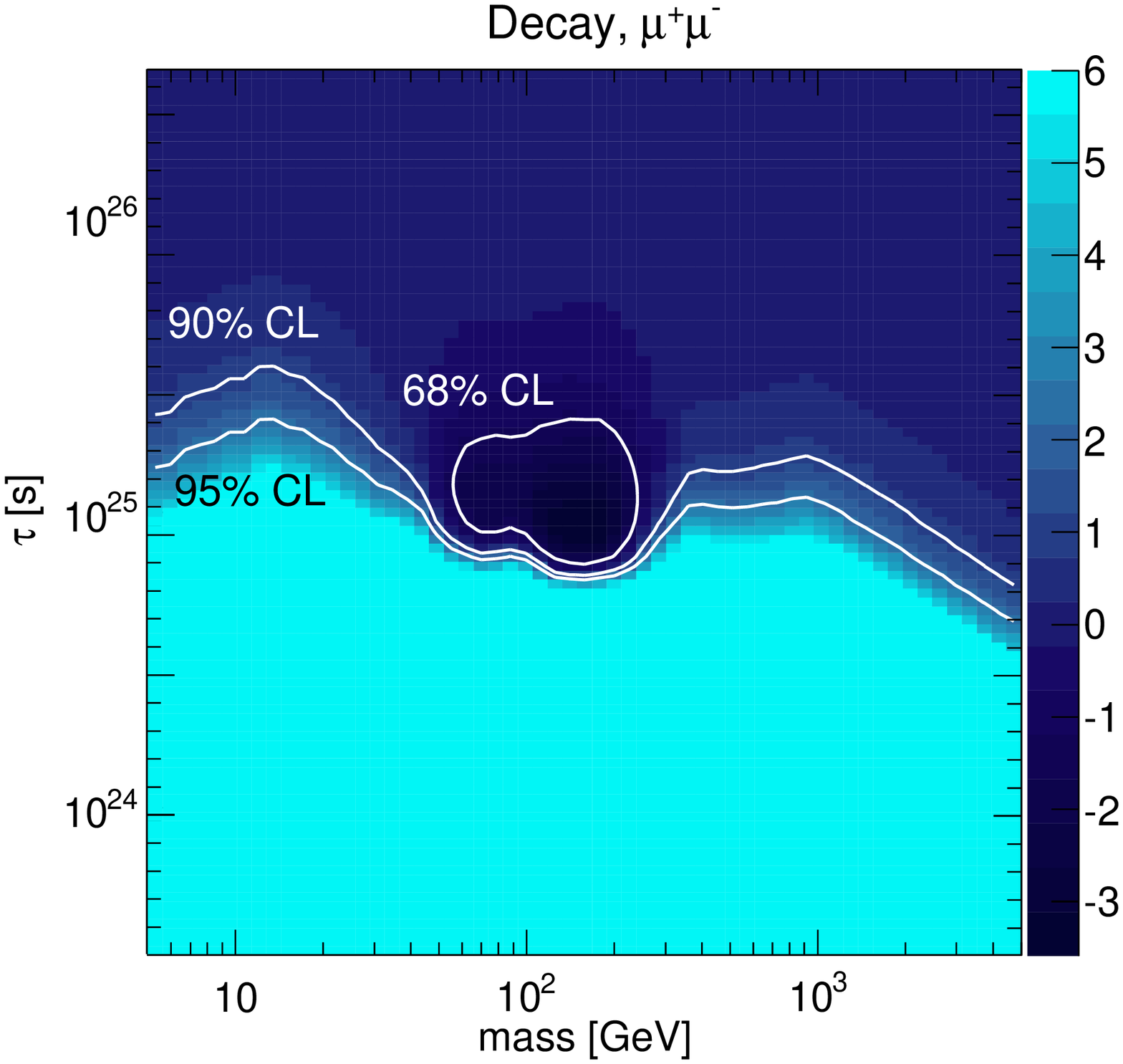}
\caption{\label{fig:table_likelihood} $\Delta\chi^2$ between the best-fit solution for the 2-component scenario and the best fit of the null hypothesis. Results presented here refer to the REF scenario with $M_{\rm min}=10^{-6} M_\odot/h$ and annihilation/decay into $\tau^+\tau^-$ (top panels) or $\mu^+\mu^-$ (bottom panels). The panels on the left are for annihilating DM and the ones on the right for decaying DM. Each point in the bi-dimensional parameter space is colored according to its $\Delta\chi^2$, i.e. the difference between the $\chi^2$ of the best fit to the auto- and cross-APS in terms of the 2-component model and the $\chi^2$ of the best fit of the null hypothesis (i.e. no DM). The closed white contour marks the 68\% CL region. The 90\% and 95\% CL ones in the left (right) panels contain all the region below (above) the white open curves labelled ``90\% CL'' and ``95\% CL''.}
\end{center}
\end{figure*}

The 2-component model developed in Sec.\ref{sec:fit} is used to fit the
measured auto-APS and cross-APS, for different values of DM mass, annihilation
cross section or decay lifetime. Fig.~\ref{fig:table_likelihood} shows the 
TS defined as the difference between the $\chi^2$ of the best fit for the null 
hypothesis (i.e. with no DM) and the $\chi^2$ of the best fit in the case with
the DM component. The top panels are for annihilation/decay into $\tau^+\tau^-$ 
and the bottom ones for the $\mu$-channel. The ones on the right are for an 
annihilating DM candidate and the ones on the left for decaying DM. They all 
refer to the REF scenario with $M_{\rm min}=10^{-6} M_\odot/h$. As indicated in
the labels, the white lines determine the 68\%, 90\% and 95\% CL regions.

\begin{figure*}
\begin{center}
\includegraphics[width=0.49\textwidth]{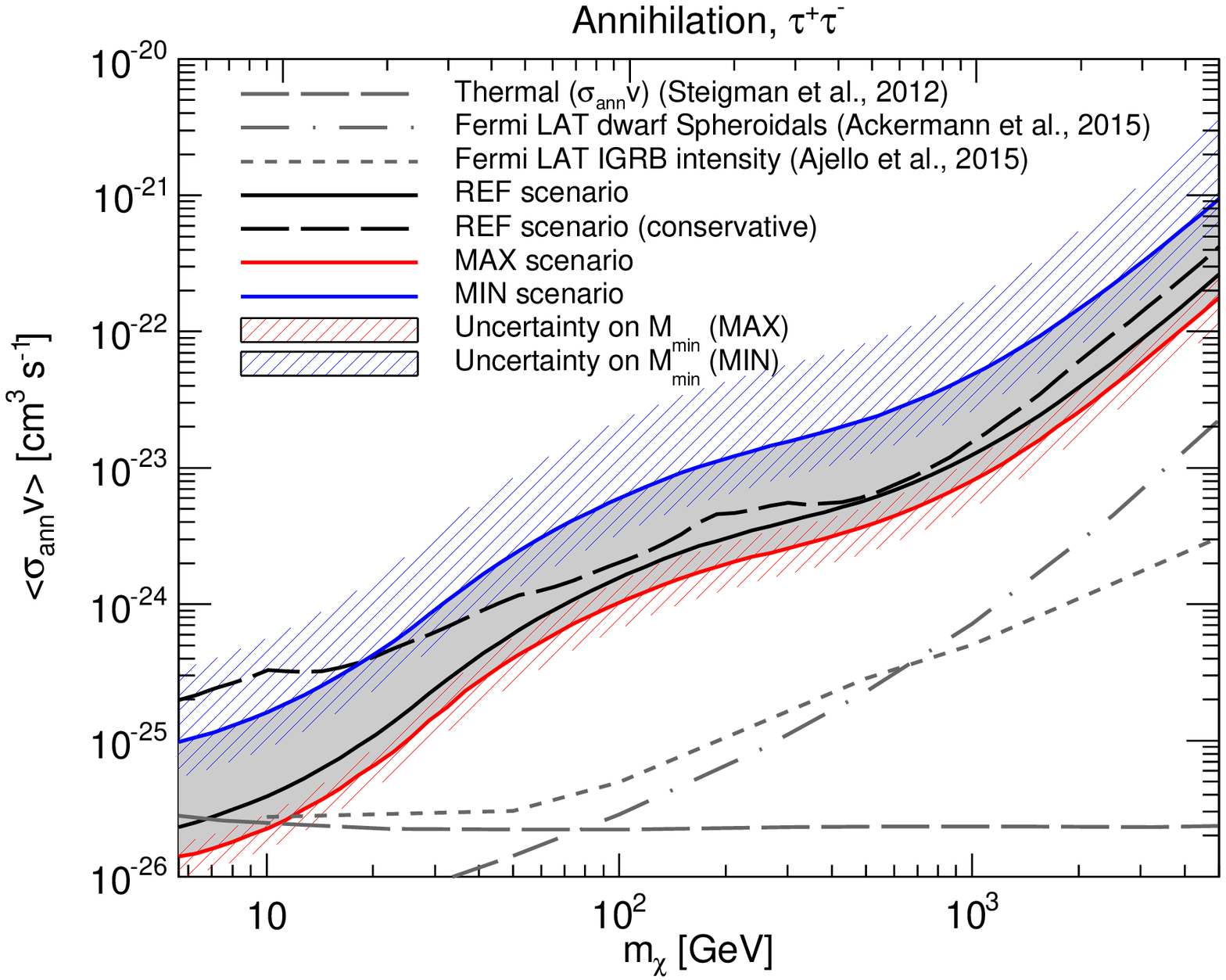}
\includegraphics[width=0.49\textwidth]{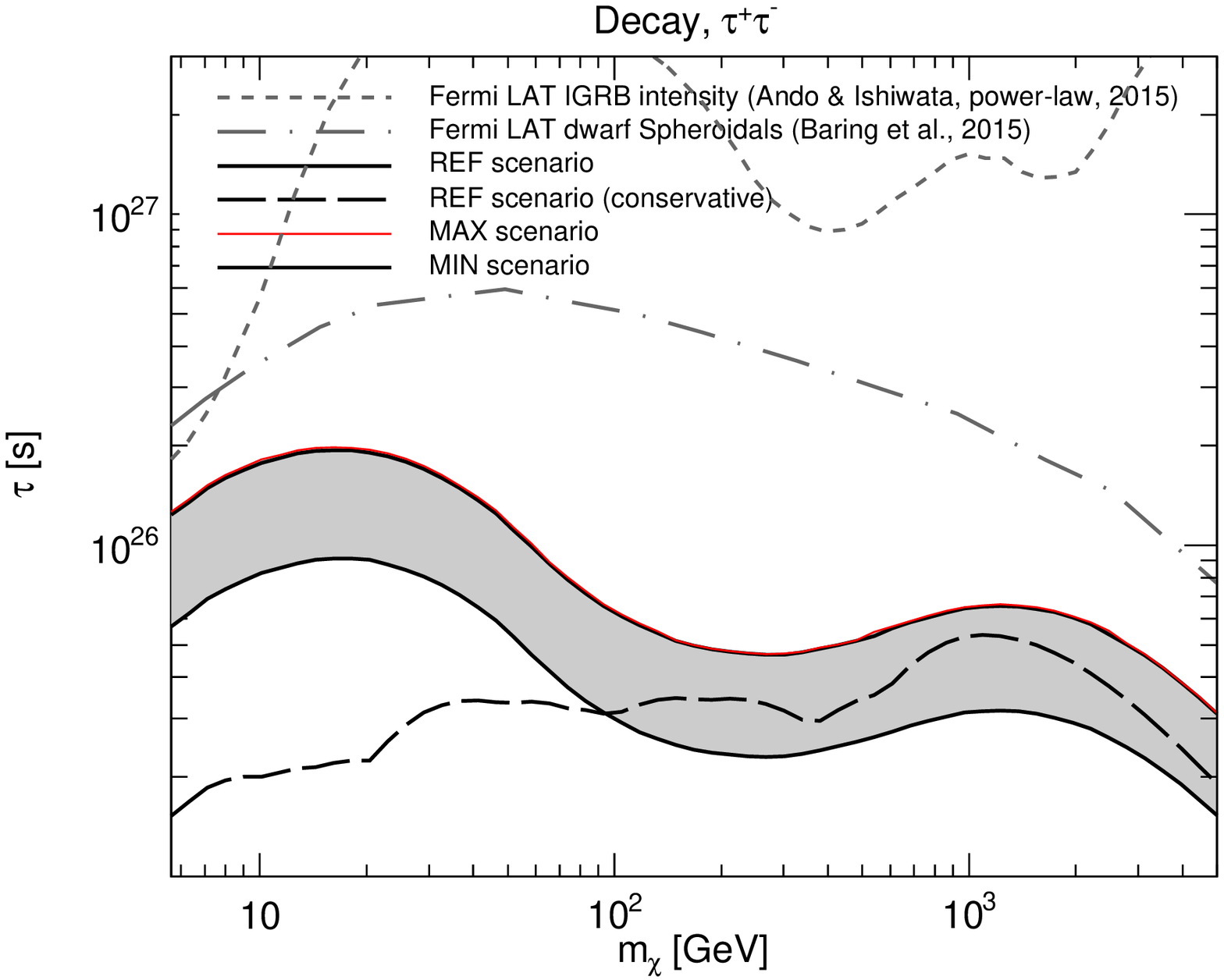}
\caption{\label{fig:limits_bestfit_tau} Exclusion limits on annihilating and decaying DM (for the $\tau$ channel) from the fit to the binned $\overline{C_\ell}$ in terms of the 2-component model. {\it Left:} The solid lines show the upper limits that can be derived on $\langle \sigma_{\rm ann}v \rangle$ as a function of $m_\chi$ (for annihilation into $\tau^+\tau^-$ quarks and $M_{\rm min}=10^{-6} M_\odot$) by fitting the \emph{Fermi} LAT data with a 2-component model that includes astrophysical sources and DM (see text for details). The black, blue and red lines correspond to the REF, MIN and MAX scenario. The blue and red shaded areas indicate how the MIN and MAX upper limits change when leaving $M_{\rm min}$ free to vary between $10^{-12} M_\odot$ and $1 \mbox { } M_\odot$. The black dashed line is the REF upper limit in the conservative case, from Fig.~\ref{fig:tau_conservative}, while the long-dashed grey line is the thermal annihilation cross section from Ref.~\cite{Steigman:2012nb}. The dot-dashed line is the upper limits derived in Ref.~\cite{Ackermann:2015zua} from the combined analysis of 15 dwarf spheroidals, while the short-dashed grey line comes from the analysis of the IGRB intensity performed in Ref.~\cite{Ajello:2015mfa}. {\it Right}: The same as in the left panel but for the lower limits on $\tau$, in the case of decaying DM. The short-dashed grey line represents the lower limit obtained in Ref.~\cite{Ando:2015qda} from the IGRB intensity. The line is taken from Fig.~5 of Ref.~\cite{Ando:2015qda}, where the IGRB is interpreted in terms of a component with a power-law emission spectrum and a DM contribution. Finally, the dot-dashed grey line is the upper limit from the analysis of 15 dwarf spheroidal galaxies performed in Ref.~\cite{Baring:2015sza}.}
\end{center}
\end{figure*}

\begin{figure*}
\begin{center}
\includegraphics[width=0.49\textwidth]{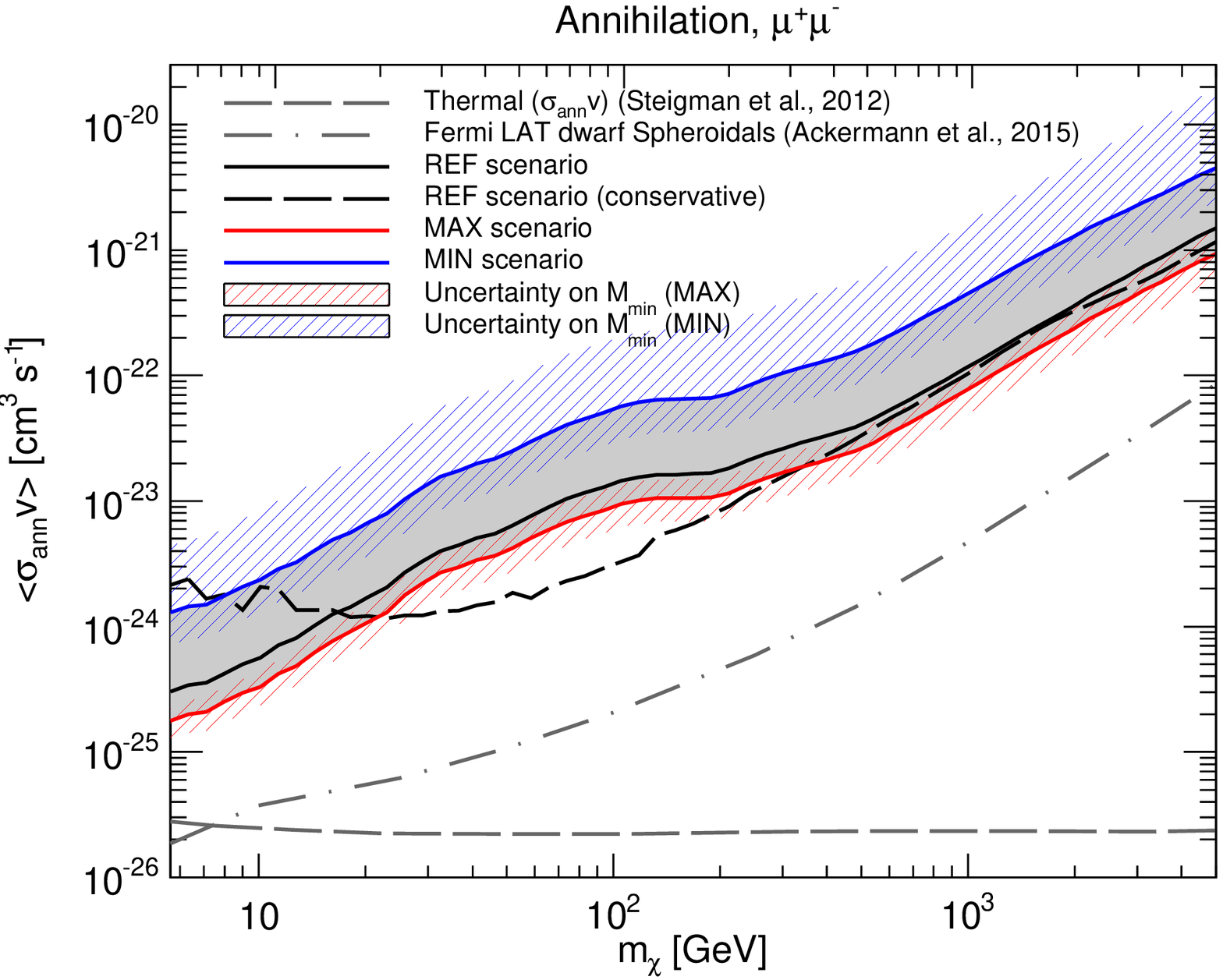}
\includegraphics[width=0.49\textwidth]{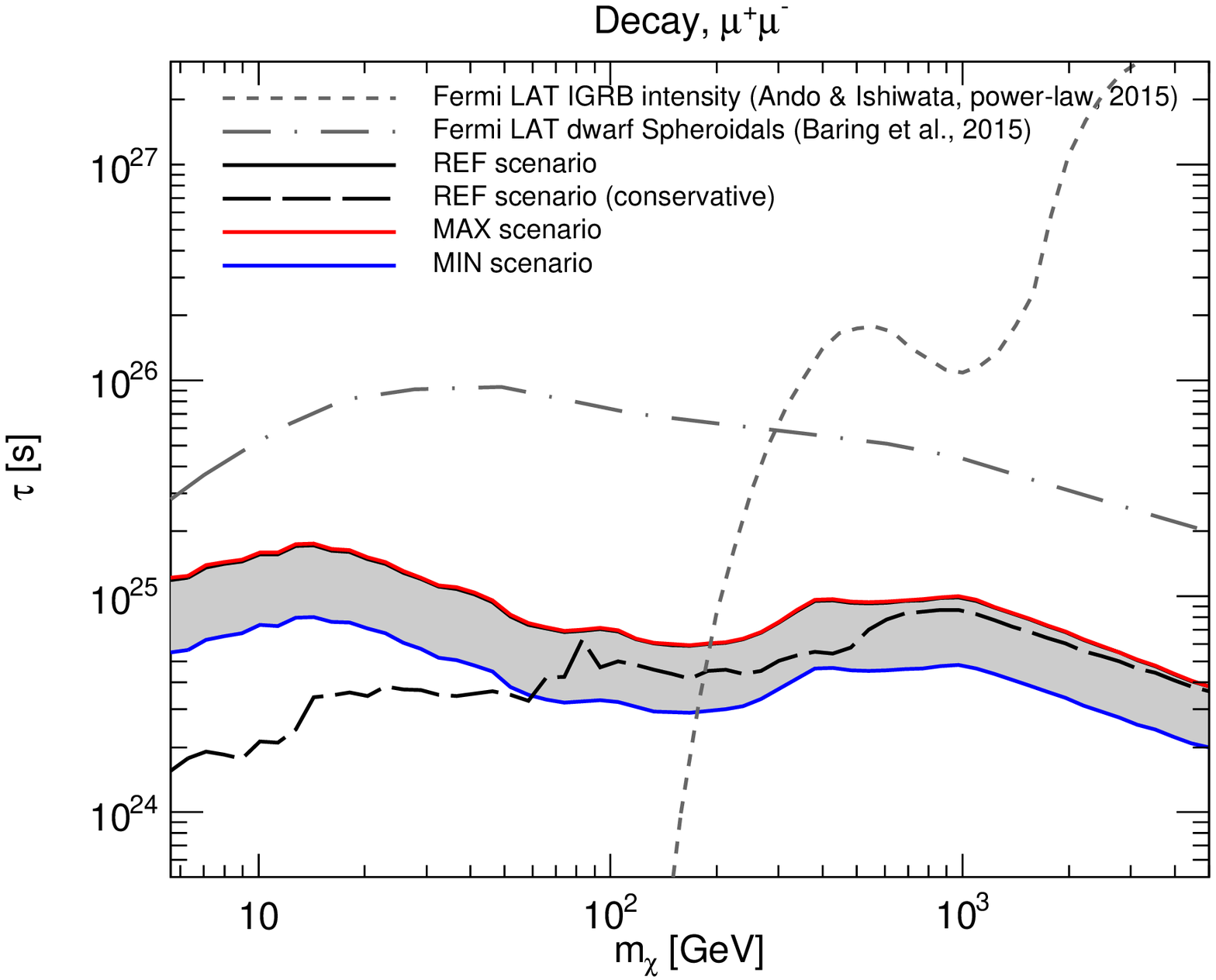}
\caption{\label{fig:limits_bestfit_mu} Same as in Fig.~\ref{fig:limits_bestfit_tau} but for annihilations/decays into $\mu^+\mu^-$.}
\end{center}
\end{figure*}

Assuming that the measured auto- and cross-APS are well described simply by a
Poissonian component, the 2-component model is used to derive exclusion limits
on DM as done in Sec.~\ref{sec:fit} but in the case of annihilations/decays
into $\tau^+\tau^-$ (Fig.~\ref{fig:limits_bestfit_tau}) and into $\mu^+\mu^-$
(Fig.~\ref{fig:limits_bestfit_mu}). In both figures the left panel is for 
annihilating DM and the right one for decaying DM. The solid black, red and 
blue line show the REF, MAX and MIN scenario for $M_{\rm min}=10^{-6} M_\odot/h$, 
respectively and the blue and shaded areas around the corresponding solid 
lines indicate how the limits change when $M_{\rm min}$ is left free to vary. 
The dashed black is the exclusion limit in the conservative scenario, from 
Fig.~\ref{fig:tau_conservative} and \ref{fig:mu_conservative}. In the left 
panels, the long-dashed grey line is the thermal annihilation cross section 
from Ref.~\cite{Steigman:2012nb} and the dot-dashed line is the upper limit 
derived in Ref.~\cite{Ackermann:2015zua} from the combined analysis of 15 
dwarf spheroidals. Also, the short-dashed grey line comes from the analysis 
of the IGRB intensity performed in Ref.~\cite{Ajello:2015mfa}. On the other 
hand, in the right panels, the short-dashed grey line represents the lower 
limit from Fig.~5 of Ref.~\cite{Ando:2015qda} from the IGRB intensity. The 
dot-dashed grey line is the lower limit from the analysis of 15 dwarf 
spheroidal galaxies performed in Ref.~\cite{Baring:2015sza}.

%

\end{document}